\documentclass[aps,onecolumn,amsmath,amsfonts,amssymb,groupedaddress,nofootinbib,nobalancelastpage,floatfix,superscriptaddress,secnumarabic,preprintnumbers]{revtex4-2}
\usepackage[dvipsnames]{xcolor}
\usepackage[utf8]{inputenc}
\usepackage[T1]{fontenc}
\usepackage{graphicx, hyperref, listings, newtxtext, newtxmath, nicefrac, relsize, xspace}
\hypersetup{colorlinks=true, linkcolor={red!50!black}, citecolor={red!50!black}, filecolor=BurntOrange, urlcolor=RawSienna}

\usepackage{listings, relsize}
\usepackage{bbm, lineno, multirow, slashed}

\DeclareUnicodeCharacter{2212}{-}
\def\lag{{\cal L}}
\def\met{E_T^\mathrm{miss}}
\def\mht{H_T^\mathrm{miss}}
\def\ptmiss{\vec{p}_T^\mathrm{miss}}

\newcommand{\be}{\begin{equation}}
\newcommand{\ee}{\end{equation}}
\def\bsp#1\esp{\begin{split}#1\end{split}}
\def\bpm{\begin{pmatrix}}
\def\epm{\end{pmatrix}}
\renewcommand{\d}{\mathrm{d}}

\newcommand{\ie}{\textit{i.e.}}
\newcommand{\eg}{\textit{e.g.}}
\newcommand{\etc}{\textit{etc.}}

\begin{document}

\title{$t$-channel dark matter models -- a whitepaper}
\preprint{CERN-LPCC-2025-001, IRMP-CP3-25-07, MITP-25-012, TTK-25-07}

\author{Chiara Arina$^\star$}
\affiliation{Centre for Cosmology, Particle Physics and Phenomenology (CP3), Universit\'e catholique de Louvain, B-1348 Louvain-la-Neuve, Belgium\\[.1cm]}

\author{Benjamin~Fuks$^\star$\footnote[1]{\href{mailto:fuks@lpthe.jussieu.fr}{fuks@lpthe.jussieu.fr}\\ \hspace*{-.275cm} $\star$ Editor\\ \hspace*{-.275cm}  $\dagger$ Convener}}
\affiliation{Laboratoire de Physique Th\'{e}orique et Hautes \'{E}nergies (LPTHE), UMR 7589, Sorbonne Universit\'{e} \& CNRS, 4 place Jussieu, 75252 Paris Cedex 05, France\\[.1cm]}

\author{Luca~Panizzi$^\star$}
\affiliation{Dipartimento di Fisica, Università della Calabria, I-87036 Arcavacata di Rende, Cosenza, Italy\\[.1cm]}
\affiliation{INFN-Cosenza, I-87036 Arcavacata di Rende, Cosenza, Italy\\[.1cm]}

\author{Michael~J.~Baker$^\dagger$}
\affiliation{Department of Physics, University of Massachusetts, Amherst, MA 01003 USA\\[.1cm]}

\author{Alan~S.~Cornell$^\dagger$}
\affiliation{Department of Physics, University of Johannesburg, PO Box 524, Auckland Park 2006, South Africa\\[.1cm]}

\author{Jan~Heisig$^\dagger$}
\affiliation{Institute for Theoretical Particle Physics and Cosmology, RWTH Aachen University, Sommerfeldstr.~16, 52074 Aachen, Germany\\[.1cm]}

\author{Benedikt~Maier$^\dagger$}
\affiliation{Imperial College, London SW7 2AZ, United Kingdom\\[.1cm]}

\author{Rute~Pedro$^\dagger$}
\affiliation{Laboratório de Instrumentação e Física Experimental de Partículas, Portugal\\[.1cm]}

\author{Dominique~Trischuk$^\dagger$}
\affiliation{Department of Physics, Brandeis University, Waltham MA, United States of America\\[.1cm]}

\author{Diyar~Agin}
\affiliation{Laboratoire de Physique Th\'{e}orique et Hautes \'{E}nergies (LPTHE), UMR 7589, Sorbonne Universit\'{e} \& CNRS, 4 place Jussieu, 75252 Paris Cedex 05, France\\[.1cm]}

\author{Alexandre~Arbey}
\affiliation{Université Claude Bernard Lyon 1, CNRS/IN2P3, IP2I UMR 5822, 4 rue Enrico Fermi, F-69100 Villeurbanne, France\\[.1cm]}

\author{Giorgio~Arcadi}
\affiliation{Dipartimento di Scienze Matematiche e Informatiche, Scienze Fisiche e Scienze della Terra, Universita degli Studi di Messina, Via Ferdinando Stagno d'Alcontres 31, I-98166 Messina, Italy\\[.1cm]}
\affiliation{INFN Sezione di Catania, Via Santa Sofia 64, I-95123 Catania, Italy\\[.1cm]}

\author{Emanuele~Bagnaschi}
\affiliation{INFN, Laboratori Nazionali di Frascati, Via E. Fermi 40, 00044 Frascati (RM), Italy\\[.1cm]}

\author{Kehang Bai}
\affiliation{University of Oregon, 1585 E 13th Ave, Eugene, OR 97403, United States\\[.1cm]}

\author{Disha~Bhatia}
\affiliation{Instituto de Física, Universidade de São Paulo, Brasil\\[.1cm]}

\author{Mathias~Becker}
\affiliation{PRISMA\textsuperscript{+} Cluster of Excellence \& Mainz Institute for Theoretical Physics, Johannes Gutenberg-Universit\"{a}t Mainz, 55099 Mainz, Germany\\[.1cm]}
\affiliation{Dipartimento di Fisica e Astronomia, Universit\`a degli Studi di Padova, Via Marzolo 8, 35131 Padova, Italy \\[.1cm]}
\affiliation{INFN, Sezione di Padova, Via Marzolo 8, 35131 Padova, Italy \\[.1cm]}

\author{Alexander~Belyaev}
\affiliation{School of Physics and Astronomy, University of Southampton, Southampton SO17 1BJ, United Kingdom\\[.1cm]}
\affiliation{Particle Physics Department, Rutherford Appleton Laboratory, Chilton, Didcot, Oxon OX11 0QX, UK\\[.1cm]}

\author{Ferdinand~Benoit}
\affiliation{Laboratoire de Physique Th\'{e}orique et Hautes \'{E}nergies (LPTHE), UMR 7589, Sorbonne Universit\'{e} \& CNRS, 4 place Jussieu, 75252 Paris Cedex 05, France\\[.1cm]}

\author{Monika~Blanke}
\affiliation{Institut f\"ur Astroteilchenphysik, Karlsruhe Institute of Technology, Hermann-von-Helmholtz-Platz 1, D-76344 Eggenstein-Leopoldshafen, Germany\\[.1cm]}
\affiliation{Institut f\"ur Theoretische Teilchenphysik, Karlsruhe Institute of Technology, Engesserstra\ss e 7, D-76128 Karlsruhe, Germany\\[.1cm]}

\author{Jackson~Burzynski}
\affiliation{Simon Fraser University, 8888 University Dr W, Burnaby, BC V5A 1S6, Canada\\[.1cm]}

\author{Jonathan~M.~Butterworth}    
\affiliation{Department of Physics and Astronomy, University College London, London, UK\\[.1cm]}

\author{Antimo~Cagnotta}
\affiliation{Texas A\&M University, College Station, USA\\[.1cm]}

\author{Lorenzo~Calibbi}
\affiliation{School of Physics, Nankai University, Tianjin 300071, China\\[.1cm]}

\author{Linda~M.~Carpenter}
\affiliation{Department of Physics, The Ohio State University, Columbus, OH 43210, U.S.A.\\[.1cm]}

\author{Xabier~Cid~Vidal}
\affiliation{Instituto Galego de Física de Altas Enerxías (IGFAE), Universidade de Santiago de Compostela, Santiago de Compostela, Spain\\[.1cm]}

\author{Emanuele~Copello}
\affiliation{PRISMA\textsuperscript{+} Cluster of Excellence \& Mainz Institute for Theoretical Physics, Johannes Gutenberg-Universit\"{a}t Mainz, 55099 Mainz, Germany\\[.1cm]}

\author{Louie~Corpe}
\affiliation{Universit\'e Clermont Auvergne, CNRS/IN2P3, LPCA, 63000 Clermont-Ferrand, France\\[.1cm]}

\author{Francesco~D'Eramo}
\affiliation{Dipartimento di Fisica e Astronomia, Universit\`a degli Studi di Padova, Via Marzolo 8, 35131 Padova, Italy \\[.1cm]}
\affiliation{INFN, Sezione di Padova, Via Marzolo 8, 35131 Padova, Italy \\[.1cm]}

\author{Aldo~Deandrea}
\affiliation{Department of Physics, University of Johannesburg, PO Box 524, Auckland Park 2006, South Africa\\[.1cm]}
\affiliation{Université Claude Bernard Lyon 1, CNRS/IN2P3, IP2I UMR 5822, 4 rue Enrico Fermi, F-69100 Villeurbanne, France\\[.1cm]}

\author{Aman~Desai}
\affiliation{Department of Physics, The University of Adelaide, Adelaide, SA 5005, Australia\\[.1cm]}

\author{Caterina~Doglioni}
\affiliation{School of Physics and Astronomy, University of Manchester, Manchester, United Kingdom\\[.1cm]}

\author{Sunil~M.~Dogra}
\affiliation{Kyungpook National University, Daegu, South Korea\\[.1cm]}

\author{Mathias~Garny}
\affiliation{Technical University of Munich, TUM School of Natural Sciences, Department of Physics, James-Franck-Str. 1, 85748 Garching, Germany\\[.1cm]}

\author{Mark~D.~Goodsell}
\affiliation{Laboratoire de Physique Th\'{e}orique et Hautes \'{E}nergies (LPTHE), UMR 7589, Sorbonne Universit\'{e} \& CNRS, 4 place Jussieu, 75252 Paris Cedex 05, France\\[.1cm]}

\author{Sohaib~Hassan}
\affiliation{University of Bergen, Norway\\[.1cm]}

\author{Philip~Coleman~Harris}
\affiliation{Massachusetts Institute of Technology, 77 Massachusetts Avenue, Cambridge, MA 02139-4307, United States\\[.1cm]}

\author{Julia~Harz}
\affiliation{PRISMA\textsuperscript{+} Cluster of Excellence \& Mainz Institute for Theoretical Physics, Johannes Gutenberg-Universit\"{a}t Mainz, 55099 Mainz, Germany\\[.1cm]}

\author{Alejandro~Ibarra}
\affiliation{Technical University of Munich, TUM School of Natural Sciences, Department of Physics, James-Franck-Str. 1, 85748 Garching, Germany\\[.1cm]}

\author{Alberto~Orso~Maria~Iorio}
\affiliation{Dipartimento di Fisica E.~Pancini, Università di Napoli Federico II, Complesso Universitario di Monte Sant'Angelo, Via Cintia, Napoli, Italy\\[.1cm]}
\affiliation{INFN, Sezione di Napoli, Italy\\[.1cm]}

\author{Felix~Kahlhoefer}
\affiliation{Institut f\"ur Astroteilchenphysik, Karlsruhe Institute of Technology, Hermann-von-Helmholtz-Platz 1, D-76344 Eggenstein-Leopoldshafen, Germany\\[.1cm]}

\author{Deepak~Kar}
\affiliation{School of Physics, University of Witwatersrand, Johannesburg, South Africa\\[.1cm]}
\affiliation{Royal Society Wolfson Visiting Fellow at the University of Glasgow, United Kingdom\\[.1cm]}

\author{Shaaban~Khalil}
\affiliation{Center for Fundamental Physics, Zewail City of Science and Technology, 6th of October City, Giza 12578, Egypt\\[.1cm]}

\author{Valery~Khoze}
\affiliation{IPPP, Durham University\\[.1cm]}

\author{Pyungwon~Ko}
\affiliation{School of Physics, Korea Institute for Advanced Study (KIAS), 85 Hoegi-ro, Dongdaemun-gu, Seoul 02455, Republic of Korea\\[.1cm]}

\author{Sabine~Kraml}
\affiliation{Laboratoire de Physique Subatomique et de Cosmologie (LPSC), Universit\'e Grenoble-Alpes, CNRS/IN2P3, 53 Avenue des Martyrs, F-38026 Grenoble, France\\[.1cm]}

\author{Greg~Landsberg}
\affiliation{Brown University, Dept.~of Physics, 182 Hope St., Providence, RI 02912, USA\\[.1cm]}

\author{Andre~Lessa}
\affiliation{Centro de Ci\^encias Naturais e Humanas, Universidade Federal do ABC, Santo Andr\'e, 09210-580 SP, Brazil\\[.1cm]}

\author{Laura~Lopez-Honorez}
\affiliation{Service de Physique Th\'eorique, C.P. 225, Universit\'e Libre de Bruxelles, Boulevard du Triomphe, B-1050 Brussels, Belgium\\[.1cm]}
\affiliation{Theoretische Natuurkunde \& The International Solvay Institutes, Vrije Universiteit Brussel, Pleinlaan 2, B-1050 Brussels, Belgium\\[.1cm]}

\author{Alberto~Mariotti}
\affiliation{IIHE/ELEM, Vrije Universiteit Brussel, Pleinlaan 2, B-1050 Brussels, Belgium\\[.1cm]}
\affiliation{Theoretische Natuurkunde \& The International Solvay Institutes, Vrije Universiteit Brussel, Pleinlaan 2, B-1050 Brussels, Belgium\\[.1cm]}

\author{Vasiliki~A.~Mitsou}
\affiliation{Instituto de F\'isica Corpuscular (IFIC), CSIC -- Universitat de Val\`encia, C/ Catedr\'atico Jos\'e Beltr\'an 2, 46980 Paterna, Spain\\[.1cm]}

\author{Kirtimaan~Mohan}
\affiliation{Department of Physics and Astronomy, Michigan State University, 567 Wilson Road, East Lansing, Michigan 48824, U.S.A.\\[.1cm]}

\author{Chang-Seong~Moon}
\affiliation{Kyungpook National University, Daegu, South Korea\\[.1cm]}

\author{Alexander~Moreno~Brice\~no}
\affiliation{Universidad Antonio Nari\~no, Bogota, Colombia\\[.1cm]}

\author{María~Moreno~Llácer}
\affiliation{Instituto de F\'isica Corpuscular (IFIC), CSIC -- Universitat de Val\`encia, C/ Catedr\'atico Jos\'e Beltr\'an 2, 46980 Paterna, Spain\\[.1cm]}

\author{Léandre~Munoz-Aillaud}
\affiliation{Laboratoire de Physique Th\'{e}orique et Hautes \'{E}nergies (LPTHE), UMR 7589, Sorbonne Universit\'{e} \& CNRS, 4 place Jussieu, 75252 Paris Cedex 05, France\\[.1cm]}

\author{Taylor~Murphy}
\affiliation{Laboratoire de Physique Th\'{e}orique et Hautes \'{E}nergies (LPTHE), UMR 7589, Sorbonne Universit\'{e} \& CNRS, 4 place Jussieu, 75252 Paris Cedex 05, France\\[.1cm]}
\affiliation{Department of Physics, Miami University, 500 E. Spring St., Oxford, OH 45056, U.S.A.\\[.1cm]}

\author{Anele~M.~Ncube}
\affiliation{Department of Physics, University of Johannesburg, PO Box 524, Auckland Park 2006, South Africa\\[.1cm]}

\author{Wandile Nzuza}
\affiliation{School of Physics, University of Witwatersrand, Johannesburg, South Africa\\[.1cm]}

\author{Clarisse~Prat}
\affiliation{School of Physics, University of Witwatersrand, Johannesburg, South Africa\\[.1cm]}

\author{Lena Rathmann}
\affiliation{Institute for Theoretical Particle Physics and Cosmology, RWTH Aachen University, Sommerfeldstr.~16, 52074 Aachen, Germany\\[.1cm]}

\author{Thobani~Sangweni}
\affiliation{Department of Physics, University of Cape Town, Cape Town, South Africa\\[.1cm]}

\author{Dipan~Sengupta}
\affiliation{Sydney Consortium for Particle Physics and Cosmology, School of Physics, The University of New South Wales, Sydney NSW 2052, Australia\\[.1cm]}

\author{William~Shepherd}
\affiliation{Department of Physics and Astronomy, Sam Houston State University, Huntsville, Texas\\[.1cm]}

\author{Sukanya~Sinha}
\affiliation{School of Physics and Astronomy, University of Manchester, Manchester, United Kingdom\\[.1cm]}

\author{Tim~M.P.~Tait}
\affiliation{University of California, Irvine\\[.1cm]}

\author{Andrea~Thamm}
\affiliation{Department of Physics, University of Massachusetts, Amherst, MA 01003 USA\\[.1cm]}

\author{Michel~H.G.~Tytgat}
\affiliation{Service de Physique Th\'eorique, C.P. 225, Universit\'e Libre de Bruxelles, Boulevard du Triomphe, B-1050 Brussels, Belgium\\[.1cm]}

\author{Zirui~Wang}
\affiliation{Fudan University, Shanghai 200443, People’s Republic of China\\[.1cm]}

\author{David~Yu}
\affiliation{Department of Physics, State University of New York at Buffalo, Buffalo, NY 14260, USA\\[.1cm]}

\author{Shin-Shan~Yu}
\affiliation{Department of Physics, The Catholic University of America, Washington, DC 20064, USA\\[.1cm]}

\begin{abstract}
  This report, summarising work achieved in the context of the LHC Dark Matter Working Group, investigates the phenomenology of $t$-channel dark matter models, spanning minimal setups with a single dark matter candidate and mediator to more complex constructions closer to UV-complete models. For each considered class of models, we examine collider, cosmological and astrophysical implications. In addition, we explore scenarios with either promptly decaying or long-lived particles, as well as featuring diverse dark matter production mechanisms in the early universe. By providing a unified analysis framework, numerical tools and guidelines, this work aims to support future experimental and theoretical efforts in exploring $t$-channel dark matter models at colliders and in cosmology.
\end{abstract}

\maketitle

\tableofcontents
\newpage

\section{Introduction}\label{sec:intro}
Despite compelling evidence for its existence~\cite{Bertone:2004pz, Bertone:2010zza, Bertone:2016nfn, Cirelli:2024ssz}, dark matter (DM) remains elusive in  direct searches using nuclear and electronic recoil experiments, indirect probes via cosmic-ray and gamma-ray spectra analyses, and collider searches for missing transverse energy signatures. Understanding the nature of dark matter is therefore among the top priorities in particle physics, astrophysics and cosmology. Currently, observational evidence confirms the existence of a gravitationally-interacting form of matter that constitutes approximately 25\% of the Universe's energy budget.  However, the fundamental properties of DM such as its particle nature, its interactions with Standard Model (SM) particles if any, its mass and its spin, are still unknown despite decades of intense efforts. From a theoretical perspective, a vast array of models predicting the existence of DM has been developed over the years. The null results from direct, indirect, and collider DM search experiments have imposed increasingly stringent constraints on these models, reducing the viable regions in the associated parameter spaces. 

These bounds are usually derived through two complementary approaches. First, a model-specific approach in which dark matter is embedded within a comprehensive theoretical framework built from first principles could be considered. Here, despite current data and related constraints, viable scenarios span orders of magnitude in terms of dark matter masses and couplings. The goal of this report is not to provide an exhaustive overview of all theoretical aspects of these DM models, for which excellent reviews are already available (see for instance the recent work~\cite{Cirelli:2024ssz} and references therein). Alternatively, experimental results could be interpreted following a phenomenological approach based on simplified models which capture essential features of diverse classes of SM extensions while remaining agnostic to the underlying high-energy theory. Simplified models are particularly useful as they allow for a systematic and model-independent exploration of viable DM scenarios, facilitating the comparison of theoretical predictions with experimental data from varied experiments and guiding the design of future experiments. In this report, we thus adopt the simplified model paradigm~\cite{Alwall:2008ag, LHCNewPhysicsWorkingGroup:2011mji}, generically assuming the existence of a DM particle that interacts minimally with the SM through a limited number of well-defined couplings and additional new physics states.

In the most simplified models for dark matter, the dark sector is assumed to consist of a single massive DM particle that interacts with the Standard Model via a single new `mediator' particle. In addition, the stability of the DM candidate is ensured by imposing a discrete or continuous symmetry under which the DM state transforms differently from the SM states. The interactions of the new physics sector are then determined by the quantum numbers of the mediator and its representations under both the new symmetry and the SM gauge group. In simplified $s$-channel DM model configurations (with thus a single DM state and a single mediator state), introduced about a decade ago~\cite{Fox:2012ru, Haisch:2013ata, Backovic:2015soa, Abercrombie:2015wmb, Boveia:2016mrp, Albert:2017onk, Albert:2022xla}, the mediator transforms under the new symmetry like the states from the SM sector. In contrast, in the $t$-channel models central to this report, the mediator's properties relative to the new symmetry are similar to those of the DM particle. For instance, under a discrete $\mathbb{Z}_2$ symmetry, an $s$-channel setup might involve a $\mathbb{Z}_2$-odd DM particle, with the SM and mediator states being $\mathbb{Z}_2$-even. In this case, the mediator couples both to a pair of SM particles and to a pair of DM particles. 

In a $t$-channel scenario, however, the mediator would be $\mathbb{Z}_2$-odd and thus couple to one DM particle and one SM particle. Furthermore, in the simplest cases, $t$-channel mediators must decay into DM and SM particles on timescales shorter than cosmological ones, as otherwise, they would appear stable and subsequently violate observational constraints as carrying colour or electric charge. While specific incarnations of minimal $t$-channel models have been largely explored in the past, the corresponding studies often focused on particular representations and spin configurations for the new states. For instance, this class of model was initially introduced in the context of Majorana or Dirac dark matter coupling to quarks~\cite{Garny:2011ii, Garny:2013ama, Bai:2013iqa, DiFranzo:2013vra, An:2013xka, Papucci:2014iwa}. Complementarily, some studies also embedded the minimal $t$-channel construction sketched above within a UV-complete framework like supersymmetry.

One of the primary goals of this report is to provide guidance for collider searches targeting final states with missing transverse energy originating from $t$-channel DM scenarios. This includes not only the minimal framework described above but also non-minimal setups that bring the models closer to UV-complete theories, and that incorporate additional features not captured by the minimal models that could be highly relevant for collider searches. We classify the considered $t$-channel models according to their degree of minimality. We begin with the most simplified frameworks, extending the SM by a single $\mathbb{Z}_2$-odd DM particle and a single $\mathbb{Z}_2$-odd mediator. Within this setup, we allow for different spin configurations and interactions with the SM, considering all possible mediator-SM couplings involving quarks and leptons (but one at a time). This approach builds on previous studies~\cite{Chang:2013oia, Giacchino:2014moa, Hisano:2015bma, ElHedri:2017nny, Arina:2020tuw, Arina:2023msd}, where efforts were made to simultaneously simplify the new physics parameter space and systematically explore variations in the quantum numbers of the new fields. Next, we extend our analysis to more complex models that involve multiple mediators and DM candidates, while still relying on $t$-channel exchanges for DM annihilation. These models include scenarios with dark sector flavour symmetries~\cite{Agrawal:2014aoa}, constructions inspired by compositeness~\cite{Colucci:2018vxz, Cornell:2021crh, Cornell:2022nky} and frustrated DM frameworks~\cite{Carpenter:2022lhj}. Another approach to non-minimality is to retain a minimal field content while allowing the fields to transform under non-trivial representations of either the SM gauge group, or an extended gauge group. Examples include the so-called Minimal DM model of~\cite{Cirelli:2005uq} or the Minimal Consistent DM scenarios of~\cite{Belyaev:2022qnf}. In this report, we focus on the latter and explore a setup featuring a non-minimal gauged dark sector~\cite{Belyaev:2022shr}. Within all these models, a wide range of DM coupling values leads to cosmologically viable scenarios, which correspond to varied possibilities for the mediator decay width. We analyse cases where mediators decay promptly into DM and SM particles, as well as scenarios with small mediator decay widths, resulting in long-lived particle (LLP) signatures at the LHC. The latter case is particularly motivated by DM production mechanisms in the early universe beyond the standard freeze-out paradigm, such as conversion-driven freeze-out~\cite{Garny:2017rxs,DAgnolo:2017dbv}, freeze-in~\cite{McDonald:2001vt,Hall:2009bx}, or superWIMP production~\cite{Covi:1999ty,Feng:2003uy}.

In all these scenarios, we discuss the implications for collider searches, cosmology and astrophysics. By leveraging the complementarity of these observations, we aim to provide a comprehensive framework for probing the parameter spaces of a wide range of $t$-channel models in the near future. To this end, we additionally provide numerical tools, including those necessary for high-precision Monte Carlo simulations for LHC studies, along with detailed instructions on how to use them. Furthermore, simulated Monte Carlo samples for simplified models can also be obtained upon request, in the form of grids of mediator and dark matter masses, in order to facilitate reinterpretation and signal modelling. We hope that these efforts will subsequently enable a systematic reinterpretation of both existing and future experimental results in terms of more complex DM models.

This report, that summarizes work achieved in the framework of the LHC Dark Matter working Group, is structured as follows. In the first part (section~\ref{sec:models}), we provide a comprehensive overview of the $t$-channel dark matter models under consideration. This includes simplified scenarios discussed in section~\ref{sec:model_minimal}, as well as more complex and non-minimal theories explored in sections~\ref{sec:DMFV} (flavoured dark matter), \ref{sec:compositeDM} (composite dark matter), \ref{sec:fDM} (frustrated dark matter) and \ref{sec:nonAbDS} (non-Abelian gauged dark sectors). Subsequently, we address the phenomenology of these models, focusing on collider studies in section~\ref{sec:colliders} and cosmological implications in section~\ref{sec:cosmology}. The collider phenomenology is discussed separately for scenarios involving promptly decaying mediators and those with long-lived mediators, while similarly, the cosmological phenomenology distinguishes between canonical dark matter freeze-out and alternative DM production scenarios, such as conversion-driven freeze-out, the superWIMP mechanism, and freeze-in production. We conclude with a summary of our findings and perspectives for future studies in section~\ref{sec:conclusion}.

\newpage\section{Benchmark models for LHC phenomenology}\label{sec:models}
In this section, we introduce various classes of models to be explored in this whitepaper. We begin with minimal setups featuring $t$-channel dark matter (DM), that we discuss in section~\ref{sec:model_minimal}. Additionally, we delve into non-minimal models incorporating diverse phenomenological features not addressed in the minimal case. These encompass flavoured dark matter models, composite realisations, frustrated dark matter scenarios and non-Abelian gauged dark sectors, which are detailed in sections~\ref{sec:DMFV}, \ref{sec:compositeDM}, \ref{sec:fDM} and \ref{sec:nonAbDS} respectively.

\subsection{Minimal options and their implementation in high-energy physics software}\label{sec:model_minimal}
In a minimal realisation of a $t$-channel simplified model for dark matter, the field content of the Standard Model (SM) is extended to include one DM candidate $X$, assumed to be a colourless electroweak singlet, and one mediator state $Y$. Additionally, to ensure DM stability, an \textit{ad hoc} $\mathbb{Z}_2$ symmetry is imposed such that all SM fields are even while the dark matter and the mediator are odd. Following the \lstinline{DMSimpt} framework introduced in \cite{Arina:2020udz}, no assumptions are made regarding the spin of the DM and the mediator, the flavour structure of their interactions, and the quantum numbers of the mediator that depend on how it couples dark matter to the SM sector. Under these conditions, the model presents three possible spin configurations for the $X$ particle: it could be a scalar (either the complex state $S$ or the real state $\tilde{S}$), a fermion (either the Dirac fermion $\chi$ or the Majorana fermion $\tilde{\chi}$), or a vector (either the complex state $V$ or the real state $\tilde{V}$). For bosonic dark matter, the mediator is then a fermionic object, that we denote as $\psi$, whereas for fermionic dark matter, the mediator is a scalar field $\varphi$. We leave the case of vector mediators to future work, as achieving next-to-leading order QCD simulations in models with coloured vector states is far from trivial, even with existing automated tools such as those discussed in this report.

Expanding upon the setup introduced in \cite{Arina:2020udz}, the full Lagrangian incorporating the interactions of these fields with the SM can be expressed as follows:
\begin{equation}
  \lag = \lag_{\rm SM} + \lag_{\rm kin} + \lag_{XY}\,.
\end{equation}
Here, $\lag_{\rm SM}$ represents the SM Lagrangian and $\lag_{\rm kin}$ encompasses gauge-invariant kinetic and mass terms for all new fields. The final term $\lag_{XY}$ incorporates the interactions between the mediator and the DM with the SM. This term involves a significant number of free coupling parameters, particularly in the absence of assumptions regarding the flavour structure of the couplings. Separating the six possibilities for the properties of the dark matter state, the Lagrangian $\lag_{XY}$ takes the following form:
\be\label{eq:lagXY}\bsp
  \lag_{XY}^\text{\lstinline{F3S}}  = \sum_{f=u,d} \lambda_{\scriptscriptstyle f}\ \bar \psi_{\scriptscriptstyle f} f_R\ \tilde S 
     + \lambda_{\scriptscriptstyle Q}\ \bar \psi_{\scriptscriptstyle Q} Q_L\  \tilde S
     + {\rm H.c.}\,, \qquad &   \lag_{XY}^\text{\lstinline{F1S}}  = \lambda_{\scriptscriptstyle \ell}\ \bar \psi_{\scriptscriptstyle \ell} \ell_R\ \tilde S 
     + \lambda_{\scriptscriptstyle L}\ \bar \psi_{\scriptscriptstyle L} L_L\  \tilde S
     + {\rm H.c.}\,,\\
  \lag_{XY}^\text{\lstinline{F3C}}  = \sum_{f=u,d} \lambda_{\scriptscriptstyle f}\ \bar \psi_{\scriptscriptstyle f} f_R\ S
     + \lambda_{\scriptscriptstyle Q}\ \bar \psi_{\scriptscriptstyle Q} Q_L\ S
     +  {\rm H.c.} \,, \qquad & \lag_{XY}^\text{\lstinline{F1C}}  = \lambda_{\scriptscriptstyle \ell}\ \bar \psi_{\scriptscriptstyle \ell} \ell_R\ S
     + \lambda_{\scriptscriptstyle L}\ \bar \psi_{\scriptscriptstyle L} L_L\ S
     +  {\rm H.c.} \,,\\
  \lag_{XY}^\text{\lstinline{S3M}}= \sum_{f=u,d} \lambda_{\scriptscriptstyle f}\ \overline{\tilde{\chi}} f_R\ \varphi_{\scriptscriptstyle f}^\dag
     + \lambda_{\scriptscriptstyle Q}\ \overline{\tilde{\chi}} Q_L\ \varphi_{\scriptscriptstyle Q}^\dag
     + {\rm H.c.} \,, \qquad & \lag_{XY}^\text{\lstinline{S1M}}= \lambda_{\scriptscriptstyle \ell}\ \overline{\tilde{\chi}} \ell_R\ \varphi_{\scriptscriptstyle \ell}^\dag
     + \lambda_{\scriptscriptstyle L}\ \overline{\tilde{\chi}} L_L\ \varphi_{\scriptscriptstyle L}^\dag
     + {\rm H.c.} \,,\\
  \lag_{XY}^\text{\lstinline{S3D}} = \sum_{f=u,d} \lambda_{\scriptscriptstyle f}\ \overline{\chi} f_R\ \varphi_{\scriptscriptstyle f}^\dag
     + \lambda_{\scriptscriptstyle Q}\ \overline{\chi} Q_L\ \varphi_{\scriptscriptstyle Q}^\dag
     + {\rm H.c.} \,, \qquad & \lag_{XY}^\text{\lstinline{S1D}} = \lambda_{\scriptscriptstyle \ell}\ \overline{\chi} \ell_R\ \varphi_{\scriptscriptstyle \ell}^\dag
     + \lambda_{\scriptscriptstyle L}\ \overline{\chi} L_L\ \varphi_{\scriptscriptstyle L}^\dag
     + {\rm H.c.} \,,\\
  \lag_{XY}^\text{\lstinline{F3V}}  = \sum_{f=u,d} \lambda_{\scriptscriptstyle f}\ \bar \psi_{\scriptscriptstyle f} \gamma^\mu f_R\ \tilde V_\mu 
     + \lambda_{\scriptscriptstyle Q}\ \bar \psi_{\scriptscriptstyle Q} \gamma^\mu Q_L\  \tilde V_\mu
     + {\rm H.c.} \,, \qquad & \lag_{XY}^\text{\lstinline{F1V}}  = \lambda_{\scriptscriptstyle \ell}\ \bar \psi_{\scriptscriptstyle \ell} \gamma^\mu \ell_R\ \tilde V_\mu 
     + \lambda_{\scriptscriptstyle L}\ \bar \psi_{\scriptscriptstyle L} \gamma^\mu L_L\  \tilde V_\mu
     + {\rm H.c.} \,,\\
  \lag_{XY}^\text{\lstinline{F3W}} = \sum_{f=u,d} \lambda_{\scriptscriptstyle f}\ \bar \psi_{\scriptscriptstyle f} \gamma^\mu f_R\ V_\mu
     + \lambda_{\scriptscriptstyle Q}\ \bar \psi_{\scriptscriptstyle Q} \gamma^\mu Q_L\ V_\mu
     +  {\rm H.c.}\,,\qquad & \lag_{XY}^\text{\lstinline{F1W}} = \lambda_{\scriptscriptstyle \ell}\ \bar \psi_{\scriptscriptstyle \ell} \gamma^\mu \ell_R\ V_\mu
     + \lambda_{\scriptscriptstyle L}\ \bar \psi_{\scriptscriptstyle L} \gamma^\mu L_L\ V_\mu
     +  {\rm H.c.}
\esp\ee
In these Lagrangians, the symbols \lstinline{F3S}, \lstinline{S3M}, \lstinline{F3V} (\lstinline{F1S}, \lstinline{S1M}, \lstinline{F1V}) refer to models with real scalar, fermionic and vector dark matter through the symbol endings \lstinline{S}, \lstinline{M} and \lstinline{V}, while the beginning of the symbol names \lstinline{F3} and \lstinline{S3} (\lstinline{F1} and \lstinline{S1}) refer to colour-triplet (colour-singlet) fermionic and scalar mediators. Similarly, the designations \lstinline{F3C}, \lstinline{S3D} and \lstinline{F3W} (\lstinline{F1C}, \lstinline{S1D} and \lstinline{F1W}) are utilised for the complex dark matter cases through the ending \lstinline{C}, \lstinline{D} and \lstinline{W}. In our notation, $Q_L$ and $L_L$ represent the weak doublets of left-handed SM quarks and leptons, respectively, while $u_R$, $d_R$ and $\ell_R$ denote the corresponding weak singlets.  The fermionic mediators $\psi_{\scriptscriptstyle Q}$, $\psi_{\scriptscriptstyle L}$, $\psi_{\scriptscriptstyle u}$, $\psi_{\scriptscriptstyle d}$ and $\psi_{\scriptscriptstyle \ell}$ interact exclusively with the $Q_L$ , $L_L$, $u_R$, $d_R$ and $\ell_R$ fields, respectively. Furthermore, we introduce one fermionic mediator for each generation of SM fermions. Consequently, they lie in the same representation as their SM fermionic partners, and they are vector-like. Similarly, the scalar mediators $\varphi_{\scriptscriptstyle Q}$, $\varphi_{\scriptscriptstyle L}$, $\varphi_{\scriptscriptstyle u}$, $\varphi_{\scriptscriptstyle d}$ and $\varphi_{\scriptscriptstyle \ell}$ only interact with $Q_L$ , $L_L$, $u_R$, $d_R$ and $\ell_R$ respectively.

The scalar and fermionic mediators all possess a flavour index. Therefore, while in the Lagrangians of eq.~\eqref{eq:lagXY} all flavour indices are omitted for clarity, it is important to note that the coupling matrices ${\bf \lambda_{\scriptscriptstyle Q}}$, ${\bf \lambda_{\scriptscriptstyle L}}$, ${\bf \lambda_{\scriptscriptstyle u}}$, ${\bf \lambda_{\scriptscriptstyle d}}$ and ${\bf \lambda_{\scriptscriptstyle \ell}}$ are $3\times 3$ matrices in the flavour space. Unless stated otherwise, these matrices are considered real and flavour-diagonal to prevent loop-induced mixing between the SM fermions. Each mediator is hence associated with a given generation of SM partners.

\begin{table}
  \centering
  \renewcommand{\arraystretch}{1.4} \setlength\tabcolsep{8pt}
  \resizebox{.96\textwidth}{!}{%
    \begin{tabular}{c | c c c c c}
      & Field & \text{\lstinline{FeynRules}} & PDG & Spin & Repr.\\[.1cm]
      \multirow{2}{*}{DM} & $\tilde{S}$ & \text{\lstinline{Xs}} & $51$ & $0$ & $({\bf 1}, {\bf 1}, 0)$\\[.1cm]
      & $S$ & \text{\lstinline{Xc}} & $56$ & $0$ & $({\bf 1}, {\bf 1}, 0)$\\[.1cm]
      \hline\\[-.6cm]
      \multirow{9}{*}{Mediators} & Field & \text{\lstinline{FeynRules}} & PDG & Spin & Repr.\\[.2cm]
      & $\psi_{\scriptscriptstyle Q} =  \bpm\psi^{(u)}_{\scriptscriptstyle Q}\\ \psi^{(d)}_{\scriptscriptstyle Q}\epm$ & $\text{\lstinline{YF3Q}} = \bpm \text{\lstinline{YF3Qu}} \\ \text{\lstinline{YF3Qd}}\epm$ & $\begin{array}{l l l}
         5910002 & 5910004 & 5910006\\
         5910001 & 5910003 & 5910005\end{array}$ & $1/2$ & $({\bf 3}, {\bf 2},  \frac16)$\\[.2cm]
      & $\psi_{\scriptscriptstyle u}$ & $\text{\lstinline{YF3u}}$ & $\begin{array}{l l l}
         5920002 & 5920004 & 5920006\end{array}$ & $1/2$ & $({\bf 3}, {\bf 1},  \frac23)$\\[.2cm]
      & $\psi_{\scriptscriptstyle d}$ & $\text{\lstinline{YF3d}}$ & $\begin{array}{l l l}
         5920001 & 5920003 & 5920005\end{array}$ & $1/2$ & $({\bf 3}, {\bf 1},  -\frac13)$\\[.2cm]
       & $\psi_{\scriptscriptstyle L} =  \bpm\psi^{(\nu)}_{\scriptscriptstyle L}\\ \psi^{(\ell)}_{\scriptscriptstyle L}\epm$ & $\text{\lstinline{YF1L}} = \bpm \text{\lstinline{YF1Lv}} \\ \text{\lstinline{YF1Le}}\epm$ & $\begin{array}{l l l}
         5910012 & 5910014 & 5910016\\
         5910011 & 5910013 & 5910015\end{array}$ & $1/2$ & $({\bf 1}, {\bf 2},  -\frac12)$\\[.2cm]
      & $\psi_{\scriptscriptstyle \ell}$ & $\text{\lstinline{YF1e}}$ & $\begin{array}{l l l}
         5920011 & 5920013 & 5920015\end{array}$ & $1/2$ & $({\bf 1}, {\bf 1},  -1)$\\[.1cm]
      \hline\\[-.6cm]
      \multirow{6}{*}{Couplings} & Parameter & {\lstinline{FeynRules}} & Les Houches Block & &\\
      &$\lambda_{\scriptscriptstyle Q}$    & \lstinline{lamF3Q} & \lstinline{DMF3Q} & &\\
      &$\lambda_{\scriptscriptstyle u}$    & \lstinline{lamF3u} & \lstinline{DMF3U} & &\\
      &$\lambda_{\scriptscriptstyle d}$    & \lstinline{lamF3d} & \lstinline{DMF3D} & &\\
      &$\lambda_{\scriptscriptstyle L}$    & \lstinline{lamF1L} & \lstinline{DMF1L} & &\\
      &$\lambda_{\scriptscriptstyle \ell}$ & \lstinline{lamF1e} & \lstinline{DMF1E} & &\\
    \end{tabular}
  }
  \caption{Model information for the \lstinline{F3S} and \lstinline{F3C} models. We provide the name of the DM particle and the mediators used in the \lstinline{FeynRules} implementation, together with the associated PDG identifiers, spin quantum number and representation under $SU(3)_c\times SU(2)_L\times U(1)_Y$. We recall that three generations of mediators are included, which requires sets of three PDG identifiers. We additionally provide the new physics couplings linking the DM, the mediators and the SM sector. Each coupling is given together with the associated \lstinline{FeynRules} symbol and the reference Les Houches block to be used in the parameter card.\label{tab:info_S}}
\end{table}

Most of the results presented in this whitepaper are obtained through the joint usage of various standard high-energy physics packages, following the tool chain outlined in \cite{Arina:2020udz}. This process involved implementing a \textit{single} model file that incorporates simultaneously all Lagrangians described above in \lstinline{FeynRules}~\cite{Christensen:2009jx,Alloul:2013bka}\footnote{Specifically, we use the \lstinline{feynrules-dev-bsm} branch of \lstinline{FeynRules} available from the \lstinline{GitHub} repository \mbox{(\url{https://github.com/FeynRules/FeynRules})}.}, which we utilised alongside \lstinline{MoGRe}~\cite{Frixione:2019fxg}, \lstinline{NLOCT}~\cite{Degrande:2014vpa} and \lstinline{FeynArts}~\cite{Hahn:2000kx} to generate a next-to-leading order (NLO) UFO~\cite{Degrande:2011ua, Darme:2023jdn} model with five flavours of massless quarks. The resulting UFO model, termed the \lstinline{DMSimpt 2.0} model, is an extension of the implementation designed in \cite{Arina:2020udz}, now incorporating leptonic fields. This implementation is available from the \lstinline{FeynRules} model database (see \url{https://feynrules.irmp.ucl.ac.be/wiki/DMsimpt} or the \lstinline{GitHub} repository \url{https://github.com/BFuks/DMSimpt}). Like the previous implementation, the available NLO UFO model includes $\lambda$ couplings that are flavour-diagonal and real. However, the \lstinline{FeynRules} model allows for a more general structure, which we relied on to generate a leading-order (LO) UFO model which is used for some searches currently conducted by the ATLAS and CMS collaborations. While an NLO extension of this more general model is feasible, it necessitates a more intricate renormalisation procedure to be implemented within \lstinline{MoGRe}. We leave this task to future work.

\begin{table}
  \centering
  \renewcommand{\arraystretch}{1.2} \setlength\tabcolsep{8pt}
  \resizebox{.96\textwidth}{!}{%
    \begin{tabular}{c | c c c c c}
      & Field & \text{\lstinline{FeynRules}} & PDG & Spin & Repr.\\[.1cm]
      \multirow{2}{*}{DM} & $\tilde{\chi}$ & \text{\lstinline{Xm}} & $52$ & $1/2$ & $({\bf 1}, {\bf 1}, 0)$\\[.1cm]
      & $\chi$ & \text{\lstinline{Xd}} & $57$ & $1/2$ & $({\bf 1}, {\bf 1}, 0)$\\[.1cm]
      \hline\\[-.6cm]
      \multirow{9}{*}{Mediators} & Field & \text{\lstinline{FeynRules}} & PDG & Spin & Repr.\\[.2cm]
      & $\varphi_{\scriptscriptstyle Q} =  \bpm\varphi^{(u)}_{\scriptscriptstyle Q}\\ \varphi^{(d)}_{\scriptscriptstyle Q}\epm$ & $\text{\lstinline{YS3Q}} = \bpm \text{\lstinline{YS3Qu}} \\ \text{\lstinline{YS3Qd}}\epm$ & $\begin{array}{l l l}
         1000002 & 1000004 & 1000006\\
         1000001 & 1000003 & 1000005\end{array}$ & $0$ & $({\bf 3}, {\bf 2},  \frac16)$\\[.2cm]
      & $\varphi_{\scriptscriptstyle u}$ & $\text{\lstinline{YS3u}}$ & $\begin{array}{l l l}
         2000002 & 2000004 & 2000006\end{array}$ & $0$ & $({\bf 3}, {\bf 1},  \frac23)$\\[.2cm]
      & $\varphi_{\scriptscriptstyle d}$ & $\text{\lstinline{YS3d}}$ & $\begin{array}{l l l}
         2000001 & 2000003 & 2000005\end{array}$ & $0$ & $({\bf 3}, {\bf 1},  -\frac13)$\\[.2cm]
       & $\varphi_{\scriptscriptstyle L} =  \bpm\varphi^{(\nu)}_{\scriptscriptstyle L}\\ \varphi^{(\ell)}_{\scriptscriptstyle L}\epm$ & $\text{\lstinline{YS1L}} = \bpm \text{\lstinline{YS1Lv}} \\ \text{\lstinline{YS1Le}}\epm$ & $\begin{array}{l l l}
         1000012 & 1000014 & 1000016\\
         1000011 & 1000013 & 1000015\end{array}$ & $0$ & $({\bf 1}, {\bf 2},  -\frac12)$\\[.2cm]
      & $\varphi_{\scriptscriptstyle \ell}$ & $\text{\lstinline{YS1e}}$ & $\begin{array}{l l l}
         2000011 & 2000013 & 2000015\end{array}$ & $0$ & $({\bf 1}, {\bf 1},  -1)$\\[.1cm]
      \hline\\[-.6cm]
      \multirow{6}{*}{Couplings} & Parameter & {\lstinline{FeynRules}} & Les Houches Block & &\\
      &$\lambda_{\scriptscriptstyle Q}$    & \lstinline{lamS3Q} & \lstinline{DMS3Q} & &\\
      &$\lambda_{\scriptscriptstyle u}$    & \lstinline{lamS3u} & \lstinline{DMS3U} & &\\
      &$\lambda_{\scriptscriptstyle d}$    & \lstinline{lamS3d} & \lstinline{DMS3D} & &\\
      &$\lambda_{\scriptscriptstyle L}$    & \lstinline{lamS1L} & \lstinline{DMS1L} & &\\
      &$\lambda_{\scriptscriptstyle \ell}$ & \lstinline{lamS1e} & \lstinline{DMS1E} & &\\
    \end{tabular}
  }
  \caption{Same as in table~\ref{tab:info_S} but for the \lstinline{S3M} and \lstinline{S3D} models.\label{tab:info_F}}\vspace*{.2cm}
  \resizebox{.96\textwidth}{!}{%
    \begin{tabular}{c | c c c c c}
      & Field & \text{\lstinline{FeynRules}} & PDG & Spin & Repr.\\[.1cm]
      \multirow{2}{*}{DM} & $\tilde{V}$ & \text{\lstinline{Xv}} & $53$ & $1$ & $({\bf 1}, {\bf 1}, 0)$\\[.1cm]
      & $V$ & \text{\lstinline{Xw}} & $58$ & $1$ & $({\bf 1}, {\bf 1}, 0)$\\[.1cm]
      \hline\\[-.6cm]
      \multirow{9}{*}{Mediators} & Field & \text{\lstinline{FeynRules}} & PDG & Spin & Repr.\\[.2cm]
      & $\psi_{\scriptscriptstyle Q} =  \bpm\psi^{(u)}_{\scriptscriptstyle Q}\\ \psi^{(d)}_{\scriptscriptstyle Q}\epm$ & $\text{\lstinline{YF3Q}} = \bpm \text{\lstinline{YF3Qu}} \\ \text{\lstinline{YF3Qd}}\epm$ & $\begin{array}{l l l}
         5910002 & 5910004 & 5910006\\
         5910001 & 5910003 & 5910005\end{array}$ & $1/2$ & $({\bf 3}, {\bf 2},  \frac16)$\\[.2cm]
      & $\psi_{\scriptscriptstyle u}$ & $\text{\lstinline{YF3u}}$ & $\begin{array}{l l l}
         5920002 & 5920004 & 5920006\end{array}$ & $1/2$ & $({\bf 3}, {\bf 1},  \frac23)$\\[.2cm]
      & $\psi_{\scriptscriptstyle d}$ & $\text{\lstinline{YF3d}}$ & $\begin{array}{l l l}
         5920001 & 5920003 & 5920005\end{array}$ & $1/2$ & $({\bf 3}, {\bf 1},  -\frac13)$\\[.2cm]
       & $\psi_{\scriptscriptstyle L} =  \bpm\psi^{(\nu)}_{\scriptscriptstyle L}\\ \psi^{(\ell)}_{\scriptscriptstyle L}\epm$ & $\text{\lstinline{YF1L}} = \bpm \text{\lstinline{YF1Lv}} \\ \text{\lstinline{YF1Le}}\epm$ & $\begin{array}{l l l}
         5910012 & 5910014 & 5910016\\
         5910011 & 5910013 & 5910015\end{array}$ & $1/2$ & $({\bf 1}, {\bf 2},  -\frac12)$\\[.2cm]
      & $\psi_{\scriptscriptstyle \ell}$ & $\text{\lstinline{YF1e}}$ & $\begin{array}{l l l}
         5920011 & 5920013 & 5920015\end{array}$ & $1/2$ & $({\bf 1}, {\bf 1},  -1)$\\[.1cm]
      \hline\\[-.6cm]
      \multirow{6}{*}{Couplings} & Parameter & {\lstinline{FeynRules}} & Les Houches Block & &\\
      &$\lambda_{\scriptscriptstyle Q}$    & \lstinline{lamF3Q} & \lstinline{DMF3Q} & &\\
      &$\lambda_{\scriptscriptstyle u}$    & \lstinline{lamF3u} & \lstinline{DMF3U} & &\\
      &$\lambda_{\scriptscriptstyle d}$    & \lstinline{lamF3d} & \lstinline{DMF3D} & &\\
      &$\lambda_{\scriptscriptstyle L}$    & \lstinline{lamF1L} & \lstinline{DMF1L} & &\\
      &$\lambda_{\scriptscriptstyle \ell}$ & \lstinline{lamF1e} & \lstinline{DMF1E} & &\\
    \end{tabular}
  }
  \caption{Same as in table~\ref{tab:info_S} but for the \lstinline{F3V} and \lstinline{F3W} models.\label{tab:info_V}}
\end{table}

We present information on the six simplified models in tables~\ref{tab:info_S}, \ref{tab:info_F} and \ref{tab:info_V} for models with scalar, fermionic and vector DM respectively. For each model, we list the new included fields, their representation under the SM gauge group, their spins, the particle names used in the \lstinline{FeynRules} implementation, and the Particle Data Group (PDG) identifiers~\cite{ParticleDataGroup:2022pth}. Additionally, we detail the conventions for the coupling parameters, including their implementation names and the corresponding Les Houches blocks~\cite{Skands:2003cj} where numerical values are stored.

The generated UFO model can be used within the \lstinline{MG5aMC} platform~\cite{Alwall:2014hca} for achieving computations relevant for to collider phenomenology. Both LO and NLO simulations are feasible for models with a flavour-diagonal structure. However, for models with off-diagonal couplings in the flavour space, only LO simulations are achievable. Additionally, we have produced LO model files in which all flavours of quarks are massive. They are available both in UFO and \lstinline{CalcHEP}~\cite{Belyaev:2012qa} formats, enabling their use with \lstinline{micrOMEGAs}~\cite{Belanger:2018ccd, Alguero:2023zol} and \lstinline{MadDM}~\cite{Ambrogi:2018jqj, Arina:2020kko, Arina:2021gfn} for evaluating cosmological observables. The implemented \lstinline{FeynRules} model and the corresponding generated UFO libraries encompass all possibilities for the DM candidate, thus incorporating all Lagrangians of eq.~\eqref{eq:lagXY}. In order to reduce the number of free parameters compared to the general case, we include specific restrictions tailored for minimal $t$-channel DM simplified models. One particular restriction involves the activation of either a single mediator or a set of mass-degenerate mediators. Consequently, all new physics states are decoupled and non-interacting, except for a single DM candidate and specific mediators. 

We first consider `\textit{universal}' possibilities (referred to as \lstinline{XYZ_uni} with \lstinline{XYZ} being one of the six symbols introduced in eq.~\eqref{eq:lagXY} to represent the various DM scenarios), where all mediators of the model are active but assumed to be degenerate, with interaction strengths that are both flavour-conserving and universal. The coupling parameters of such a scenario therefore satisfy the condition: 
\be
  (\lambda_{\scriptscriptstyle Q})_{ij} = (\lambda_{\scriptscriptstyle L})_{ij} =  (\lambda_{\scriptscriptstyle u})_{ij} =  (\lambda_{\scriptscriptstyle d})_{ij} = (\lambda_{\scriptscriptstyle \ell})_{ij} = \lambda\, \delta_{ij}\,,
\ee
where $\lambda$ represents the sole, universal, free coupling parameters and $i,j = 1,2,3$ are flavour indices. Alongside the universal mediator mass ($M_Y$) and the DM mass ($M_X$), the model is defined by three new physics parameters. Furthermore, we can construct a second class of universal scenarios in the context of leptophilic dark matter. These scenarios, termed \lstinline{XYZ_lR} models, exclusively involve DM couplings to all three SM right-handed leptons, generalising the framework studied in \cite{Baker:2018uox} to accommodate various spin configurations. In this case, the coupling parameters are defined by 
\be
  (\lambda_{\scriptscriptstyle Q})_{ij} = (\lambda_{\scriptscriptstyle L})_{ij} =  (\lambda_{\scriptscriptstyle u})_{ij} =  (\lambda_{\scriptscriptstyle d})_{ij} = 0
  \qquad\text{and}\qquad
  (\lambda_{\scriptscriptstyle \ell})_{ij} = \lambda\, \delta_{ij}\,,
\ee
and the model is again defined by three free parameters, the DM and (universal) mediator masses $M_X$ and $M_Y$, as well as the coupling parameter $\lambda$. Such a class of models are however not examined in this report. In all these universal setups, the presence of several mediators associated with a specific generation of SM fermion prevents strong constraints from flavour-changing-neutral-current processes, both in the quark and lepton sectors. This contrasts to models featuring mediators coupling to several generations of fermions. 

Next, we focus on models in which a single class of mediator is considered. One such scenario is exemplified by the \lstinline{XYZ_uR} possibilities investigated in \cite{Arina:2020tuw, Arina:2023msd}, wherein the mediator couples exclusively to the right-handed up quark. In this setup, the coupling parameters are constrained to 
\be
  (\lambda_{\scriptscriptstyle u})_{11} = \lambda
  \qquad\text{and}\qquad
  (\lambda_{\scriptscriptstyle f})_{ij} = 0 \ \text{for all other couplings (with $f=Q,L,u,d,\ell$).}
\ee
In contrast to the universal case, $\lambda$ represents the coupling strength between the mediator and the right-handed up quark, and this restricted scenario is characterised by three free parameters: the masses of the dark matter and the mediator $M_X$ and $M_Y$, and the coupling $\lambda$. In the present work, we will examine scenarios in which the new particles couple to any flavour of right-handed quark (referred to as \lstinline{XYZ_dR}, \lstinline{XYZ_cR}, \lstinline{XYZ_sR}, \lstinline{XYZ_bR} and \lstinline{XYZ_tR} models) and the right-handed muon (referred to as \lstinline{XYZ_muR} models), motivated by the specificities inherent to each generation. In particular, we keep in mind the special role that the second generation could play with respect to new physics (to solve, for instance, the anomalous magnetic moment of the muon puzzle or some of the remaining flavour anomalies), as well as that of the third generation due the heavier corresponding fermion masses and their possible connection to the electroweak symmetry breaking mechanism. In contrast, we only outline in our cosmological analysis the complementarity of astroparticle searches to LHC-based ones, relying on previous studies of models related to the first and third generations of quarks. The reasons is that in this case, the flavour of the quark is not relevant for the constraints that can be imposed on the model so that a smooth extrapolation of the results for models relevant to first-generation quarks can be applied to models relevant to second-generation and bottom quarks.

\subsection{Flavoured dark matter: Dark Minimal Flavour Violation}\label{sec:DMFV}
While minimal models serve as a valuable parametrisation for a broad range of $t$-channel DM scenarios, they often fail to provide a comprehensive description of non-minimal features. Therefore, it is crucial to explore models beyond the minimal setups, to ensure thorough theoretical and experimental investigations without overlooking potential loopholes. 

One way to expand upon the minimal framework outlined in section~\ref{sec:model_minimal} involves introducing multiple flavours of DM (\ie\ a field $X_i$ carrying a flavour index $i$) and assuming that they transform under a certain flavour symmetry. Early studies of such flavoured DM models~\cite{Kile:2011mn, Kamenik:2011nb, Batell:2011tc, Agrawal:2011ze, Batell:2013zwa, Kile:2013ola, Kile:2014jea, Lopez-Honorez:2013wla} have applied the concept of Minimal Flavour Violation (MFV)~\cite{Chivukula:1987py, Hall:1990ac, Buras:2000dm, DAmbrosio:2002vsn, Buras:2003jf, Cirigliano:2005ck} to the dark sector. In MFV models, an approximate flavour symmetry $U(3)^5 \equiv U(3)_Q \times U(3)_u \times U(3)_d \times U(3)_L \times U(3)_e$ is imposed, treating the five SM fundamental fermion fields $Q_L$, $u_R$, $d_R$, $L_L$, and $\ell_R$ as flavour triplets whose components rotate into each other via five separate $U(3)$ groups. This flavour symmetry is then broken by Yukawa interactions. In minimally flavour-violating dark matter models, the DM field $X\equiv (X_1,X_2,X_3)$ is assumed to transform as a triplet under one of these $U(3)$ symmetries. The coupling matrix $\lambda$ is then determined following the MFV assumption, and its structure follows from an expansion in terms of the SM Yukawa couplings. Additionally, $X$ is required to be a complex field. Unlike the simplified model discussed in section~\ref{sec:model_minimal}, the mediator field in flavoured DM models does not carry a flavour index. However, in less minimal `skew-flavoured' constructions~\cite{Agrawal:2015kje}, both the DM and the mediator carry flavour indices.

Models featuring Dark Minimal Flavour Violation (DMFV)~\cite{Agrawal:2014aoa} extend this concept further. The flavour symmetry group is promoted to $U(3)^6$ (or $U(3)^5 \times O(3)$ for real DM fields), with additional symmetry transformations acting on the DM field $X$ ($\chi$ in the notation of \cite{Agrawal:2014aoa}), rendering it a flavour triplet. For models with a single mediator, we introduce a $t$-channel mediator $Y$ ($\phi$ in the notation of \cite{Agrawal:2014aoa}) that couples the dark flavour triplet $X$ to one of the representations of SM fermions $f$. The Lagrangian governing the interaction between the dark matter and the mediator is generically expressed as
\begin{equation}\label{eq:lag_dmfv}
\mathcal{L}_{XY} = \lambda_{ij} \bar f_i X_j Y\,,
\end{equation}
where $i$ and $j$ represent flavour indices. Adhering to the DMFV principle, the complex $3\times 3$ coupling matrix $\lambda$ serves as the sole source of flavour symmetry breaking apart from the Standard Model Yukawa couplings. In contrast to the Lagrangians introduced in eq.~\eqref{eq:lagXY}, the second flavour index of the coupling  $\lambda_{ij}$ is here associated with the dark matter, rather than the mediator for which only a single state is considered. In accordance with the DMFV principle, the mass matrix $M_X$ of the DM states is not an arbitrary parameter in the theory, but is instead determined by a spurion expansion based on the flavour-violating coupling $\lambda$. For example, for $X$ being a Dirac fermion it reads
\begin{equation}\label{eq:flavDM_eta}
  M_X = m_X \Big[\mathbbm{1}+\eta \lambda^\dagger\lambda +\mathcal{O}(\lambda^4)\Big]\,,
\end{equation}
where $\eta$ is a real expansion parameter. This relationship ties the dark mass spectrum directly to the flavour structure of the coupling $\lambda$ that connects the visible and dark sectors. Moreover, the approximate $U(3)$ flavour symmetry in the dark sector reduces the complexity of the coupling matrix $\lambda$ that can, for complex DM, be characterised by three diagonal coupling strengths, three mixing angles, and three complex phases. It is worth noting that in quark-flavoured DMFV models (\ie\ models in which $X$ and $Y$ couple to quarks) featuring Dirac DM, the stability of the lightest DM flavour is ensured by a residual $\mathbb{Z}_3$ symmetry, which persists after the flavour symmetry is broken. 

This has motivated previous studies that have investigated the phenomenology of DMFV models with Dirac dark matter coupling to a specific quark flavour~\cite{Agrawal:2014aoa, Jubb:2017rhm, Blanke:2017tnb, Blanke:2017fum}. Scenarios with right-handed up, down, charm and bottom quarks have hence been explored, as well as setups in which the DM state predominantly couples to the doublet of left-handed top and bottom quarks. This corresponds to $f_i = u_R$, $d_R$, $c_R$, $b_R$ and $(t_L, b_L)^T$ in eq.~\eqref{eq:lag_dmfv}. Furthermore, the possibility of Majorana DM has been examined for the case of $f=u_R$~\cite{Acaroglu:2021qae, Acaroglu:2023phy}. Here, the presence of a smaller dark flavour symmetry group $O(3)$ results in a coupling matrix $\lambda$ containing more free parameters. Additionally, the absence of the DM-stabilising $\mathbb{Z}_3$ symmetry necessitates the introduction of an \textit{ad hoc} $\mathbb{Z}_2$ symmetry to ensure DM stability.

The correspondence between the general DMFV parametrisation discussed above and the simplified model parametrisation outline in section~\ref{sec:model_minimal} can be established in two cases: either when all physical dark matter states are nearly mass-degenerate, or when one DM state is much lighter than the others. In such scenarios, the DMFV model effectively comprises one dark matter state $X$ and one mediator state $Y$ featuring couplings to all generation of SM fermions. Consequently, the coupling matrices $\lambda$ appearing in eq.~\eqref{eq:lagXY} reduce to vectors in the flavour space and only one mediator state is active. Various DMFV-inspired simplified models can then be constructed, depending on the SM fermion representation $f$ to which the DM state couples, and the particle nature of the new physics fields $X$ and $Y$. While the \lstinline{DMSimpt 2.0} UFO model incorporates such a flexibility, we employ dedicated LO UFO libraries available from the \lstinline{DMSimpt GitHub} repository \url{https://github.com/lena-ra/Flavored-Dark-Matter}.

Non-minimally flavoured dark matter can also be introduced to couple to the lepton sector~\cite{Chen:2015jkt, Acaroglu:2022hrm, Acaroglu:2023cza}, as studied in the context of Dirac and complex scalar DM coupling to right-handed charged leptons. In these constructions, a discrete $\mathbb{Z}_2$ symmetry is always necessary to stabilise dark matter. Additionally, less minimal models~\cite{Acaroglu:2022boc} featuring two mediators have been studied. Such a setup can provide an explanation to the longstanding deviations between theoretical predictions and experimental measurements relevant to the anomalous magnetic moment of the muon. While the matrices of couplings $\lambda_1$ and $\lambda_2$ associated with the two mediators are related by DMFV, the DM spectrum becomes instead free. Finally, flavoured DM models beyond DMFV have also been found to offer a promising avenue for achieving successful baryogenesis~\cite{Heisig:2024mwr}.

\subsection{Dark matter simplified models inspired by compositeness}\label{sec:compositeDM}
Another compelling possibility for exploring $t$-channel dark matter models beyond minimal frameworks emerges from composite scenarios of new physics~\cite{Panico:2015jxa, Cacciapaglia:2020kgq, Cacciapaglia:2022zwt}. These models typically postulate the existence of additional coloured and non-coloured resonances, arising as bound states from an underlying fermionic construction beyond the SM featuring a new strong dynamics. While some of these resonances are closely tied to the top quark, elucidating the question of its large mass via mechanisms like partial compositeness~\cite{Kaplan:1991dc}, others could be stable, electrically neutral, and colourless, thereby potentially serving as candidates for dark matter~\cite{Colucci:2018vxz, Cornell:2021crh, Cornell:2022nky}. In such frameworks, DM stability is ensured by a residual discrete $\mathbb{Z}_2$ symmetry stemming from the breaking of the new strong dynamics. The DM state is assigned an odd parity under this new $\mathbb{Z}_2$ symmetry, while all SM fields are assigned an even parity. Meanwhile, the particle spectrum comprises both even and odd new states. 

An example of next-to-minimal setup, just slightly more complex than the simplified models introduced in section~\ref{sec:model_minimal}, involves the inclusion of one real scalar DM state together with two fermionic top partners, or mediators as per the terminology used in this manuscript. Specifically, this setup encompasses one $\mathbb{Z}_2$-even state denoted as $Y'_{\scriptscriptstyle t}$ (or $T'$ in the notation of \cite{Cornell:2021crh}), and one $\mathbb{Z}_2$-odd state labelled as $Y_{\scriptscriptstyle t}$ (or $T$ in the notation of \cite{Cornell:2021crh}). Both these states share the same quantum numbers as the SM right-handed top quark field $t_R$, and their interactions with the real scalar DM state $X$ (or $S$ in the notation of \cite{Cornell:2021crh}) can be captured within an effective Lagrangian resembling those introduced in eq.~\eqref{eq:lagXY}. This Lagrangian can be expressed as
\be\label{eq:lagtopcompoDM}
  \lag_{XY} = 
        \lambda_{\scriptscriptstyle t}\ \bar Y_{\scriptscriptstyle t} t_R\ X 
     + \lambda'_{\scriptscriptstyle t}\ \bar Y_{\scriptscriptstyle t} Y'_{\scriptscriptstyle t}\ X 
     + {\rm H.c.}
\ee
Here, the parameters $\lambda_{\scriptscriptstyle t}$ and $\lambda'_{\scriptscriptstyle t}$ represent two new physics couplings, that were noted $\tilde{y}_{t}$ and $\tilde{y}_{T'}$ in the notation of \cite{Cornell:2021crh}. The list of free parameters additionally includes the dark matter mass $M_X$ and the two mediator masses $M_Y$ and $M_{Y'}$. The \lstinline{FeynRules} model and associated UFO libraries are available from a dedicated \lstinline{GitHub} repository, located at \url{https://github.com/BFuks/CompositeDM}.

A less minimal effective DM model, still inspired by composite constructions but closer to UV completions compared the simplified setup of section~\ref{sec:model_minimal}, could entail a sufficient number of mediator fields to generate the top quark mass via partial compositeness. In such a scenario, both weak singlet and doublet mediators are necessary~\cite{Belyaev:2021zgq}. Assuming all these mediators to be $\mathbb{Z}_2$-odd for simplicity, an effective interaction Lagrangian incorporating their coupling to dark matter would be given by
\be\renewcommand{\arraystretch}{1.25}\label{eq:lagchacal}
  \lag_{XY} = 
        \lambda_{\scriptscriptstyle t}\ \bar Y_{\scriptscriptstyle t} t_R\ X 
     + \lambda_{\scriptscriptstyle Q}\ \bar Y_{\scriptscriptstyle Q} Q_L\ X 
     + {\rm H.c.}
    \qquad\text{with}\qquad
    Y_{\scriptscriptstyle Q} = \bpm Y_{\scriptscriptstyle Q,t}\\ Y_{\scriptscriptstyle Q,b}\epm\quad\text{and}\quad
    Q_L = \bpm t_L\\b_L\epm\,.
\ee
This model introduces a certain number of free parameters, including two new physics couplings $\lambda_{\scriptscriptstyle Q}$ and $\lambda_{\scriptscriptstyle t}$, one dark matter mass $M_X$, and three mediator masses $M_{Y_t}$, $M_{Y_{Q,t}}$ and $M_{Y_{Q,b}}$. Although these masses are predictable in effective setups incorporating partial compositeness, for simplicity we treat them as free parameters. Consequently, such a configuration can be mapped to the \lstinline{DMSimpt 2.0} simplified model, which is provided with an associated restriction termed \lstinline{F3S-VLQ}, recalling that the choice of scalar DM and fermionic mediators is the relevant one for composite constructions.

\subsection{Towards UV completions - frustrated dark matter}\label{sec:fDM}
Simple UV-complete dark matter models, which avoid the theoretical shortcomings or arbitrary motivations of certain simplified models, can often be integrated into larger and well-motivated theoretical frameworks~\cite{Belyaev:2022qnf}. Consequently, they could serve as non-minimal simplified models for dark matter that offer the advantage of a more complex phenomenology including a variety of astrophysical and collider signatures. Frustrated dark matter (fDM) models~\cite{Carpenter:2022lhj} exemplify these simple UV-complete setups. They consist of models where interactions between dark matter and the visible sector arise at one-loop order, rather than at tree level. Such loop-level couplings can emerge in many UV-complete theories, and they yield signatures that are highly sensitive to the details of the mediator-SM interactions. The fDM framework encompasses a broad class of models where the mediators carry SM charges, but the interactions of dark matter are `frustrated' in the sense that the specific mediator assignments preclude tree-level interactions with the SM. 

In this framework, as is customary, the dark matter field $X$ is assumed to transform as a singlet under the SM gauge symmetry. However, the SM gauge charges of all mediator fields coupling both to $X$ and to the Standard Model are chosen to forbid renormalisable gauge-invariant contact interactions between $X$ and any SM fermion. For fermionic dark matter, these models necessitate a pair of mediators to couple the dark state $X$ to the SM. A schematic representation of this family of models, featuring SM-singlet Dirac dark matter coupled to a pair of mediators, takes the form:
\begin{align}
    \text{SM}\ \longleftrightarrow\ \text{mediators}\ [ \text{$\varphi$, $\psi$} ] \ \longleftrightarrow\ \text{DM [$X$]}\,, 
\end{align}
where, as in section~\ref{sec:model_minimal}, $\varphi$ denotes a scalar mediator and $\psi$ a fermionic one. Since the mediators carry SM charges, they can generally decay into SM particles which has the extra benefit to circumvent cosmological and phenomenological issues associated with new stable particles. Therefore, one or both mediators should have renormalisable interactions with the SM, hence allowing for mediator decays.

Within the scope of this whitepaper, we focus on fDM models where mediators carry $SU(3)_C$ quantum numbers, as particles with non-trivial colour charges are anticipated to exhibit the highest production cross sections at the LHC. Depending on the details of the mediator sector, these mediators may also carry non-trivial $SU(2)_L$ charges and $U(1)_Y$ hypercharges. Among the numerous possible $SU(3)_C$ charge assignments for the mediating sector, only certain choices permit direct renormalisable interaction between the mediators and the SM. For instance, only the colour triplet, sextet, and octet options allow for renormalisable interactions between mediators and pairs of quarks. Notably, colour-octet mediators can enjoy such couplings without the necessity of non-trivial electroweak charges.  

We consider a particular fDM realisation where the mediators are $SU(3)_C$ sextets and weak singlets, this last condition being imposed for simplicity, and where the DM is a Dirac fermion. In this scenario, only the scalar mediator directly couples to the SM at tree level, specifically to a pair of quarks of possibly the same electric charge. The hypercharge $Y$ of the mediators depends on the structure of these quark-messenger couplings. While a comprehensive examination of all possible low-dimensional couplings is provided in \cite{Carpenter:2021rkl}, only one of them is renormalisable. This interaction couples the scalar sextet to a $uu$, $ud$, or $dd$ quark pair pursuant to its hypercharge, and its phenomenological consequences have been explored in~\cite{Shu:2009xf, Han:2009ya, Han:2010rf}. For simplicity, we only consider a sextet scalar with $Y = 4/3$ so that the corresponding Lagrangian $\mathcal{L}_{XY}$\footnote{We use the notation $\mathcal{L}_{XY}$ for any Lagrangian including DM-mediator interactions, regardless of the model, for the purpose of uniformity.} is given by
\begin{equation}\label{eq:fdmlag}
  \mathcal{L}_{XY} = \lambda_X \varphi^{\dagger} \bar{X}\psi
    + \lambda_{qq} \bar{K}_6 \varphi^\dagger \bar{u}^c_R u_R  + \text{H.c.}\,,
\end{equation}
where all indices have been omitted for clarity. Here, $\lambda_X$ represents the  mediator-DM coupling, and $\lambda_{qq}$ is the mediator-SM coupling matrix in the flavour space. Additionally, $\bar{K}_6$ denotes the sextet Clebsch-Gordan coefficient tensor, \ie\ an elementary colour tensor with one antisextet colour index and two fundamental colour indices. Furthermore, we impose a $\mathbb{Z}_2$ symmetry on the DM and messenger sector to ensure DM stability. The dark matter state $X$ and one mediator (in this case, $\psi$) are odd, while all other fields, including the SM ones and the other mediator ($\varphi$), are even. Consequently, only one of the mediators needs be heavier than the DM state as its decay involves the DM particle. In contrast, the other mediator decays directly into SM fields so that its mass is unconstrained by DM stability requirements.

The free parameters of this model include the three particle masses $M_{X}$, $M_{\psi}$, and $M_\varphi$, the DM-mediator Yukawa coupling $\lambda_X$, and the mediator-quark coupling matrix $\lambda_{qq}$. In our analysis, we restrict the entries of $\lambda_{qq}$ to be real, while allowing flexibility in choosing the flavour structure. This flexibility is constrained by stringent limits on flavour-changing neutral currents (FCNCs), which can be enhanced by the presence of a colour-sextet scalar field coupling to up-type quarks~\cite{Babu:2008rq}. Additionally, certain couplings to charm quarks are tightly constrained~\cite{Babu:2013yca}, leading to the following bounds:
\be\label{eq:fcnclim}\bsp
  &(\lambda_{qq})_{11}(\lambda_{qq})_{22} \leq 9.3\ \Bigg(\frac{M_\varphi}{1~\mathrm{TeV}}\Bigg)^2 \times 10^{-7}\,,\\
  &\sum_{I=1}^3(\lambda_{qq})_{I2}(\lambda_{qq})_{I1} \leq 2.5\ \Bigg(\frac{M_\varphi}{1~\mathrm{TeV}}\Bigg) \times 10^{-3}\,.
\esp\ee
A straightforward way to comply with these constraints, as explored in~\cite{Carpenter:2022lhj}, is to assume a flavour-diagonal coupling scheme with $(\lambda_{qq})_{22} = 0$, effectively yielding a charm-phobic scenario. However, completely excluding charm couplings, while convenient, is neither necessary nor symmetry-driven. On the other hand, as discussed in section~\ref{sec:collider_frustrated}, large couplings involving third-generation quarks lead to intriguing heavy-flavour phenomenology. To balance these considerations, we adopt a coupling scheme that reflects the Standard Model quark mass hierarchy while respecting the FCNC constraints in~\eqref{eq:fcnclim}.

In the investigations carried out in this white paper, we generate a UFO version of the above model with \lstinline{FeynRules}, to be used with \lstinline{MG5aMC} and \lstinline{MadDM} for collider and cosmology phenomenology, respectively. This model is available upon request.

\subsection{New gauge interactions to connect the dark sector to the Standard Model}\label{sec:nonAbDS}

Models with a vector DM field, particularly those involving a non-Abelian gauge sector, remain among the least explored extensions of the SM, despite being well-motivated. Here, gauge principles provide natural constraints and guidance limiting possible theoretical constructions (see, \eg, \cite{Hubisz:2004ft, Hambye:2008bq, Chen:2009ab, Diaz-Cruz:2010czr, Bhattacharya:2011tr, Lebedev:2011iq, Farzan:2012hh, Baek:2012se, Koorambas:2013una, Fraser:2014yga, Ko:2014gha, Huang:2015wts, Gross:2015cwa, DiFranzo:2015nli, Ko:2016fcd, Barman:2017yzr, Huang:2017bto, Barman:2018esi, Barman:2019lvm, Buttazzo:2019mvl, Abe:2020mph, Gross:2020zam, Chowdhury:2021tnm, Baouche:2021wwa, Hu:2021pln, Babu:2021hef} for discussions of non-Abelian DM in different setups, including scenarios with non-renormalisable kinetic mixing terms or Higgs portal couplings). In this section, we highlight a recently proposed minimal framework that extends the SM gauge sector by introducing a non-Abelian gauge group under which all SM particles are singlets and for which no renormalisable kinetic mixing terms are allowed.\footnote{Gauge kinetic mixing terms may arise at loop level, depending on the Higgs sector structure, but these correspond to suppressed higher-dimensional operator contributions.} For further details, we refer the reader to~\cite{Belyaev:2022shr}, while here we summarise the construction and key properties of this framework.

We consider the simplest non-Abelian group, denoted as $SU(2)_D$, to connect the SM to the dark sector, and we label the associated gauge bosons as 
\be\renewcommand{\arraystretch}{1.3}
  V_\mu^D \equiv \bpm V^0_{D+\mu}\\ V^0_{D0\mu}\\ V^0_{D-\mu}\epm\,,
\ee
where the superscripts refer to the field electric charges and the subscripts denote their isospin under $SU(2)_D$ ($D$-isospin). The spontaneous breaking of the $SU(2)_L \times U(1)_Y$ and $SU(2)_D$ gauge symmetries is achieved through two scalar doublets,
\be
  \Phi_H = \begin{pmatrix} \phi^+ \\ \phi^0 \end{pmatrix}\,\qquad\qquad 
  \Phi_D = \begin{pmatrix} \varphi^0_{D+{\frac{1}{2}}} \\ \varphi^0_{D-{\frac{1}{2}}} \end{pmatrix} \,
\ee
whose lower components get the vacuum expectation values (vevs) $v$ and $v_D$. The scalar potential, that was introduced in~\cite{Hambye:2008bq}, is given by
\be
  V(\Phi_H, \Phi_D) = - \mu^2 \Phi_H^\dagger \Phi_H - \mu_D^2 \Phi_D^\dagger \Phi_D 
   + \lambda (\Phi_H^\dagger \Phi_H)^2 + \lambda_D (\Phi_D^\dagger \Phi_D)^2 
   + \lambda_{\Phi_H\Phi_D} \Phi_H^\dagger \Phi_H \Phi_D^\dagger \Phi_D\;,
\ee
and involves various bilinear ($\mu_i^2$) and quartic ($\lambda_i$) couplings. This potential ensures the degeneracy and stability of the $SU(2)_D$ gauge bosons due to the custodial symmetry inherent to the scalar Lagrangian. Moreover, the interaction between the scalar fields via the portal term $\lambda_{\Phi_H\Phi_D}$ induces scalar mixing and modifies the couplings of the Higgs boson to the SM states, which therefore provides strong constraints on the model~\cite{Arcadi:2020jqf}. 

A new mechanism for communication between the dark and visible sectors is introduced through a vector-like fermion doublet,
\be 
  \Psi = \bpm \psi_D \\ \psi\epm \,,
\ee
which is charged under $SU(2)_D$ but singlet under $SU(2)_L$, and where both components of $\Psi$ share the same hypercharge quantum numbers as an SM right-handed fermion.\footnote{Vector-like portals have been studied for scalar DM candidates in~\cite{Baek:2017ykw, Colucci:2018vxz}, and for vector DM states in~\cite{Hisano:2020qkq, Babu:2021hef}. In these contexts, simplifying assumptions include neglecting new Yukawa couplings~\cite{Hisano:2020qkq} or introducing an extended particle content~\cite{Babu:2021hef}.} The mass and interaction terms for the fermion $\Psi$ are given by
\begin{equation}\label{eq:FPVDM_yukawa}
  \mathcal{L} = - M_\Psi \bar{\Psi} \Psi - \left( y^\prime \bar{\Psi}_L \Phi_D f_R^{\rm SM} + \text{H.c.} \right)\,, 
\end{equation}
where $f_R^{\rm SM}$ denotes a generic SM right-handed fermion, and $y^\prime$ is a new Yukawa coupling connecting the SM fermion to the dark fermion $\Psi$ via the scalar dark doublet $\Phi_D$. The stability of the DM is ensured by the absence of an additional Yukawa term $y^{\prime\prime} \bar{\Psi}_L \Phi_D^c f_R^{\rm SM}$, and is protected by an unbroken global $U(1)_D \equiv e^{i\Lambda Y_D}$ symmetry. Without this symmetry, the term involving the $y^{\prime\prime}$ coupling would be unavoidable since the scalar doublet $\Phi_D$ lies in a pseudo-real representation. The symmetry-breaking pattern in this framework is $SU(2)_D \times U(1)_D \to U(1)_D^d$. Assigning $U(1)_D$ charges as $Y_D = 1/2$ for the dark scalar and fermion doublets and $Y_D = 0$ for the vector triplet, there remains an invariance under the discrete subgroup $\mathbb{Z}_2 \equiv (-1)^{Q_D}$, where $Q_D = T^3_D + Y_D$.

Among all new particles, the lightest $\mathbb{Z}_2$-odd particle is stable. Our construction features two potential candidates, namely the $V^0_{D\pm}$ or $\psi_D$ states, with different implications for cosmology~\cite{Belyaev:2022shr}. In the present work, we focus on the scenario where the lightest $\mathbb{Z}_2$-odd particle is the $V^0_{D\pm}$ boson, which we designate as the Fermion Portal Vector Dark Matter (FPVDM) framework. The theory predicts six massive gauge bosons (the $Z$, $W^\pm$, $V^0_{D0}$, and $V^0_{D\pm}$ vector bosons), whose longitudinal components correspond to six Goldstone bosons. The remaining two scalar degrees of freedom include the Standard Model Higgs boson and an additional CP-even scalar. In the unitary gauge, the scalar mass terms in the Lagrangian take the form
\begin{equation}
  \mathcal{L}_m^{\mathcal{S}} =  \bpm h_1 & \varphi_1\epm 
     \bpm \lambda v^2 & \frac{\lambda_{\Phi_H\Phi_D}}{2} v v_D \\ 
     \frac{\lambda_{\Phi_H\Phi_D}}{2} v v_D & \lambda_D v_D^2 \epm
     \bpm h_1 \\ \varphi_1 \epm \,,
\end{equation}
where $h_1$ and $\varphi_1$ are defined from $\phi^0 =(v + h_1)/\sqrt{2}$ and $\varphi^0_{D-1/2} = (v_D + \varphi_1)/\sqrt{2}$. Diagonalising the above mass matrix yields the scalar mass eigenvalues:
\begin{equation}
  m_{h,H}^2 = \lambda v^2 + \lambda_D v_D^2 \mp \sqrt{(\lambda_D v_D^2 - \lambda v^2)^2 + \lambda_{\Phi_H\Phi_D}^2 v^2 v_D^2}\,,
\end{equation}
with a mixing angle given by
\begin{equation}
  \sin\theta_S = \sqrt{2 \frac{m_H^2 v^2 \lambda - m_h^2 v_D^2 \lambda_D}{m_H^4 - m_h^4}} \,.
\end{equation}

In the dark sector, the $\Psi$ component with $T_{{3D}} = +1/2$ ($Q_D = +1$) does not mix with any other fermion, and its mass, $m_{\psi_{D}} = M_\Psi$, is determined solely by the vector-like mass term in \eqref{eq:FPVDM_yukawa}. In contrast, the masses of the other fermions depend on the vevs of both scalars, as well as on the SM Yukawa interactions and the one introduced in \eqref{eq:FPVDM_yukawa}. The corresponding fermionic mass Lagrangian is given by
\begin{equation}\renewcommand{\arraystretch}{1.3}
  \mathcal{L}_m^f =  \bpm \bar f^{\rm SM}_L & \bar\psi_L\epm 
  \bpm y \frac{v}{\sqrt{2}} & 0 \\ y^\prime \frac{v_{D}}{\sqrt{2}} & M_\Psi  \epm 
  \bpm  f^{\rm SM}_R \\ \psi_R  \epm\,.
\end{equation}
The corresponding mass eigenvalues are easily obtained, and read
\begin{eqnarray}
 m_{f,F}^2 &=& \frac{1}{4} \bigg[\Delta \mp \sqrt{\Delta^2 - 8 y^2 v^2 M_\Psi^2}\bigg]\,,
\end{eqnarray}
with $\Delta = y^2 v^2 + y^{\prime2} v_{D}^2 + 2 M_\Psi^2$. Here, $f$ represents the SM fermion connected to the dark sector, while $F$ denotes its heavier partner, and they always satisfy the hierarchy $m_f < m_{\psi_{D}} \leq m_F$. Let us note that in principle, a vector-like fermion may interact with one or more SM flavours, and multiple vector-like fermions may be included in the theory. In addition, the $SU(2)_D$ gauge bosons are degenerate in mass at tree level, with $m_{V_D} \equiv m_{V^0_{D\pm}} = m_{V^0_{D0}} = g_D v_D/2$ (and $g_D$ being the $SU(2)_D$ gauge coupling). However, this degeneracy is lifted by fermionic loop corrections, which account for the differing $\mathbb{Z}_2$ parities of the $SU(2)_D$ gauge bosons. For simplicity, we label from now on the states $V_D \equiv V^0_{D\pm}$ of mass $m_{V_D}$, and $V^\prime \equiv V^0_{D0}$ of mass $m_{V^\prime}$. The leading contribution to the radiative mass splitting $\Delta m_V = m_{V_D} - m_{V^\prime}$ is driven by $F$ and $\psi_D$ loops, 
\begin{equation}
\Delta m_V = \frac{\varepsilon^2 g_D^2 m_F^2}{32 \pi^2 m_{V_D}} + \mathcal{O}(\varepsilon^2), \quad 
\text{where } 
\varepsilon = \frac{m_F^2 - m_{\psi_D}^2}{m_F^2}.
\end{equation}
Notably, $\Delta m_V$ vanishes in the limit $y^\prime \to 0$. 

To analyse the simplest realisation of the FPVDM framework, we assume that the new vector-like fermions interact exclusively with one SM flavour, that we take to be the top quark. The new physics sector is thus described by six independent input parameters,
\begin{equation}
    g_D, \quad  
    m_{V_{D}}, \quad  
    m_{H}, \quad  
    \sin\theta_S, \quad  
    m_T\equiv m_F, \quad  
    m_{t_{D}} \equiv m_{\psi_{D}}.
\label{eq:fpvdm_prm}\end{equation}
We further simplify the parameter space by enforcing no mixing between the two scalars $h$ and $H$ (\ie\ $\theta_S = 0$), so that the SM Higgs sector remains unaffected by new physics at tree level and the dark side of the potential mirrors the structure of the SM potential. It is important to note that for $y^\prime \to 0$, the quartic coupling $\lambda_{\Phi_H\Phi_D}$ cannot be generated radiatively. This implies that the scalar mixing is induced solely by $y^\prime$ effects through fermionic loops, and can thus be safely neglected. In this configuration, the fermion sector satisfies the mass hierarchy $m_t < m_{t_D} \leq m_T$, while the mass of the heavy Higgs boson $H$ can take any value consistent with experimental bounds, including values below that of the SM Higgs boson. In this study we test this realisation of the model against multiple observables from cosmology, DM direct and indirect detection and LHC searches. For this purpose the Lagrangian has been implemented in \lstinline{LanHEP}~\cite{Semenov:2008jy} and \lstinline{FeynRules}~\cite{Alloul:2013bka} to generate model files for \lstinline{CalcHEP}~\cite{Belyaev:2012qa}, in the \lstinline{UFO}~\cite{Degrande:2011ua, Darme:2023jdn} format, as well as for \lstinline{FeynArts}~\cite{Hahn:2000kx}. They are available from the \lstinline{HEPMDB} database~\cite{Brooijmans:2012yi}. 

\newpage\section{\texorpdfstring{$t$}--channel dark matter at the LHC}\label{sec:colliders}
\noindent \textit{Contributions from D. Agin, C. Arina, E. Bagnaschi, K. Bai, M.J. Baker, M. Becker, A. Belyaev, F. Benoit, M. Blanke, J.~Burzynski, J.M. Butterworth, A.~Cagnotta, L. Calibbi, L.M. Carpenter, A.S. Cornell, L. Corpe, F. D’Eramo, A. Deandrea, A. Desai, B. Fuks, M.D. Goodsell, J. Harz, J. Heisig, A.O.M. Iorio, D. Kar, S. Kraml, A. Lessa, L. Lopez-Honorez, A. Mariotti, A.~Moreno~Briceño, \mbox{L.~Munoz-Aillaud}, T.~Murphy, A.M.~Ncube, W. Nzuza, L. Panizzi, R. Pedro, C. Prat, L. Rathmann, T. Sangweni, D. Sengupta, W.~Shepherd, A.~Thamm, D. Trischuk}\vspace{.2cm}

This section explores the phenomenology of a wide class of $t$-channel DM models at colliders. We begin, in section~\ref{sec:collider_simpl}, by examining the set of simplified models introduced in section~\ref{sec:model_minimal}. In section~\ref{sec:collider_generalities} we describe the general features of collider signals typical of these $t$-channel simplified models, and explain why a straightforward naive approach is generally insufficient. Additionally, we provide details on leveraging standard Monte Carlo event generators, such as \lstinline{MadGraph5_aMC@NLO}, to simulate these signals by matching fixed-order matrix elements at NLO accuracy in QCD with parton showers. We also discuss the use of public tools to extract current experimental constraints on these models. Next, we consider in section~\ref{sec:collider_1stgen} simplified models where the DM candidate couples to right-handed quarks of the first generation (\ie\ $u_R$ and $d_R$) and present updated results for the existing LHC constraints obtained using state-of-the-art simulations. This analysis is extended in sections~\ref{sec:bounds_2nd} and~\ref{sec:bounds_3rd}, where we investigate scenarios involving couplings to quarks of the second generation (\ie\ $c_R$ and $s_R$) and third generation (\ie\ $t_R$ and $b_R$), respectively. The unique features of the signals are examined in section~\ref{sec:distr}, where we demonstrate again the importance of considering all signal components across different mass spectra. Finally, we briefly discuss leptophilic models in section~\ref{sec:leptoph}.

In section~\ref{sec:collider_nonmin}, we leave the minimal assumption and investigate a few non-minimal setups. In particular, sections~\ref{sec:collider_flavour}, \ref{sec:collider_composite}, \ref{sec:collider_frustrated}, and~\ref{sec:collider_nonabelian} delve into flavoured dark matter constructions, composite $t$-channel DM models, frustrated $t$-channel DM models, and models where the dark sector is linked through a new non-Abelian gauge interaction, respectively. 

We close the discussion on $t$-channel collider phenomenology by considering, in section~\ref{sec:LLP}, scenarios where the mediator is long-lived, highlighting their distinctive signatures.

\subsection{Minimal simplified models -- prompt decays}\label{sec:collider_simpl}
\subsubsection{Generalities about the signal of quark-philic dark matter}\label{sec:collider_generalities}

The phenomenology of the simplified $t$-channel DM models under study must account for a description of the signal kinematics that is as accurate as possible. It should therefore include all contributions. At tree level and for models in which DM couples to quarks, this involves all diagrams leading to the production of any pair of new physics states. Specifically, this includes the production of a pair of dark matter states ($p p \to X X$, with a squared matrix element proportional to $\lambda^4$), the associated production of a DM particle and a mediator ($p p \to X Y + X\bar{Y}$, with a squared matrix element proportional to $\lambda^2\alpha_s$), and the production of a pair of mediator particles or antiparticles ($p p \to Y Y + Y\bar{Y} + \bar{Y}\bar{Y}$). 

In the following, we label the contributions from the first two processes by $XX$ and $XY$ respectively, while for the last process we must distinguish the different contributing components. The pair production of a mediator and an anti-mediator can originate from QCD diagrams (labelled by $Y\bar{Y}_\mathrm{QCD}$, with a squared matrix element proportional to $\alpha_s^2$), $t$-channel DM exchange diagrams (labelled by $Y\bar{Y}_t$ with a squared matrix element proportional to $\lambda^4$), and the corresponding interference (labelled by $Y\bar{Y}_i$, with a matrix element proportional to $\alpha_s\lambda^2$). Conversely, the production of two mediators or two anti-mediators can only be induced by $t$-channel dark matter exchanges in models where the DM state is real, with the corresponding matrix elements being proportional to $\lambda^4$. The total new physics cross section $\sigma_\mathrm{BSM}$ is thus given, with the dependence on the new physics coupling $\lambda$ factorised and as a function of the new physics masses $M_X$ and $M_Y$, by:
\begin{equation}\begin{split}    
  \sigma_\mathrm{BSM} = &\
      \lambda^2\, \sigma_{XY}(M_X,M_Y) + \lambda^4\, \sigma_{XX}(M_X,M_Y)
    + \sigma_{Y\bar{Y}_\mathrm{QCD}}(M_Y) + \lambda^4\, \sigma_{Y\bar{Y}_t}(M_X,M_Y) + \lambda^2\, \sigma_{Y\bar{Y}_i}(M_X,M_Y)\\ &\ 
    + \lambda^4\, \sigma_{YY_t}(M_X,M_Y) + \lambda^4\, \sigma_{\bar{Y}\bar{Y}_t}(M_X,M_Y) \,.
\end{split}\end{equation}
Considering a signal region of an analysis sensitive to the signal, the total efficiency $\varepsilon$ depends not only on the mass spectrum but also on the coupling, as the latter can alter the relative contributions of the various signal components. The fiducial new physics cross section $\hat{\sigma}_\mathrm{BSM}$ corresponding to that region is thus given, once again with the dependence on the new physics masses $M_X$ and $M_Y$ introduced explicitly, by:
\begin{equation}\label{eq:fiducial_sigma}\begin{split}    
  & \hat{\sigma}_\mathrm{BSM} =
      \lambda^2\, {\sigma}_{XY}(M_X,M_Y)\ \varepsilon_{XY}(M_X,M_Y) 
    + \lambda^4\, \sigma_{XX}(M_X,M_Y)\ \varepsilon_{XX}(M_X,M_Y) \\
 &\ + {\sigma}_{Y\bar{Y}_\mathrm{QCD}}(M_Y)\      \varepsilon_{Y\bar{Y}_\mathrm{QCD}}(M_X,M_Y) 
    + \lambda^4\, {\sigma}_{Y\bar{Y}_t}(M_X,M_Y)\ \varepsilon_{Y\bar{Y}_t}(M_X,M_Y) \\
 &\ + \lambda^2\, {\sigma}_{Y\bar{Y}_i}(M_X,M_Y)\ \varepsilon_{Y\bar{Y}_i}(M_X,M_Y) 
    + \lambda^4\, \sigma_{YY_t}(M_X,M_Y)\ \varepsilon_{YY_t}(M_X,M_Y)\\
 &\ + \lambda^4\, \sigma_{\bar{Y}\bar{Y}_t}(M_X,M_Y)\ \varepsilon_{\bar{Y}\bar{Y}_t}(M_X,M_Y) \,.
\end{split}
\end{equation}
In this expression, fiducial cross sections include mediator decays, $Y\to X q$, where $q$ represents the relevant SM quark species for the model considered. In the simulation framework used throughout this work, we always assume a small mediator width, ensuring that the narrow-width approximation (NWA) is valid, allowing mediator production and decay to be considered in a factorised way~\cite{Berdine:2007uv}. Consequently, as shown by the above equation, the kinematics originating from each component to the signal depend solely on the masses $M_X$ and $M_Y$, with the coupling $\lambda$ serving only to globally rescale each contribution.

Specifically, the $XX$ component has a strong dependence on the new physics coupling $\lambda$, to the fourth power, but its modelling requires considering an additional hard visible object in the final state to be detectable. On the other hand, the various contributions involving two mediators or anti-mediators in the final state ($YY$, $Y\bar{Y}$ and $\bar{Y}\bar{Y}$) exhibit different dependencies on $\lambda$. Moreover, these are generally phase-space disfavoured due to the large mediator mass required to satisfy LHC constraints on coloured particles and the fact that we deal with pair production. Eq.~\eqref{eq:fiducial_sigma} also shows that, even in feebly-coupled scenarios, the QCD contribution serves as a baseline component independent of the DM mass. Conversely, other (anti-)mediator pair channels become increasingly relevant with higher $\lambda$ values, and some of them can benefit from parton density enhancements. This is specially relevant for scenarios where DM couples to up or down quarks. In such cases, processes like $uu\to YY$ and $dd\to YY$ can involve a pair of initial valence quarks. Finally, $XY$ production has a weaker $\lambda$ dependence, only to the second power, but is phase-space favoured owing to the production of a lighter DM state. Consequently, there is no clear model-independent hierarchy between the different contributions, necessitating their inclusion in signal simulations.

In this work, hard-scattering simulations are performed with \lstinline{MadGraph5_aMC@NLO}~\cite{Alwall:2014hca} (version 2.9.18) using the \lstinline{MadSTR} plugin~\cite{Frixione:2019fxg}\footnote{The \lstinline{MadSTR} plugin can be downloaded from \url{https://code.launchpad.net/~maddevelopers/mg5amcnlo/MadSTRPlugin}.} to handle resonant contributions appearing at NLO, where an intermediate mediator is produced on-shell and subsequently decays into a DM state and a SM quark. This approach ensures that such contributions are not double-counted across the three new physics processes $XX$, $XY$, and $YY$. For instance, at NLO, the process $pp \to XY \to X (Xj)$ includes configurations where the intermediate mediator is on-shell (to be included in $XY$ simulations) as well as off-shell contributions (to be included in the real-emission corrections to $XX$ production). Furthermore, LO and NLO matrix elements are determined (with \lstinline{MadGraph5_aMC@NLO}) using the tree-level Feynman rules, $R_2$ contributions, and counterterms relevant to the models introduced in section~\ref{sec:models} and encoded in the UFO format~\cite{Degrande:2011ua, Darme:2023jdn} as detailed above. These matrix elements are convolved with the LO and NLO sets of NNPDF4.0 parton densities~\cite{NNPDF:2021njg, Buckley:2014ana}, respectively. Additionally, decays of heavy unstable particles are handled with \lstinline{MadSpin}~\cite{Artoisenet:2012st} and \lstinline{MadWidth}~\cite{Alwall:2014bza} to retain off-shell propagation and spin correlation effects. Finally, parton showering (PS) and hadronisation effects are simulated with \lstinline{Pythia} (version 8.306)~\cite{Bierlich:2022pfr} and matched with fixed-order calculations according to the MC@NLO procedure\cite{Frixione:2002ik}.

In practice, the event generation procedure is divided into dedicated simulation runs, each associated with a specific component of the new physics signal. Except for the QCD production of a pair of mediator particles and antiparticles ($Y\bar{Y}_\mathrm{QCD}$) and its interference with the $t$-channel diagrams ($Y\bar{Y}_i$), all simulations rely on the \lstinline{MadSTR} plugin. To start the \lstinline{MadGraph5_aMC@NLO} package with \lstinline{MadSTR}, we need to type the following command in a shell:
\begin{lstlisting}
  <MG folder>/bin/mg5_aMC --mode=MadSTR
\end{lstlisting}
Here, \lstinline{<MG folder>} represents the directory where \lstinline{MadGraph5_aMC@NLO} is installed. We then define the new physics process in the command line interface of \lstinline{MadGraph5_aMC@NLO} by typing:
\begin{lstlisting}[mathescape]
  import model DMSimpt_v2_0-<restriction> --modelname
  define xx = <X state>
  define yy = <Y state>
  define yy~ = <Ybar state>
  define yyy = yy yy$\sim$
  <generate command>
\end{lstlisting}
In this snippet of script, \lstinline{<restriction>} specifies the model restriction considered (see section~\ref{sec:models}), and generic labels are introduced for the dark matter (\lstinline{xx}) and mediator states (\lstinline{yy}, \lstinline{yy~}, and \lstinline{yyy}). The placeholders \lstinline{<X state>}, \lstinline{<Y state>}, and \lstinline{<Ybar state>} refer to the \lstinline{FeynRules} names of the relevant dark matter state and mediator particles and antiparticles, as listed in Tables~\ref{tab:info_S}, \ref{tab:info_F}, and \ref{tab:info_V}. 

The event generation commands for the $XX$, $XY$, $Y\bar{Y}_t$, $YY_t$, and $\bar{Y}\bar{Y}_t$ processes are as follows:
\begin{lstlisting}[mathescape]
  generate p p > xx xx / <excluded states> [QCD]
  generate p p > xx yyy / <excluded states> [QCD]
  generate p p > yy yy$\sim$ DMT=2 QCD=0 QED=0 / <excluded states> [QCD]
  generate p p > yy yy DMT=2 QCD=0 QED=0 / <excluded states> [QCD]
  generate p p > yy$\sim$ yy$\sim$ DMT=2 QCD=0 QED=0 / <excluded states> [QCD]
\end{lstlisting}
Here, \lstinline{<excluded states>} is a sequence listing all particles irrelevant to the chosen model restriction. We emphasise once again that it is important to address these production modes individually, as their relative importance depends on the model (particularly, the $YY$ and $\bar{Y}\bar{Y}$ modes are relevant only for setups with self-conjugate dark matter) and the benchmark scenario chosen (\ie\ the values of the mass and coupling parameters). Event generation is further performed normally, as stated in the \lstinline{MadSTR} documentation~\cite{Frixione:2019fxg}. Technically, we remove all resonant diagram contributions squared potentially arising at NLO, which corresponds to the \lstinline{istr=2} option of the \lstinline{MadSTR} configuration that we set in the \lstinline{run_card.dat} configuration file of \lstinline{MadGraph5_aMC@NLO}. 

Additionally, event generation for QCD-induced mediator pair production ($Y\bar{Y}_\mathrm{QCD}$) is performed by starting \lstinline{MadGraph5_aMC@NLO} normally, without using the \lstinline{MadSTR} plugin as there is no resonant diagrams appearing at NLO. This is achieved by typing the following command:
\begin{lstlisting}[mathescape]
  generate p p > yy yy$\sim$ / <excluded states> [QCD]
\end{lstlisting}
Furthermore, the interference between the QCD diagrams and the $t$-channel ones for $pp\to Y\bar{Y}$ is performed at LO with the command:
\begin{lstlisting}[mathescape]
  generate p p > yy yy$\sim$ DMT^2==2 / <excluded states>
\end{lstlisting}
As it is technically not possible to handle event generation for this contribution at NLO, the corresponding predictions are rescaled with a constant $K$-factor $K_{Y\bar{Y}_i} $. This $K$-factor is defined as the geometric mean of the $Y\bar{Y}_\mathrm{QCD}$ and $Y\bar{Y}_t$ $K$-factors,
\begin{equation}
  K_{Y\bar{Y}_i} 
    \equiv \sqrt{K_{Y\bar{Y}_t}\ K_{Y\bar{Y}_\mathrm{QCD}}} = \sqrt{ \frac{\hat{\sigma}_{Y\bar{Y}_t}^{\rm NLO}}{\hat{\sigma}_{Y\bar{Y}_t}^{\rm LO}}\ \frac{\hat{\sigma}_{Y\bar{Y}_\mathrm{QCD}}^{\rm NLO}}{\hat{\sigma}_{Y\bar{Y}_\mathrm{QCD}}^{\rm LO}} } \,.
\end{equation}
where $\hat{\sigma}^{\rm LO}$ and $\hat{\sigma}^{\rm NLO}$ respectively refer to cross sections evaluated at the LO and NLO accuracy in QCD.

For correct pole cancellation between the virtual and real emission contributions appearing at NLO in the above processes, it is crucial to allow non-coloured DM states to run in virtual diagrams. By default, this is forbidden in the default settings and internal mechanisms of \lstinline{MadGraph5_aMC@NLO}. To fix this, a few core files of the code need to be modified, as detailed in appendix~A of \cite{Borschensky:2021hbo}. First, we add the following lines in the function \lstinline{is_perturbating} implemented in the file \lstinline{base_objects.py}:\footnote{The exact location of this function in the files generated by \lstinline{MadGraph5_aMC@NLO} depends on the version of the program.}
\begin{lstlisting}
  is_dm = abs(self.get('pdg_code')) in [51,52,53,56,57,58]
  if order in int.get('orders').keys() and is_dm:
      return True
\end{lstlisting}
Next, we modify the function \lstinline{user_filter} in the file \lstinline{loop_diagram_generation.py} as follows,
\begin{lstlisting}
  dm_ids = [51,52,53,56,57,58]
  loop_pdgs = [abs(x) for x in diag.get_loop_lines_pdgs()]
  is_loop_dm = (len([x for x in loop_pdgs if x in dm_ids])>0)
  is_loop_gluon = (21 in loop_pdgs)
  if is_loop_dm and not is_loop_gluon:
      valid_diag=False
  connected = diag.get_pdgs_attached_to_loop(structs)
  isnot_dmcorrection = [x for x in connected if not abs(x) in dm_ids]
  if not len(isnot_dmcorrection)>0:
      valid_diag=False
\end{lstlisting}

As mentioned above, mediator decays into a dark matter and a quark are handled using \lstinline{MadSpin} and \lstinline{MadWidth}. This is achieved by updating the \lstinline{MadSpin} card at runtime, specifying explicitly that the mediator must decay through the only open channel,
\begin{lstlisting}
  set max_weight_ps_point 400
  decay <Y state> > <X state> <q state> 
  decay <Ybar state> > <X state> <qbar state>
  launch
\end{lstlisting}
As before, the placeholders  \lstinline{<Y state>}, \lstinline{<Ybar state>}, \lstinline{<X state>} and \lstinline{<q state>} refer to the model's labels for the mediator, anti-mediator, DM particle and the relevant SM quark for the considered model restriction (see Tables~\ref{tab:info_S}, \ref{tab:info_F} and \ref{tab:info_V}). In the case of the interference contribution between the $t$-channel and the QCD diagrams, the command \lstinline{set spinmode none} must be additionally included. Parton showering and hadronisation are then performed using the default \lstinline{Pythia} cards generated by \lstinline{MadGraph5_aMC@NLO}.

Constraints on the new physics signal emerging from the considered models can be obtained by reinterpreting the results of various experimental searches for new physics signatures comprising missing transverse energy ($\met$) and jets. This procedure, in which constraints from existing LHC analyses are derived for new models not originally considered in experimental publications by implementing the analysis logic in some \textit{ad-hoc} code, is commonly referred to as \textit{recasting}.\footnote{This term will be used throughout this report.} Specifically, we focus on recent exclusive LHC Run~2 searches that impose stringent requirements on a small number of jets and more inclusive searches with looser requirements enforcing the presence of a larger number of jets: ATLAS-EXOT-2018-06~\cite{ATLAS:2021kxv}, ATLAS-SUSY-2018-17~\cite{ATLAS:2020xgt}, ATLAS-CONF-2019-040~\cite{ATLAS:2019vcq}, CMS-SUS-19-006~\cite{CMS:2019zmd} and CMS-EXO-20-004~\cite{CMS:2021far}. The selection criteria defining these analyses have been designed to observe a signal of new physics characterised by a substantial amount of missing transverse energy, energetic jets, and no leptons. While the cuts across the different analyses are largely similar, their differences define signal regions yielding varying sensitivities depending on the signal details. In other words, for a given new physics theoretical framework, the various searches are expected to be sensitive in different parts of the model’s parameter space. For instance, the requirements on the number of jet candidates in the final state  ($N_j$) along with the definition of such jet candidates in terms of pseudo-rapidity ($\eta(j)$) and transverse momentum ($p_T(j)$) slightly differ, as illustrated by table~\ref{tab:ana_defs} where we summarise the \textit{main} requirements included in the different analyses\footnote{We refer to the relevant publications for the exact definition of all signal regions of the considered analyses. In particular, different regions of a given analysis may involve different cut thresholds on a given observable, table~\ref{tab:ana_defs} indicating in this case the typical softest threshold.}.

\begin{table}
  \centering\renewcommand{\arraystretch}{1.3}\setlength{\tabcolsep}{12pt}
  \resizebox{.96\textwidth}{!}{%
    \begin{tabular}{l | c c c c c}
      \multirow{2}{*}{Cuts} & ATLAS & ATLAS & ATLAS & CMS & CMS\\
      & EXOT-2018-06 & SUSY-2018-17 & CONF-2019-040 & SUS-19-006 & EXO-20-004 \\
      \hline
      $N_j$ & $\in [1, 4]$& $\in [8, 12]$ & $\geq 2$ & $\geq 2$ &  $\geq 1$\\
      $|\eta(j)|$ & $<2.8$ & $<2.0$ & $<2.8$ & $<2.4$ & $<2.4$ \\
      $p_T(j_1)$ & $>150$~GeV & \multirow{2}{*}{$>50$~GeV} & $>200$~GeV & \multirow{2}{*}{$>30$~GeV} & $>100$~GeV  \\
      $p_T(j_2,...,j_{N_j})$ & $>30$~GeV &  & $>50$~GeV & & $>20$~GeV\\
      \hline
      $\met$ & $>200$~GeV & $-$ & $>300$~GeV & $-$ & $>250$~GeV \\
      $\mht$ & $-$ & $-$ & $-$ & $>300$~GeV & $-$ \\
      $\met/\sqrt{H_T}$ & $-$ & $>5~\sqrt{\mathrm{GeV}}$ & $>10~\sqrt{\mathrm{GeV}}$ & $-$ & $-$ \\
      $\Delta \Phi(j_i, \ptmiss)$ & $>0.4$ & $-$ & $>0.2$& $>0.3$ & $>0.5$ \\
      \hline
      $m_{\text{eff}}$ & $-$ & $-$ & $>800$~GeV & $-$ & $-$ \\
         $H_T$ & $-$ & $-$ & $-$ & $>300$~GeV & $-$ \\
    \end{tabular}}
  \caption{Summary of the typical event selection cuts included in the five jets+$\met$ analyses considered. Detailed information can be found in the experimental publications~\cite{ATLAS:2021kxv, ATLAS:2020xgt, ATLAS:2019vcq, CMS:2019zmd, CMS:2021far}.}
  \label{tab:ana_defs}
\end{table}

The CMS-EXO-20-004 analysis is hence sensitive to a softer monojet-like signature compared to the ATLAS-EXOT-2018-06 or ATLAS-CONF-2019-040 analyses by virtue of a milder cut on the $p_T$ of the leading jet, thus offering complementary sensitivity to signals with less hadronic activity. Additionally, all analyses involve different cuts on the missing transverse momentum (or the missing hadronic activity $\mht$, which is the norm of the vector sum of the transverse momenta $\vec{p}_T$ of all jets in the event, in the case of CMS-SUS-19-006) and its properties, such as its separation from the jets or its significance $\met/\sqrt{H_T}$ (with $H_T$ being defined below). For instance, the ATLAS-EXOT-2018-06 analysis features the smallest threshold on the $\met$ requirement, albeit with a rather hard selection on the leading jet transverse momentum. In addition, the more inclusive searches include cuts on global observables like the effective mass $m_{\text{eff}}$ or the hadronic activity $H_T$ defined by 
\begin{equation}
    m_{\text{eff}} = \met + \sum_j p_T(j)\qquad\text{and}\qquad
    H_T = \sum_j p_T(j)\,.
\end{equation}

We derive constraints on the quark-philic $t$-channel models introduced in section~\ref{sec:models} by relying on the implementation of the above analyses in the \lstinline{MadAnalysis 5} framework~\cite{Conte:2012fm, Conte:2014zja, Conte:2018vmg} (version 1.10.12), which uses \lstinline{FastJet}~\cite{Cacciari:2011ma} (version 3.3.4) and its implementation of the anti-$k_T$ algorithm~\cite{Cacciari:2008gp}, as well as \lstinline{Delphes 3}~\cite{deFavereau:2013fsa} (version 3.5.0) and the \lstinline{SFS} framework~\cite{Araz:2020lnp} for the simulation of the LHC detectors. Details regarding the integration of these implementation into \lstinline{MadAnalysis 5}, along with corresponding validation notes\footnote{The implementation of the ATLAS-EXOT-2018-06 search in \lstinline{MadAnalysis 5} has been carried out in the context of the work done for the present whitepaper, and has therefore not been documented in any peer-reviewed publication. Validation details are consequently provided in appendix~\ref{app:atlas_exot_2018_06}.}, can be found on the public analysis database of \lstinline{MadAnalysis 5}~\cite{Dumont:2014tja}, as well as in~\cite{Araz:2019otb, Kim:2020nrg, Mrowietz:2020ztq, CMS:2021far, Fuks:2021zbm, Carpenter:2021vga}.\footnote{The different codes can be obtained from the \lstinline{MadAnalysis 5} dataverse~\cite{DVN/NW3NPG_2021, DVN/REPAMM_2023, DVN/4DEJQM_2020, DVN/IRF7ZL_2021, DVN/I2CZWU_2021}.}

Bounds on new physics can also be obtained by means of detector-corrected particle-level measurements at the LHC instead of detector-level searches, which thus does not require an approximation of the detector response. \lstinline{Contur}~\cite{Butterworth:2016sqg,Buckley:2021neu} is currently the only tool which uses this, exploiting that measurement publications from the collaborations systematically provide associated \lstinline{Rivet}~\cite{Bierlich:2024vqo} routines. The library of reusable measurements is thus always growing organically, without the need for further work from the phenomenological community to exploit the results. The workflow for \lstinline{Contur} starts with a UFO file~\cite{Degrande:2011ua, Darme:2023jdn} which encodes a chosen new physics model, that is then passed to an event generator to produce \lstinline{HEPMC}~\cite{Verbytskyi:2020sus} files. These events are next analysed with \lstinline{Rivet}, which outputs a set of histograms displaying where the signal events would have shown up in the bank of LHC measurements having \lstinline{Rivet} routines available. Finally, one can stack the predicted signal (properly scaled relative to cross sections) on top of the predictions from the SM, and compare to the observed data in each measurement to derive exclusions, without the need for smearing since everything is done at particle-level. In practice, measurements are grouped into orthogonal pools (defined by final state, experiment and centre-of-mass energy) to avoid double counting: only the best exclusion from a given pool is conserved. In the context of the $t$-channel DM models discussed in this work, the relevant pools of measurements focus on the $\ell^+\ell^-\gamma$ final state~\cite{ATLAS:2019gey}, the $\met+\mathrm{jet}$ final state~\cite{ATLAS:2024vqf, ATLAS:2017txd}, the hadronic $t\bar{t}$ final state~\cite{ATLAS:2022mlu} and the $\ell+\met+\mathrm{jet}$ final state~\cite{CMS:2018htd, CMS:2016oae}. 

\subsubsection{First generation simplified models}\label{sec:collider_1stgen}
We begin our phenomenological analysis of DM $t$-channel models at colliders by exploring the LHC constraints that can be imposed on models belonging to the \lstinline{XYZ_uR} and \lstinline{XYZ_dR} classes. We hence extend the studies of~\cite{Arina:2020tuw, Arina:2023msd} to scenarios where DM couples to any first-generation right-handed quark, and we additionally incorporate reinterpretations of a broader set of LHC analyses. Given that complex dark matter models are disfavoured by cosmological observations, at least when DM couplings to light generations of SM fermions are considered (see~\cite{Arina:2023msd} and section~\ref{sec:CosmConstMinQuark}), we focus exclusively on scenarios involving a real, self-conjugate DM particle (\ie\ models of type \lstinline{F3S}, \lstinline{F3V}, and \lstinline{S3M}). The collider analyses that we recast are discussed in section~\ref{sec:collider_generalities}, and include ATLAS-EXOT-2018-06, ATLAS-SUSY-2018-17, ATLAS-CONF-2019-040, CMS-SUS-19-006, and CMS-EXO-20-004.

\begin{figure}
  \centering
  \includegraphics[width=.48\textwidth]{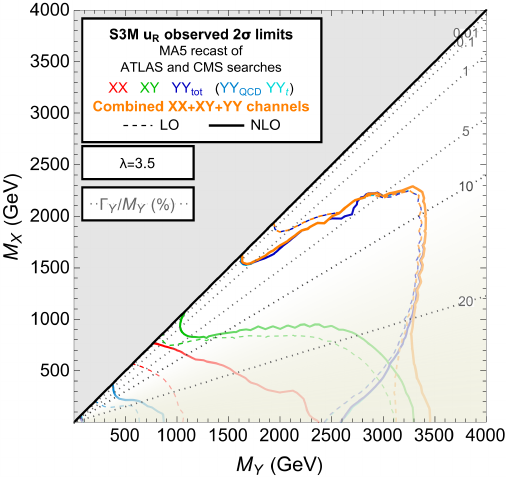}\hfill
  \includegraphics[width=.48\textwidth]{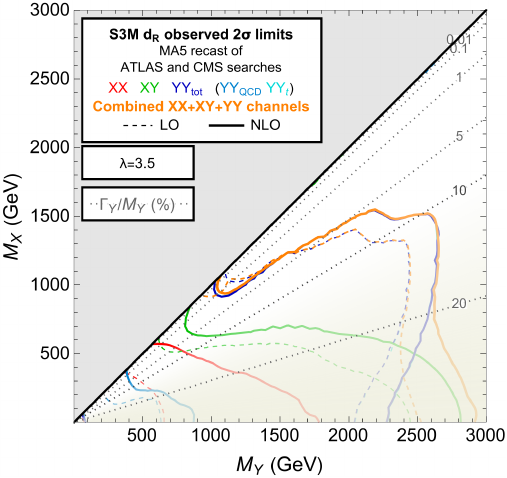}\\ \vspace{.3cm}
  \includegraphics[width=.48\textwidth]{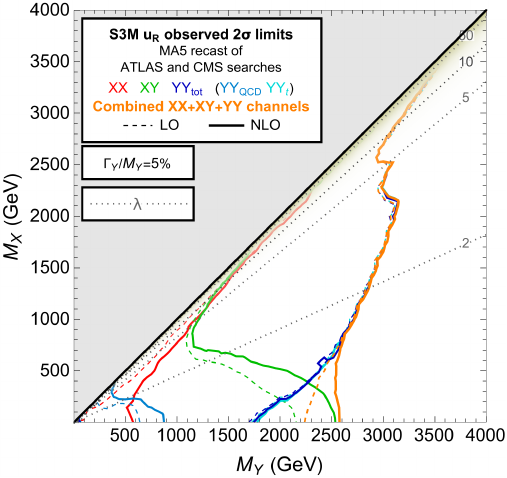}\hfill
  \includegraphics[width=.48\textwidth]{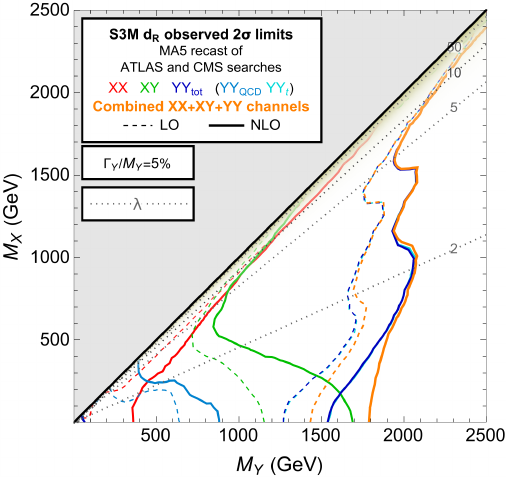}
  \caption{Exclusion limits at 95\% confidence level from the reinterpretation of several Run~2 ATLAS and CMS searches~\cite{ATLAS:2021kxv, ATLAS:2020xgt, ATLAS:2019vcq, CMS:2019zmd, CMS:2021far}. The results are shown for the \lstinline{S3M_uR} (left) and \lstinline{S3M_dR} (right) real dark matter scenarios described in section~\ref{sec:model_minimal}, considering two configurations: $\lambda = 3.5$ (top row) and $\Gamma_Y / M_Y = 0.05$ (bottom row). For scenarios with $\lambda = 3.5$, dotted grey lines represent isolines of constant $\Gamma_Y / M_Y$ value. Conversely, for scenarios with $\Gamma_Y / M_Y = 0.05$, these lines correspond to isolines of fixed $\lambda$ value. Individual contributions to the bounds are displayed for processes $XX$ (red), $XY$ (green), and $YY$ (dark blue), with the $YY$ process further decomposed into its purely QCD part ($YY_{\rm QCD}$, teal) and its $t$-channel part ($YY_t$, turquoise). The yellow gradient highlights regimes where either the perturbative approach becomes increasingly invalid due to large coupling values or the narrow-width approximation loses validity due to a large mediator width-to-mass ratio.}
  \label{fig:1stgen}
\end{figure}

We first present the bounds obtained for various choices of the \lstinline{S3M_uR} and \lstinline{S3M_dR} model's free parameters in figure~\ref{fig:1stgen}. These scenarios are supersymmetry-inspired, featuring a scalar mediator in the fundamental representation of $SU(3)_C$ (analogous to a right-handed up or down squark) and a fermionic Majorana DM particle (resembling a neutralino). The results are shown in the mediator and DM mass plane, $(M_Y, M_X)$, for several different setups. The top row of the figure corresponds to a fixed coupling value of $\lambda = 3.5$, such a large value being motivated by cosmological considerations (see section~\ref{sec:cosmology}), while the bottom row assumes a mediator width-to-mass ratio of $\Gamma_Y / M_Y = 0.05$, which guarantees that the narrow-width-approximation is valid (a necessary condition for using \lstinline{MadSpin} for mediator decays). Exclusion limits at the 95\% confidence level are displayed as solid and dashed orange lines for NLO and LO simulations respectively. The left panel of the figure shows results for scenarios where DM couples to the first-generation right-handed up quark $u_R$, while its right panel displays exclusions for DM couplings to the first-generation right-handed down quark $d_R$. Additionally, grey dotted isolines are included to represent constant $\lambda$ values (in the case where $\Gamma_Y / M_Y = 0.05$) or constant $\Gamma_Y / M_Y$ values (for fixed $\lambda = 3.5$). Regions of the parameter space where $\lambda$ becomes too large for a perturbative treatment (although our calculations are only at first-order in $\lambda$) or where the size of $\Gamma_Y / M_Y$ challenges the validity of the narrow-width approximation are shaded to indicate theoretical and/or technical limitations.

In all the figures, mass configurations to the left of the solid (dashed) orange exclusion line are ruled out at the 95\% confidence level, based on state-of-the-art simulations at NLO+PS (LO+PS). For scenarios with $\lambda = 3.5$ (top row), the mediator mass is constrained to be larger than approximately $3.2-3.5$~TeV for up-like setups and $2.5-2.7$~TeV for down-like setups, provided that the spectrum compression, defined as $r = M_Y / M_X - 1$, is greater than $0.6-0.7$. This corresponds to an LHC sensitivity to DM masses $M_X$ smaller than approximately $2$~TeV and $1.5$~TeV for the up-like and down-like cases, respectively. The loss in sensitivity for decreasing $r$ values (\ie\ scenarios closer to the diagonal in the figures) has a twofold origin. First, the $XX$ contribution (red contours), despite being enhanced by the large $\lambda^4$ factor as shown in \eqref{eq:fiducial_sigma}, is only significant for very light new physics spectra. Consequently, for heavier mass spectra, such as those close to the current bounds, mediator production (both pair and associated) and subsequent decay drives the constraints. The final-state jets originating from mediator decays thus become softer for increasingly compressed spectra, leading to a reduction in signal selection efficiencies and consequently weaker bounds. Specifically, our predictions show that the LHC currently has no sensitivity to models where the DM mass $M_X \gtrsim 2.2$~TeV for up-like setups, and $M_X \gtrsim 1.5$~TeV for down-like setups. 

A key feature emerging from our results is that, in general, the LHC sensitivity is primarily driven by mediator pair production ($YY$), as indicated by the near-overlap of the blue and orange exclusion lines. Only in scenarios with lighter DM masses (and heavy mediator mass typical in the vicinity of the current bounds) does the associated production channel ($XY$) begin to contribute significantly, as shown by the green contours. This contribution arises from phase-space enhancements relevant for light dark matter, combined with the large imposed value of the new physics coupling. This large coupling value $\lambda = 3.5$ also induces a hierarchy among the different components of the $YY$ channel. Contributions from QCD diagrams are negligible, as evidenced by the teal contours, which exclude only a small portion of the parameter space. In contrast, $t$-channel diagrams, with their $\lambda^4$ dependence and the existence of a $q_v q_v\to YY$ sub-process driven by two valence quarks, dominate and yield exclusions equivalent to those of the full $YY$ channel. However, caution is required for scenarios with a split spectrum, or equivalently featuring large $r$ values such as $r > 2$, where the mediator width-to-mass ratio $\Gamma_Y / M_Y$ generally exceeds 20\%. Such a high ratio signals an ill-defined new physics setup, and challenges the validity of calculations within the narrow-width approximation.

In the bottom row of the figure, we address this last issue by considering a second class of scenarios. Here, the two new physics masses $M_X$ and $M_Y$ remain free parameters, but the coupling $\lambda$ is dynamically computed to enforce a fixed mediator width-to-mass ratio of $\Gamma_Y / M_Y = 0.05$. Consequently, this approach eliminates any issues related to the use of \lstinline{MadSpin} for modelling mediator decays. The resulting exclusion contours exhibit a distinct shape compared to the case where $\lambda$ was fixed to $3.5$. For scenarios where the mediator width-to-mass ratio is fixed, the exclusion bounds (orange line) are nearly independent of the dark matter mass, reaching $2.5-3$~TeV and $1.8-2.1$~TeV in the \lstinline{S3M_uR} and \lstinline{S3M_dR} cases, respectively. As before, the exclusion is primarily driven by the $YY$ channel, except for scenarios involving lighter dark matter masses and a heavy mediator, where associated $XY$ production begins to contribute. This behaviour is consistent with the dependence of the mediator pair production rates on the DM mass, as detailed in the analysis of related matrix elements and cross sections in \cite{Arina:2023msd}. Interestingly, the bounds become stronger for increasingly compressed spectra, but this trend arises from the progressively larger values of $\lambda$ required to maintain $\Gamma_Y / M_Y = 0.05$. In such regions of the parameter space, the perturbativity assumption underlying the entire calculation no longer holds so that the exclusion bounds should be interpreted with caution.

Another notable feature of our results is the impact of higher-order QCD corrections on the exclusion bounds. As outlined in \cite{Arina:2020udz}, NLO QCD corrections influence not only the overall rate of the new physics signal (\ie\ the combined contributions from the $XX$, $XY$, and $YY$ channels) but also significantly alter the shapes of key observables, such as the missing transverse momentum $\ptmiss$ and the missing energy $\met$ used in all LHC analyses considered (see also section~\ref{sec:distr}). As a result, the exclusion contours at LO and NLO are not merely related by a simple translation of each other, but exhibit more complex differences. This is evident in figure~\ref{fig:1stgen} for all scenarios considered, where the LO bounds (dashed lines) and NLO bounds (solid lines) can be compared. For \lstinline{S3M_uR} models, global NLO effects are generally mild, except for scenarios involving a heavy mediator ($M_Y \in [2.5, 3.5]$~TeV) and a light dark matter particle ($M_X \lesssim 1$~TeV). In this region of parameter space and for decreasing $M_X$ values, the $XY$ channel begins to contribute significantly, and it turns out that it exhibits a strong sensitivity to NLO corrections due to the strong coupling already entering in the LO matrix elements. Moreover, while NLO contributions are highly relevant for the $YY_\mathrm{QCD}$ and $XX$ channels, these channels contribute negligibly given the mass ranges probed by current LHC analyses. Conversely, QCD corrections to the dominant $YY_t$ channel are mild, except for setups with light DM masses, where the dependence of the matrix elements on the DM mass becomes non-trivial~\cite{Arina:2023msd}. Consequently, NLO contributions influence only a specific portion of the parameter space for \lstinline{S3M_uR} models. The situation is markedly different for \lstinline{S3M_dR} models. Here, the distinct parton distribution functions (PDFs) involved in the $YY_t$ and $XY$ matrix elements lead to much stronger NLO effects for these channels, resulting in exclusion bounds that are significantly more sensitive to higher-order corrections than in the \lstinline{S3M_uR} case. This is clearly illustrated in the figures, where shifts in the mediator mass limits approach $500$~GeV for a fixed DM mass below approximately $1$~TeV.

As previously outlined, significant differences are observed between the bounds applicable on \lstinline{S3M_uR} and \lstinline{S3M_dR} models, despite the overall similarity in the shapes of their exclusion contours. The primary distinction lies in the mass range covered. In the region of parameter space where the current bounds reside, the signal is in both cases predominantly driven by contributions from the $YY_t$ and $XY$ channels. These two channels are directly influenced by the flavour of the initial-state quark, with the subprocesses $uu \to YY$ and $ug \to XY$ dominating in the \lstinline{S3M_uR} case, and $dd \to YY$ and $dg \to XY$ dominating in the \lstinline{S3M_dR} case. These subprocesses are strongly affected by partonic luminosities and potential enhancements from valence quarks. Since the up-quark content in the proton is substantially higher than the down-quark content, the bounds on \lstinline{S3M_dR} models are consequently weaker by approximately $500-700$~GeV for a given dark matter mass.

\begin{figure}
  \centering
  \includegraphics[width=.48\textwidth]{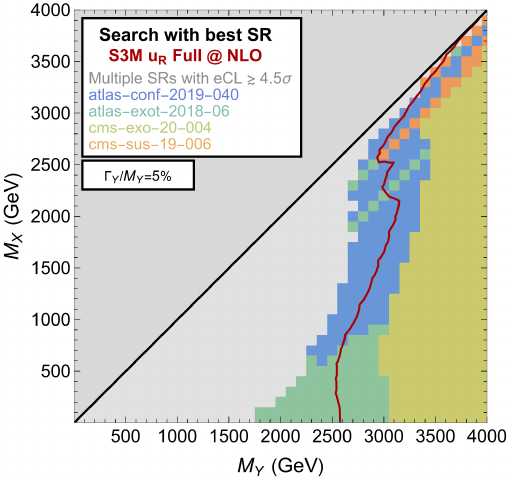}\hfill
  \includegraphics[width=.48\textwidth]{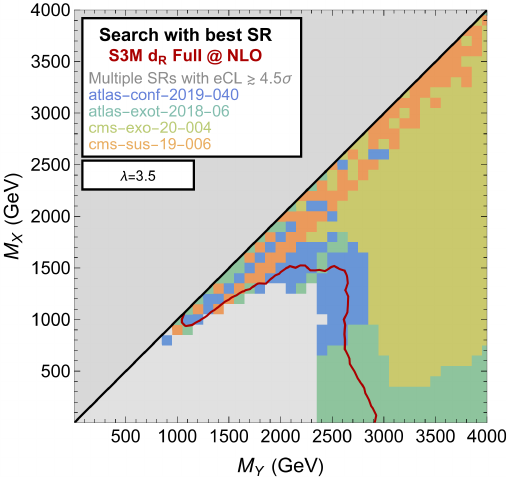}
  \caption{Search including the most sensitive signal region for the new physics signal originating from the combination of all processes at NLO for \lstinline{S3M_uR} models with a fixed width/mass ratio (left), and \lstinline{S3M_dR} models with a fixed coupling (right). \label{fig:BestSR_S3M_Full_NLO_ud}}
\end{figure}

In figure~\ref{fig:BestSR_S3M_Full_NLO_ud}, we explore the dependence of the total exclusion for each mass point, accounting for all signal contributions evaluated at NLO, on the analysis driving it. Two illustrative setups are considered: \lstinline{S3M_uR} benchmark scenarios featuring a fixed mediator width-to-mass ratio $\Gamma_Y / M_Y = 0.05$ (left panel), and \lstinline{S3M_dR} scenarios characterised by a fixed new physics coupling value $\lambda = 3.5$ (right panel). Scenarios excluded by signal regions from more than one analysis, with exclusion levels equivalent to $4.5$ standard deviations or greater, are displayed in grey. These scenarios typically exhibit a much lighter new particle spectrum compared to the current exclusion limits, represented by the solid red line. For scenarios near or beyond these limits, we use a colour code to indicate the most sensitive analysis among those considered. The figure highlights the strengths of leveraging multiple LHC analyses, each providing an optimal sensitivity to different mass configurations.

The CMS monojet analysis (CMS-EXO-20-004) is particularly sensitive to split spectra, where the dark matter state is at least a few hundred GeV lighter than the mediator. These configurations are shown in yellow in the figure. Additionally, the ATLAS-EXOT-2018-06 analysis, which has a similar selection (as depicted in table~\ref{tab:ana_defs}), is equally sensitive to such configurations and is, in some cases, even more constraining (the relevant spectrum configuration being shown in green). Notably, these two analyses together provide the best sensitivity to heavier first-generation \lstinline{S3M} scenarios compare to existing bounds, making them strong candidates for follow-up investigations during the LHC Run~3 and its high-luminosity phase. Moreover, the more inclusive ATLAS-CONF-2019-040 analysis proves to be the most sensitive for scenarios where both the mediator and the dark matter are relatively heavy, but with a significant mass gap, as indicated by the blue regions in the figure. Conversely, the highly inclusive CMS-SUS-19-006 analysis demonstrates strong sensitivity to compressed spectra, as highlighted by the orange regions displayed in the figure. The combined sensitivity of these four analyses helps to clarify the shape and sharp features of the exclusion contour that had been found in figure~\ref{fig:1stgen}. Importantly, it is known that some signal regions across these analyses are uncorrelated in light of the targeted signal, despite all these analyses targeting a final state comprising jets and missing transverse momentum~\cite{Araz:2022vtr, Feike:2024zfz}. There thus exists some potential for their combination, as demonstrated by the \lstinline{TACO} approach introduced in~\cite{Araz:2022vtr} and applied to a supersymmetric scenario resembling the \lstinline{S3M} model in~\cite{Feike:2024zfz}, or the approach of \cite{MahdiAltakach:2023bdn, Altakach:2023tsd}. A comprehensive reassessment of the current exclusion via analysis combination is, however, left for future work.

\begin{figure}
  \centering
  \includegraphics[width=.48\textwidth]{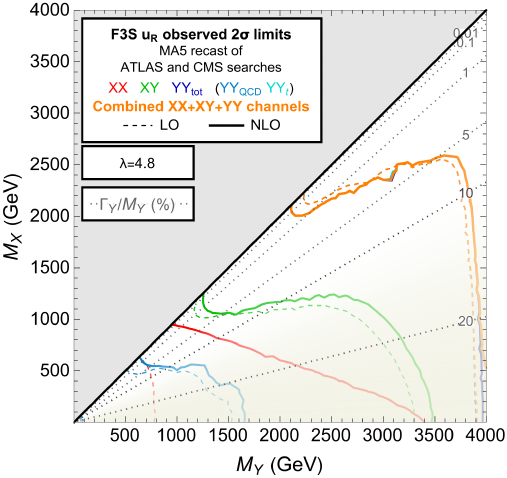}\hfill
  \includegraphics[width=.48\textwidth]{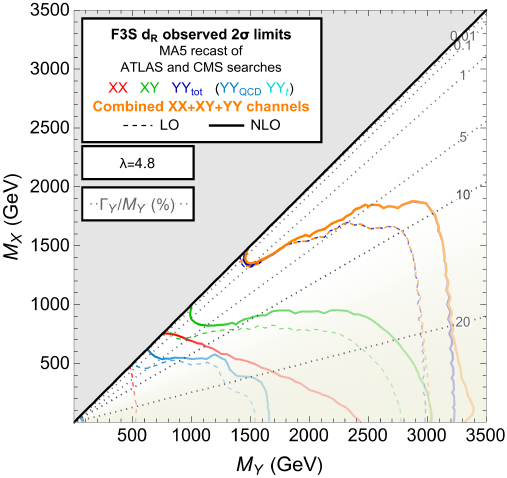}\\ \vspace{.3cm}
  \includegraphics[width=.48\textwidth]{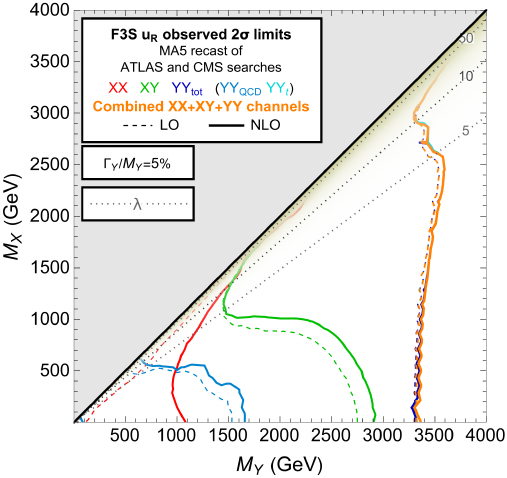}\hfill
  \includegraphics[width=.48\textwidth]{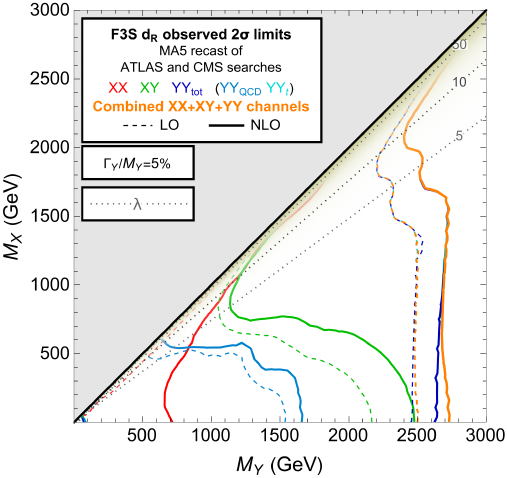}
  \caption{Same as in figure~\ref{fig:1stgen}, but for the \lstinline{F3S_uR} (left) and \lstinline{F3S_dR} (right) real dark matter scenarios. For scenarios with fixed $\lambda$ values, we adopt $\lambda = 4.8$.}
  \label{fig:1stgenbis}
\end{figure}

\begin{figure}
  \centering
  \includegraphics[width=.48\textwidth]{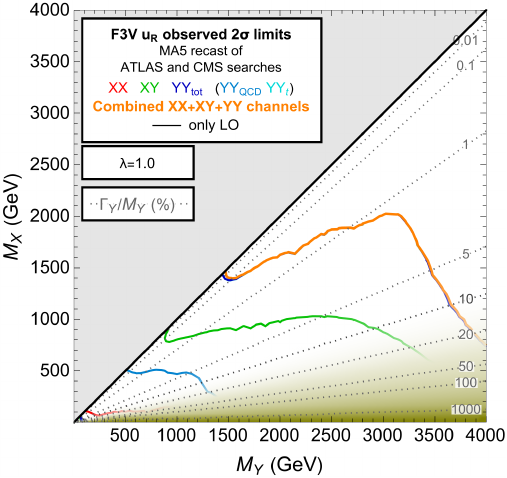}\hfill
  \includegraphics[width=.48\textwidth]{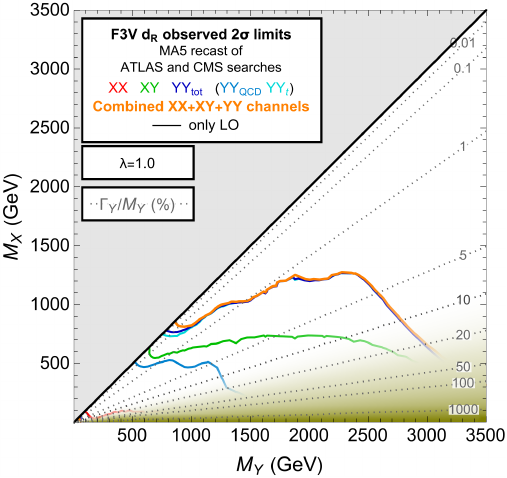}\\ \vspace{.3cm}
  \includegraphics[width=.48\textwidth]{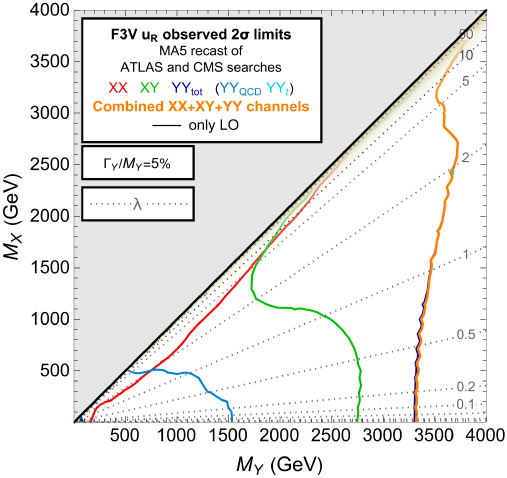}\hfill
  \includegraphics[width=.48\textwidth]{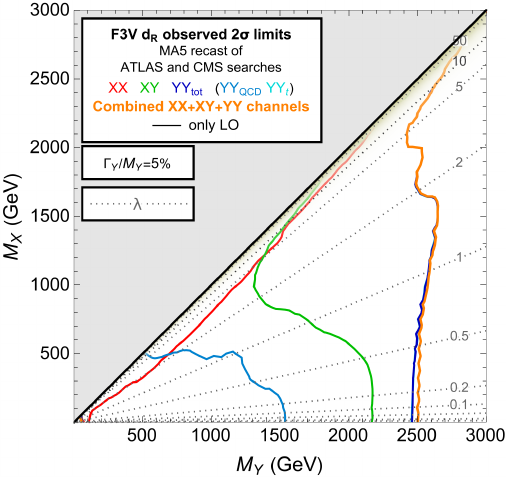}
  \caption{Same as in figure~\ref{fig:1stgen}, but for the \lstinline{F3V_uR} (left) and \lstinline{F3V_dR} (right) real dark matter scenarios. For scenarios with fixed $\lambda$ values, we adopt $\lambda = 1$. Moreover, results are given at LO only.}
  \label{fig:1stgenter}
\end{figure}

In figure~\ref{fig:1stgenbis}, we turn our attention to the \lstinline{F3S} class of models, in which the dark matter particle is a real scalar state and the mediator is a vector-like fermion lying in the fundamental representation of $SU(3)_C$. We remind that such a setup is typical of composite constructions in theories beyond the Standard Model. Similar to our study of the \lstinline{S3M} models above, the dark matter and mediator masses are treated as free parameters, while two complementary approaches are used to define the coupling $\lambda$ of dark matter to quarks. In the first approach, $\lambda$ is fixed to a value of $4.8$, motivated by cosmological considerations (see again section~\ref{sec:cosmology}). In the second approach, the mediator width-to-mass ratio is set to $\Gamma_Y / M_Y = 0.05$, and the corresponding $\lambda$ value is then dynamically determined. The obtained exclusion contours exhibit a similar shape to those of the \lstinline{S3M} models, although they are shifted toward significantly higher mediator mass values. This shift originates from the larger cross sections associated with vector-like quark pair production compared to coloured scalar pair production for a given mediator mass and coupling value~\cite{Fuks:2016ftf,Frixione:2019fxg}. Additionally, scenarios with a fixed coupling value naturally result in higher exclusion limits than in the \lstinline{S3M} case due to the increased chosen value for $\lambda$. Consequently, mediator masses as high as $3.5-4$~TeV are excluded for \lstinline{F3S_uR} models, and $2.7-3.5$~TeV for \lstinline{F3S_dR} models. These bounds are almost insensitive to the dark matter mass, as the relevant matrix elements only depend on $M_X$ for much larger dark matter masses.  As a result, the $q q \to YY$ subprocess (with $q$ representing either an up- or down-type quark) plays an even more dominant role compared to the \lstinline{S3M} models when consider the parameter space as a whole. Enhanced by the high partonic luminosity associated with valence quarks, this subprocess alone is indeed sufficient to drive the exclusion bounds across the entire parameter space, with the only mild exception occurs for dark matter masses $M_X \lesssim 500$~GeV.

Finally, in figure~\ref{fig:1stgenter}, we examine the \lstinline{F3V} class of models. The mediator is, as in the \lstinline{F3S} models, a vector-like quark, but the dark matter particle is this time a vector resonance. Such a setup is again characteristic of composite constructions beyond the Standard Model. Unlike the other exclusion bounds computed in this subsection, the results are based solely on simulations at LO+PS, this limitation arising from practical constraints in treating massive vector states at NLO in our computational toolchain. As in previous analyses, we consider two options for the choice of the new physics coupling $\lambda$. In the first, $\lambda$ is fixed to $1$, consistent with cosmological constraints, while in the second, $\lambda$ is dynamically determined by requiring $\Gamma_Y / M_Y = 0.05$. Overall, the findings align with those obtained for the other two model classes. In particular, the exclusion is dominated by the $YY_t$ channel across the entire relevant parameter space. Consequently, bounds are stronger for up-type scenarios compared to down-type scenarios due to the higher partonic luminosity associated with up quarks. In addition, the $XY$ channel contributes marginally in scenarios with a heavy mediator and light dark matter. This is evident in the results shown for \lstinline{F3V_dR} scenarios with $\Gamma_Y / M_Y = 0.05$ in the parameter space region defined by $M_X \lesssim 500$~GeV. However, for fixed $\lambda$ scenarios, only the $YY_t$ channel contributes for all represented contours. This is due to an artifact of the plots, where exclusion limits have been removed from regions where the particle width becomes extremely large, and thus unphysical (that we have arbitrarily chosen to correspond to $\Gamma_Y / M_Y \gtrsim 0.30$). Such large widths challenge not only the validity of the narrow-width approximation, but also the very definition of a particle. Consequently to all these considerations, for scenarios with $\lambda = 1$ we obtain exclusion bounds for mediator masses up to $4$~TeV and $3$~TeV in the \lstinline{F3V_uR} and \lstinline{F3V_dR} cases, respectively, and for dark matter masses below $2$~TeV and $1$~TeV. Conversely, in scenarios enforcing a narrow mediator resonance (with $\Gamma_Y / M_Y = 0.05$), the bounds are nearly independent of $M_X$, this independence being related to the functional dependence of the associated matrix elements on the new physics masses~\cite{Arina:2023msd}.

\begin{figure}
  \centering
  \includegraphics[width=.48\textwidth]{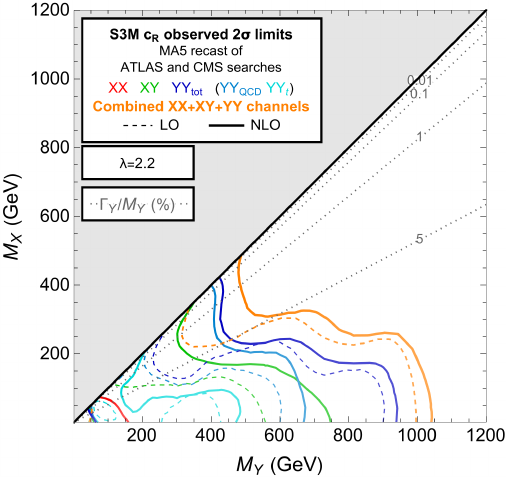}\hfill
  \includegraphics[width=.48\textwidth]{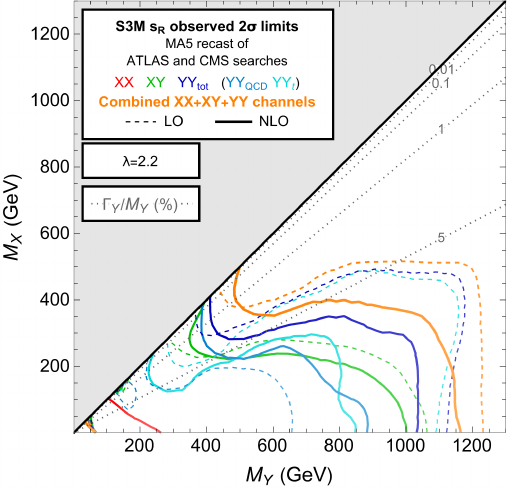}\\
  \includegraphics[width=.48\textwidth]{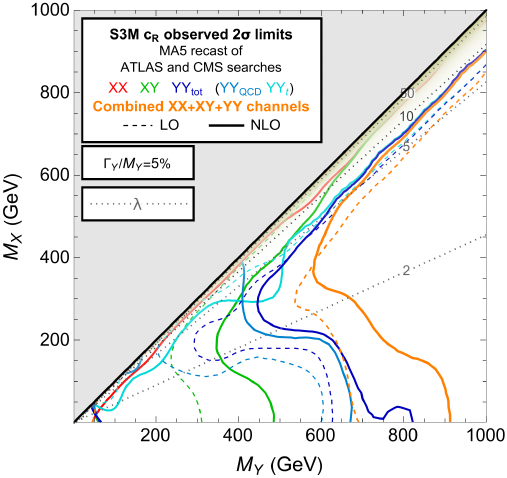}\hfill
  \includegraphics[width=.48\textwidth]{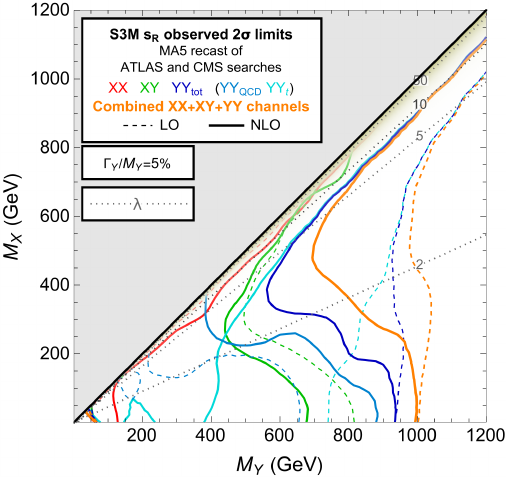}
  \caption{Same as in figure~\ref{fig:1stgen}, but for the \lstinline{S3M_cR} (left) and \lstinline{S3M_sR} (right) classes of models. For scenarios with a fixed coupling value, we adopt $\lambda = 2.2$.}
  \label{fig:2ndgenS3M}
\end{figure}

\subsubsection{Second generation simplified models}\label{sec:bounds_2nd}

\begin{figure}
  \centering
  \includegraphics[width=.48\textwidth]{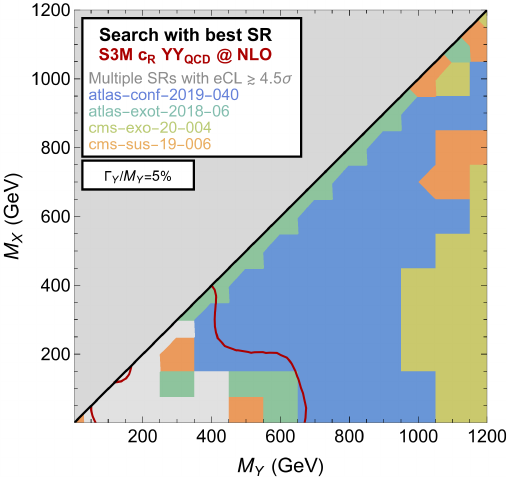}\hfill
  \includegraphics[width=.48\textwidth]{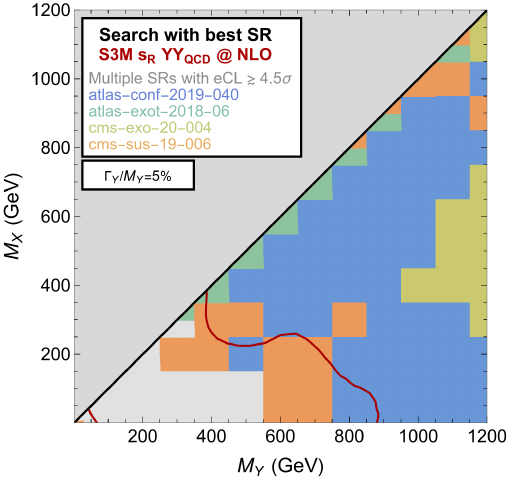}
  \caption{Search including the most sensitive signal region for the new physics signal emerging from the $YY_{\mathrm{QCD}}$ processes at NLO, for the \lstinline{S3M_cR} (left) and \lstinline{S3M_sR} (right) models with a fixed width-to-mass ratio. The 95\% CL exclusion limit is also reported.\label{fig:BestSR_S3M_YYQCD_NLO_cs}} \vspace{.6cm}  
  \includegraphics[width=.485\textwidth]{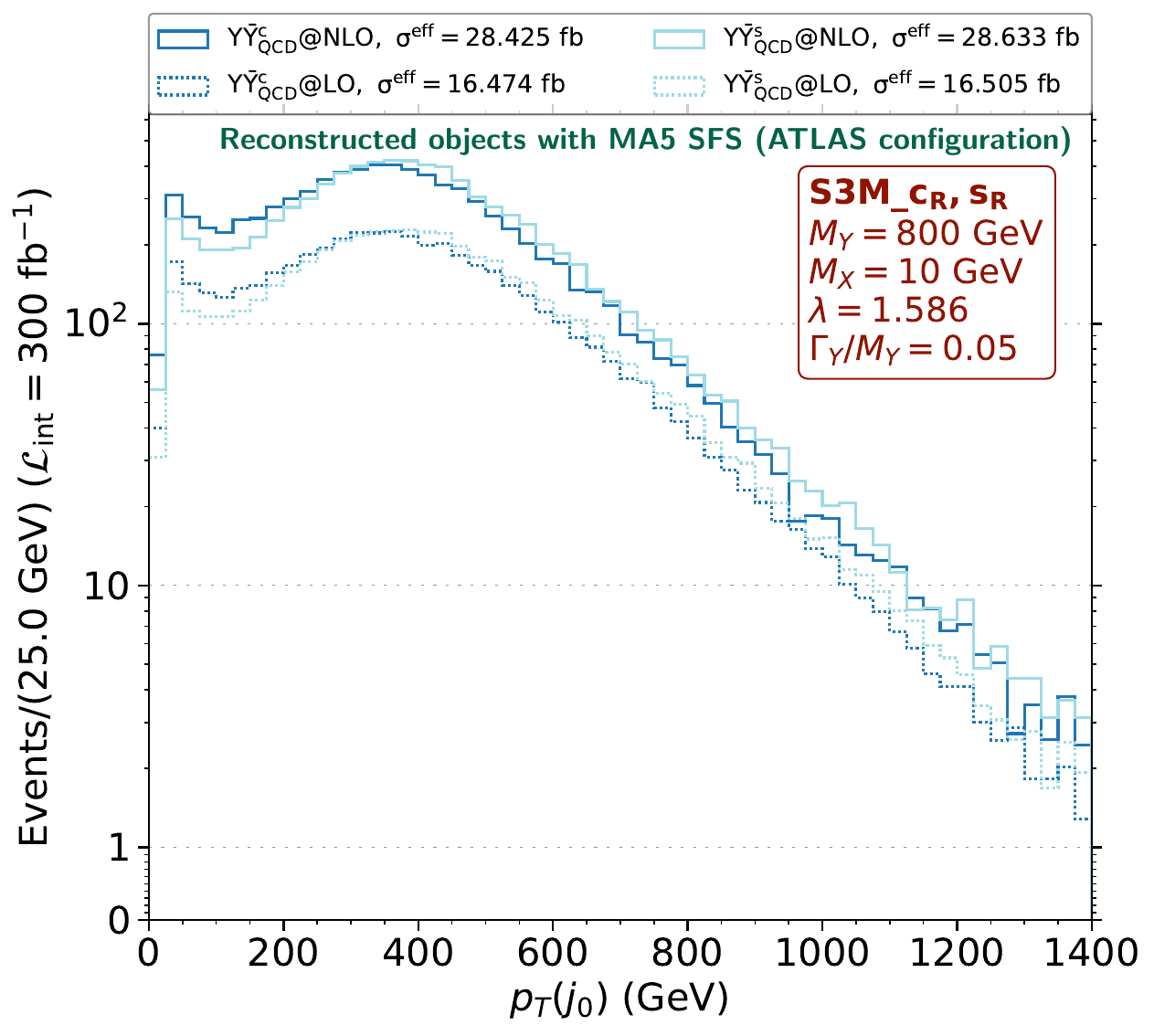}\hfill
  \includegraphics[width=.47\textwidth]{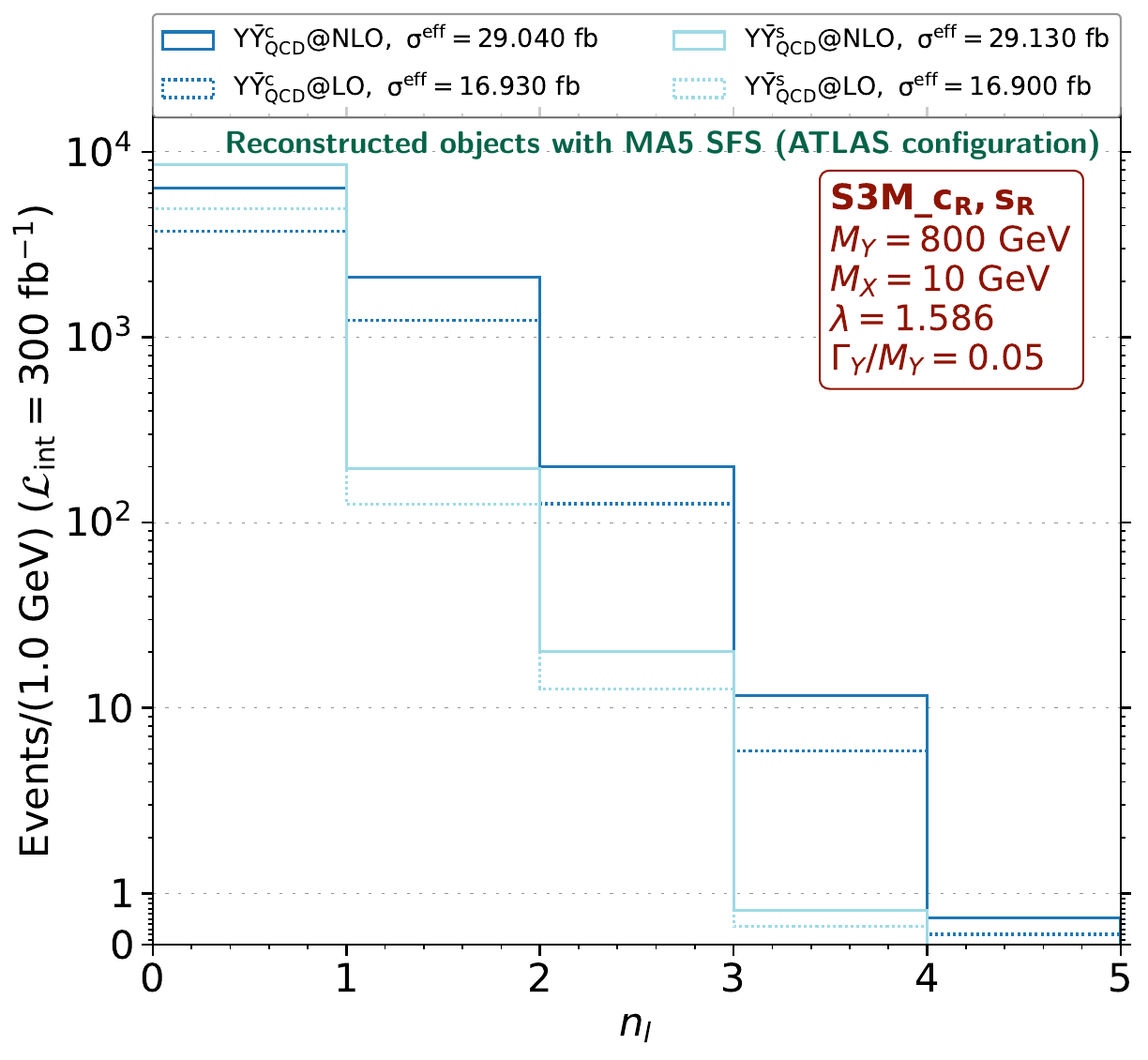}
  \caption{LO and NLO distributions of the transverse momentum of the leading jet (left) and of the number of leptons (right) for the $YY_{\rm QCD}$ channel. We show the number of events obtained at reconstruction level, after hadronisation, detector simulation and before any selection, and for an integrated luminosity of 300~fb$^{-1}$. The distributions correspond to the \lstinline{S3M_cR} and \lstinline{S3M_sR} scenarios with $M_Y=800$~GeV, $M_X=10$~GeV and $\lambda$ fixed to obtain $\Gamma_Y/M_Y=0.05$. The cross sections reported in the legend are effective to reflect the reduction due to selecting specific final state particles.\label{fig:MY800_MX10_distributions}}
\end{figure}

In this section, we present the results obtained when the DM and mediator particles interact with SM quarks of the second generation, namely charm and strange quarks. Following the approach used for the first-generation case, figure~\ref{fig:2ndgenS3M} shows the bounds that are applicable to the \lstinline{S3M_cR} and \lstinline{S3M_sR} models where the dark matter is a Majorana fermion and the mediator a coloured scalar state. In fixed-coupling scenarios, the interplay between astrophysical and cosmological constraints favours a different range of $\lambda$ values compared to the first-generation case, by virtue of the existence of a large valence up and down quark content in nuclei compared to other quarks: we consequently here focus on benchmarks with $\lambda = 2.2$, for which the mediator remains narrow across the entire allowed parameter space.

For both the \lstinline{S3M_cR} and \lstinline{S3M_sR} classes of models, the bounds are weaker than those obtained for the corresponding first-generation models. For light DM, the mediator mass must exceed approximately 1~TeV for \lstinline{S3M_cR} scenarios and 1.2~TeV for \lstinline{S3M_sR} scenarios. On the other hand, in the small mass-splitting region, the bounds on the mediator mass reduce to around 500~GeV in both scenarios, while the DM mass is constrained to be larger than 300--400~GeV for lower values of $M_Y$, and $M_Y/M_X-1 \gtrsim 0.2$. Unlike in the first-generation case, the bounds result from a more intricate interplay between the different production channels due to the absence of a dominant same-charge $YY$ contribution driven by valence-quark PDFs. In second-generation scenarios, the total $YY$ channel receives comparable contributions from the $t$-channel diagrams ($YY_\mathrm{t}$) and from the QCD ones ($YY_\mathrm{QCD}$), while the $XY$ channel is additionally sensitive to the signal in a similar way as the $YY$ contributions. Consequently, no single contribution entirely dominates, and the location of the overall bounds in the parameter space originates truly from the combination of the various subprocesses contributing to the signal.

A key distinction between scenarios where the DM couples to charm and strange quarks is the strength of the bounds. Charm scenarios are found to exhibit bounds that are typically weaker by 100--200~GeV, particularly in the light DM parameter space region although this holds for most of the parameter space. Moreover, this feature applies both to results obtained on the basis of LO and NLO simulations. In order to understand the origin of this difference, we examine as a representative example the $YY_\mathrm{QCD}$ channel, which is independent of $\lambda$ and dominated by gluon-initiated topologies. Associated cross sections are therefore approximately independent of the nature of the quark to which the DM couples (which also holds for scenarios with couplings to up and down quarks). For benchmark scenarios relevant with the position of the bounds in the framework of light DM, the most sensitive signal region (to a new physics signal only comprising the $YY_\mathrm{QCD}$ contribution) is the \lstinline{SR2j_1600} region of the ATLAS-CONF-2019-040 analysis~\cite{ATLAS:2019vcq}, as shown in figure~\ref{fig:BestSR_S3M_YYQCD_NLO_cs} where we display information on the most sensitive analysis for all considered mass configurations. A deep investigation of the associated cut-flow reveals that the preselection criteria disproportionately impact charm scenarios. The corresponding requirements include zero leptons, a leading and sub-leading jet with a transverse momentum imposed to be above 200~GeV  and 50~GeV respectively, missing transverse energy larger than 300~GeV, and an effective mass (defined as the scalar sum of the transverse momenta of all visible objects and the missing transverse momentum) greater than 800~GeV. However, charm quarks tend to produce more leptons and softer jets during hadronisation than strange quarks, subsequently leading to a stronger signal event rejection by the preselection. This is illustrated in figure~\ref{fig:MY800_MX10_distributions} where we show the $p_T$ spectrum of the leading jet (left) and the distribution in the number of leptons (right), both at LO and NLO. These predictions are obtained with the standard tool chain described in section~\ref{sec:collider_generalities}, with a detector parametrisation set to the default ATLAS-SFS configuration included in \lstinline{MadAnalysis 5}. The charm signal is therefore correspondingly associated with a smaller selection efficiency then the strange signal, thus explaining the weaker exclusion bounds. Additionally, we emphasise that the contribution related to the quark-initiated processes leads to stronger bounds for \lstinline{S3M_sR} scenarios than for \lstinline{S3M_cR} scenarios across the entire parameter space, primarily due to the larger PDF contributions for strange quarks relative to charm quarks.

\begin{figure}
  \includegraphics[width=.48\textwidth]{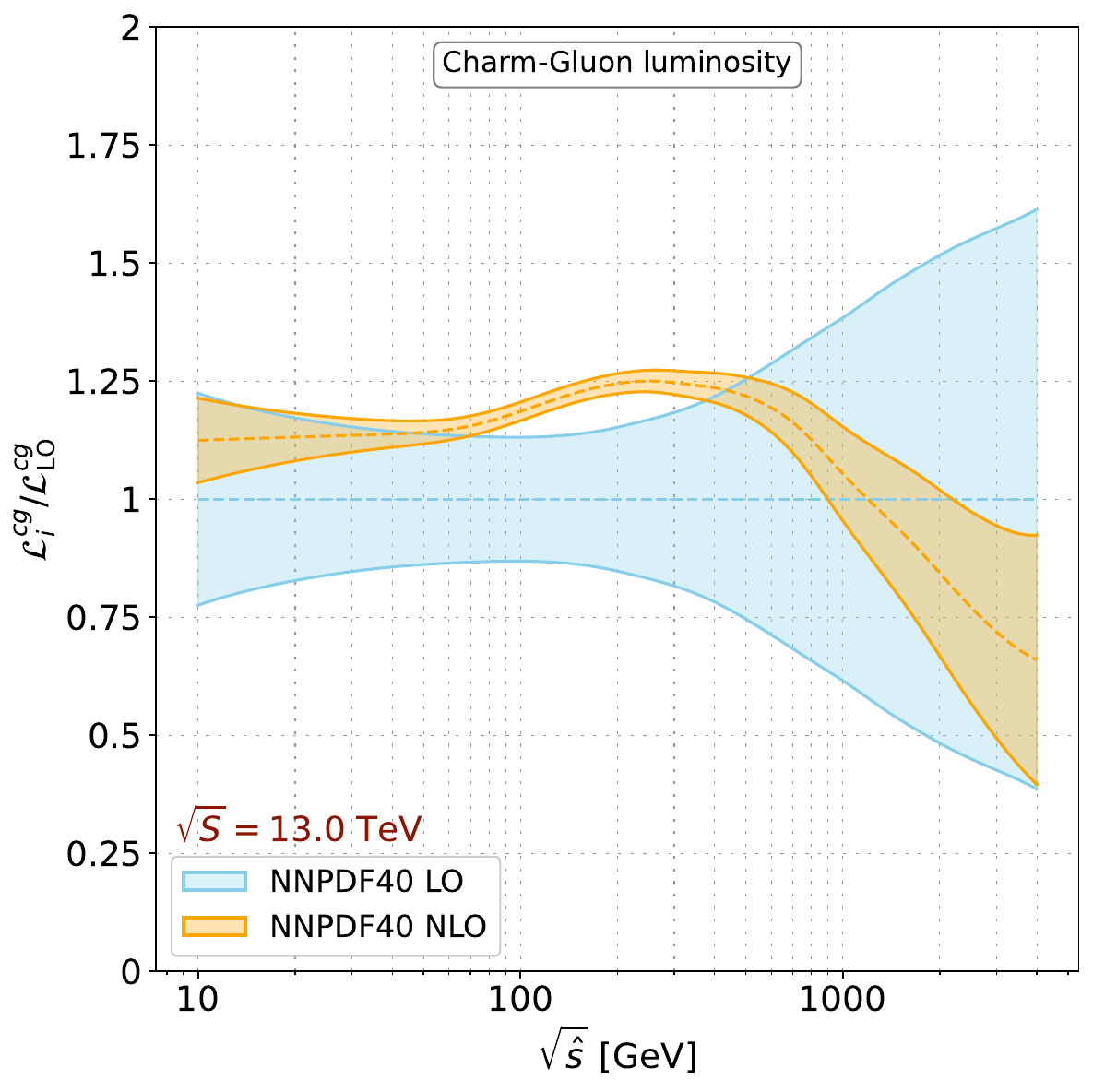}\hfill
  \includegraphics[width=.48\textwidth]{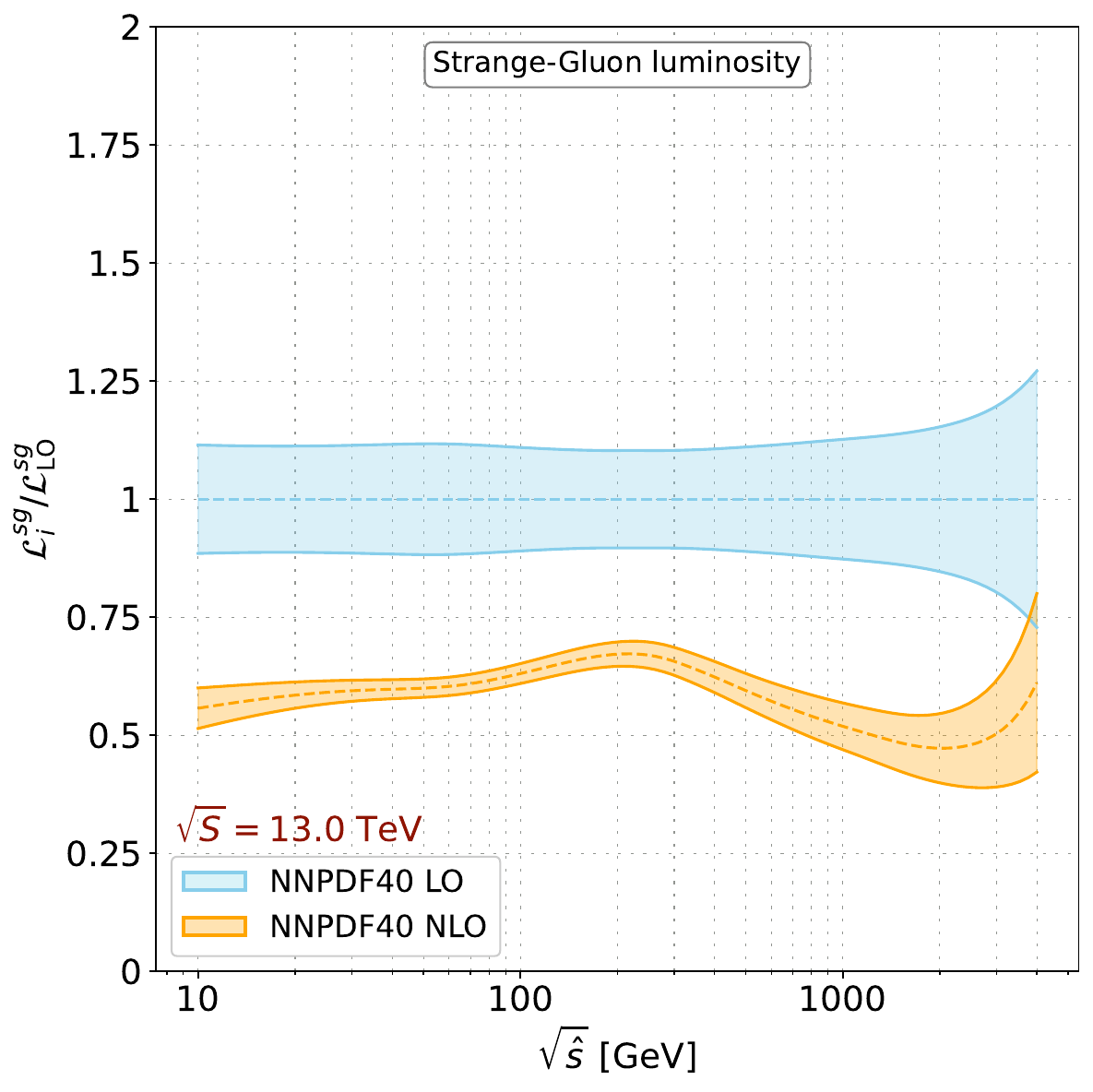} \\  \vspace{.3cm}
  \includegraphics[width=.48\textwidth]{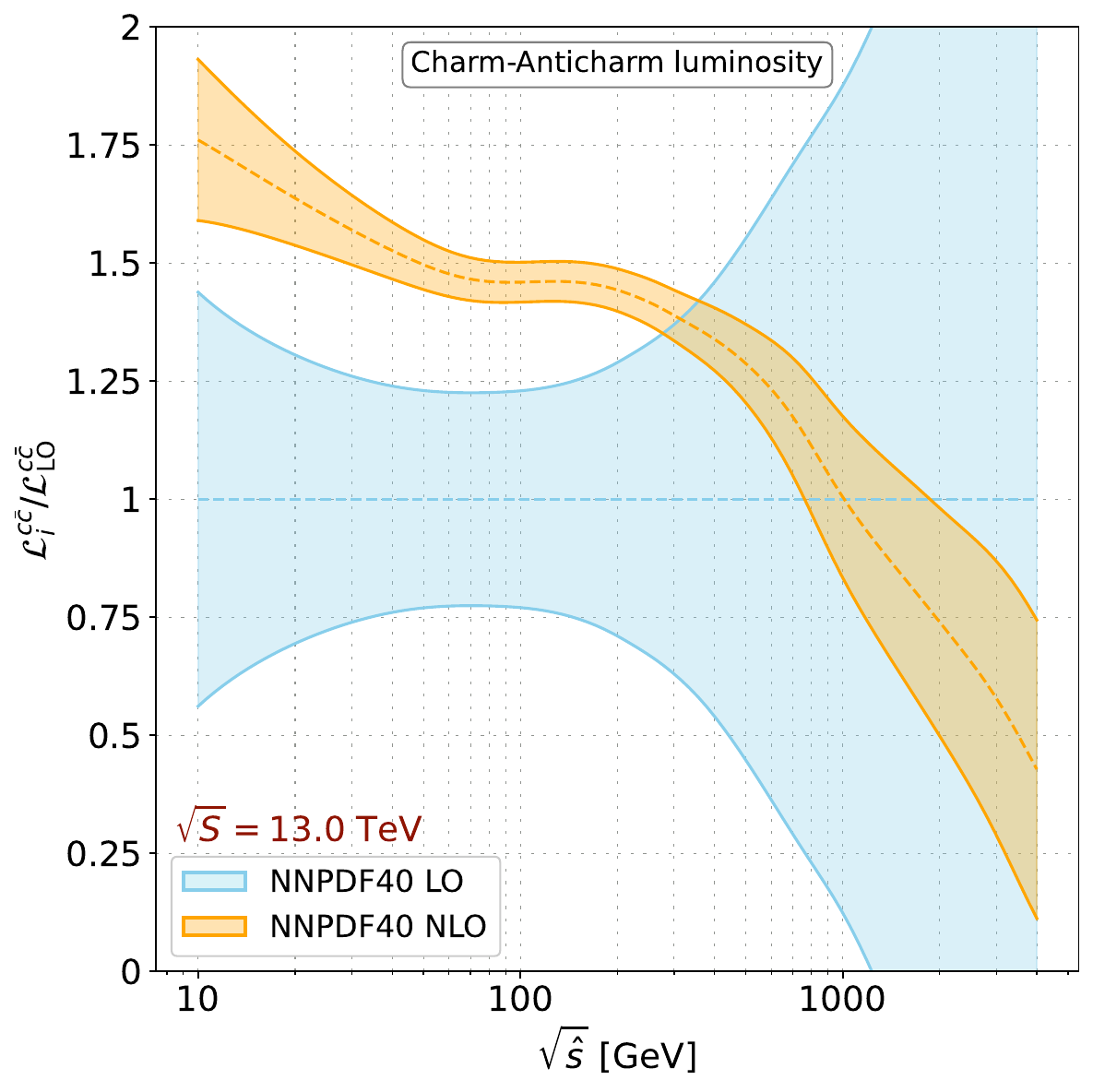}\hfill
  \includegraphics[width=.48\textwidth]{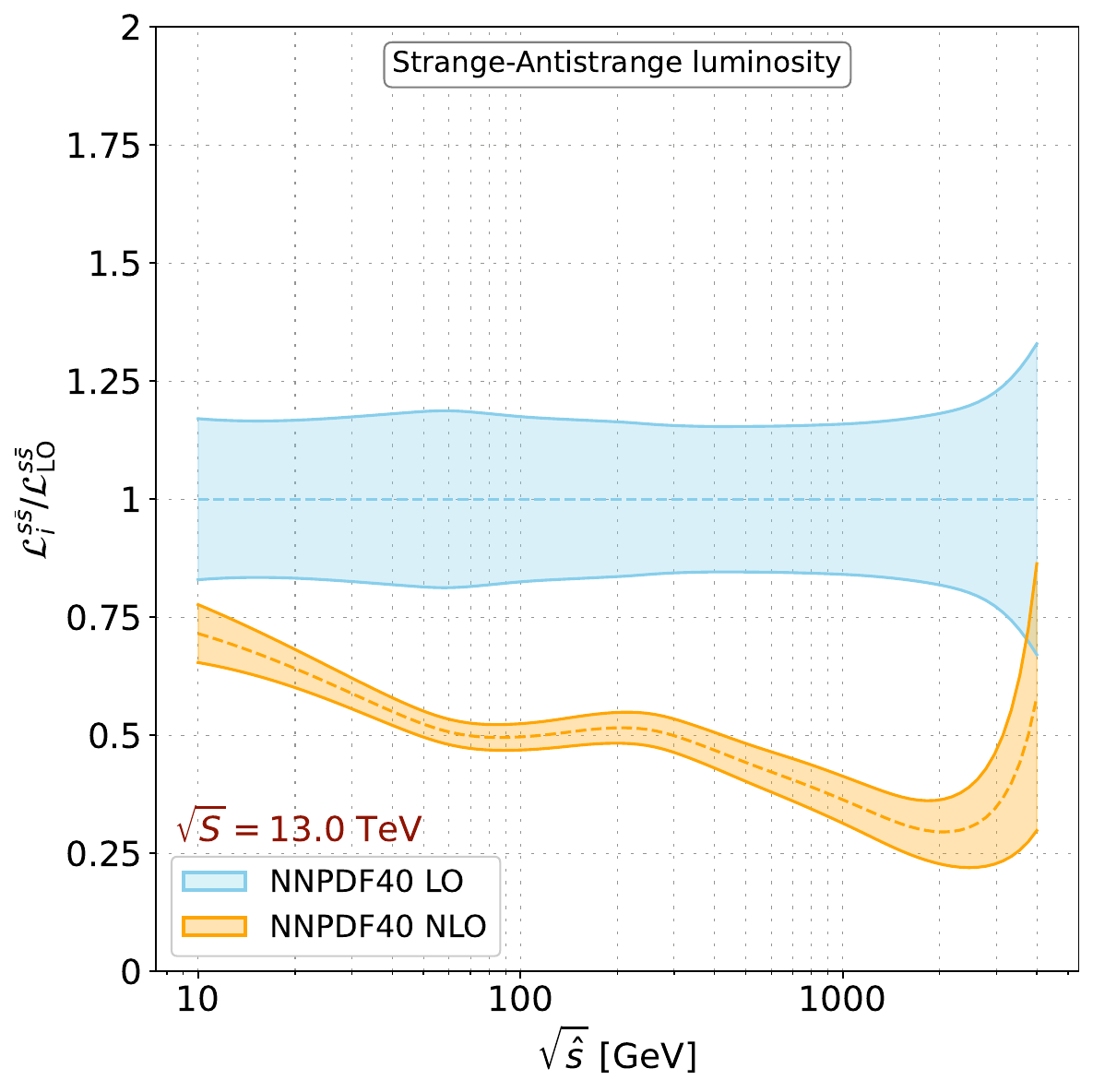}
  \caption{NLO/LO ratios of parton luminosities for the NNPDF40 set of parton densities. We consider first the charm-gluon (top left) and strange-gluon (top right) initial states relevant for the $XY$ process, and next the charm-anticharm (bottom left) and strange-antistrange (bottom right) initial states relevant for the $Y\bar Y_t$ process. Band thickness refers to PDF uncertainties.\label{fig:NLOLOPDFratio_cs}}
\end{figure}

A second key difference between charm and strange scenarios lies in the relationship between NLO and LO bounds for the $XY$ and $YY_\mathrm{t}$ contributions. For scenarios where the DM couples to charm quarks, NLO bounds are stronger than LO bounds, while the opposite holds for strange quarks. This behaviour is again attributed to PDF effects: as shown in figure~\ref{fig:NLOLOPDFratio_cs}, strange-initiated partonic luminosities are higher at LO than NLO across the entire range of partonic centre-of-mass energies $\sqrt{\hat{s}}$, whereas for charm-initiated processes, the NLO/LO ratio exceeds 1 only for $\sqrt{\hat{s}} \lesssim 1$~TeV. Despite larger PDF uncertainties for charm quarks, this effect significantly impacts the interpretation of results, particularly when combined with the different distributions in the final-state jet multiplicity at NLO. Consequently, it is clear that understanding the impact of the PDFs on the bounds is of utmost importance for the interpretation of the results.

\begin{figure}
  \centering
  \includegraphics[width=.48\textwidth]{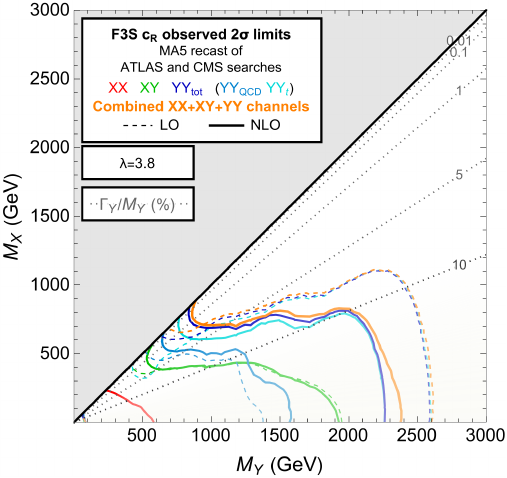}\hfill
  \includegraphics[width=.48\textwidth]{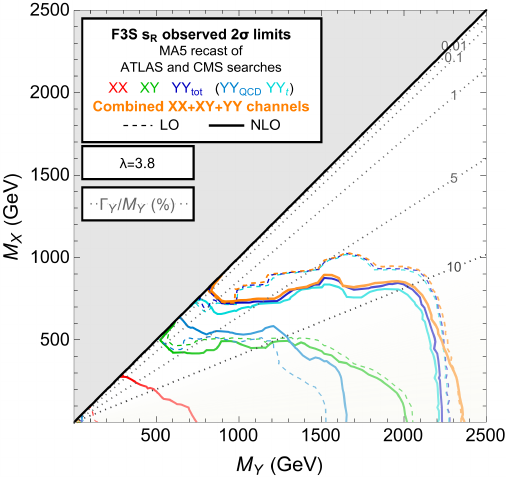}\\ \vspace{.3cm}
  \includegraphics[width=.48\textwidth]{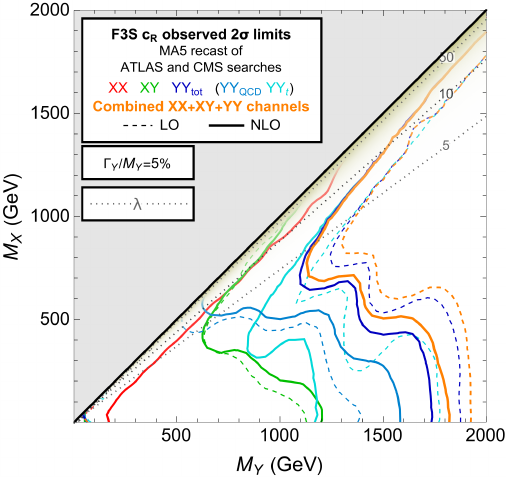}\hfill
  \includegraphics[width=.48\textwidth]{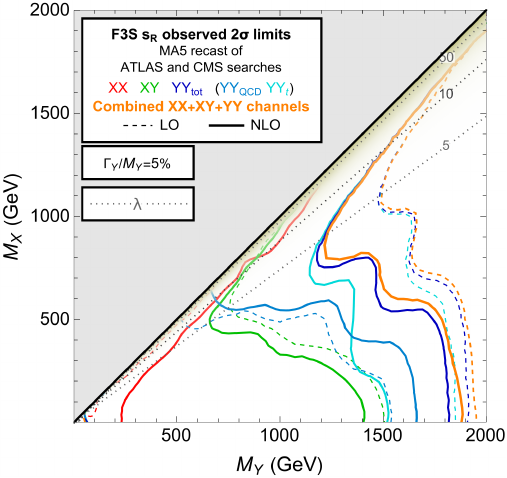}
  \caption{Same as figure \ref{fig:2ndgenS3M}, but for the \lstinline{F3S_cR} (left) and \lstinline{F3S_sR} (right) classes of scenarios.}
  \label{fig:2ndgenF3S}
\end{figure}

We now move on with the \lstinline{F3S_cR} and \lstinline{F3S_sR} classes of scenarios in which the mediator is a coloured fermion and the dark matter a scalar state. The bounds on the simplified models are shown in figure~\ref{fig:2ndgenF3S}. As expected, the overall bounds are stronger compared to \lstinline{S3M} setups due to the higher number of spin degrees of freedom of the mediator. The latter impacts the rates of the $YY_\mathrm{t}$ and $XY$ processes, that get larger than for the production of a scalar mediator, and then contribute dominantly to the combined signal. For light DM, the bounds on the mediator  reach 2--2.5~TeV, depending on the scenario. For benchmarks with a fixed coupling value, caution is nevertheless necessary because of the total mediator width, which easily exceeds 10\% of the mediator mass and thus makes the narrow-width approximation less reliable. Moreover, in both the charm and strange cases, the DM mass is constrained to be above approximately 800~GeV as long as the mediator is light enough to get significant bounds from its production at the LHC. In scenarios featuring a fixed width-over-mass ratio (and for DM couplings to either the strange or the charm quark), the bounds are almost entirely driven by the $YY_\mathrm{QCD}$ channel, at least until the $\lambda$-dependent $YY_\mathrm{t}$ contributions become dominant for setups where the spectrum approaches the kinematic limit where the DM and mediator masses are equal. Finally, we also note that the LO bounds are stronger than the NLO ones for all considered scenarios, which is compatible with the PDF behaviour shown in figure~\ref{fig:NLOLOPDFratio_cs}.

\begin{figure}
  \centering
  \includegraphics[width=.48\textwidth]{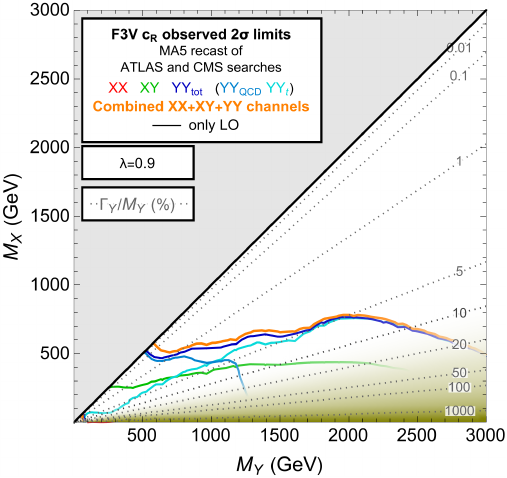}\hfill
  \includegraphics[width=.48\textwidth]{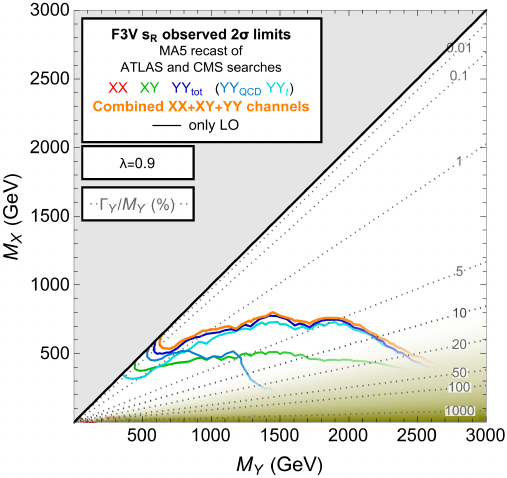}\\
  \includegraphics[width=.48\textwidth]{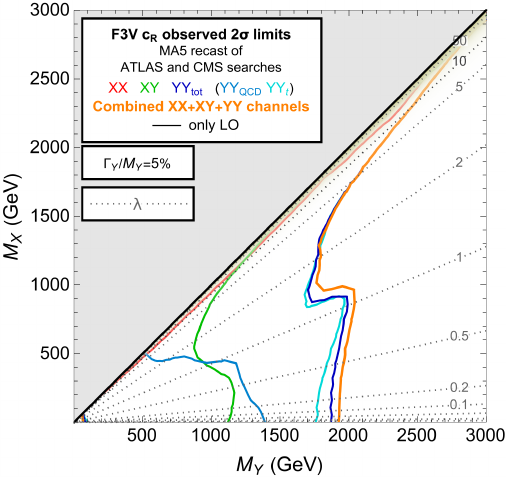}\hfill
  \includegraphics[width=.48\textwidth]{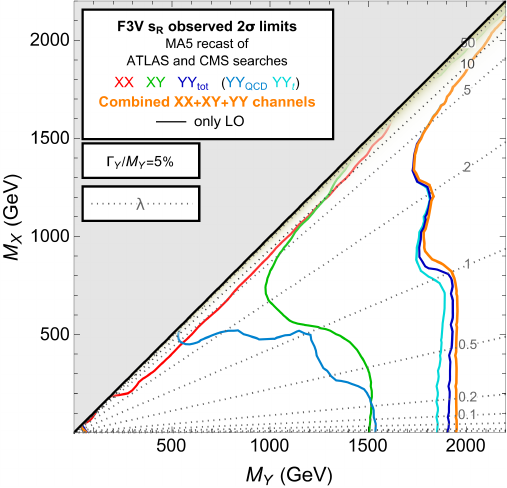}
  \caption{Same as figure \ref{fig:2ndgenS3M}, but for the \lstinline{F3V_cR} (left) and \lstinline{F3V_sR} (right) scenarios.}\label{fig:2ndgenF3V}
\end{figure}

Finally, we examine in figure~\ref{fig:2ndgenF3V} the LHC bounds imposed on the two \lstinline{F3V} classes of models where the dark matter is a vector state, and the mediator a coloured fermion. Here, the $YY_\mathrm{t}$ channel dominates almost everywhere in the parameter space, except in regions where its width-to-mass ratios is around or below 1\%, for scenarios with a fixed coupling value. There, the $YY_\mathrm{QCD}$ contribution takes over. For the two classes of models featuring a coupling to charm and strange quarks, the combined bounds exclude DM masses below 500--800~GeV, depending on the mediator mass. Moreover, in scenarios with a fixed width-over-mass ratio, a visible positive interference effect between the $YY_\mathrm{t}$ and $YY_\mathrm{QCD}$ channels emerges when the DM is light. Numerically, mediator masses are then constrained to exceed 2~TeV almost independently of DM mass until $M_Y/M_X-1 \gtrsim 0.3$, where the bounds asymptotically approach the kinematic limit for increasing mediator and DM masses.

\subsubsection{Third generation simplified models}\label{sec:bounds_3rd}

Third-generation scenarios differ significantly from those in which the mediator couples to first- or second-generation quarks. The relevant quark parton densities are either entirely absent (for top quarks) or highly suppressed (for bottom quarks). As a result, mediator pair production through QCD interactions ($YY_\mathrm{QCD}$) becomes the dominant contribution to the entire new physics signal. At leading order and in scenarios where dark matter couples to top quarks, $YY_\mathrm{QCD}$ hence constitutes the sole contribution to the signal, and dark matter pair production ($XX$) becomes accessible at NLO. For setups where the dark matter couples to bottom quarks, the situation is largely similar, with two notable exceptions. First, associated $XY$ production plays a sub-leading role, particularly for light dark matter masses. Second, mediator pair production via $t$-channel exchange of a dark matter particle ($YY_t$) can contribute in regions with a very compressed mass spectrum.

\begin{figure}
  \centering
  \includegraphics[width=.325\textwidth]{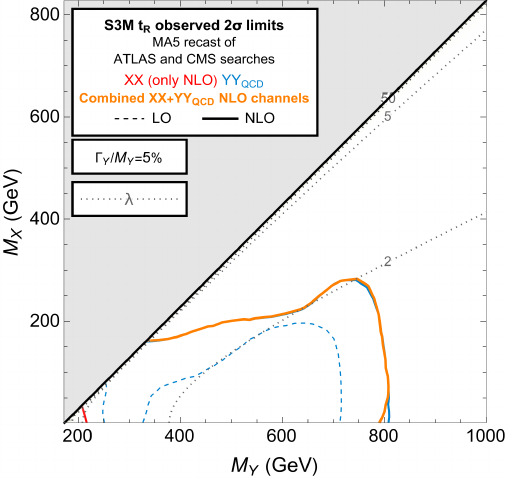}\hfill
  \includegraphics[width=.325\textwidth]{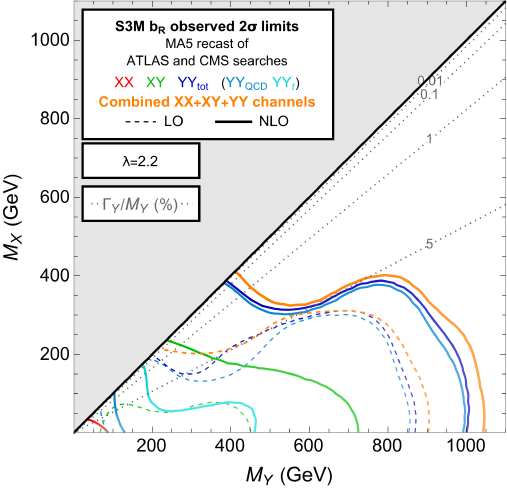}\hfill
  \includegraphics[width=.325\textwidth]{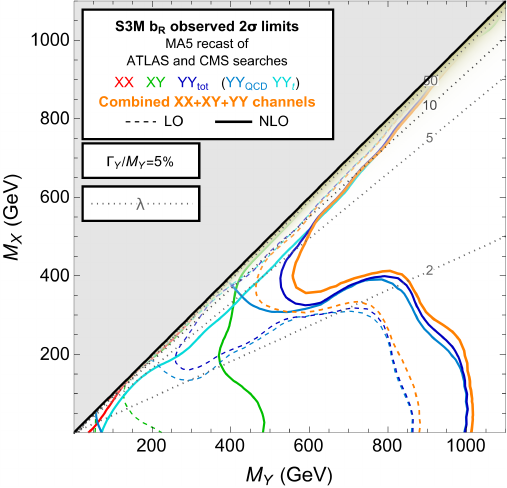}
  \caption{Same as in figure~\ref{fig:1stgen}, but for the \lstinline{S3M_tR} real dark matter scenarios with $\Gamma_Y/M_Y=0.05$ (left), and for the \lstinline{S3M_bR} scenarios with either $\lambda=2.2$ (central) or $\Gamma_Y/M_Y=0.05$ (right).}
  \label{fig:3rdgen}
\end{figure}

These features are illustrated in figure~\ref{fig:3rdgen}, which presents exclusion contours for three classes of third-generation simplified models with fermionic dark matter in the $(M_Y, M_X)$ plane. First, the left panel of the figure focuses on supersymmetry-like top-philic \lstinline{S3M_tR} scenarios, where the mediator is a coloured scalar similar to a top squark, the dark matter is a Majorana fermion, and the new physics coupling is dynamically set to ensure $\Gamma_Y / M_Y = 0.05$. Since the dependence on the coupling is mild due to the fact that the signal is dominated by its $\lambda$-independent QCD component, using a fixed coupling value of $\lambda = 2.2$ as motivated by cosmology would not significantly alter the results. The corresponding figure is therefore omitted from this report. The constraints obtained are relatively weak. The LHC Run~2 is indeed sensitive only to scenarios with mediator masses below $800$~GeV and dark matter masses lighter than about $200$~GeV. These weaker bounds, compared to those for other scenarios with Majorana DM, stem from the distinct decay patterns of the top quarks produced in mediator decays. The latter hence lead to a variety of final states differing from the simpler (hard-scattering-level) one-jet $+ \met$  or two-jet $+ \met$ signatures. In principle, the limits could be significantly improved by incorporating dedicated $t\bar{t} + \met$ searches in the analysis, as suggested in past studies of composite constructions~\cite{Colucci:2018vxz, Cornell:2021crh}. However, the absence of validated implementations of such full Run~2 searches in public recasting tools limits this possibility. While we could extrapolate predictions from partial Run~2 results available for some time~\cite{Fuks:2018yku}, it is unclear whether such a naive extrapolation would provide meaningful new insights beyond the jet+$\met$ searches already considered, as pointed out in~\cite{Cornell:2021crh}. We therefore refrain from doing so in this report. Finally, as anticipated, the exclusion limits are entirely driven by the $YY_\mathrm{QCD}$ channel. While $XX$ contributions could in principle play a role, they are found being negligible and only relevant for scenarios with a mass only slightly larger than that of the top quark.  

The central and right panels of figure~\ref{fig:3rdgen} explore bottom-philic \lstinline{S3M_bR} scenarios. In the central panel, the new physics coupling is fixed to $\lambda = 2.2$, while in the right panel, $\lambda$ is dynamically determined to satisfy $\Gamma_Y / M_Y = 0.05$. The $Y \to X b$ decay, which produces this time always one jet and missing energy (unlike for top-philic DM), impacts the sensitivity of the recast analyses compared to the \lstinline{S3M_tR} case and strengthen the bounds, owing to a larger mediator branching ratio into the relevant final states. Additionally, the non-zero bottom-quark parton PDF introduces non-negligible (albeit suppressed) contributions from processes beyond $YY_\mathrm{QCD}$. Specifically, in scenarios with a fixed coupling and light dark matter, $XY$ production slightly tightens the bounds, while in cases with a fixed mediator width-to-mass ratio and a compressed spectrum, the larger new physics coupling leads to an increasing contribution from the $YY_t$ channel. However, caution is in order when interpreting predictions in this regime, as it approaches the limits of validity for perturbative treatments. For both \lstinline{S3M_bR} scenarios, mediator masses up to $1$~TeV are found excluded for dark matter masses up to approximately $400$~GeV. Similar to the \lstinline{S3M_tR} case, these bounds are conservative, as dedicated $b\bar{b} + \met$ searches have not been included in the analysis due to the lack of validated implementations in public tools.

\begin{figure}
  \centering
  \includegraphics[width=.325\textwidth]{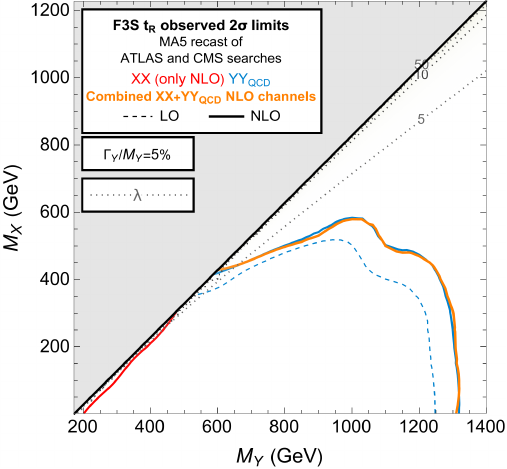}\hfill
  \includegraphics[width=.325\textwidth]{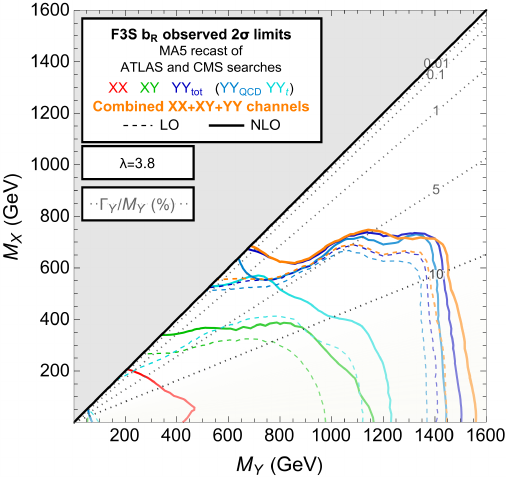}\hfill
  \includegraphics[width=.325\textwidth]{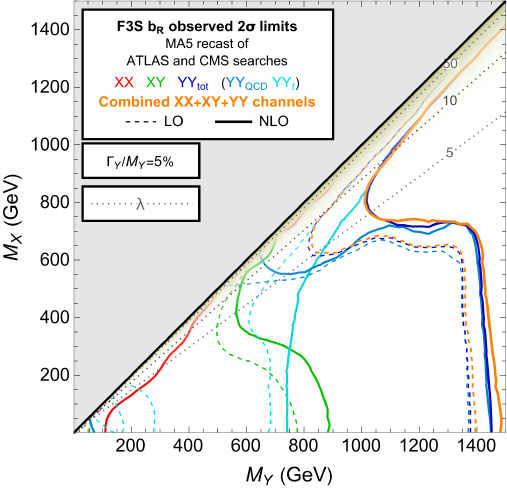}\\ \vspace{.3cm}
  \includegraphics[width=.325\textwidth]{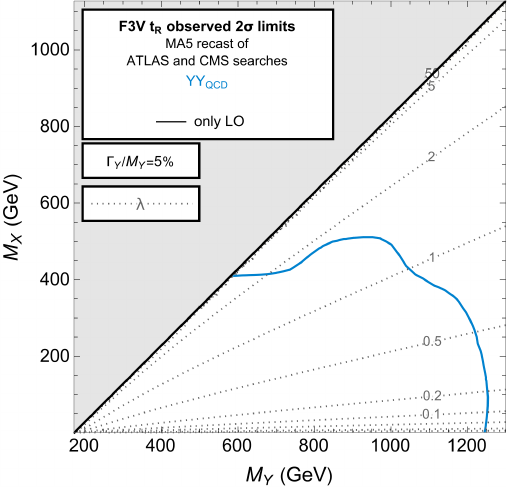}\hfill
  \includegraphics[width=.325\textwidth]{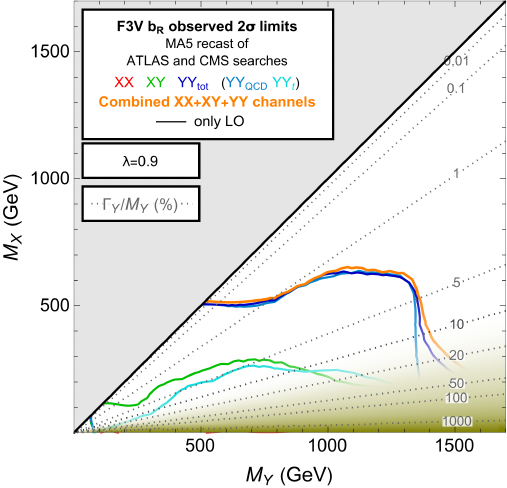}\hfill
  \includegraphics[width=.325\textwidth]{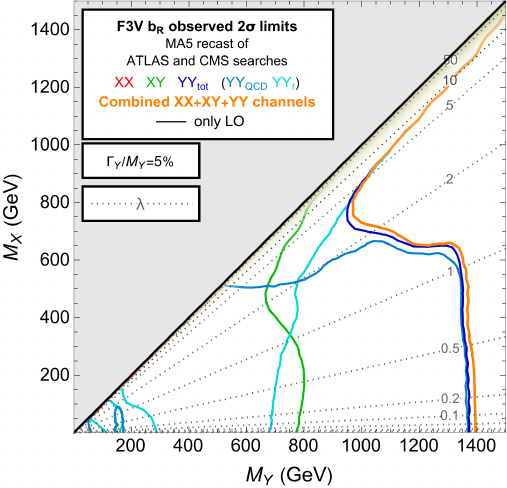}\\
  \caption{Same as in figure~\ref{fig:1stgen}, but for the \lstinline{F3S_tR} (top left), \lstinline{F3V_tR} (bottom left) scenarios with a fixed mediator width-to-mass ratio $\Gamma_Y / M_Y = 0.05$, and for the \lstinline{F3S_bR} (top row, central and right panels) and \lstinline{F3V_bR} (bottom row, central and right panels) setups. We either fix the new physics coupling to $\lambda = 3.8$ (top central) or $0.9$ (bottom central), or derive it from $\Gamma_Y / M_Y = 0.05$ (right).}
  \label{fig:3rdgenbis}
\end{figure}

In figure~\ref{fig:3rdgenbis}, we present exclusion contours for scenarios with a fermionic mediator, specifically for the \lstinline{F3S} (top row) and \lstinline{F3V} (bottom row) classes of models. For the top-philic cases (leftmost figures), we focus on scenarios where the coupling is dynamically set to satisfy $\Gamma_Y / M_Y = 0.05$. As in the \lstinline{S3M_tR} case, the exclusion contours in the $(M_Y, M_X)$ plane are nearly identical to those obtained using a fixed coupling value motivated by cosmology, so the latter are omitted for brevity. No new features emerge compared to models with scalar mediators, except for an increase in the exclusion limits driven by the larger mediator pair production cross section, a direct consequence of the fermionic nature of the mediator. As a result, mediator masses up to $1.3$~TeV and $1.2$~TeV are excluded for the \lstinline{F3S_tR} and \lstinline{F3V_tR} models, respectively, with corresponding dark matter masses constrained to be larger than approximately $800$~GeV and $500$~GeV. It is worth noting that the weaker exclusions for the \lstinline{F3V_tR} model arise from the reliance on LO+PS simulations, lacking thus of important $K$-factor enhancement in the signal rates that cannot be computed due to technical limitations in our toolchain.

The remaining panels in figure~\ref{fig:3rdgenbis} explore bottom-philic setups. As with the \lstinline{S3M_bR} scenarios, we distinguish between cases where the coupling $\lambda$ is fixed based on cosmological considerations (central figures) and those where $\Gamma_Y / M_Y = 0.05$ (rightmost figures). For the fixed coupling case, we use $\lambda = 3.8$ for \lstinline{F3S_bR} scenarios and $\lambda = 0.9$ for \lstinline{F3V_bR} scenarios. Again, no new features emerge compared to scalar mediator cases. The limits remain dominated by the $YY_\mathrm{QCD}$ channel, with sub-leading contributions from $XY$ production for spectrum featuring light dark matter, and from $YY_t$ production for more compressed spectra. Subsequently, mediator masses are excluded up to $1.5$--$1.6$~TeV in \lstinline{F3S_bR} scenarios with dark matter masses below approximately $600$--$700$~GeV. Strong constraints are also obtained for compressed spectra in cases where $\lambda$ is derived from $\Gamma_Y / M_Y = 0.05$, but these correspond to baroque setups where perturbative methods are unreliable, so that predictions could not be trusted. For the \lstinline{F3V_bR} model, the bounds are similar but slightly weaker, reflecting the use of LO predictions without (differential) $K$-factors.

\subsubsection{Simplified models: considerations on signal modelling}\label{sec:distr}

Designing searches optimised to probe $t$-channel DM scenarios requires a detailed understanding of the distinct kinematic features of the signal final state. This involves weighing the relative importance of individual contributions and examining the effects of NLO corrections at the differential level. Such an approach enables the targeting of dominant contributions in specific signal regions, and determines whether NLO corrections are impactful enough to be experimentally observable. In the following, we present representative kinematic distributions for the different scenarios discussed in the previous sections. The aim is to highlight differences and similarities that can aid in designing new searches. All distributions are computed using the simplified fast detector simulation (\lstinline{SFS}) built within \lstinline{MadAnalysis 5}~\cite{Araz:2020lnp, Araz:2021akd}. For these illustrative case studies, we utilised the ATLAS default settings shipped with the code.

\begin{figure}
  \centering
  \includegraphics[width=.48\textwidth]{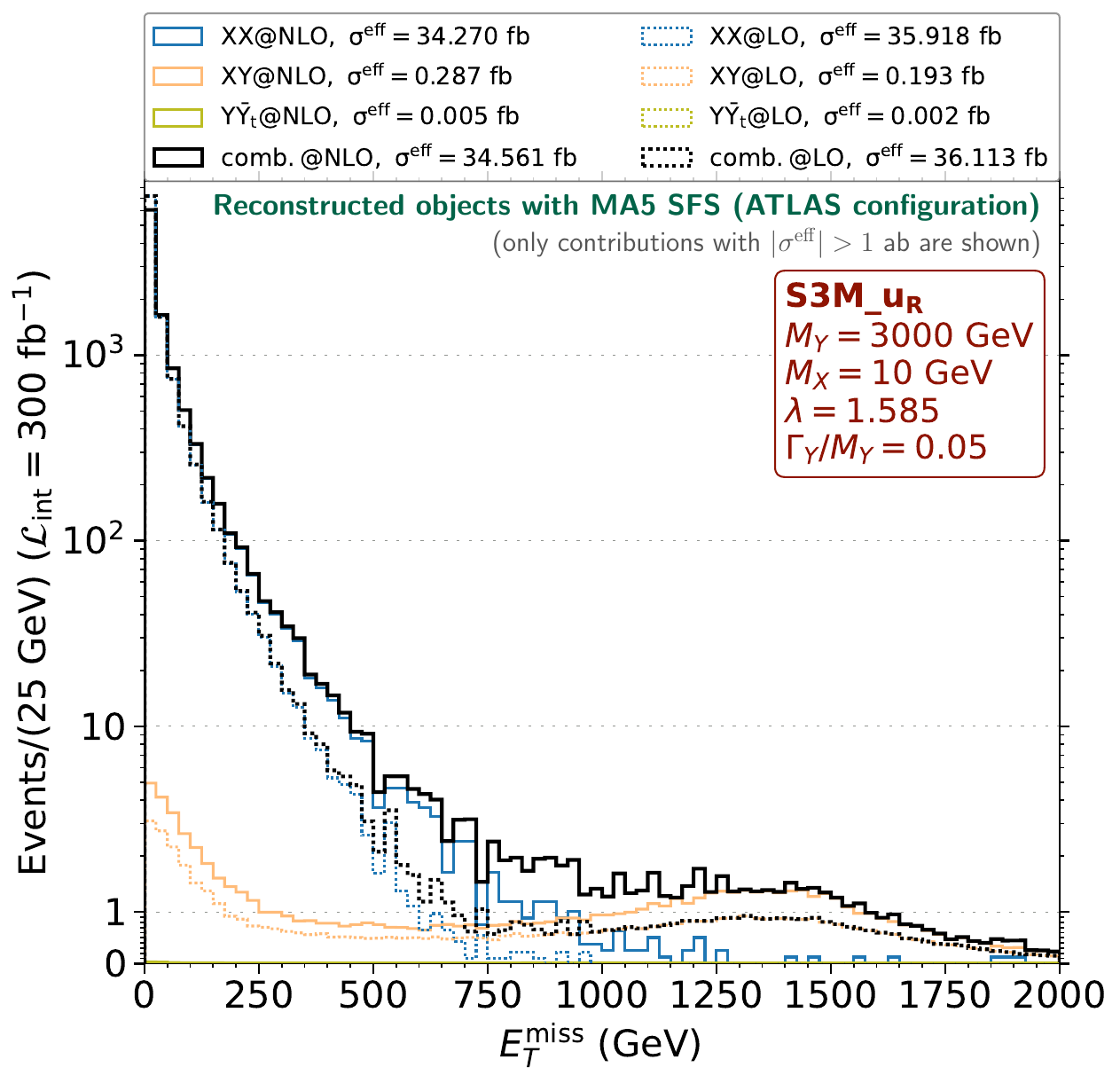}
  \includegraphics[width=.48\textwidth]{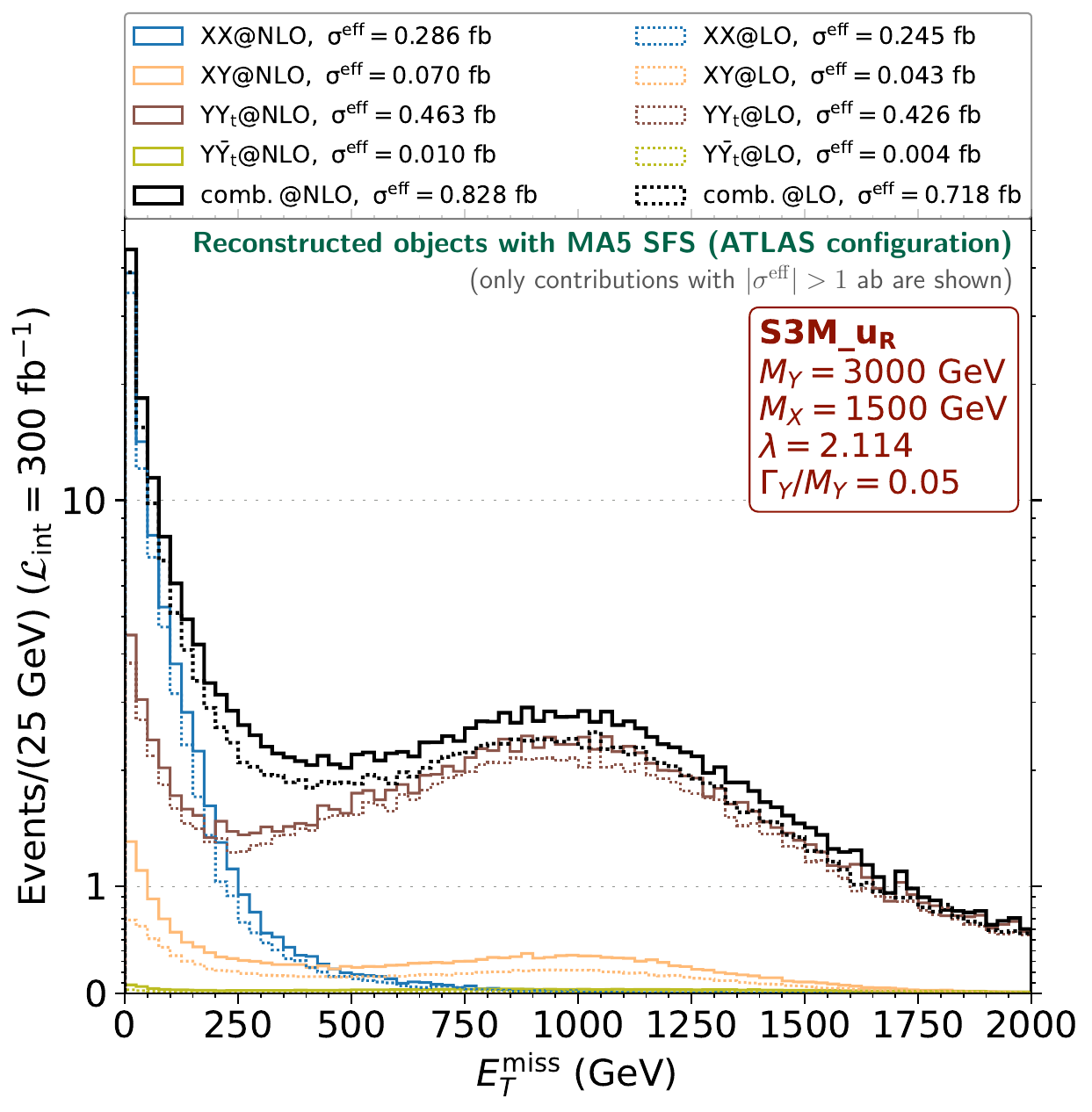}\\
  \includegraphics[width=.48\textwidth]{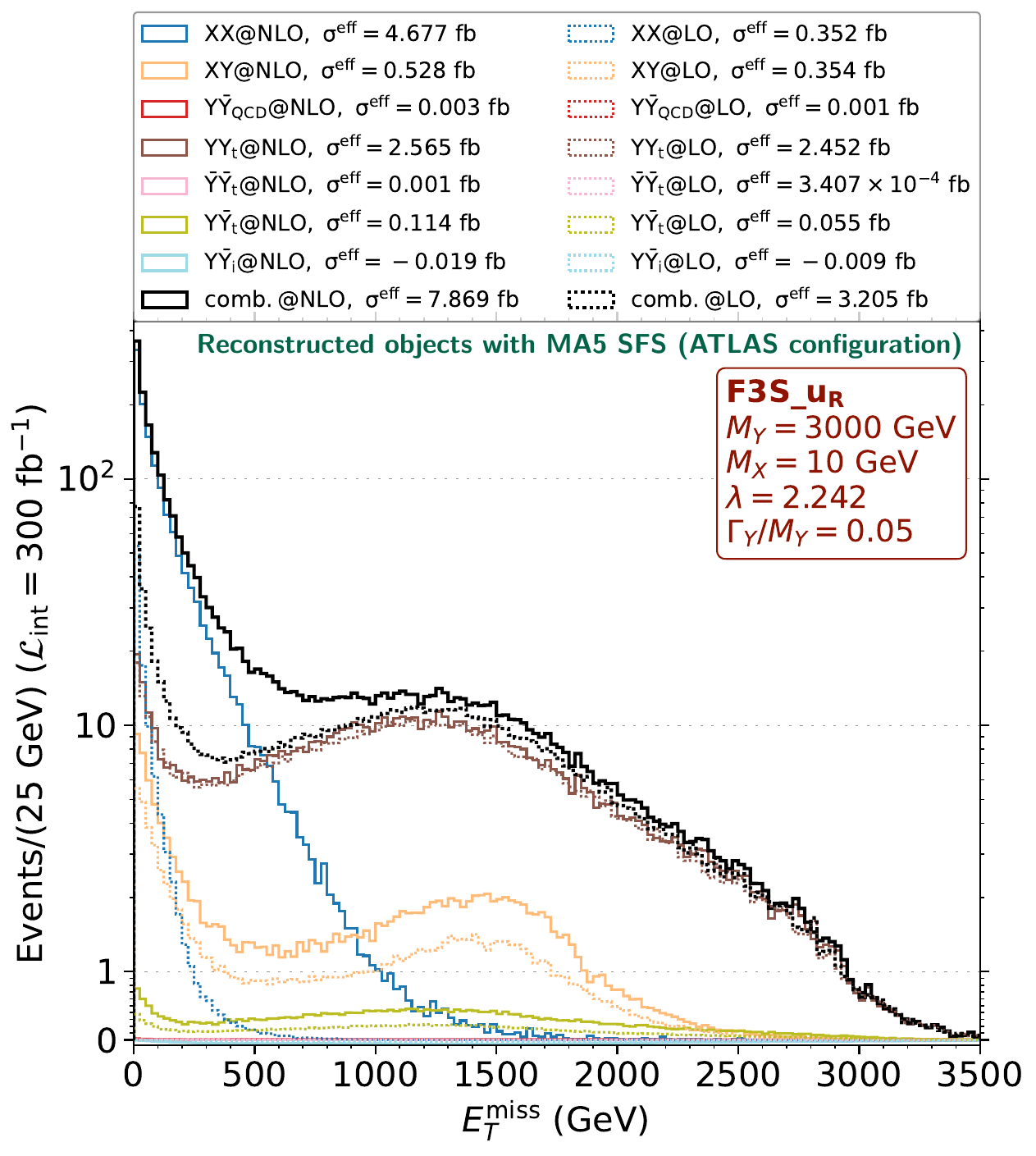}
  \includegraphics[width=.48\textwidth]{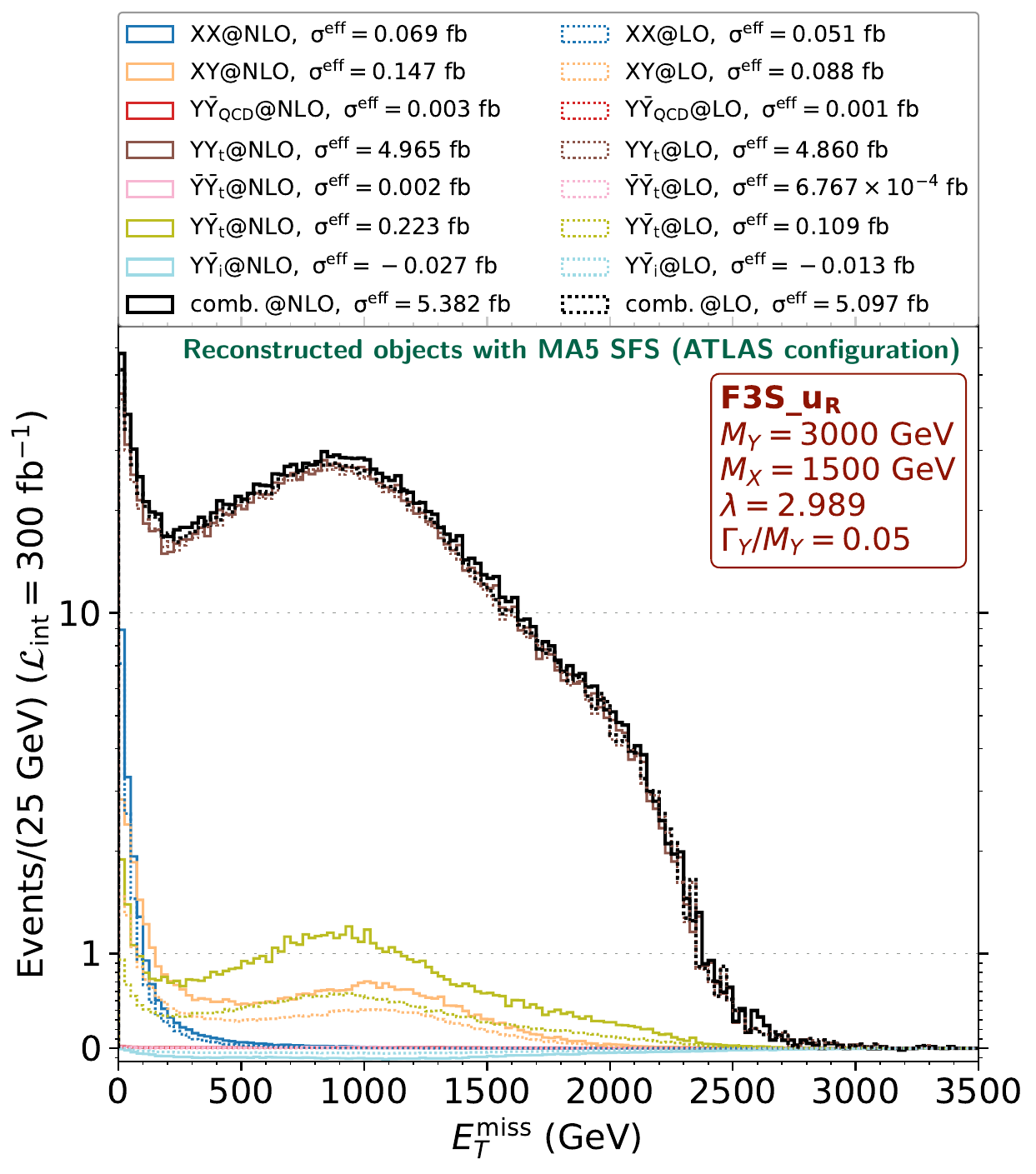}
  \caption{\label{fig:dist_S3M_u} Differential distributions for the missing transverse energy $\met$ in the case of the \lstinline{S3M_uR} (top row) and \lstinline{F3S_uR} (bottom row) scenarios, with mass configurations $\{M_Y,M_X\}=\{3000,10\}$ GeV (left column) and $\{M_Y,M_X\}=\{3000,1500\}$ GeV (right column). The distributions represent the number of signal events expected for an integrated luminosity of 300 fb$^{-1}$ of proton-proton collisions at 13~TeV, and have been obtained using the \lstinline{SFS} detector simulation module included in \lstinline{MadAnalysis 5}  with the default ATLAS configuration. The contributions of each channel with a cross section larger than 1~ab at both NLO and LO are shown, together with the resulting combined distribution.}
\end{figure}

In figure~\ref{fig:dist_S3M_u}, we consider the \lstinline{S3M_uR} and \lstinline{F3S_uR} scenarios, and we show the relative contributions of the various production channels to the $\met$ distribution for two different mass configurations close to the recast bounds at NLO: $M_Y=3$ TeV and $M_X=10$ or $1500$ GeV. In both cases, the total width of the mediator is set to 5\% of its mass (see figure \ref{fig:1stgen}), with the $\lambda$ coupling determined accordingly. The differing mass gaps between the mediator and the DM result in significantly distinct shapes for the combined distributions. As already pointed out in \cite{Arina:2020udz}, at both LO and NLO the $XX$ contribution dominates in the low $\met$ region, but the size of its contribution decreases rapidly with increasing $\met$ values, making other contributions relatively more relevant. In the \lstinline{S3M_uR} setup with a large mass gap, the $XY$ contribution becomes dominant around $\met \sim 1$ TeV, peaks between $1.3 - 1.4$ TeV, and then decreases with less than one event per 25 GeV bin above approximately $1.5$ TeV for an integrated luminosity of $300\,\text{fb}^{-1}$. In contrast, for the \lstinline{F3S_uR} scenario, the PDF-enhanced $YY_t$ contribution (in which same-charge mediators are produced via $t$-channel DM exchange) dominates above 600 GeV and strongly contributes to shape the $\met$ distribution in the high $\met$ range, as described in \cite{Arina:2023msd} (see also, \eg, ~\cite{Garny:2013ama}). When the DM state is half the mediator mass, the $YY_t$ contribution for \lstinline{S3M_uR} again dominates, but this time already for $\met \sim 200$ GeV, with a peak around $1$ TeV, and a fall below one event per bin above $1.7$ TeV for an integrated luminosity of $300\,\text{fb}^{-1}$. In contrast, for the \lstinline{F3S_uR} setup, $YY_t$ is always, and by far, the dominant contribution.

The total signal cross section in the considered \lstinline{S3M_uR} cases varies significantly between the two mass configurations. Including all contributions to the signal, it is approximately $40$ ($50$) times larger at NLO (LO) for $M_X=10$~GeV compared to $M_X=1500$ GeV. However, we must keep in mind that experimental selections for DM searches often impose strong cuts on the missing transverse energy. In contrast, for the considered \lstinline{F3S_uR} scenarios the $YY_t$ contribution scales with a weak dependence on the DM mass (for a fixed width-to-mass ratio of the mediator), while the $XX$ cross section significantly depends on the benchmark point. The latter however peaks in a kinematic regime featuring low $\met$ and low-$p_T$ jets, and the bulk of the related events are thus usually cut away by experimental searches. The impact of the $XX$ channel on the overall number of selected signal events is therefore mild. The markedly different shapes depicted in figure~\ref{fig:dist_S3M_u} motivate a deeper investigation into whether scenarios can be distinguished with sufficient accuracy once appropriate selection criteria and cuts are imposed, particularly for benchmarks with similar effective cross sections. We explore this aspect in detail in the rest of this section.

\begin{figure}
  \centering
  \includegraphics[width=.49\textwidth]{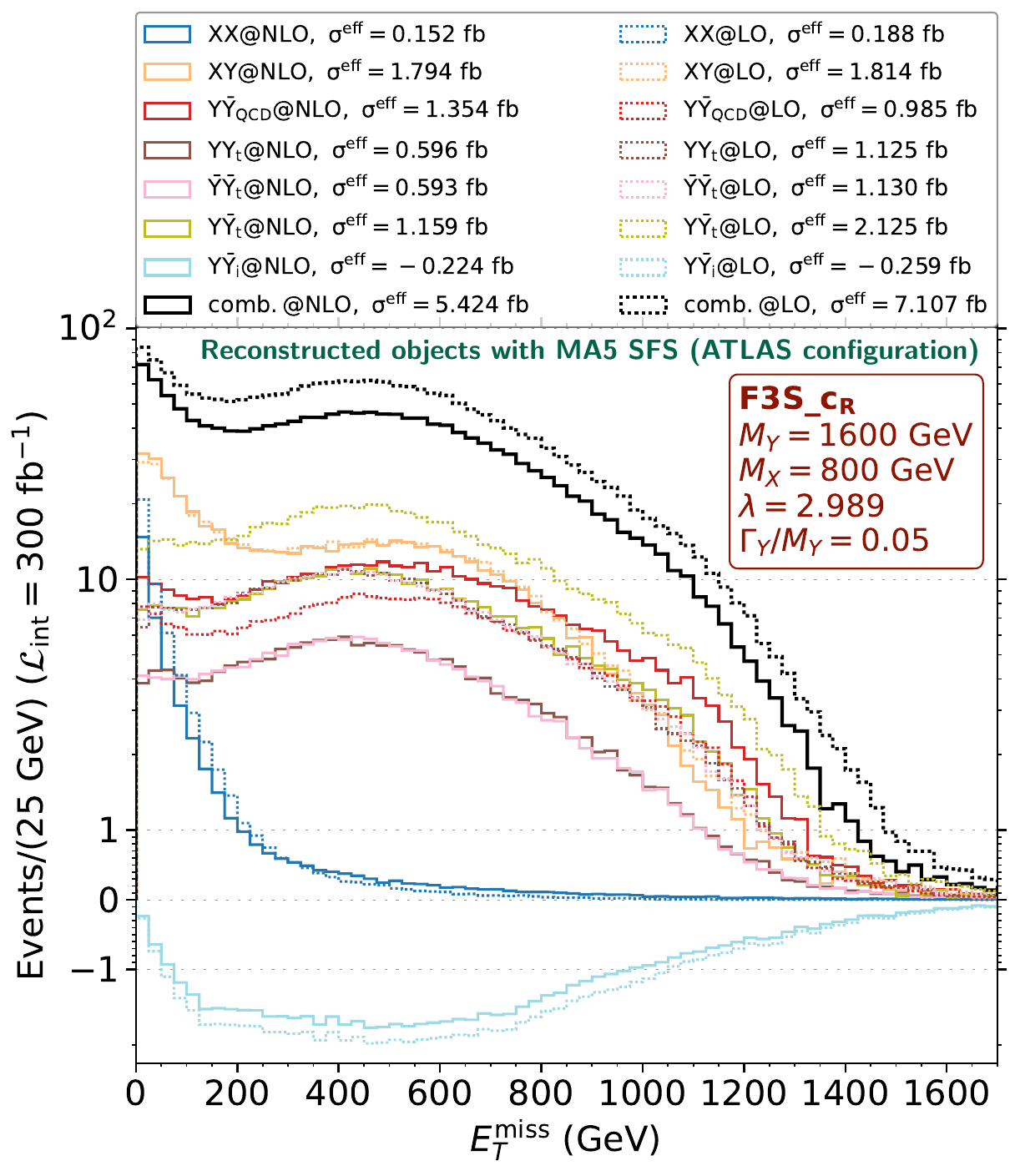}\hfill
  \includegraphics[width=.49\textwidth]{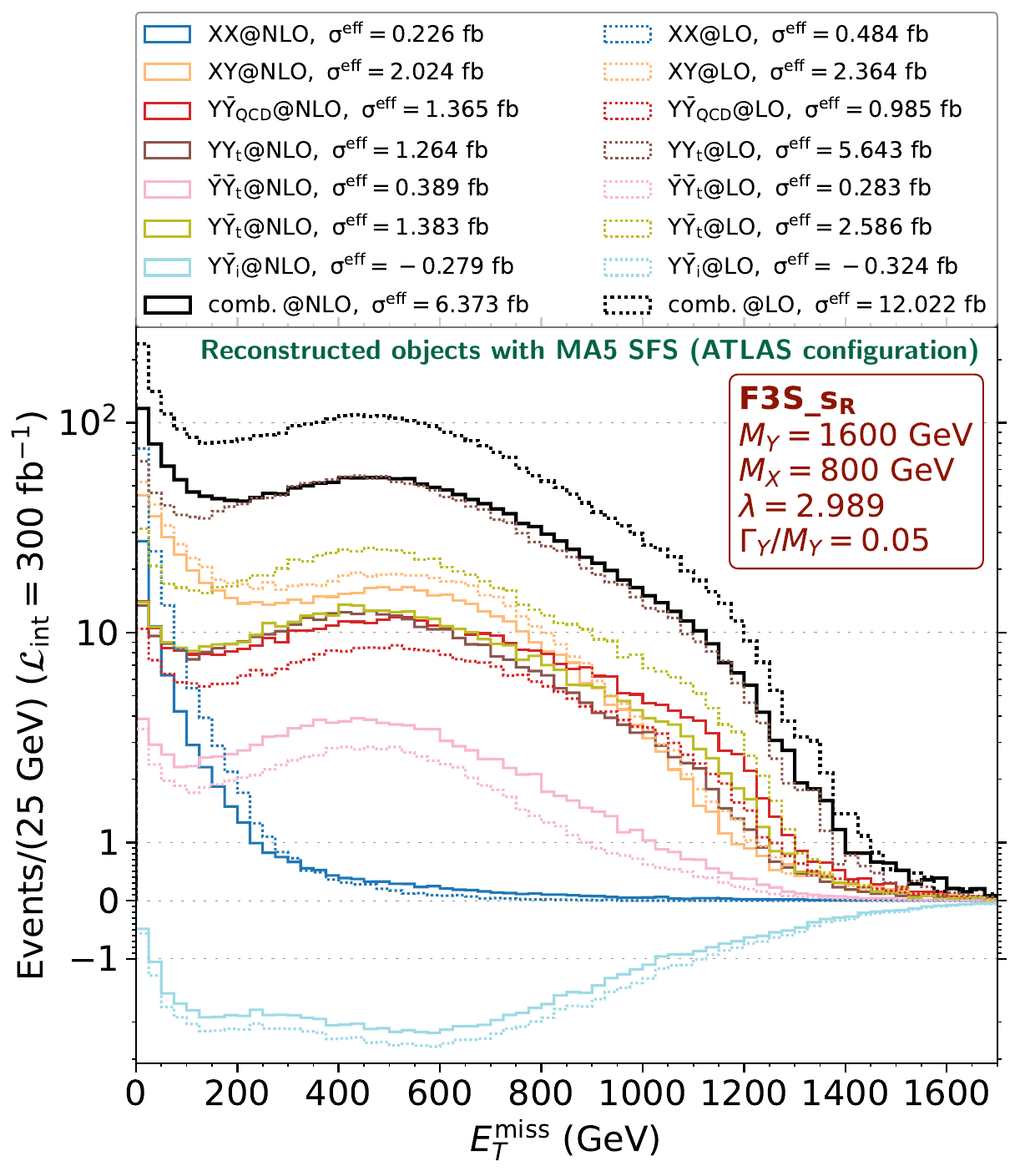}
  \caption{\label{fig:dist_F3S_cs} Same as in figure~\ref{fig:dist_S3M_u} but for the \lstinline{F3S_cR} (left panel) and \lstinline{F3S_sR} (right panel) scenarios, and the mass spectrum $\{M_Y,M_X\}=\{1600,800\}$~GeV.}\vspace{.3cm}
 \includegraphics[width=.49\textwidth]{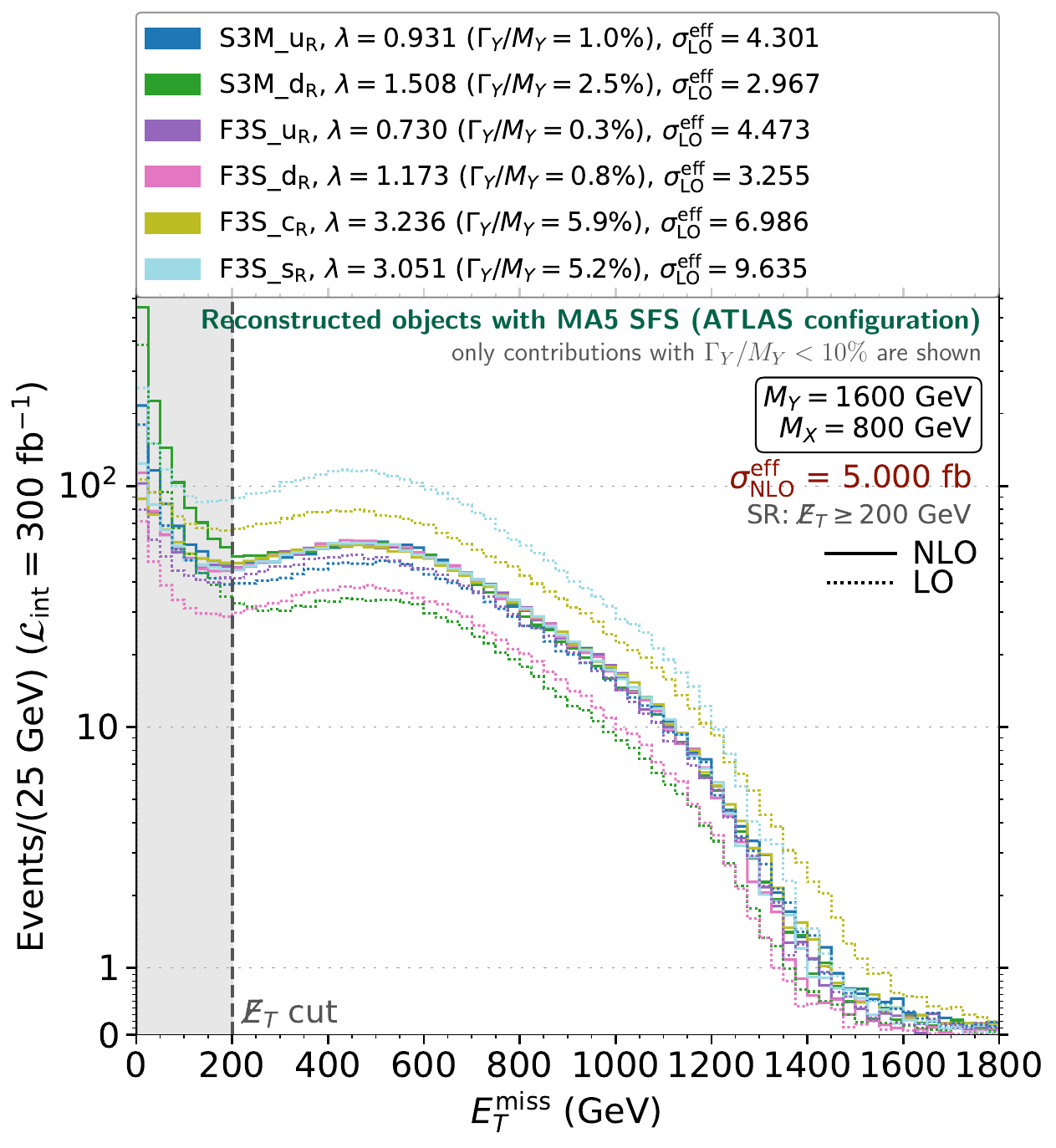}\hfill
  \includegraphics[width=.49\textwidth]{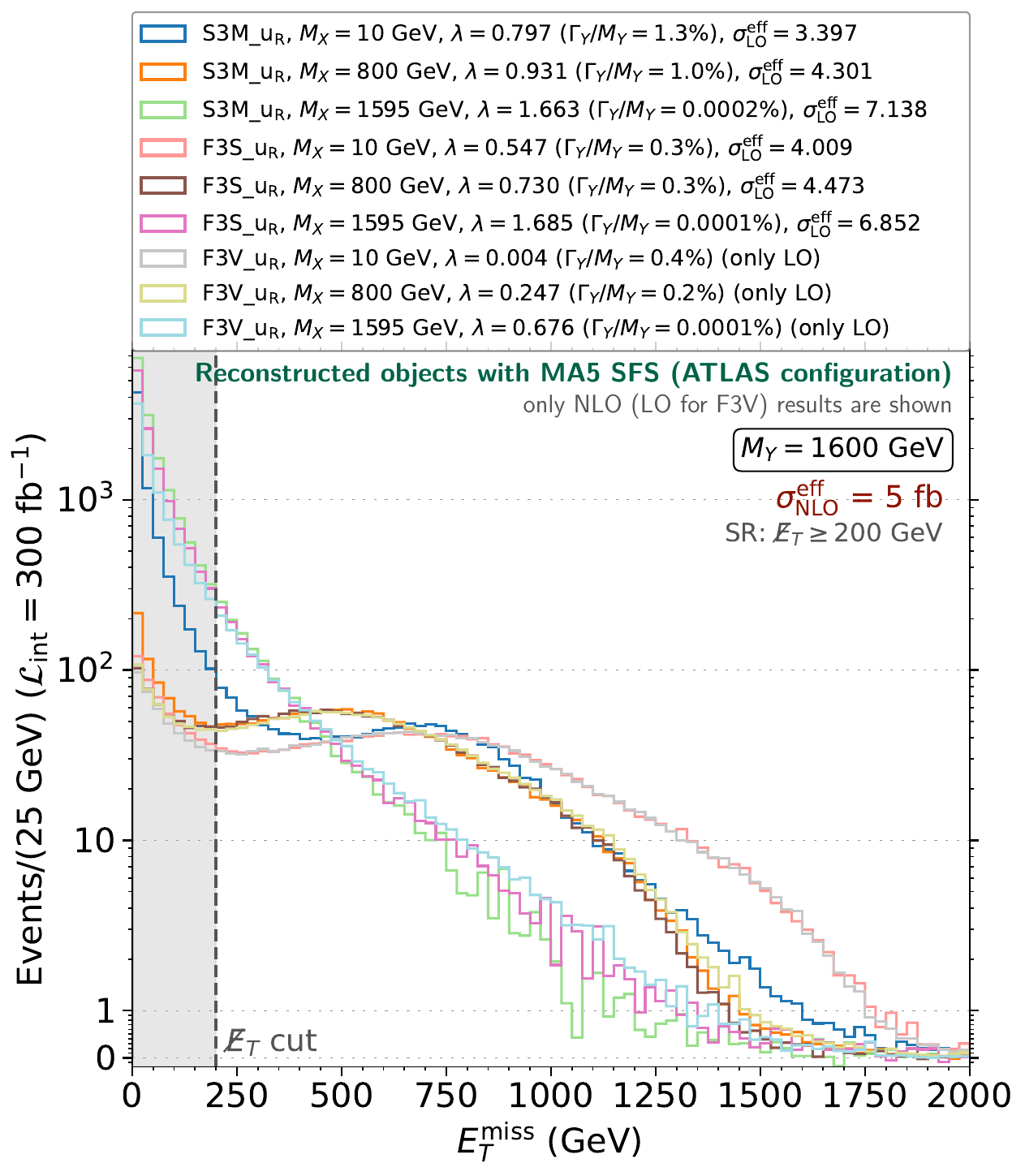}
  \caption{\label{fig:dist_effxs} Differential distributions for the missing transverse energy $\met$ for different $t$-channel DM scenarios in which the mediator width is less than 10\% of its mass. The coupling $\lambda$ is determined to obtain an NLO total signal cross section of 5 fb after the selection cut $\met>200$ GeV. We either consider NLO and LO predictions for mediator and DM masses fixed to 1600 GeV and 800 GeV respectively (left panel), or we compute NLO predictions for a scenario in which we assume a mediator mass of 1600~GeV, interactions with the up quark only, and DM masses of 10, 800 and 1595~GeV (right panel). In this case, the LO rates are indicated in the figure's legend.}
\end{figure}

Figure~\ref{fig:dist_F3S_cs} presents the missing transverse energy distributions obtained in a scenario where $\{M_Y, M_X\} = \{1600, 800\}$ GeV, but this time for the second-generation models \lstinline{F3S_cR} (left panel) and \lstinline{F3S_sR} (right panel). As discussed in section \ref{sec:bounds_2nd}, PDFs play a critical role in determining the differential $K$-factors and the bounds on the mediator and the DM states. At the differential level, the impact arises from the varying weights of the different contributions to the signal. While the total cross sections and the NLO distributions are similar in the two scenarios, the \lstinline{F3S_sR} LO cross section is about $1.7$ times larger than that obtained in the \lstinline{F3S_cR} case, primarily due to the same-charge $YY_t$ contribution. Additionally, we can note that the interference between $YY_{\rm{QCD}}$ and $Y\bar Y_t$ has a minimal effect of the order of a few percent, reducing for example the total event count by a few units per bin at the peak.

Figure~\ref{fig:dist_effxs} explores the impact of a simple kinematic cut of $\met > 200$ GeV. We consider various $t$-channel DM scenarios in which the $\lambda$ coupling value is tuned to yield an effective cross section including all signal contributions of $\sigma^{\rm eff} \equiv \sigma \times \varepsilon = 5$~fb, with $\varepsilon$ representing the imposed missing energy cut efficiency. In the left panel, we focus on split spectra in which we fix the mediator mass to $M_Y = 1600$~GeV and the dark matter mass to $M_X = 800$ GeV, scanning across a variety of \lstinline{S3M} and \lstinline{F3S} models featuring couplings to first-generation and second-generation quarks. Other scenarios, for such a choice of masses, would require large couplings leading to mediator widths incompatible with the NWA, and these are thus not represented in the figure. The resulting NLO distributions show identical shapes in the $\met$ region above the cut, within statistical fluctuations, making them indistinguishable. Conversely, the LO distributions differ significantly, underscoring the importance of including NLO modelling for accurately describing final-state kinematics and total cross sections.

The right panel of the figure focuses on interactions involving the up quark, with a scenario in which we have fixed $M_Y$ to 1600~GeV and varied $M_X$ across the light regime (1 GeV), intermediate regime (800 GeV), and compressed regime (1595 GeV). For NLO results (with LO cross sections provided in the legend for information), scenarios yielding an identical total number of signal events exhibit visibly distinct kinematic features, already in regions with substantial event counts. This clearly reflects the different weights of the various signal contributions, which depend on the size of the $\lambda$ coupling (with the exception of $YY_{\rm{QCD}}$) and its impact on the determination of the final-state properties. For instance, in the $M_X = 1595$~GeV case, the $XX$ contribution dominates, irrespective of the mediator and DM spins. Similarly, the $M_X = 800$ GeV distributions show consistent shapes across models. In the light DM case, the scalar mediator (\lstinline{S3M}) exhibits a significantly larger $YY_t$ contribution compared to fermion mediators (\lstinline{F3S} and \lstinline{F3V}), reflecting the different cross section scaling with the DM mass, as also seen in figure~\ref{fig:dist_S3M_u}.

\subsubsection{Leptophilic models}\label{sec:leptoph}
\begin{figure}
  \centering
  \includegraphics[width=0.48\textwidth]{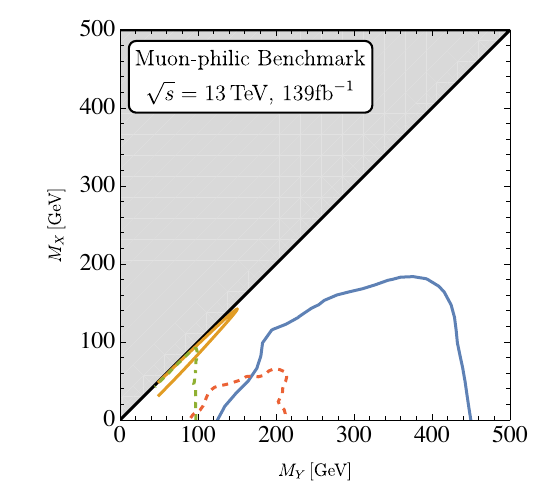}
  \includegraphics[width=0.48\textwidth]{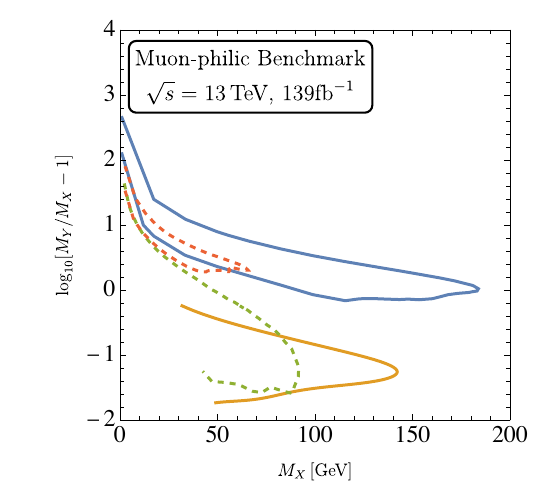}
  \caption{Current LHC exclusions on the muon-philic benchmark models considered. The exclusion curves correspond to  the reinterpretation of the results of the run~2 ATLAS searches from~\cite{ATLAS:2019lff} (blue) and~\cite{ATLAS:2019lng} (orange), as well as from the LHC Run~1 ATLAS results from~\cite{ATLAS:2014zve} (dashed red). The dashed green curve represents the limits from the \href{https://lepsusy.web.cern.ch/lepsusy/www/sleptons_summer04/slep_final.html}{LEP2 SUSY Working Group}.  \label{fig:current-exclusions}}
\end{figure}

Before discussing non-minimal models, we briefly consider signals originating from prompt mediator decays in leptophilic scenarios, focusing on a model where Majorana dark matter couples exclusively to right-handed muons (as discussed in section~\ref{sec:model_minimal}). In the region of the parameter space where the observed relic abundance is achieved via thermal freeze-out (including potential co-annihilation effects), mediators produced via Drell-Yan processes at the LHC always decay promptly. Consequently, the Yukawa coupling $\lambda$ does not directly influence the LHC phenomenology. 

The main collider signature in this scenario consists of two opposite-sign leptons accompanied by missing transverse energy. If the mass splitting between the $X$ and $Y$ states is small, the resulting leptons tend to be soft, making their reconstruction and identification challenging. This parameter space is best probed using dedicated strategies directly targeting these soft leptons, or through searches exploiting initial-state radiation (ISR) which boosts the system and produces harder leptons. In figure~\ref{fig:current-exclusions}, we present the most stringent ATLAS and CMS constraints on promptly decaying mediators within this benchmark scenario. The ATLAS search for slepton pair production in the decoupled region~\cite{ATLAS:2019lff} constrains dark matter masses up to approximately 200\,GeV (corresponding to mediator masses around 450\,GeV) in the regime where the mediator mass is roughly twice that of the dark matter. Additionally, the ATLAS ISR-based search for soft leptons~\cite{ATLAS:2019lng} can probe dark matter masses up to about 150\,GeV when the mass splitting is around 10\%. These constraints significantly improve upon LEP limits; however, viable regions of the parameter space remain, particularly at higher masses and for intermediate mass splittings ($M_Y/M_X-1 \sim 0.1-2$), which remain challenging to probe.

\subsection{Non-minimal models -- prompt decays}\label{sec:collider_nonmin}

\subsubsection{Flavoured dark matter at the LHC}\label{sec:collider_flavour}

Compared to the minimal models discussed previously, DM models incorporating DM Flavour Violation (see section~\ref{sec:DMFV}) feature a significantly richer phenomenology at the LHC due to their more complex flavour structure. Similar to non-flavoured models, mediator pair-production remains the dominant production channel across much of the parameter space. However, both the production cross section and the mediator decay depend on the model's flavour structure. In this section, we focus on quark-flavoured DM within the DMFV framework, while for LHC constraints on lepton-flavoured DM models we refer to~\cite{Chen:2015jkt, Acaroglu:2022hrm, Acaroglu:2022boc, Acaroglu:2023cza}. Moreover, for simplicity, we assume that the DM flavours $X_i$ are approximately mass-degenerate. In contrast, non-degenerate scenarios lead to cascade decays of new particles, which opens the possibility for LLP signatures as those discussed in section~\ref{sec:LLP}.

Mediator pair-production at the LHC proceeds both via QCD interactions and through $t$-channel exchange of the DM flavour triplet $X = (X_1, X_2, X_3)$. The latter process becomes particularly relevant when the coupling of $X$ to the first-generation quarks is sizeable. In models with real DM representations, this $t$-channel exchange also enables same-sign mediator pair production~\cite{Acaroglu:2021qae} that could be enhanced by valence quark-pair contributions as discussed in section~\ref{sec:collider_1stgen}. On the other hand, the mediator $Y$ decays into an SM quark and a dark flavour state $X_i$ via the coupling matrix $\lambda$, the different branching ratios depending on the relative sizes of the elements of $\lambda$. This typically leads to several relevant signatures of mediator pair production. Consequently, LHC constraints on mediator and DM masses are generally weaker than in the single-generation scenarios discussed previously,  where only a single decay model prevails.

\begin{figure}
  \includegraphics[width=0.49\textwidth]{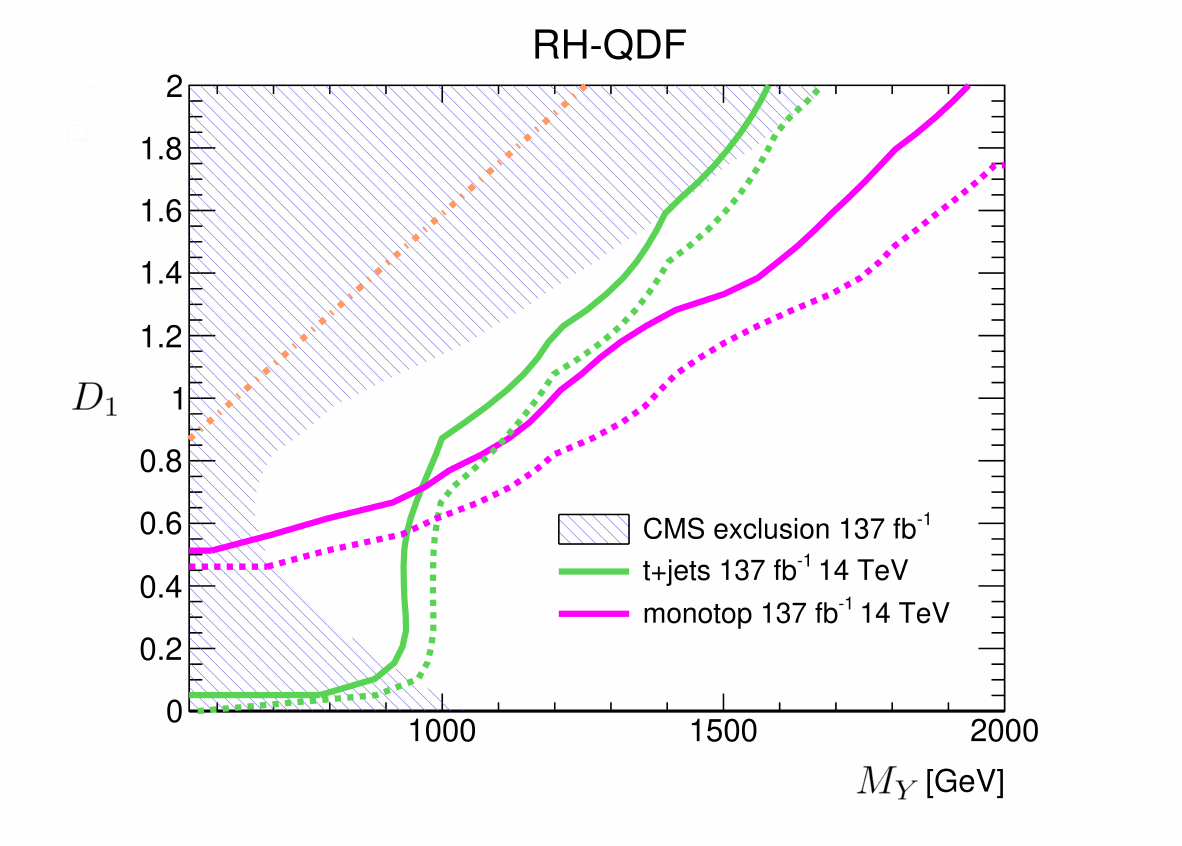}\hfill
  \includegraphics[width=0.49\textwidth]{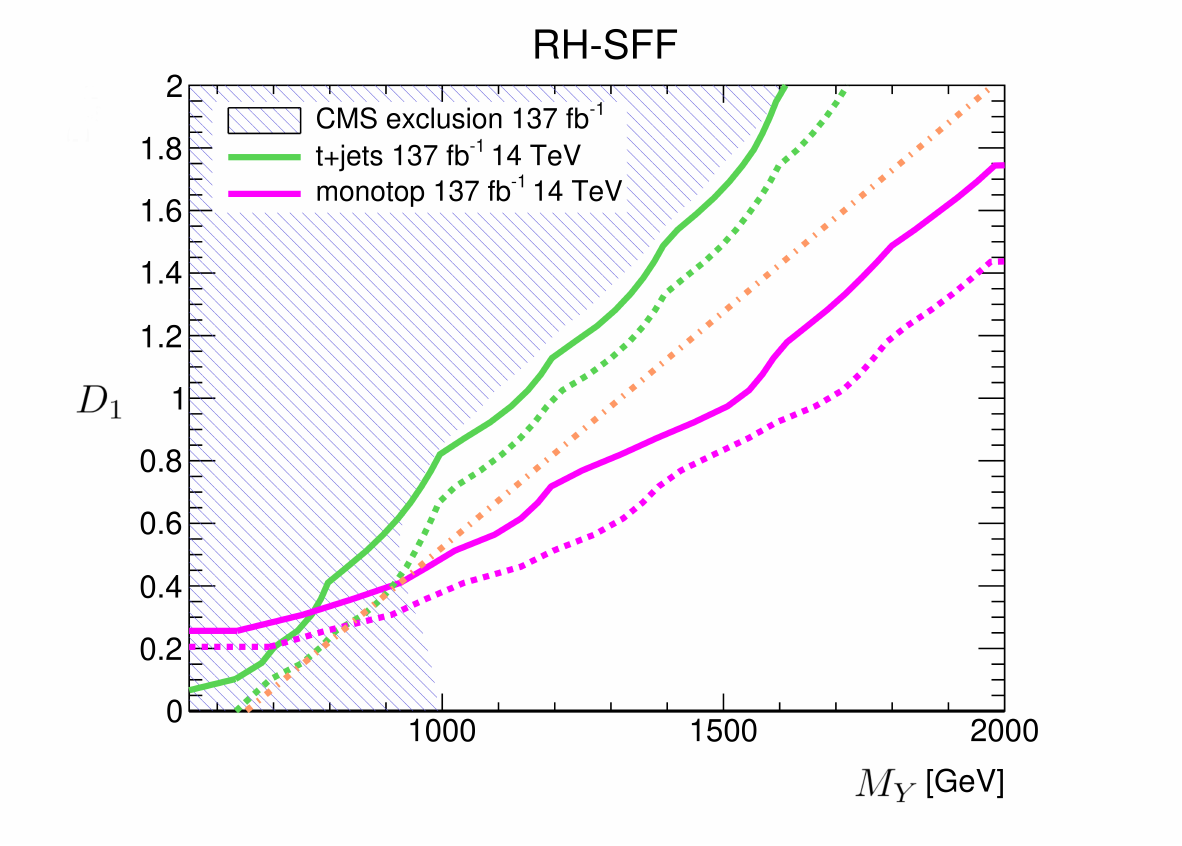}
  \caption{Expected limits from single-top final states on a model with Dirac flavoured DM coupling to right-handed up-type quarks, shown for the QDF (left) and FFS (right) benchmark scenarios of DM freeze-out. Solid lines indicate the expected reach in the $(M_Y, D_1)$ plane for $tj+\met$ and $t+\met$ analyses, assuming an integrated luminosity of 137\,fb$^{-1}$ at the 14\,TeV LHC. The excluded regions lie to the left of the curves. Dashed lines show the corresponding projections for 300\,fb$^{-1}$, and results for 3000\,fb$^{-1}$ can be found in \cite{Blanke:2020bsf}. The shaded region represents the exclusion derived from a recast of the CMS search~\cite{CMS:2019zmd}, and the orange dash-dotted lines indicate parameter values that yield the correct relic abundance. Figure adapted from \cite{Blanke:2020bsf}.\label{fig:single-top}}
\end{figure}

\begin{figure}
  \centering{\includegraphics[width=0.7\textwidth]{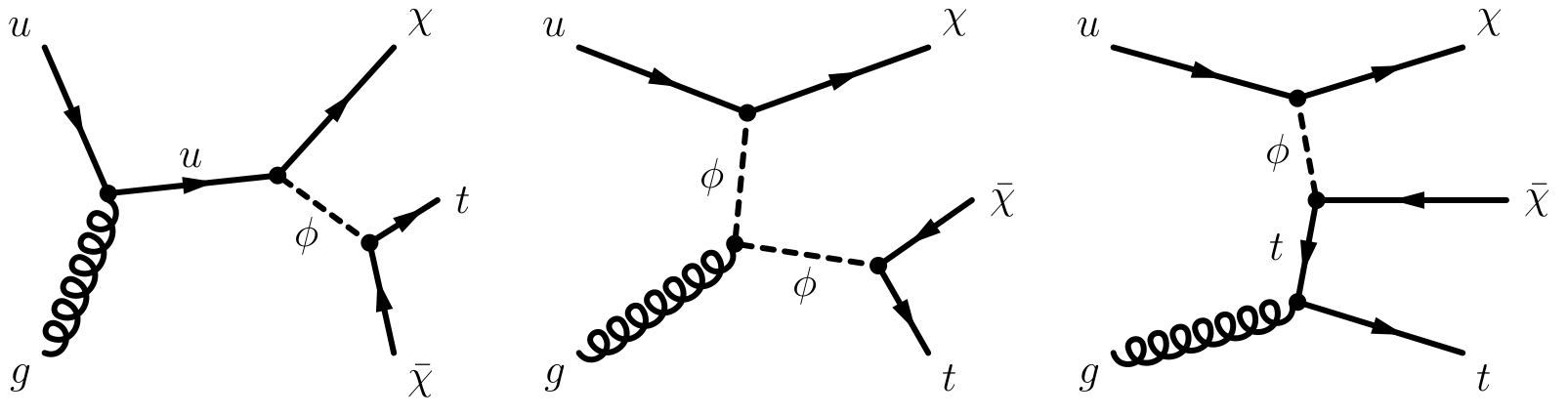}}
  \caption{Representative Feynman diagrams illustrating the production of two dark matter particles in association with a single top quark. Adapted from \cite{Blanke:2020bsf}.\label{fig:monotop} }
\end{figure}

The case of Dirac flavoured DM coupling to right-handed down-type quarks has been explored in~\cite{Agrawal:2014aoa}. The strongest LHC constraints arise from searches for supersymmetric squarks in final states with either two light jets or two $b$-tagged jets, both accompanied by missing transverse energy. Additionally, the compressed region is constrained by monojet searches. On the other hand, \cite{Blanke:2017tnb, Jubb:2017rhm, Blanke:2020bsf} examined scenarios with couplings to right-handed up-type quarks. Here, the most stringent bounds stem from recasts of supersymmetric squark searches in the $jj + \met$ and $t\bar t + \met$ final states, with the precise exclusion limits depending on the flavour structure of the matrix $\lambda$. The blue shaded regions in figure~\ref{fig:single-top} show the resulting constraints, derived from the experimental search~\cite{CMS:2019zmd}, on the mediator mass $M_Y$ as a function of the first-generation coupling strength $D_1$. The limits are shown for two benchmark DM freeze-out scenarios, a first one featuring quasi-degenerate freeze-out (QDF) and a second one predicting single-flavour freeze-out (SFF), their cosmology being discussed in section~\ref{sec:CosmConstNonMin_flav}. QDF setups consist in scenarios in which all three dark flavours are quasi-degenerate, hence participating all together in the thermal freeze-out process through their combined (co-)annihilations. On the other hand, SFF configurations refer to scenarios where only one particle flavour remains thermally active and undergoes freeze-out, while other flavour states are too heavy to contribute. In addition to these flavour-conserving final states, flavoured DM models also give rise to flavour-violating signatures. For instance, \cite{Blanke:2020bsf} proposed search strategies dedicated to single-top final states, which arise in flavoured DM models coupling to up-type quarks and where mediator pair-production can lead to final states such as $tj + \met$. Similarly, the monotop signature~\cite{Andrea:2011ws}, $t + \met$, can also arise, as illustrated in the Feynman diagrams in figure~\ref{fig:monotop}. The green and magenta lines in figure~\ref{fig:single-top} show that these search strategies significantly extend the LHC reach beyond the limits obtained from the flavour-conserving channels. Notably, in the case of the SFF benchmark, the monotop search could probe the thermal freeze-out hypothesis, as indicated by the orange dash-dotted line. Finally, Dirac flavoured DM coupling to left-handed quarks has been investigated in~\cite{Blanke:2017fum, Blanke:2020bsf}. Here, constraints from $jj + \met$ are significantly stronger due to the larger multiplicity of possible partonic final states. Nevertheless, the LHC reach can be further extended through dedicated searches for single-top final states~\cite{Blanke:2020bsf}. Interestingly, this model predicts a unique $tb + \met$ signature for which the dedicated search strategy developed in~\cite{Blanke:2020bsf} was found to only have limited sensitivity.

\begin{figure}
  \includegraphics[height=0.45\textwidth]{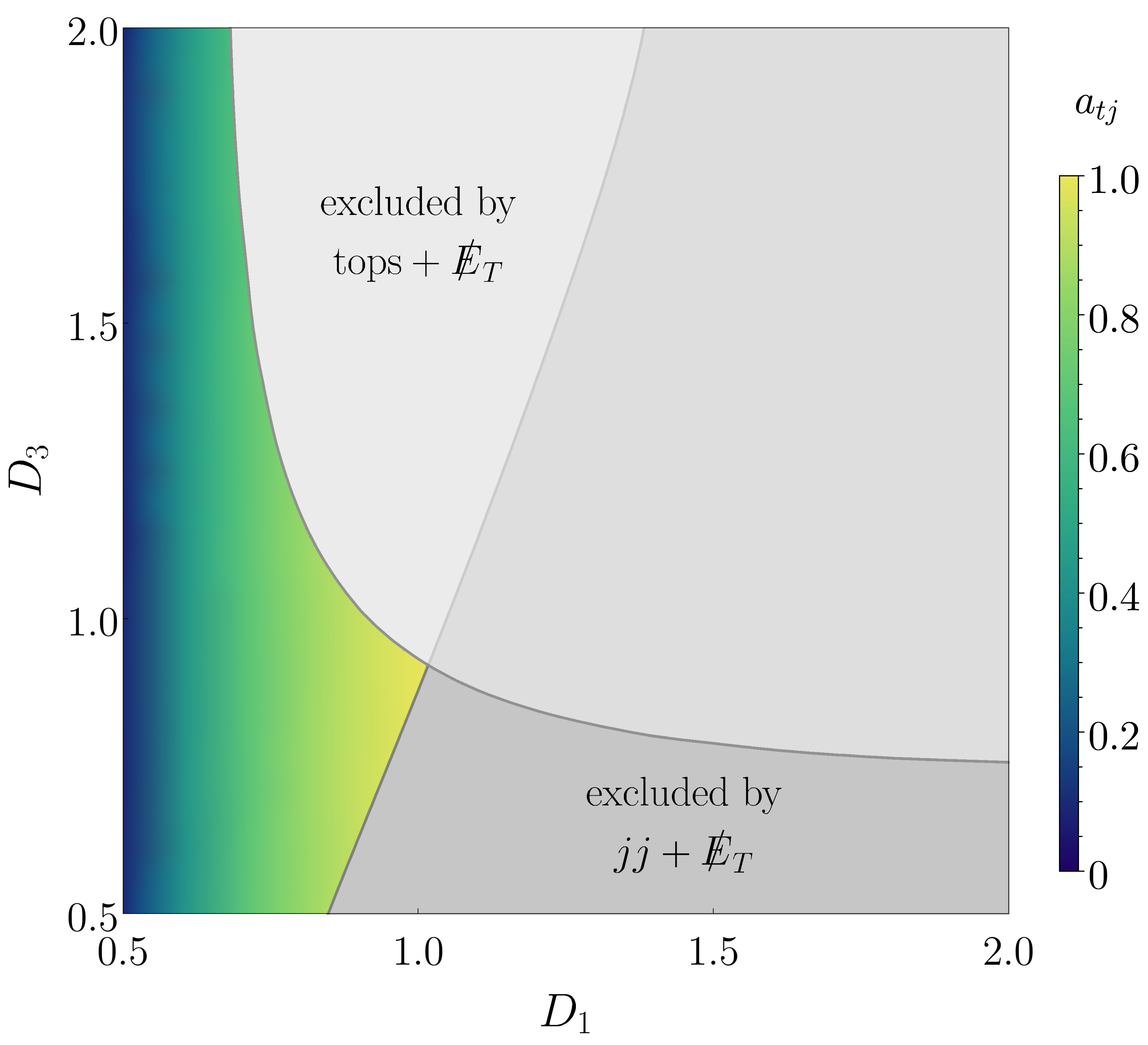}\hfill
  \includegraphics[height=0.45\textwidth]{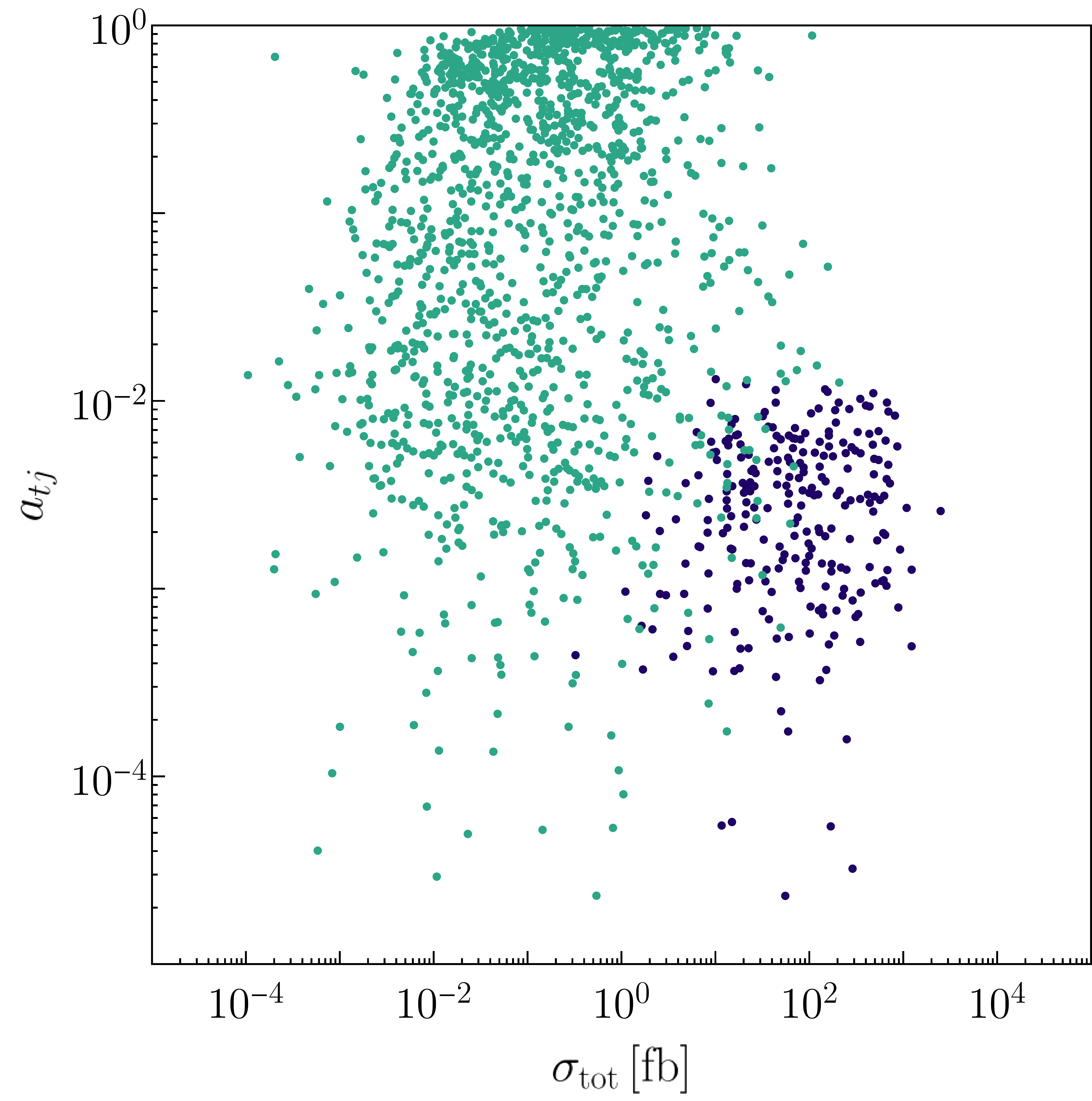}
  \caption{Predictions for the single-top charge asymmetry $a_{tj}$ in a model of Majorana flavoured DM coupling to the three flavours of right-handed up-type quarks.  We present predictions for $a_{tj}$ (colour-coded) at the 14\,TeV LHC, compared with current constraints from tops$+\met$ (light grey) and jets$+\met$ (dark grey) searches, for $D_2 = 0$, $M_Y = 1200$\,GeV, and $M_X = 400$\,GeV (left); we also show the correlations between predictions for $a_{tj}$ and the total single-top cross section $\sigma_\text{tot}$ as defined in the denominator of \eqref{eq:atj}, for viable scenarios featuring canonical DM freeze-out (green) and conversion-driven freeze-out (blue).  Here, $D_i$ represent the DM coupling strengths to the $i^\mathrm{th}$ generation, and the figures have been adapted from \cite{Acaroglu:2023phy}. \label{fig:atj}
  }
\end{figure}

Turning to scenarios with Majorana flavoured DM, models with coupling to right-handed up-type quarks have been studied in~\cite{Acaroglu:2021qae, Acaroglu:2023phy}. The key difference between a Majorana scenario and the Dirac case is the possibility of same-sign mediator pair production via $t$-channel exchange of $X$ states with a mass insertion. If $X$ has significant coupling to first-generation quarks, this process is strongly enhanced due to the large up-quark PDF in the proton. As a result, constraints from charge-insensitive searches, such as those exploiting the $jj + \met$ and $t\bar t + \met$ signatures with zero or one lepton in the final state, are considerably stronger (at least for non-zero DM masses $M_X$). Additionally, in final states containing top quarks where the charge can be reconstructed, same-sign mediator pair production leads to unexplored smoking-gun signatures. One such process is the production of two positively charged top quarks in association with missing transverse energy, $tt + \met$, that could be probed through dileptonic decays of the $tt$ system~\cite{Acaroglu:2021qae}. However, due to the small leptonic branching fraction of the top quark, a naive estimate suggested that this channel is not competitive with standard searches, necessitating a more refined analysis~\cite{Acaroglu:2023phy}. A promising alternative could rely on the charge asymmetry $a_{tj}$ of single-top final states~\cite{Acaroglu:2023phy}, defined as
\begin{equation}\label{eq:atj}
    a_{tj} = \frac{\sigma(tj+\met)-\sigma(\bar tj+ \met)}{\sigma(tj+\met)+\sigma(\bar tj+\met)}\,.
\end{equation}
The single-top search strategy developed in~\cite{Blanke:2020bsf} can be directly applied here, as it relies on leptonic top decays. In Dirac models, $a_{tj}$ is then expected to be close to zero, making this asymmetry a powerful discriminator between Majorana and Dirac flavoured DM. As shown in figure~\ref{fig:atj}, large positive values, $a_{tj} \sim 1$, are possible in regions of parameter space where the total single-top cross section reaches up to 100\,fb. These regions are not excluded by standard LHC searches, as well as by the cosmological probes discussed in section~\ref{sec:CosmConstNonMin_flav}. In addition, they can still accommodate the observed DM relic density through canonical freeze-out. Conversely, in conversion-driven freeze-out scenarios, $a_{tj}$ is found to be close to zero due to the specific coupling structure of this mechanism. Such cases should instead be probed through LLP signatures, as discussed in section~\ref{sec:LLP}.

\subsubsection{Composite dark matter at the LHC}\label{sec:collider_composite}
The analysis presented so far has focused on the minimal $t$-channel dark matter scenarios described in section~\ref{sec:model_minimal}. In the current section, we explore the impact of non-minimality by considering the composite constructions detailed in section~\ref{sec:compositeDM}. We begin this exploration with the class of models described by the Lagrangian~\eqref{eq:lagtopcompoDM}, which introduces two mediators: a $\mathbb{Z}_2$-odd $t$-channel mediator $Y \equiv Y_{\scriptscriptstyle t}$ and a $\mathbb{Z}_2$-even vector-like quark $Y' \equiv Y^\prime_{\scriptscriptstyle t}$. This setup hence corresponds to an \lstinline{F3S_tR} scenario extended by the addition of the $Y'$ mediator, whose presence is subsequently expected to modify the bounds derived in section~\ref{sec:bounds_3rd}. As previously determined, $YY$ pair production, followed by the decay $Y \to X t$, dominates the total new physics signal relevant for DM production at the LHC, with sub-leading contributions from $XX$ dark matter pair production and $XY$ associated production. As a result, scenarios with light dark matter ($M_X \lesssim 400$~GeV) are excluded if the mediator mass is smaller than approximately $1.3$~TeV. For intermediate dark matter masses ($M_X \in [500, 600]$~GeV), the constraints that could be imposed on the mediator are significantly weakened. Finally, heavier dark matter scenarios are in principle reachable; for instance, it was shown in \cite{Cornell:2022nky} that for $M_X \sim 700$~GeV, top partners with masses up to $1$~TeV could be excluded, provided that the spectrum is not too compressed (\ie\ $M_Y > M_X + M_t$, where $M_t$ is the mass of the top quark). The analysis of the impact of searches for dark matter in the $t\bar t + \met$ channel is however left for future work, due to the absence of validated implementations of corresponding full Run~2 ATLAS and CMS searches in public recasting tools.

The introduction of $Y'$ mediators must comply with ATLAS and CMS searches for vector-like top partners decaying into a SM top quark and an electroweak boson. These searches generally target signatures of the process $pp \to Y'Y' \to ttVV$, where $V$ denotes a $Z$ boson, a $W$ boson, or a Higgs boson $h$, and where the two produced vector-like quarks decay similarly. Consequently, we enforce $M_{Y'} \geq 1.3$~TeV. The presence of the $Y'$ mediator can, in principle, affect dark matter production in association with a jet (\ie\ the $XX$ channel) through additional box diagrams involving both $Y$ and $Y'$ mediators in the loops. However, within the parameter space favoured by cosmological considerations, these diagrams contribute negligibly, ensuring that new contributions to monojet or multijet + missing energy signals remain minimal. For scenarios in which $M_{Y'} > M_Y + M_X$, the decay channel $Y' \to Y X$ becomes kinematically open, leading to new contributions to the $t\bar{t} + \met$ signal through the process $pp \to Y'\bar{Y}' \to YX\bar{Y}X \to tXX\bar{t}XX$. This can potentially modify the constraints existing on both the $Y$ and $Y'$ states. However, for light dark matter, the mass of the $Y$ mediator is already independently constrained to be at least $1.3$~TeV, leaving no room for constraint weakening. For heavier dark matter, the condition $M_Y + M_X < M_{Y'}^\mathrm{min} = 1.3$~TeV (the minimum allowed mass for a vector-like quark) is never satisfied for scenarios reachable at the LHC Run~2. As a result, collider bounds on composite dark matter models remain unaffected by this first exploration of non-minimality.

We extend our exploration of non-minimality by considering the Lagrangian~\eqref{eq:lagchacal}, which incorporates top partial compositeness into the \lstinline{F3S_tR} class of models. This results in a scenario featuring three vector-like quark mediators ($Y_t$, $Y_{Q,t}$, and $Y_{Q,b}$), one dark matter state ($X$), and two new couplings ($\lambda_{\scriptscriptstyle Q}$ and $\lambda_{\scriptscriptstyle t}$). To simplify the resulting six-dimensional parameter space, we assume equal masses for all mediators ($M_Y \equiv M_{Y_t} = M_{Y_{Q,t}} = M_{Y_{Q,b}}$) and equal couplings ($\lambda \equiv \lambda_{\scriptscriptstyle Q} = \lambda_{\scriptscriptstyle t}$).

\begin{figure}
    \centering
    \includegraphics[width=.45\textwidth]{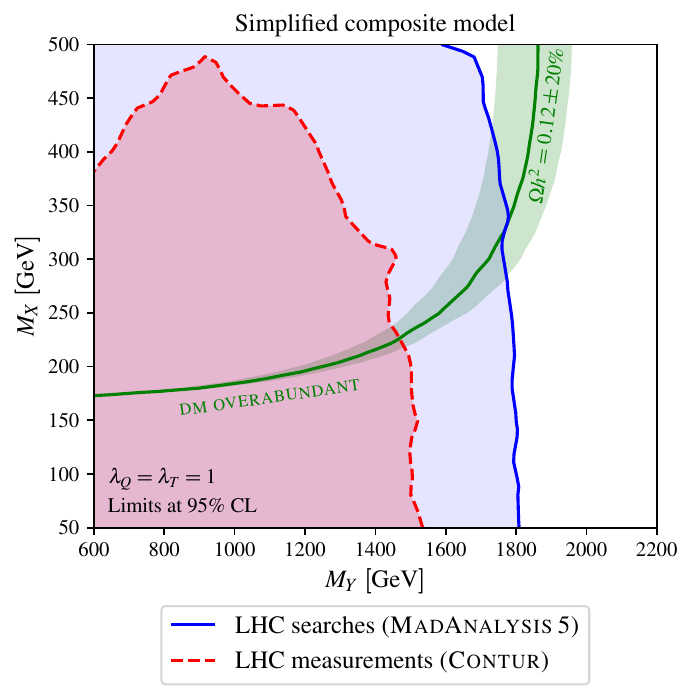} 
    \raisebox{1.1cm}{\includegraphics[width=.54\textwidth]{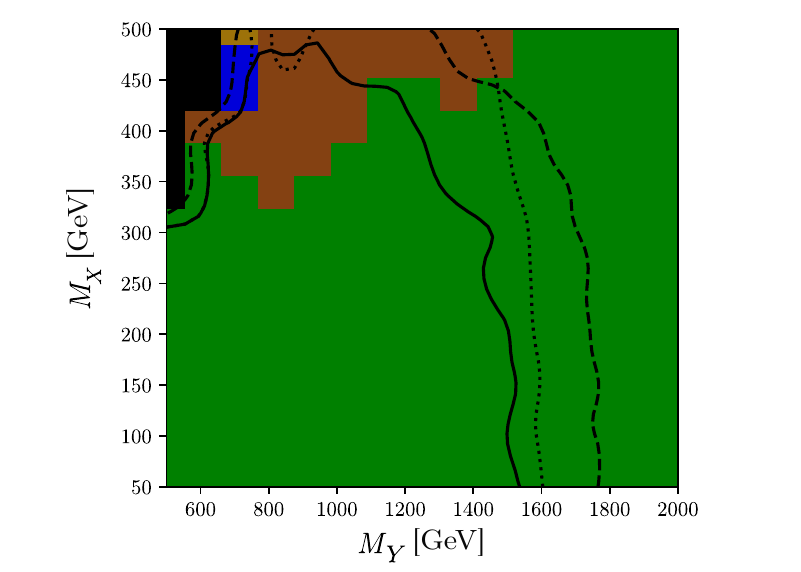}}
    \caption{Exclusion limits at 95\% confidence level (CL) for the composite DM model with partial compositeness considered. Results are shown in the $(M_Y, M_X)$ plane, where $M_Y$ is the common mediator mass, $M_X$ the DM mass, and with fixed new physics couplings of $1$. The left panel displays the region yielding the correct DM relic abundance (green), and exclusions from LHC new physics searches (blue, solid) and SM measurements (red, dashed). The right panel shows the most sensitive analysis pool at each grid point, to which we superimpose 95\% observed (solid), 68\% CL observed (dashed), and 95\% CL expected (dotted) exclusions. Analysis pools include $\ell^+\ell^-\gamma$ (light brown), $\met+$jets (green), hadronic $t\bar{t}$ (dark brown), and $\ell+\met+$jet measurements, with the black area denoting the unconstrained region.}
    \label{fig:DMcomposite}
\end{figure}

In the left panel of figure~\ref{fig:DMcomposite}, we present exclusion bounds derived from the reinterpretation of the full new physics signal in this model, which includes contributions from mediator pair production, mediator-DM associated production, and DM pair production. Each component of the signal implicitly include a sum over all possible mediator combinations, and corresponding simulations are achieved at LO for simplicity. To highlight the regions of parameter space favoured by cosmology, we additionally overlay the region in which the DM relic density predictions agree within 20\% with Planck data~\cite{Planck:2018vyg}. The constraints, based on the searches detailed in section~\ref{sec:collider_generalities} (blue contour), demonstrate that scenarios with light DM ($M_X < 500$~GeV) and mediators lighter than $1.8$~TeV are excluded. These bounds, primarily driven by the ATLAS-CONF-2019-040 search results, are significantly stronger than those reported for the \lstinline{F3S} models in figure~\ref{fig:3rdgenbis}. This enhanced sensitivity arises from the combined contributions of the three QCD production channels corresponding to $Y_t$, $Y_{Q,t}$, and $Y_{Q,b}$ pair production. Notably, while the results are obtained for a coupling value of $1$, the dominance of QCD contributions ensures that this choice has little impact on the exclusion bounds.

For comparison, we also consider constraints derived from recent detector-corrected SM measurements provided by the LHC collaborations (red exclusion). These constraints are somewhat less stringent than those resulting from searches for new physics, as the latter utilise additional kinematic variables to enhance sensitivity. In the right panel of figure~\ref{fig:DMcomposite}, we identify the most sensitive `pool’ of analyses used by \lstinline{Contur} at each point in the parameter space. The majority of the excluded region is dominated by the `missing energy plus jets' pool, which aggregates several analyses relying on this final-state signature in 13~TeV LHC data. In particular, a 13 TeV ATLAS measurement of the differential cross section for missing energy production~\cite{ATLAS:2024vqf} is responsible for most of the exclusion. The discrepancy between the expected and observed exclusions in this figure arises from a small excess in the unfolded $\met$ spectrum around 1200~GeV. Consequently, any signal favouring this kinematic regime slightly improves the data/prediction agreement, leading to a marginally weaker exclusion limit.

\subsubsection{Frustrated dark matter at the LHC}\label{sec:collider_frustrated}

\begin{figure}
    \centering
    \includegraphics[width=.35\textwidth]{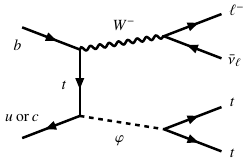} 
    \caption{Representative Feynman diagram for scalar sextet mediator production in association with a $W$ boson in the fDM model, followed by a leptonic $W$ decay and a $\varphi$ decay to same-sign top quarks. \label{fig:fDMfeyn}}
\end{figure}

In this section, we continue with our study of non-minimal models of $t$-channel dark matter. In the fDM model introduced in section~\ref{sec:fDM}, the only efficient mechanism for DM annihilation involves the $t$-channel exchange of the fermionic mediator $\psi$, where DM particles annihilate into a pair of scalar mediators, $X\bar{X} \to \varphi^\dagger\varphi \to qq\bar{q}\bar{q}$. However, this $2 \to 4$ process, while relevant for the DM relic abundance, is not directly relevant for the LHC. Instead, the model can be probed at colliders in a number of ways, ranging from jets $+\met$ analyses to direct searches for the colour-charged mediators. The latter category is particularly versatile because colour-charged particles appear in a wide variety of constructions beyond the SM, so many existing searches could be used to constrain this fDM model through reinterpretations. One promising channel is the $t$-channel single production of the scalar mediator $\varphi$ in association with a $W$ boson, as illustrated by the representative Feynman diagram of figure~\ref{fig:fDMfeyn} (that includes specific $W$ and $\varphi$ decays). Whereas this $t$-channel process leads to diverse final states depending on the decays of the $\varphi$ mediator and the $W$ boson, the figure illustrates a scenario where the $W$ boson decays hadronically, and the scalar sextet mediator decays into like-sign top quarks ($\varphi \to tt$). 

\begin{figure}
    \begin{center}
        \includegraphics[width=.7\textwidth]{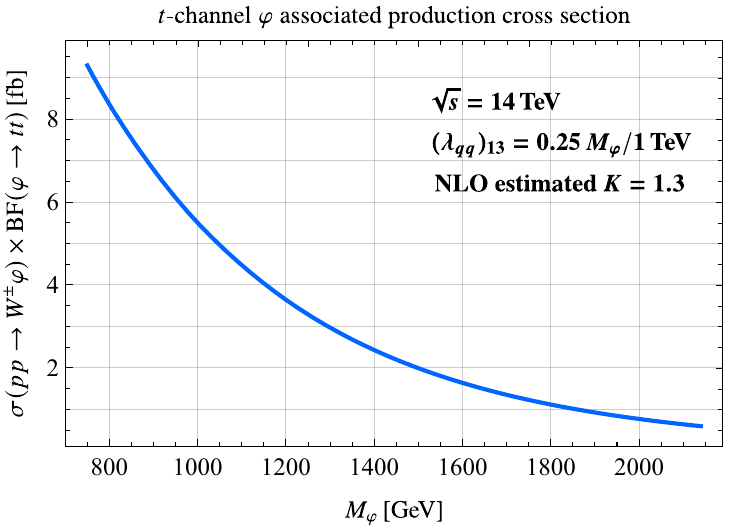}
    \end{center}
    \caption{\label{fig:fDMsignalxsec}Cross section for scalar mediator production in association with a $W$ boson at the LHC, for the fDM scenario considered. Results are inclusive with respect to the $W$-boson decay, but the mediator $\varphi$ is enforced to decay into a same-sign top pair. Calculation applies to the benchmark scenario defined in~\eqref{eq:lambda_benchmark}, and for a hadronic centre-of-mass energy of $\sqrt{S} = 14$~\text{TeV}. A flat $K$ factor of 1.3 is included to estimate the NLO yields~\cite{Han:2009ya}.}
\end{figure}

The production of same-sign top pairs is quite distinctive compared to conventional SM $t\bar{t}$ production, and it has therefore been previously searched for at the LHC~\cite{ATLAS:2012iws}. This channel is particularly relevant to the considered fDM model, since top-philic sextet scalars yield very large associated rates. This enhancement arises from a chirality flip due to the quark exchanged in the $t$-channel (see the Feynman diagram in figure~\ref{fig:fDMfeyn}): the $W$ boson couples to left-handed quarks, while the scalar $\varphi$ couples exclusively to right-handed quarks, as shown in the Lagrangian~\eqref{eq:fdmlag}. Consequently, the resulting cross section is proportional to the mass of the exchanged up-type quark, which leads to a strong enhancement for the third generation. Specifically, the parton-level cross section $\d\hat{\sigma}$ for $W\varphi$ production via $ud_f$ fusion is given, for a specific up-type quark flavour $u_f$ in the $t$-channel (assuming a diagonal CKM matrix), by
\be
    \frac{\d\hat{\sigma}}{\d t}\, (ud \to W^- \varphi) = \frac{\alpha_2}{48}\,(\lambda_{qq})_{1f}^2 \left(\frac{M_{u_f}}{M_W}\right)^2 \bigg[M_W^2(M_{\varphi}^2 - 2s - t) - t(M_{\varphi}^2 - s - t)\bigg]\,,
\ee
where $\alpha_2 = g_2^2/4\pi$ with $g_2$ being the $SU(2)_L$ gauge coupling, and where $M_W$ denote the mass of the $W$ boson. The presence of the $M_{u_f}^2$ factor limits the sensitivity of the LHC to this process, unless the exchanged quark is a top quark ($f=3$, $M_{u_f} = M_t$). Therefore, as alluded in section~\ref{sec:fDM}, we consider a benchmark sextet-quark coupling matrix $\lambda_{qq}$ that prioritises sextet couplings to top quarks. Non-flavour-diagonal couplings are however also essential, due to the parton distribution functions involved in the production process. For this purpose, we fix the scalar sextet coupling matrix $\lambda_{qq}$ as
\be\renewcommand{\arraystretch}{1.4}\label{eq:lambda_benchmark}
  \lambda_{qq} \equiv \frac{1}{2}
  \begin{pmatrix}
     2 \times 10^{-5} & 10^{-4} & 0.25\,\frac{M_\varphi}{1~\mathrm{TeV}}\\
     10^{-4} & 2 \times 10^{-3} & 10^{-2}\\
     0.25\,\frac{M_\varphi}{1~\mathrm{TeV}} & 10^{-2} & 2
   \end{pmatrix}\,,
\ee
which maximises the production cross section for same-sign top pairs while satisfying the FCNC constraints~\eqref{eq:fcnclim} for any sextet mass. A less aggressive benchmark with a constant value for $(\lambda_{qq})_{13} \equiv \lambda_{13} = 0.325$ is also valid, provided that $M_\varphi > 1.3$~TeV. This last bound is anyway almost certainly required in view of the results of LHC searches for the production of four top quarks~\cite{Carpenter:2022lhj}. The dependence of the hadron-level cross section $\sigma(pp \to W\varphi)$ with $\varphi \to tt$ on the sextet mass $M_{\varphi}$, in which the sum with the charge-conjugate process is implicit, is shown for this benchmark in figure~\ref{fig:fDMsignalxsec}, highlighting a high signal rate that is in principle not impossible to observe at the LHC. Predictions are obtained for a hadronic centre-of-mass energy of 14~TeV, and the LO set of NNPDF2.3 parton densities~\cite{Ball:2012cx}.

The size of the signal cross section, of at most $\mathcal{O}(1)$~fb, nevertheless poses a significant challenge for its discovery at the LHC, even with the full high-luminosity dataset of $\mathcal{L} = 3~\text{ab}^{-1}$. Nonetheless, the distinctive kinematics of the process shown in figure~\ref{fig:fDMfeyn} could allow the exclusion of a colour-sextet mediator in the mid-TeV mass range, improving upon current limits from searches for pair-produced colour-charged resonances~\cite{CMS:2018mts}. This $t$-channel signal produces a pair of boosted top quarks which, if they decay hadronically, can potentially be reconstructed. The invariant mass of the di-top system should then be localised within a narrow window around the mediator mass. Additionally, we enforce the recoiling $W$ boson to decay leptonically. This provides exactly one lepton and missing transverse momentum that could be used to suppress backgrounds such as $t\bar{t}$ events. A $t\bar t$ and $tt$ pair cannot indeed be distinguished in the fully hadronic channel, so that other means are in order to control the associated background.

The final state of interest thus consists of multiple jets, including two $b$-tagged jets, a single lepton, and missing transverse momentum. The dominant SM backgrounds for this signature are opposite-sign top-quark pair production in association with a $W$ boson, and opposite-sign bottom-quark pair production in association with three electroweak bosons including at least one $W$ boson. We produce samples of $10^5$ events for each of the two background processes with \lstinline{MadGraph5_aMC@NLO}, that we normalise according to cross sections of 769~fb~\cite{Campbell:2012dh, Alwall:2014hca} and 1211.9~fb~\cite{Alwall:2014hca}. In addition, for the $ttW$ sample, we combine matrix elements featuring up to two additional partons following the MLM procedure~\cite{Mangano:2006rw, Alwall:2008qv}. Finally, for the signal, we produce samples of $10^5$ events for an array of colour-sextet scalar masses $M_{\varphi} \in [1300,2100]$~GeV.

To motivate the selection criteria for our search, we present two distributions of observables with good discriminating power. These distributions are generated using \lstinline{MadAnalysis 5}~\cite{Conte:2012fm, Conte:2014zja, Conte:2018vmg}, after performing object reconstruction with its built-in simplified fast detector simulator (\lstinline{SFS})~\cite{Araz:2020lnp}. The latest version of \lstinline{MadAnalysis 5}~\cite{Araz:2023axv} supports jet reclustering and includes an implementation of \lstinline{HepTopTagger}~\cite{Plehn:2010st, Kasieczka:2015jma}. These features enable the creation of two collections, one of narrow jes and one of fat jets. In our analysis, narrow jets are reconstructed using \lstinline{FastJet}~\cite{Cacciari:2011ma} with the anti-$k_t$ algorithm~\cite{Cacciari:2008gp} and a radius parameter of $R=0.4$. Fat jets are instead clustered using the Cambridge-Aachen algorithm~\cite{Dokshitzer:1997in, Bentvelsen:1998ug, Wobisch:1998wt} with $R=1.0$.

\begin{figure}
  \begin{center}
    \includegraphics[width=.49\textwidth]{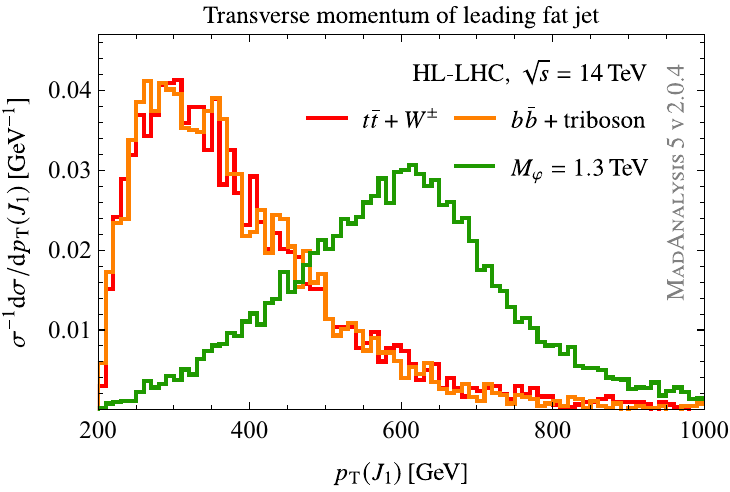}\hfill
    \includegraphics[width=.49\textwidth]{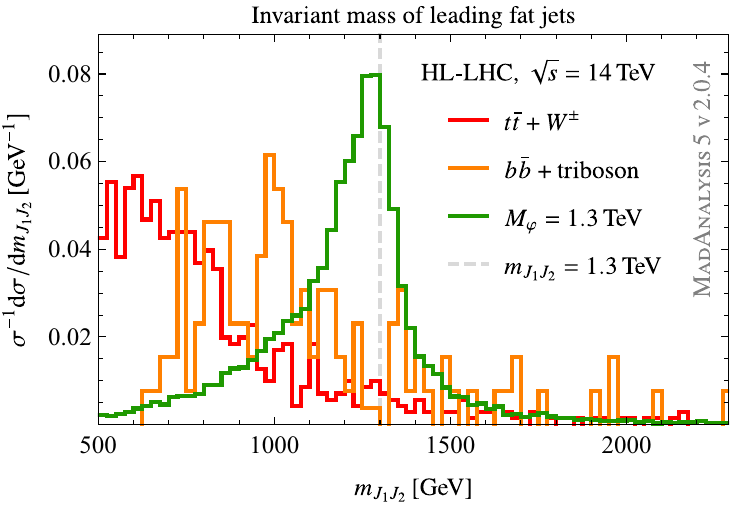}
  \end{center}
  \caption{\label{fig:fDM_distr}Distribution in the leading fat jet transverse momentum $p_T(J_1)$ (left), and in the invariant mass $m_{J_1J_2}$ of the pair of leading fat jets (right), for a signal scenario with $M_{\varphi}=1.3$~TeV (\ie\ $\varphi_Y \to tt$ production in association with a $W^{\pm}$ boson, followed by a leptonic $W$-boson decay and hadronic top decays), and the two components of the irreducible SM background.}
\end{figure}

\begin{table}
  \centering\renewcommand{\arraystretch}{1.3}\setlength{\tabcolsep}{12pt}
  \begin{tabular}{c | c}
    Selection criterion & Selection ranges\\ \hline
    \multirow{3}{*}{Narrow jets, anti-$k_{\text{T}}$ with $R = 0.4$} & $N_j \geq 2$\\[-.1cm]
      & $p_{\text{T}}(j) > 20$~GeV,\ $|\eta(j)| < 2.5$\\[-.1cm]
      & Veto any pair with invariant \\[-.15cm]
      & mass $m_{jj} \in [81,101]$~GeV\\[.15cm]
    $b$-tagged narrow jets & $N_b \geq 2$\\[.15cm]
    \multirow{2}{*}{Fat jets, Cambridge-Aachen with $R = 1.0$} & $N_J \geq 2$\\[-.1cm]
      & $p_{\text{T}}(J) > 200$~GeV,\ $|\eta(J)| < 2.5$\\[.15cm]
    Leading fat-jet transverse momentum & $p_{\text{T}}(J_1) \geq 475$~GeV\\[.15cm]
    \multirow{2}{*}{\ \ Lepton from decaying $W$ boson} & \ $N_{\ell} = 1$\\[-.1cm]
      & \ \ $p_{\text{T}}(\ell) > 25\,\text{GeV}$,\ $|\eta(\ell)| < 2.5$\\[.15cm]
    \multirow{2}{*}{Overlap removal} & Veto any narrow jet with  $\Delta R(j, \ell) < 0.1$ \\[-.1cm]
      & Veto any lepton with $\Delta R(\ell,j) < 0.4$ \\[.15cm]
    Missing transverse momentum & $\met > 15$~GeV\\[.15cm]
    Binned cut: reconstructed top & $m_{\varphi} - m_{tt}^{\text{had}}< 250$~GeV\\[-.15cm]
    pair invariant mass & \\
  \end{tabular}
  \caption{\label{tab:fDM_sel1} Selection criteria in our proposed search for TeV-scale colour-sextet scalar mediator produced in association with a leptonically decaying $W$ boson, and decaying to like-sign top quarks ($\varphi \to tt$) which themselves decay hadronically.}
\end{table}

The left panel in figure~\ref{fig:fDM_distr} shows the distribution in the transverse momentum $p_T(J_1)$ of the leading fat jet $J_1$ for the background processes and a representative signal sample with $M_{\varphi} = 1.3$~TeV. In addition, the right panel in this figure presents the invariant mass $m_{J_1J_2}$ of the system made of the two highest-momentum fat jets $J_1$ and $J_2$. These distributions are computed for events passing basic selection criteria, including the presence of at least two fat jets with $p_T> 200$~GeV. As expected (and as a consistent validation of our analysis framework), the $m_{J_1J_2}$ spectrum exhibits a pronounced peak at the mass of the colour-sextet resonance for the $M_{\varphi} = 1.3$~TeV signal, with no analogous structure in the background samples. This finding suggests that $m_{J_1J_2}$ could be a powerful variable not only for suppressing the SM backgrounds, but also for distinguishing between different new-physics scenarios with varying $M_{\varphi}$. The complete selection criteria are collected in table~\ref{tab:fDM_sel1}.

\begin{table}
  \centering
  \renewcommand{\arraystretch}{1.4} \setlength\tabcolsep{8pt}
    \begin{tabular}{l|c c|c c|c c}
      \multirow{2}{*}{Selection} & \multicolumn{2}{c|}{$M_{\varphi} = 1.3\,\text{TeV}$} & \multicolumn{2}{c|}{$b\bar{b} + \text{triboson}$} & \multicolumn{2}{c}{$t\bar{t}+W$}\\ 
        & $N_i$ & $\varepsilon_i$ & $N_i$ & $\varepsilon_i$ & $N_i$ & $\varepsilon_i$\\ \hline
      Initial & $1.00\times 10^5$ & -- & $1.00\times 10^5$ & -- & $1.00\times 10^5$ & --\\
      $N_j \geq 2$ & $9.95 \times 10^4$ & 0.995 & $9.74 \times 10^4$ & 0.974 & $9.67 \times 10^4$ & 0.967 \\
      No jets in $Z$ window & $9.24 \times 10^4$ & 0.929 & $1.63 \times 10^4$ & 0.167 & $8.02 \times 10^4$ & 0.830 \\
      $N_b \geq 2$ & $4.29 \times 10^4$ & 0.465 & $1.11 \times 10^4$ & 0.685 & $2.92 \times 10^4$ & 0.363 \\
      $N_J \geq 2$ & $3.78 \times 10^4$ & 0.881 & $2.51 \times 10^3$ & 0.226 & $6.01 \times 10^3$ & 0.206 \\
      $p_{\text{T}}(J_1) \geq 475\,\text{GeV}$\ \ & $3.02 \times 10^4$ & 0.797 & $1.98 \times 10^2$ & 0.079 & $1.14 \times 10^3$ & 0.190 \\
      $N_{\ell} = 1$ & $2.41 \times 10^4$ & 0.799 & $3.47 \times 10^1$ & 0.175 & $2.68 \times 10^2$ & 0.234\\
      $p_{\text{T}}^{\text{miss}} > 15\,\text{GeV}$ & $2.35 \times 10^4$ & 0.973 & $3.26 \times 10^1$ & 0.938 & $2.66 \times 10^2$ & 0.993 \\
      $m_{tt}^{\text{had}} >  1050\,\text{GeV}$\ \ & $2.06 \times 10^4$ & 0.878 & $1.55 \times 10^1$ & 0.475 & $1.20 \times 10^2$ & 0.451 \\
    \end{tabular}
  \caption{\label{tab:fDM_cutflow1}Illustrative cut-flow for the selection strategy given in table~\ref{tab:fDM_sel1} and a signal benchmark featuring $M_{\varphi} = 1.3$~TeV, shown together with the cut-flows related to the irreducible backgrounds. We present the results as yields $N_i$ normalised from an initial number of events set to $10^5$, as well as consecutive selection efficiencies $\varepsilon_i$.}
\end{table}

We now present the results of our mock search. For this analysis, we estimate the expected 95\% CL limit under the assumption of no signal observation, using the Asimov approximation for the median signal significance~\cite{Cowan:2010js},
\be\label{eq:significance}
  \mathcal{S} = \sqrt{2\bigg[(N_{\text{s}}+N_{\text{b}})\ln\,\bigg(1+\frac{N_{\text{s}}}{N_{\text{b}}}\bigg) - N_{\text{s}}\bigg]}\,,
\ee
where $N_{\text{s}}$ and $N_{\text{b}}$ represent the signal and background yields after selection. In the limit where $N_{\text{s}} \ll N_{\text{b}}$, this expression simplifies to the well-known formula
\be
  \mathcal{S} \to \frac{N_{\text{s}}}{\sqrt{N_{\text{b}}}} + \mathcal{O}(N_{\text{s}}^2/N_{\text{b}}^2)\,.
\ee
However, this approximation does not hold uniformly across our parameter space following the applied selection criteria. For simplicity, we take $\mathcal{S} = 2$ as an approximate threshold for 95\% CL exclusion (the correct value being $\mathcal{S} = 1.64$~\cite{Cowan:2010js}). It should be noted that our calculation assumes no uncertainties in the signal and background yields, leading to an optimistic estimate of sensitivity. We show in table~\ref{tab:fDM_cutflow1} the yields and cut-by-cut efficiencies for the background processes and the $M_{\varphi} = 1.3$~TeV signal. The most effective selection cuts are those on the number of fat jets ($N_J$), the transverse momentum of the leading fat jet, and the final binned cut on the invariant mass of the pair of leading fat jets.

\begin{figure}
  \begin{center}
    \includegraphics[width=.56\textwidth]{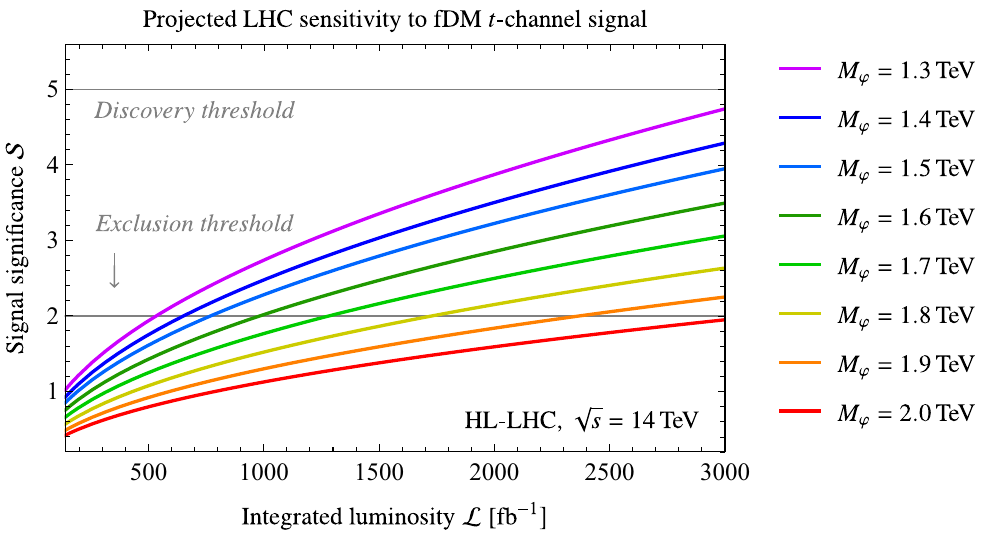}\hfill
    \includegraphics[width=.43\textwidth]{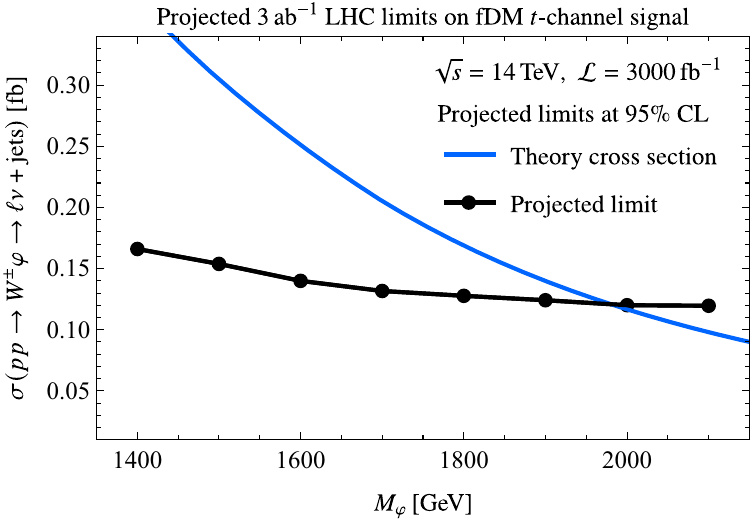}
  \end{center}
  \caption{\label{fig:fDM_sensitivity}Projected LHC sensitivity $\mathcal{S}$ (left) to the considered fDM colour-sextet scalar signal and associated exclusion (right), following the search strategy proposed. Our sensitivity predictions are given as a function of the HL-LHC integrated luminosity, the exclusion ones for $\mathcal{L} = 3\,\text{ab}^{-1}$, and we take $\mathcal{S} = 2$ as an estimate of the 95\%~CL exclusion threshold.}
\end{figure}

Using analogous results for all signal samples, the left panel of figure~\ref{fig:fDM_sensitivity} presents the sensitivity $\mathcal{S}$ as a function of the LHC integrated luminosity, $\mathcal{L} \in [139, 3000]\,\text{fb}^{-1}$. Ultimately, our simple cut-and-count strategy can probe colour-sextet scalars with masses up to $m_{\varphi} \sim 1.95$~TeV using the full planned HL-LHC dataset. This represents a best-case scenario for the method that we introduced, as our significance estimate does not account for background uncertainties or the presence of reducible backgrounds. Nevertheless, under these idealised assumptions, as shown in the right panel of figure~\ref{fig:fDM_sensitivity}, our proposed strategy can exclude mediator masses up to $700$~GeV beyond the Run~2 limits imposed on the pair production of colour-sextet scalars in a similar benchmark scenario~\cite{Carpenter:2022lhj}, provided no excess is observed. It should be noted that the cross section in figure~\ref{fig:fDM_sensitivity} is not the inclusive cross section shown in figure~\ref{fig:fDMsignalxsec}, and instead incorporates appropriate branching ratios reflecting the final states targeted by our search. Moreover, the projections exceed the luminosity-scaled improvements in new-physics bounds from measurements of the $t\bar{t}t\bar{t}$ production cross section and associated searches for new resonances, which extend to $2$~TeV for colour-octet scalars and are expected to be slightly weaker for colour-sextet scalars~\cite{Darme:2021gtt, Darme:2024epi}. Nevertheless, a $5\sigma$ discovery in the same-sign top channel at the HL-LHC appears unlikely, although as discussed elsewhere in this report, multiple experimental approaches could be used to further explore frustrated dark matter. While direct LHC searches and astrophysical constraints may offer greater sensitivity, our strategy could serve as a complementary avenue for such studies.

\subsubsection{A fermionic portal to a non-Abelian dark sector}\label{sec:collider_nonabelian}

Moving further along non-minimality, we discuss now the collider phenomenology arising from the FPVDM construction featuring a non-Abelian dark sector described in section~\ref{sec:nonAbDS}. Our model implementation in the \lstinline{UFO} format has been used in \lstinline{MadGraph5_aMC@NLO}~\cite{Alwall:2014hca} for the determination of the LHC constraints, relying on collider simulations at LO using the NNPDF3.0 LO set of parton densities~\cite{NNPDF:2014otw, Buckley:2014ana}. Moreover, a simplified version of the model has been implemented to calculate cross sections at one loop in \lstinline{MadGraph5_aMC@NLO} and \lstinline{FormCalc}~\cite{Hahn:2016ebn}.

\begin{figure}
  \centering
  \begin{minipage}{.35\textwidth}
  \centering
  \begin{tabular}{l}
  \includegraphics[width=\textwidth]{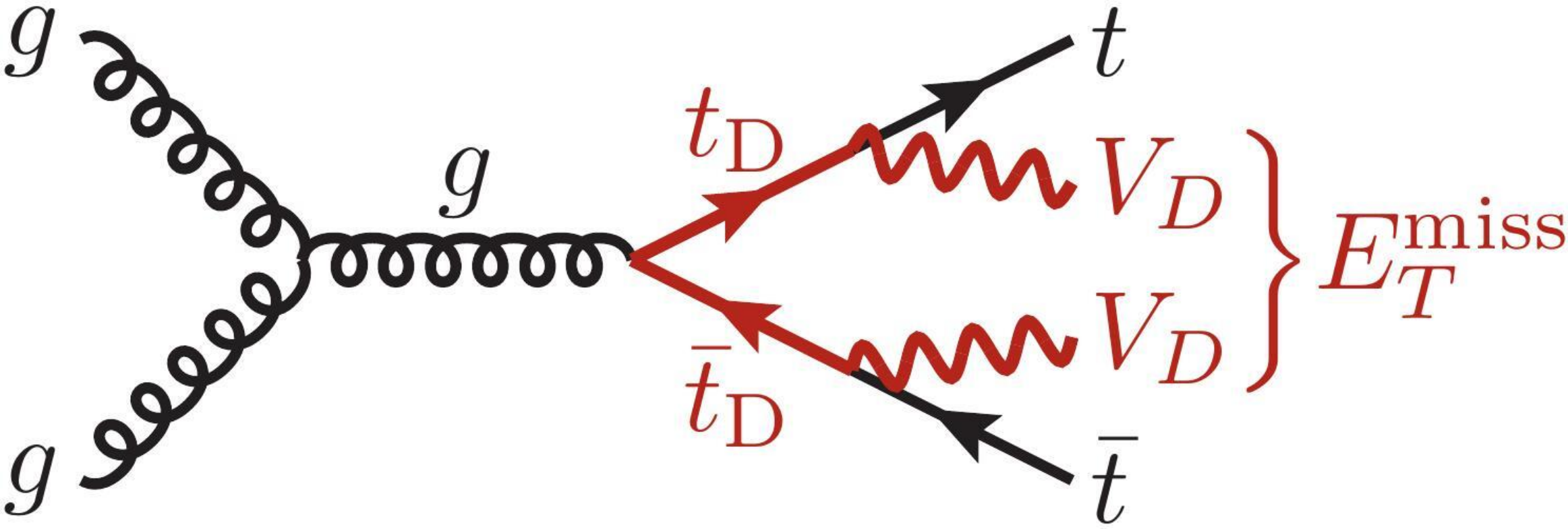}\\[10pt]
  \includegraphics[width=.79\textwidth]{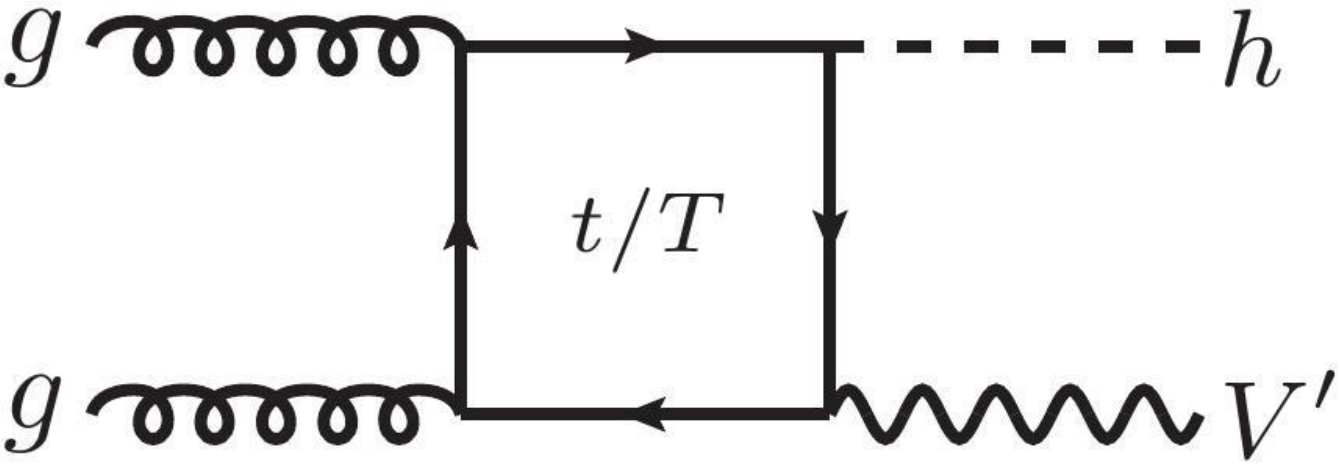}\\[10pt]
  \includegraphics[width=.9\textwidth]{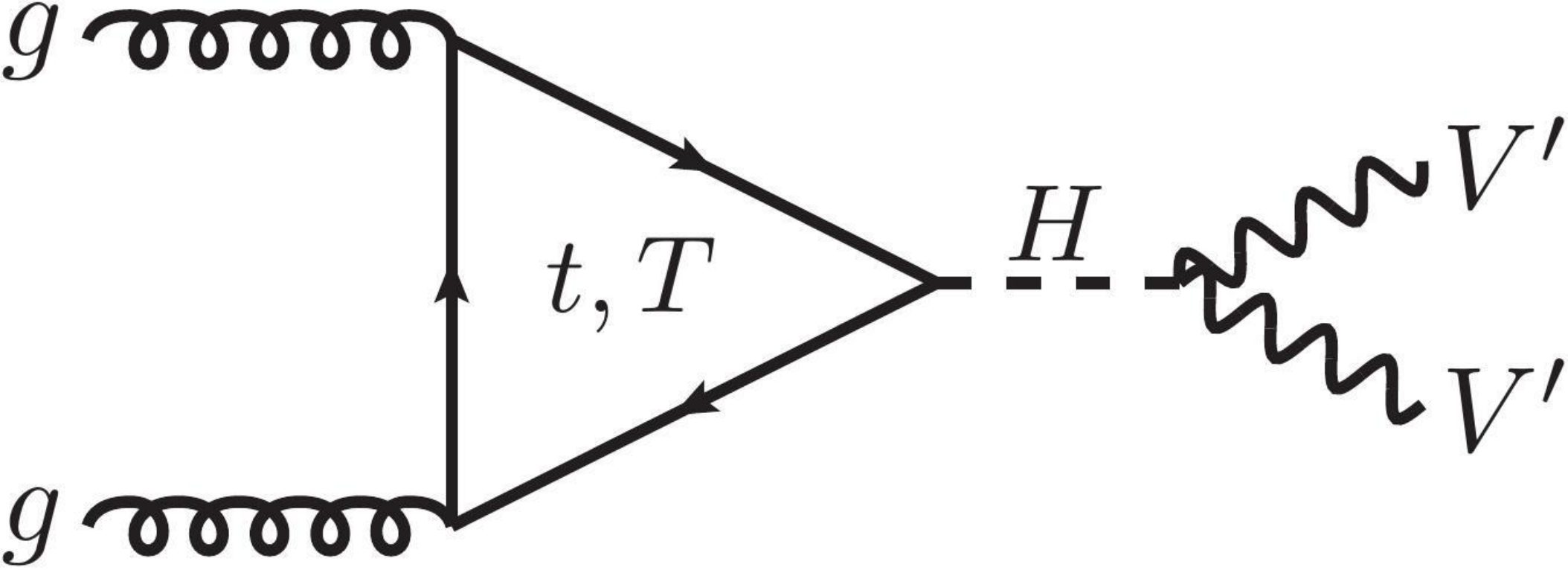}
  \end{tabular}
  \end{minipage} \hskip 10pt
  \begin{minipage}{.5\textwidth}
  \includegraphics[width=\textwidth]{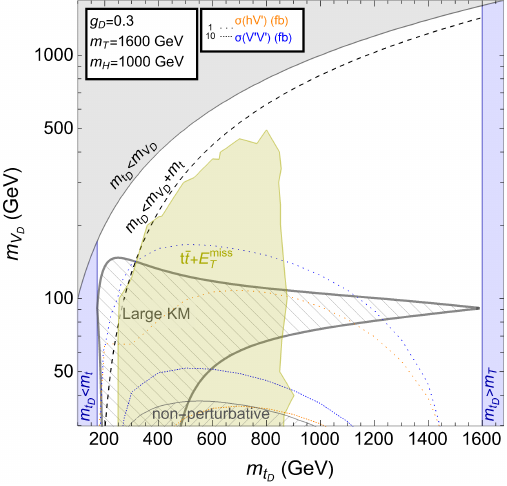}
  \end{minipage}  
  \caption{Feynman diagram representative of LHC processes relevant for an FPVDM scenario with top-partners (left), and regions of the parameter space excluded by the results of the LHC at the 95\% confidence level (right). We consider exclusion limits projected onto the $(m_{t_D}, m_{V_D})$ plane for $m_T = 1600$ GeV, $m_H = 1000$ GeV, and $g_D = 0.3$. Cross section isolines for $hV^\prime$ and $V^\prime V^\prime$ production processes are also displayed, together with the regions related to non-perturbativity and large kinetic mixing (hatched areas). The blue regions correspond to non-physical scenarios where $t_D$ is heavier than $T$ or lighter than the top quark.
\label{fig:FPVDM_bounds}}
\end{figure}

For the present analysis, we consider a scenario where the new fermions are partners of the top quark. LHC bounds have been derived by confronting the signal originating from $t_{D}$ pair production, followed by decay into the DM state $V_D$ and top quarks, against the results of CMS searches for top squark pair production decaying into DM and tops with partial Run 2 data~\cite{CMS:2017jrd}. Moreover, we have imposed limits on $T\bar{T}$ production, which are approximately given by $m_T > 1.5$~TeV~\cite{ATLAS:2021ibc, CMS:2022fck}, keeping in mind that single $T$ production is less constrained due to rate suppression by the small $T-t$ mixing. Additionally, we have estimated the relevance of $V^\prime$ pair production as well as that of the associated production of a $V^\prime$ boson with the Higgs boson, which occur at LO via fermion loops. Representative Feynman diagrams for all these processes are displayed in the left panel of figure~\ref{fig:FPVDM_bounds}.

We present in the right panel of figure~\ref{fig:FPVDM_bounds} the regions of the parameter space allowed by LHC constraints. The bounds are projected onto the $(m_{t_D}, m_{V_D})$ plane for a representative benchmark point with $g_D=0.3$, $m_T=1600$ GeV, and $m_H=1000$ GeV. In the small $g_D$ limit, where the width of $t_D$ is narrow, the $t_D$ pair production cross section depends solely on $m_{t_D}$, as it is governed by the strong interaction. Consequently, as long as the mass difference between the $t_D$ and $V_D$ states remains well above the top-quark threshold, the exclusion limits are largely independent of $m_{V_D}$ and rule out values of $m_{t_D} \lesssim 850$ GeV. However, as this threshold is approached, the amount of missing transverse energy originating from the signal decreases, reducing the sensitivity of the searches considered. This results in scenarios with a small mass gap between $t_D$ and $V_D$ remaining allowed by current data. Subsequently, throughout the entire parameter space, dark matter masses above approximately $500$ GeV are hence never excluded by the considered CMS search. In the degenerate region where $m_{t_D} \simeq m_{V_D}$, the $t_D$ state becomes increasingly long-lived, particularly for lower values of $g_D$, since its width is proportional to $g_D$. The only available decay mode allowed by the $\mathbb{Z}_2$-odd nature of the top partner, $t_D \to V_D t^{(*)}$, implies that the model is testable only through searches dedicated to long-lived particles. Notably, variations in the masses of the $T$ and $H$ states do not significantly affect this qualitative picture.  

On the other hand, we find that $V^\prime$ pair production and $V^\prime h$ associated production are only accessible in parameter regions already excluded by cosmological constraints (see section~\ref{sec:CosmConstNonMin_nonab}), rendering these channels irrelevant for our analysis. For completeness, figure~\ref{fig:FPVDM_bounds} also includes cross-section isolines for $hV^\prime$ and $V^\prime V^\prime$ production, along with the region corresponding to large kinetic mixing, and the non-perturbative region where the corrections to gauge boson masses become larger than 50\%.

\begin{figure}
  \centering
  \includegraphics[width=.3\textwidth]{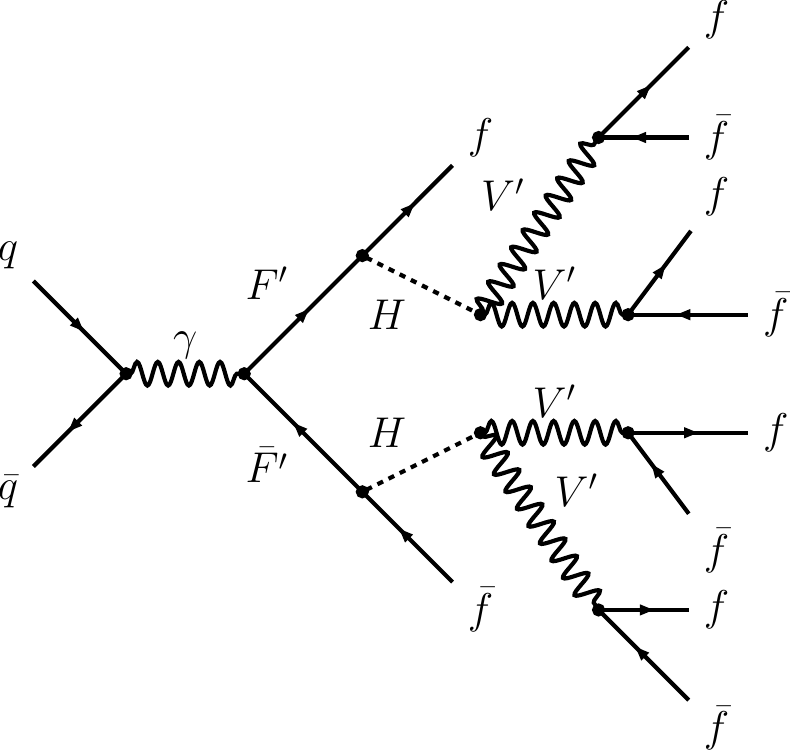}
  \caption{Representative Feynman diagram of a multi-fermion final state achievable with the FPVDM scenario.\label{fig:FPVDM_multifermion}}
\end{figure}

Finally, an outstanding distinctive feature of the FPVDM scenario, especially when the new fermions are the partners of light quarks or leptons, is the possibility of multiple SM fermion production induced by the pair production of the new fermions, as illustrated in figure~\ref{fig:FPVDM_multifermion}, where they cascade decay into $H$ and $V'$ bosons. This ultimately results in a striking final state featuring ten hard fermions that is worth analysing. This task is however left for future work.

\subsection{Long-lived \texorpdfstring{$t$}--channel mediators}\label{sec:LLP}

\subsubsection{Generalities}\label{sec:generalities_llp}
In theories beyond the Standard Model, particles with macroscopic decay lengths, known as \emph{long-lived particles} (LLPs), can arise due to phase-space suppression or small (effective) couplings that govern their decay. For a comprehensive review, see for example \cite{Alimena:2019zri}. In the simplified $t$-channel DM models considered in this work, LLPs may emerge from either or both of these mechanisms. In this section, our focus is thus on the cosmologically motivated scenarios discussed below in section~\ref{sec:cosmology}, where small couplings and/or small mass gaps in the new physics spectrum are required to explain the observed relic density. These include, in decreasing order of DM interaction strength, the conversion-driven freeze-out (CDFO)~\cite{Garny:2017rxs, DAgnolo:2017dbv}, freeze-in (FI)~\cite{McDonald:2001vt, Asaka:2005cn, Hall:2009bx}, and superWIMP (SW)~\cite{Covi:1999ty, Feng:2003uy} production mechanisms. Additionally, phase-space suppression may further extend lifetimes in these scenarios, and also lead to LLPs in specific regions of the WIMP parameter space. For instance, in a top-philic model with small mass splittings between the $X$ and $Y$ states, tree-level two-body and three-body decays can be kinematically forbidden, resulting in LLPs~\cite{Garny:2018icg} despite sizeable couplings. However, such cases only affect small regions of the WIMP parameter space and are not a primary focus of this discussion. 

It turns out that the mass splitting $\Delta m \equiv M_Y - M_X$ not only impacts the lifetime of the LLP mediator but also affects the kinematics of the visible objects produced at the LHC from its decay. To address this, we categorise LLP scenarios into two main classes: those with relatively small mass splittings ($\Delta m \ll M_X, M_Y$) and those without particularly small mass splittings. This distinction reflects not only the need for different LLP search strategies, but also corresponds to different DM genesis scenarios. In the CDFO regime, $\Delta m$ is typically less than 10\% of the DM mass, resulting in the emission of soft SM particles during the decay $Y \to X + \text{SM}$. These particles thus have relatively low momentum compared to the centre-of-mass energy of the process or the missing energy carried away by the DM particle. Consequently, such kinematics pose challenges for search strategies, as discussed below. Notably, the cosmologically viable parameter space for CDFO scenarios is constrained to mediator masses below a few TeV, exposing this entire parameter space to collider searches in the near future. 

In contrast, the superWIMP and freeze-in production mechanisms, which belong to the class of non-thermalised DM scenarios, do not require small mass splittings. Instead, they often exhibit extremely large mass hierarchies between the $Y$ and $X$ states in the bulk of the cosmologically viable parameter space~\cite{Garny:2018ali, Deshpande:2023zed, Decant:2021mhj}. The DM particle in these scenarios can be as light as $\mathcal{O}(10)\,\text{keV}$~\cite{Ballesteros:2020adh, DEramo:2020gpr, Decant:2021mhj}, effectively rendering it effectively massless with respect to a typical collider scale in large regions of the parameter space. On the other hand, testable mediator masses at the LHC are typically around the TeV scale. Nevertheless, the full cosmologically viable range of mediator masses extends far beyond the LHC reach, leaving only a small fraction of the parameter space accessible to collider searches. 

In general, both the CDFO and freeze-in/superWIMP scenarios are largely unchallenged by conventional WIMP searches via indirect or direct detection (see section~\ref{sec:cosmology}) due to the highly suppressed interaction rates resulting from their very weak couplings. Consequently, these parameter space regions often represent the only regions in certain $t$-channel models that have not yet been excluded. This highlights the critical role of collider searches in probing these scenarios and underscores the strong motivation for performing LLP searches. 

Experimental signatures featuring displaced objects, often accompanied by significant missing trasnverse energy, are key to detecting LLPs predicted by $t$-channel mediator DM models. Experimental results from collider experiments, such as ATLAS and CMS, impose valuable constraints on the mediator parameter space. Although no dedicated searches targeting $t$-channel mediator DM models have been performed to date, existing results can be recast to establish bounds. The current experimental constraints are detailed in the following sections. First, in section~\ref{sec:currentLLP}, we review existing search strategies for LLPs. Next, in section~\ref{sec:LLPscoverage}, we analyse the constraints originating from these searches within the freeze-in/superWIMP scenarios (section~\ref{sec:LLPsFISW}) and the CDFO scenario (section~\ref{sec:LLPsCDFO}). In section~\ref{sec:LLPsGaps}, we identify gaps in the current coverage of LLP searches, emphasising the need to explore new signatures. Finally, in section~\ref{sec:RecastMaterial}, we provide concluding remarks on the reinterpretability of the searches. 

\subsubsection{Current LLP searches}\label{sec:currentLLP}

\begin{table}
  \centering\renewcommand{\arraystretch}{1.3}\setlength{\tabcolsep}{8pt}
  \begin{tabular}{l l c c}
    Signature  & Analysis \& reference & ${\cal L}$ [fb$^{-1}$] &  Decay length [mm] \\
    \hline
    \multirow{2}{*}{HSCP} & \href{https://atlas.web.cern.ch/Atlas/GROUPS/PHYSICS/PAPERS/SUSY-2018-42/}{ATLAS-SUSY-2018-42}~\cite{ATLAS:2022pib} & 139  & $\gtrsim$ 300 \\
       & \href{https://cms-results.web.cern.ch/cms-results/public-results/publications/EXO-18-002/index.html}{CMS-EXO-18-002}~\cite{CMS:2024nhn} & 101 & $\gtrsim$ 1000\\[.1cm]
    \multirow{2}{*}{DT} & \href{https://cms-results.web.cern.ch/cms-results/public-results/publications/SUS-21-006/}{CMS-SUS-21-006}~\cite{CMS:2023mny} & 137 & 100\,--\,1000 \\
      & \href{https://atlas.web.cern.ch/Atlas/GROUPS/PHYSICS/PAPERS/SUSY-2018-19/}{ATLAS-SUSY-2018-19}~\cite{ATLAS:2022rme} & 136 & 100\,--\,300 \\[.1cm]  
    \multirow{2}{*}{DV plus $\met$} & \href{https://cms-results.web.cern.ch/cms-results/public-results/publications/EXO-22-020/}{CMS-EXO-22-020}~\cite{CMS:2024trg} & 137 & 0.1\,--\,20 \\
      & \href{https://atlas.web.cern.ch/Atlas/GROUPS/PHYSICS/PAPERS/SUSY-2016-08/}{ATLAS-SUSY-2016-08}~\cite{ATLAS:2017tny} & 32.8 & 4\,--\,300  \\[.1cm]
    Displaced soft tracks & \href{https://atlas.web.cern.ch/Atlas/GROUPS/PHYSICS/PAPERS/SUSY-2020-04/}{ATLAS-SUSY-2020-04}~\cite{ATLAS:2024umc} & 140 & 0.1\,--\,10  \\[.1cm]
    \multirow{2}{*}{DL} & \href{https://atlas.web.cern.ch/Atlas/GROUPS/PHYSICS/PAPERS/SUSY-2018-14}{ATLAS-SUSY-2018-14}~\cite{ATLAS:2020wjh} & 139 & 3\,--\,300  \\
      & \href{https://cms-results.web.cern.ch/cms-results/public-results/publications/EXO-18-003}{CMS-EXO-18-003}~\cite{CMS:2021kdm} & 113-118 & 0.1\,--\,100  \\[.1cm]
    DJ (Calorimeter) & \href{https://atlas.web.cern.ch/Atlas/GROUPS/PHYSICS/PAPERS/EXOT-2019-23}{ATLAS-EXOT-2019-23}\cite{ATLAS:2022zhj}  & 139 & 2000-4000  \\[.1cm]
    \multirow{2}{*}{DJ (Muon system)} & \href{https://cms-results.web.cern.ch/cms-results/public-results/publications/EXO-21-008/}{CMS-EXO-21-008}~\cite{CMS:2024bvl}  & 138 & 3000-7000  \\
      & \href{https://atlas.web.cern.ch/Atlas/GROUPS/PHYSICS/PAPERS/EXOT-2019-24/}{ATLAS-EXOT-2019-24}~\cite{ATLAS:2022gbw}  & 139 & 4000-8000   \\
\end{tabular}
\caption{Summary of recent LLP searches at the LHC Run 2 relevant for the signatures discussed in the text, and involving missing transverse energy together with heavy stable charged particles (HSCP), displaced tracks (DT), dispalced vertices (DV), displaced leptons (DL) and displaced jets (DJ).}
\label{tab:llp_recentsearches}
\end{table}

The LHC has now established a comprehensive programme dedicated to searches for LLPs. Many of these searches directly target DM production alongside objects originating from positions significantly displaced from the collision point. Moreover, other searches, particularly those selecting events with large missing transverse momentum ($\met$), may also indirectly probe the DM models considered in this report. In this section, we provide an overview of the signatures examined in these searches. We focus both on searches that explicitly involve missing transverse momentum for triggering and/or event selection, and on searches without explicit $\met$ requirements that are inclusive enough to capture signals from the considered set of models. Table~\ref{tab:llp_recentsearches} lists a selection of recent LHC Run~2 searches representing these types, and we refer to \cite{Alimena:2019zri} for a broader overview of LLP searches at the LHC.

Searches for heavy stable charged particles (HSCPs) are sensitive to DM models where the mediator carries electric charge and/or hadronises into charged hadrons, and then decays outside the tracker. These searches~\cite{ATLAS:2022pib, CMS:2024nhn} rely on high ionisation losses ($\d E/\d x$) of heavy charged particles, and/or anomalous time-of-flight measurements between their production at the collision point and their arrival in the muon system. Furthermore, triggering often relies on calorimetric missing transverse energy, originating from either new physics particles produced in association with the charged LLP or the LLP itself if it decays outside the calorimeter. The reconstructed HSCP mass is then used to suppress backgrounds. HSCP searches are usually sensitive to models in which the electrically charged and/or coloured mediators have proper decay lengths larger than approximately $0.1$~m~\cite{ATLAS:2022pib}. Since they typically employ model-independent selection criteria, they are easily recastable using tabulated efficiencies like those provided in~\cite{CMS:2015lsu}, and HSCP results have thus been reinterpreted not only within supersymmetric models~\cite{Heisig:2012zq, Heisig:2015yla}, but also in frameworks featuring heavy resonances decaying into doubly charged LLPs~\cite{Giudice:2022bpq}. 

In scenarios where charged LLPs decay within the tracker to mostly soft and/or invisible final states, disappearing track (DT) signatures arise. Disappearing track searches~\cite{CMS:2020atg, ATLAS:2022rme, CMS:2023mny} identify short tracks with a few hits in the innermost tracking layers caused by charged particles that decay invisibly or to low-momentum states not reconstructed as tracks. Events are triggered by the large amount of $\met$ generated by the invisible LLP decay, and the  searches are usually sensitive to models with small mass splittings $\Delta m$ and LLP decay lengths between $10$~mm and $100$~mm. ATLAS and CMS disappearing track searches therefore commonly target simplified supersymmetric scenarios where a long-lived charged wino or higgsino decays into a neutral particle and a low-momentum pion.

In contrast to HSCP and disappearing track analyses, searches for displaced vertices (DVs) focus on visible LLP decays within the tracker~\cite{ATLAS:2017tny, CMS:2024trg}, and they can be very inclusive as long as the LLP decays to SM particles and invisible states. While displaced vertex searches often include stringent $\met$ requirements for triggering reasons, their sensitivity further diminishes in scenarios with small $\Delta m$ due to the additional requirement for high-momentum displaced objects. Moreover, some displaced vertex searches specifically target resonant displaced hadronic decays without missing energy, although these are less relevant for the (non-resonant) mediator models considered here. Since it is not always possible to determine the sensitivity of this last set of searches to those non-resonant mediator decays, they will not be considered in this report.

In addition, ATLAS has recently performed a search for mildly displaced soft tracks~\cite{ATLAS:2024umc}. It relies on a trigger on the missing transverse energy and focuses on events with soft tracks ($2\mbox{ GeV} < p_T < 5\mbox{ GeV}$) exhibiting small displacements ($|d_0| < 10$~mm). This search specifically targets supersymmetric scenarios with compressed mass spectra, such as setups in which charginos decay to neutralinos and soft pions with small displacements. Despite being a challenging signal due to the presence of the soft objects, the search exploits the relatively small associated SM background rates (\ie\ related to the production of soft and displaced pions). It thus allows to look for an LLP signal and hence cover previously existing gaps in typical LLP parameter spaces where $\Delta m < 1$~GeV and $c\tau \sim 0.1\mbox{ mm} - 1\mbox{ cm}$.

Beyond $\met$-triggered searches, displaced leptons (DL) searches without any specific requirement on the missing energy are sufficiently inclusive to apply to long-lived mediators decaying to charged leptons and DM~\cite{ATLAS:2020wjh, ATLAS:2023ios, ATLAS:2024vnc, CMS:2021kdm, CMS:2022qej, CMS:2022fut, CMS:2024qxz, CMS:2021sch}. Such searches are typically triggered by either standard lepton triggers, or more specialised ones such as ATLAS’s Run 3 displaced lepton triggers~\cite{ATLAS:2023nze} (that focuses on the reconstruction of tracks with a large impact parameter) or CMS’s data scouting dimuon stream with low $p_T$ thresholds~\cite{CMS:2024zhe}. They are sensitive to a broad range of LLP decay lengths (0.1~mm to 10~m), and to spectra featuring large mass splittings, so that they can then be used to probe both supersymmetric models and scenarios featuring a more generic dark sector.

Finally, displaced jets (DJ) searches target LLPs decaying hadronically within different detector regions, such as the inner detector, the calorimeter, or the muon spectrometer~\cite{ATLAS:2023oti, ATLAS:2022gbw, ATLAS:2022zhj, ATLAS:2024ocv, ATLAS:2024qoo, CMS:2020iwv, CMS:2021tkn, CMS:2019qjk, CMS:2018qxv, CMS:2018bvr, CMS:2024bvl}. These searches often rely on multijet triggers or lepton triggers (when jets and/or leptons are produced in association with the LLP), and in some cases, specialised triggers designed for the unique related calorimeter and muon spectrometer signatures. Displaced jet searches primarily probe dark sector scenarios involving LLPs decaying to SM fermions, although supersymmetric models are also considered, commonly in the $R$-parity-violating case.

\subsubsection{Coverage of current searches}\label{sec:LLPscoverage}
\begin{figure}
  \centering
  \includegraphics[width=0.65\textwidth]{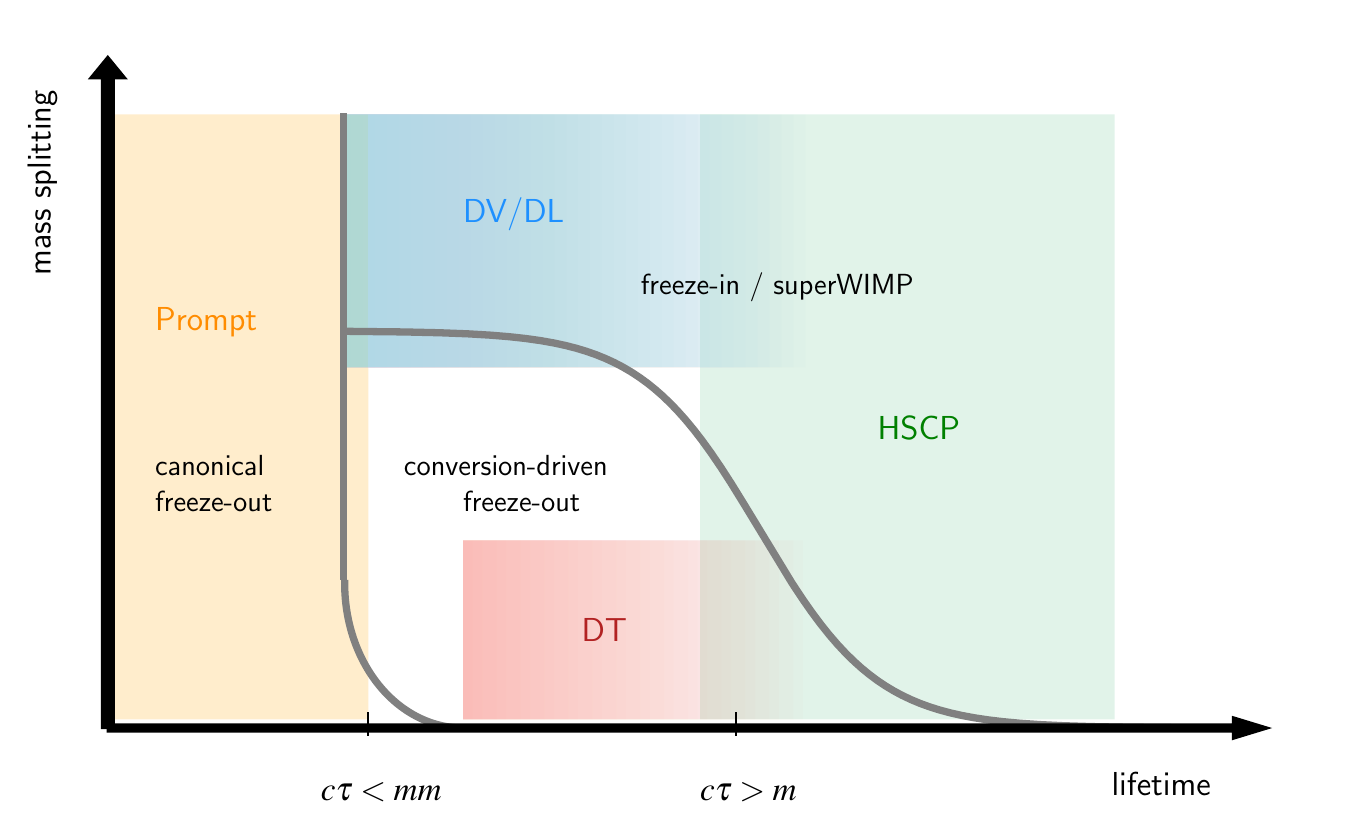}
  \caption{Schematic representation of the coverage of LLP searches in the parameter space defined by the LLP proper lifetime and the associated spectrum mass splitting. Regions corresponding to various cosmological DM production scenarios are also indicated.} \label{fig:schematic}
\end{figure}

Figure~\ref{fig:schematic} schematically illustrates the parameter space coverage of the LLP searches discussed in section~\ref{sec:currentLLP}, alongside cosmological DM production scenarios. This coverage is presented in the plane defined by the LLP proper lifetime $c\tau$ and the associated mass splitting. The parameter space of canonical freeze-out, which typically results in short lifetimes ($c\tau \lesssim 1~\mathrm{mm}$), is associated with prompt signatures. In contrast, the parameter space relevant for CDFO scenarios, predominantly characterised by $1~\mathrm{mm} \lesssim c\tau \lesssim 1~\mathrm{m}$ and relatively small mass splittings, remains only partially covered by current searches. However, these regions can still be probed from multiple angles using various LLP and even prompt search strategies. Freeze-in/superWIMP scenarios, on the other hand, allow for larger mass splittings, while the mediator lifetime spans values similar to those in the CDFO scenario and extends well into the detector-stable regime. As a result, their parameter space is also exposed to searches for displaced leptons and jets, as well as for HSCP signatures.

\begin{table}
  \centering\renewcommand{\arraystretch}{1.2}\setlength{\tabcolsep}{8pt}
  \begin{tabular}{l l c c}
    Signature & Analysis \& reference & ${\cal L}$  [fb$^{-1}$] & Reinterpretation code \\
    \hline
    \multirow{4}{*}{HSCP} & \href{http://cms-results.web.cern.ch/cms-results/public-results/publications/EXO-13-006/index.html}{CMS-EXO-13-006}~\cite{CMS:2015lsu} & 18.8  & \checkmark\\
      & \href{http://cms-results.web.cern.ch/cms-results/public-results/preliminary-results/EXO-16-036/index.html}{CMS-PAS-EXO-16-036}~\cite{CMS:2016ybj} & 12.9  & \checkmark\\ 
      & \href{https://atlas.web.cern.ch/Atlas/GROUPS/PHYSICS/PAPERS/SUSY-2016-32/}{ATLAS-SUSY-2016-32}~\cite{ATLAS:2019gqq} & 36.1  & \checkmark\\
      & \href{https://atlas.web.cern.ch/Atlas/GROUPS/PHYSICS/PAPERS/SUSY-2018-42/}{ATLAS-SUSY-2018-42}~\cite{ATLAS:2022pib} & 139  & \checkmark\\[.1cm]
    \multirow{3}{*}{DT} & \href{https://atlas.web.cern.ch/Atlas/GROUPS/PHYSICS/PAPERS/SUSY-2016-06/}{ATLAS-SUSY-2016-06}~\cite{ATLAS:2017oal}  & 36.1  & \checkmark\\
      & \href{https://cms-results.web.cern.ch/cms-results/public-results/publications/EXO-16-044/}{CMS-EXO-16-044}~\cite{CMS:2018rea}  & 38   & \checkmark\\
      & \href{https://cms-results.web.cern.ch/cms-results/public-results/publications/EXO-19-010/}{CMS-EXO-19-010}~\cite{CMS:2020atg} & 101  & \checkmark\\[.1cm]
    DV plus $\met$ & \href{https://atlas.web.cern.ch/Atlas/GROUPS/PHYSICS/PAPERS/SUSY-2016-08/}{ATLAS-SUSY-2016-08}~\cite{ATLAS:2017tny} & 32.8 & \checkmark\\[.1cm]
    DV plus $\mu$ & \href{https://atlas.web.cern.ch/Atlas/GROUPS/PHYSICS/PAPERS/SUSY-2018-33/}{ATLAS-SUSY-2018-33}~\cite{ATLAS:2020xyo} & 136 & \checkmark\\[.1cm]
    DJ plus $\met$ & \href{https://cms-results.web.cern.ch/cms-results/public-results/publications/EXO-19-001/}{CMS-EXO-19-001}~\cite{CMS:2019qjk} & 137 &  \checkmark\\[.1cm]
    \multirow{2}{*}{DL} & \href{https://cms-results.web.cern.ch/cms-results/public-results/publications/EXO-18-003}{CMS-EXO-18-003}~\cite{CMS:2021kdm} & 113-118 &   \\
      & \href{https://atlas.web.cern.ch/Atlas/GROUPS/PHYSICS/PAPERS/SUSY-2018-14}{ATLAS-SUSY-2018-14}~\cite{ATLAS:2020wjh} & 139 & \checkmark\\[.1cm]
    Monojet & \href{https://cms-results.web.cern.ch/cms-results/public-results/publications/EXO-20-004/}{CMS-EXO-20-004}~\cite{CMS:2021far} & 137  &  \checkmark\\
  \end{tabular}
  \caption{Summary of the LHC searches reinterpreted to constrain LLP scenarios in this report. Those that are accessible through the publicly available \lstinline{llprecasting} repository and the \lstinline{MadAnalysis 5}, \lstinline{CheckMATE} and \lstinline{SModelS} programs are indicated with a check mark.} \label{tab:llp_usedsearches}
\end{table}

One of the significant challenges from a theoretical perspective is the limited availability of recast analyses. In many instances, the absence of reinterpretation material or publicly available recasting codes makes it difficult to rigorously evaluate the coverage of specific searches for the models discussed in this report. We refer to section~\ref{sec:RecastMaterial} for a more detailed discussion on advancing reinterpretation methods.  Consequently, the results presented here do not always reflect searches using the full integrated luminosity available. Currently, publicly available recasting and reinterpretation tools include ten LLP searches available from the \lstinline{llprecasting} repository~\cite{llprecastingRepo}, three analyses implemented in the \lstinline{MadAnalysis 5} public analysis database~\cite{Araz:2021akd}, six analyses implemented in \lstinline{CheckMATE}~\cite{Desai:2021jsa}, and twelve analyses included in \lstinline{SModelS}~\cite{Heisig:2018kfq, Ambrogi:2018ujg, Alguero:2021dig, Altakach:2024jwk}. It should be noted that the numbers above include some overlap between tools and represent only a subset of the LLP analyses conducted by the LHC collaborations. A summary of all the searches considered in this section is provided in table~\ref{tab:llp_usedsearches}. \vspace{.5cm}

\paragraph{Freeze-in/superWIMP regime}\label{sec:LLPsFISW}
\paragraph*{}\vspace{.3cm}
If the DM-mediator coupling $\lambda$ is too small to thermalise dark matter in the early universe, the relic density as observed in data can still be achieved through freeze-in and superWIMP production. In general, both production mechanisms coexist, with their relative contributions being model-dependent. For this reason, we consider them together. As the related cosmologically viable parameter space extends to very large mediator masses, up to around $10^9\,\mathrm{GeV}$, the LHC can only probe a small fraction of it. Nonetheless, as we discuss below, current LLP search strategies offer good sensitivity.

\begin{figure}
  \centering
  \includegraphics[width=0.48\textwidth]{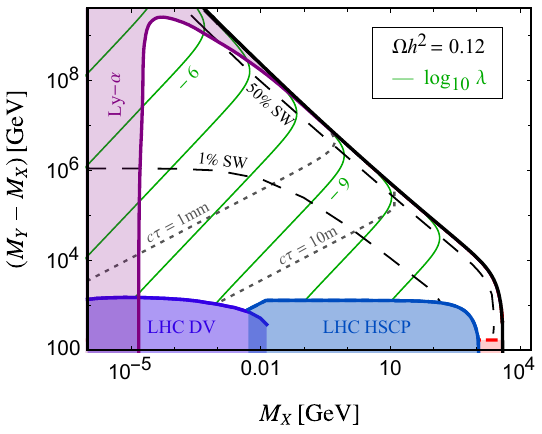}
  \includegraphics[width=0.5\textwidth]{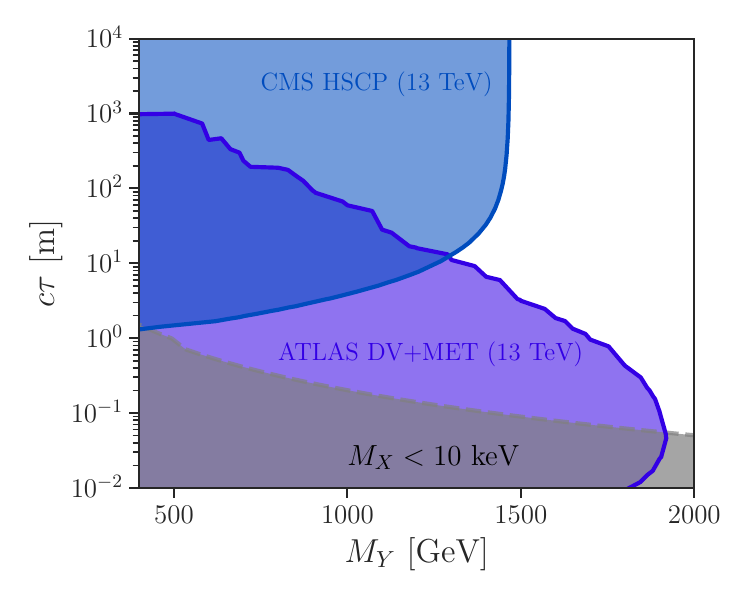}
  \caption{Constraints on $t$-channel DM models featuring quark-philic mediators in the freeze-in/superWIMP (non-thermalised) DM production regime, adapted from~\cite{Decant:2021mhj, Belanger:2018sti}. In the left panel, we display the viable region of the parameter space (\ie\ with $\Omega h^2 = 0.12$) of a scenario with a top-philic scalar mediator and a Majorana DM state. Shaded regions indicate exclusions from DV (violet, \cite{ATLAS:2017tny}) and HSCP (blue, \cite{ATLAS:2019gqq}) searches, as well as from structure formation (purple) and BBN (red) (see details in section~\ref{sec:cosmology}). In the right panel, we consider instead a model where a fermionic mediator couples to first-generation quarks and scalar DM, and display again the constraints originating from DV (blue, \cite{ATLAS:2017tny}) and HSCP (purple, \cite{CMS:2016ybj}) searches. In addition, the grey region represents setups with a hot DM candidate.} \label{fig:nonTHparamquarkphil}
\end{figure}

We begin our study with an exploration of quark-philic models. The left panel of figure~\ref{fig:nonTHparamquarkphil} shows the cosmologically viable parameter space for a class of scenarios where a top-philic scalar mediator couples to a Majorana DM particle (\ie\ which corresponds to the \lstinline{S3M_tR} class of simplified models discussed in section~\ref{sec:model_minimal}). At large mediator and DM masses, superWIMP production dominates, with the solid black line marking where this mode alone saturates the observed relic density, $\Omega h^2 = 0.12$. To the right of this line, it would thus overclose the universe, which subsequently excludes that part of the parameter space. To its left, both freeze-in and superWIMP production contribute to the relic density, with the black dashed contours indicating the relative superWIMP contribution. In addition, the figure includes two isolines in the mediator proper decay length, shown as grey dotted lines. They respectively span the prompt decay and detector-stable cases, covering decay lengths of $1$~mm and $10$~m respectively. For mediator masses around the TeV scale (\ie\ within the reach of the LHC) and DM masses allowed by structure formation ($M_X \gtrsim 10$~keV), mediator decay lengths typically exceed a few centimetres. For DM masses close to the exclusion limit, displaced vertex searches are the most relevant, and they exclude the violet-shaded region of the parameter space. For larger DM masses ($M_X \gtrsim 10$~MeV), the mediator decays predominantly outside the tracker, making HSCP searches most effective. The corresponding excluded regions of the parameter space are indicated by the blue-shaded excluded region of the parameter space. Both the DV and HSCP search strategies therefore restrict the mediator to be quite heavy. 

Similar results are expected for other models of the \lstinline{S3M} type where quarks of the first and second generation are involved, provided that $\Delta m \gg M_t$. Differences would only be due to the uniqueness of the decay modes of the displaced top quark. For instance, in top-philic models, an additional displacement arises from $B$-hadron decays, potentially complicating vertex reconstruction due to the extra associated tracks, and leptonic top decays could provide an alternative search avenue that is not available for models featuring quarks of the two lighter generations.

\begin{figure}
  \centering
  \includegraphics[width=0.5\textwidth]{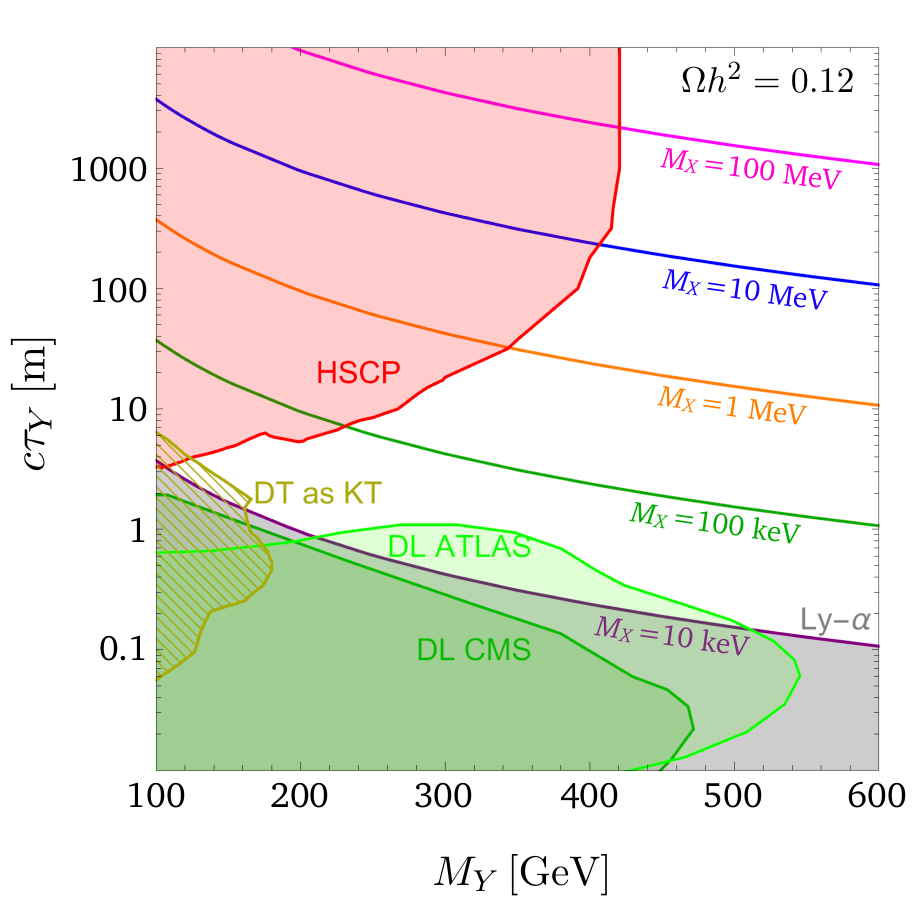}
    \caption{Isolines of constant relic density $\Omega h^2 = 0.12$ for different choices of the DM mass in a leptophilic Majorana DM model with a scalar mediator coupling to right-handed muons. We show constraints stemming from HSCP searches (red, \cite{ATLAS:2019gqq}), ATLAS and CMS displaced lepton searches (green, \cite{ATLAS:2020wjh, CMS:2021kdm}) and disappearing track searches (yellow, \cite{CMS:2020atg}). The plot is adapted from~\cite{Calibbi:2021fld, Junius:2022vzl}.} \label{fig:nonTH_leptophilic}
\end{figure}

In the right panel of figure~\ref{fig:nonTHparamquarkphil}, we perform a similar analysis, but this time for a scenario with a scalar DM state and a fermionic mediator coupling to first-generation quarks. Here, the entire displayed region of the parameter space is consistent with $\Omega h^2 = 0.12$, which is achieved through freeze-in production. In this case, as the mediator lifetime decreases, the rate for DM production increases, thus requiring smaller DM masses in order to reproduce the observed relic abundance. For $c\tau \lesssim 0.1$~m, the required DM mass even drops below 10~keV, which is excluded by structure formation constraints. Once again, displaced vertex and HSCP searches remain the most sensitive searches to this setup, constraining scenarios with mediator masses ranging up to $M_Y \sim 1.8$~TeV. Compared with the case of a scalar mediator and a fermionic DM state, the bounds are slightly weaker because of the different production cross section for coloured fermions and scalars~\cite{Calibbi:2021fld, Belanger:2018sti}. Moreover, as for Majorana DM, we expect our results to be qualitatively similar for the other quark-philic models of this class. 

We now move on with leptophilic scenarios that we illustrate by considering a setup in which a scalar mediator couples to Majorana DM and the right-handed muon, with an interaction strength relevant for freezing-in DM. Figure~\ref{fig:nonTH_leptophilic} shows contours of correct relic abundance for different choices of the DM mass, assuming a reheating temperature higher than the freeze-in scale. It turns out that low DM masses are excluded by structure formation constraints from Lyman-$\alpha$ measurements~\cite{Ballesteros:2020adh, DEramo:2020gpr, Decant:2021mhj}, and HSCP searches constrain large mediator lifetimes. For intermediate mediator lifetimes, displaced lepton searches can probe mediator masses up to approximately 500~GeV, while scenarios with a mediator decay length of $\mathcal{O}(1)$ metre are less constrained. In this case, a signature featuring a kinked charged track (made from the combination of the charged scalar track and the muon one) is the primary handle on the model. Reinterpreted disappearing track searches therefore provide the strongest bounds in this regime~\cite{Calibbi:2021fld}, due to the fact that there is no LHC analysis dedicated to kinked tracks to reinterpret.

\begin{figure}
  \centering
  \includegraphics[width=0.415\textwidth]{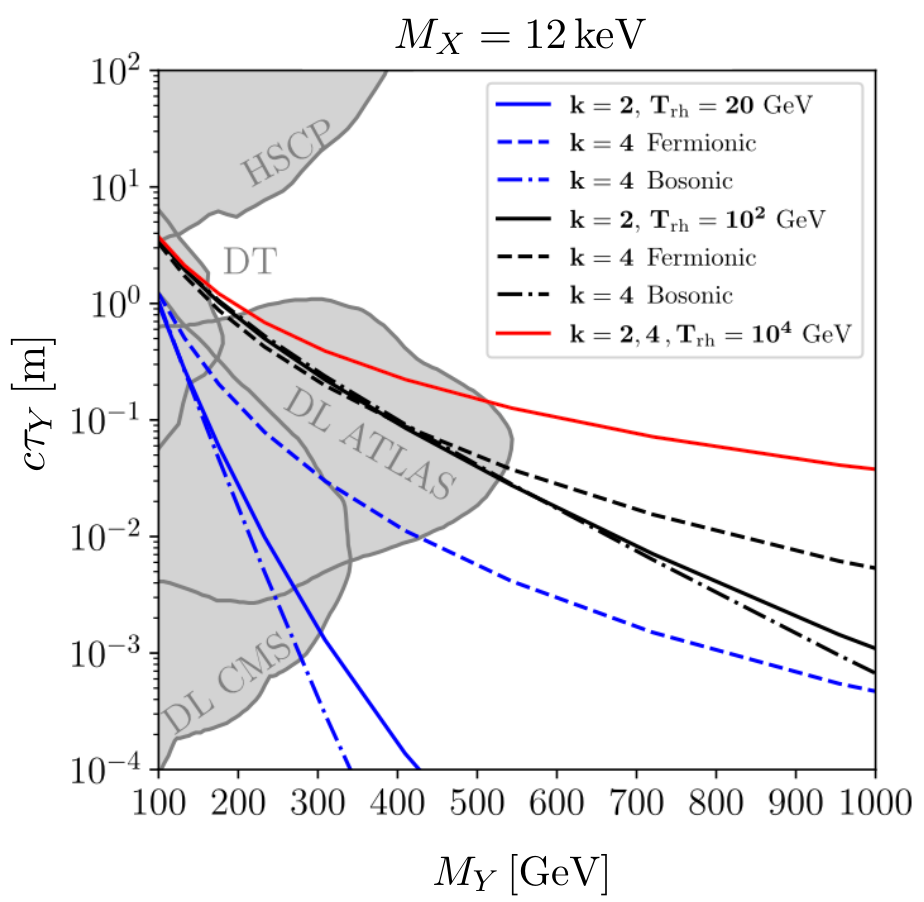}\hfill
  \includegraphics[width=0.54\textwidth]{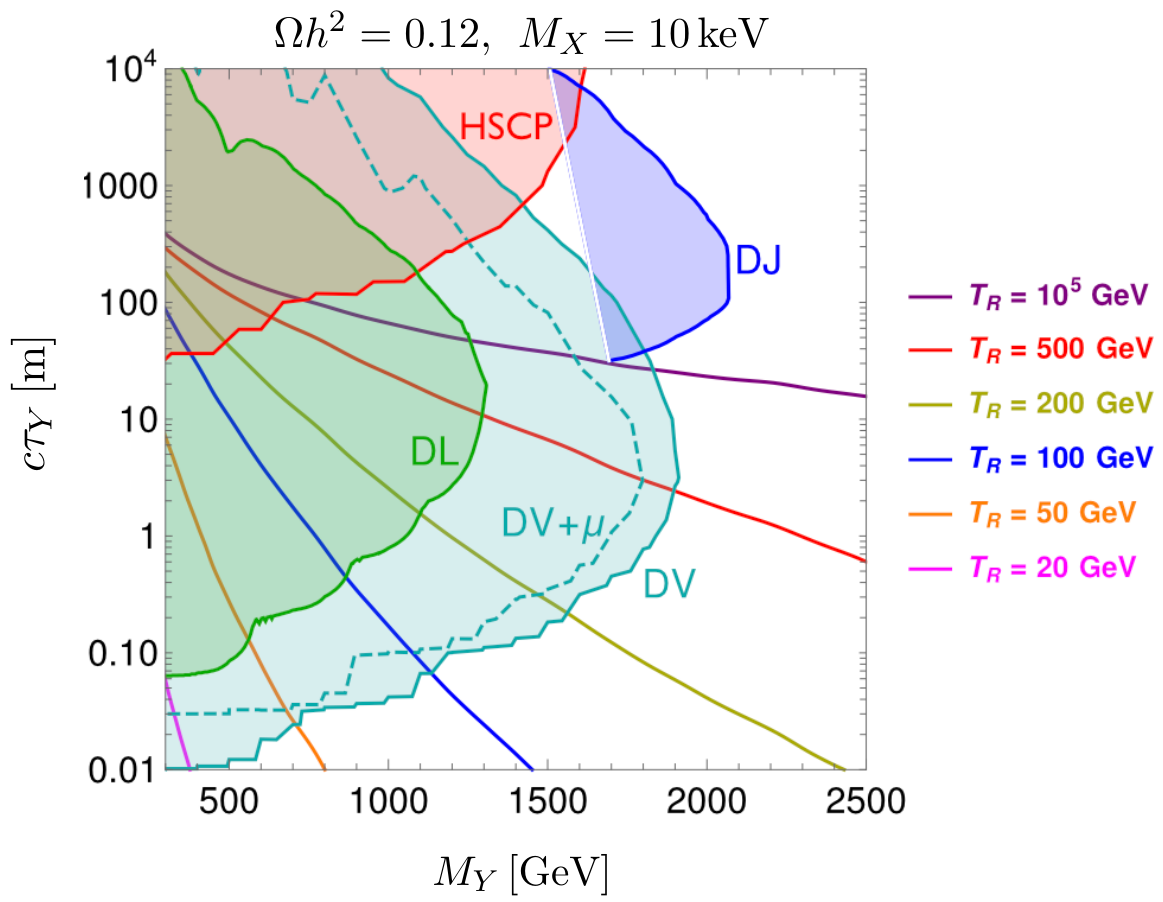}
  \caption{Isolines of constant reheating temperature $T_\text{rh}$ expressed in the $(M_Y, c\tau)$ plane, for scenarios accounting for the observed DM relic abundance and a DM mass of 12~keV for a muon-philic Majorana DM model (left), and 10~keV for a top-philic model with scalar DM (right). In the left panel, adapted from~\cite{Becker:2023tvd}, we study the impact of the reheating potential $V(\Phi) \sim \Phi^k$, and consider $T_\text{rh} = 20\,\text{GeV}$ (blue), $100\,\text{GeV}$ (black) and $10^4\,\text{GeV}$ (red, with details of reheating being no longer relevant for the considered mediator masses). Solid lines apply to $k = 2$, while dashed and dot-dashed lines illustrate fermionic and bosonic reheating scenarios for $k = 4$ potentials respectively. Various constraints from LLP searches are additionally shown as grey-shaded regions. In the right panel, adapted from~\cite{Calibbi:2021fld}, we consider $k=2$, vary $T_\text{rh}$, and show collider constraints from DJ (blue, \cite{CMS:2019qjk}), DV  (green, \cite{ATLAS:2017tny, ATLAS:2020xyo}) and DL (dark green, \cite{ATLAS:2020wjh}) searches.}\label{fig:ctaumP}
\end{figure} 

The interpretation of the LLP searches as achieved so far depends on the assumption used for the cosmological history~\cite{Co:2015pka, Belanger:2018sti, Brooijmans:2020yij, Calibbi:2021fld, Becker:2023tvd} (see also section~\ref{sec:nonstandardCosmo}). Here, we have considered reheating temperatures $T_\text{rh}$ much larger than the masses of the new physics states, which in particularly implies that $T_\text{rh} \gg M_Y$. Consequently, the collider phenomenology is insensitive to $T_\text{rh}$. However, for $T_\text{rh} \lesssim M_Y$, larger couplings would be required to match the observed relic density, expanding the regions of the parameter space accessible to LLP searches at colliders. In such cases, freeze-in occurs during reheating, with entropy injection diluting the relic abundance. The dilution depends on the reheating potential ($V(\Phi) \sim \Phi^k$) and on the fermionic or bosonic nature of the reheating. These dependencies are illustrated in figure~\ref{fig:ctaumP}, in particular for early matter domination scenarios corresponding to $k=2$. Concretely, for $T_\text{rh} \gg M_Y$, the region below the red solid line in the left panel of the figure would be excluded by structure formation constraints (\ie\ imposing $M_X \gtrsim 10$~keV)~\cite{Ballesteros:2020adh, DEramo:2020gpr, Decant:2021mhj}, thus limiting the relevance of LLP searches within the model. In case of a lower reheating temperatures $T_\text{rh} < M_Y$, this bound shifts towards smaller decay lengths, revealing new relevant regions of the parameter space that are testable at colliders with displaced signatures, as visible from the left and right panels of the figure. The observation of an associated LLP signal corresponding to a region that would be excluded in the case of a standard cosmological history could thus provide valuable insights into the dynamics of the (inflationary) reheating phase for a given DM model. Conversely, for a given reheating scenario, the measurement of the mediator mass and lifetime would give an absolute upper limit on the reheating temperature consistent with freezing-in DM. In this context, a coherent analysis of specific inflation models (thus with a particular reheating potential) reveals an interesting interplay with cosmological data from Planck~\cite{Becker:2023tvd}.

Alternatively, scenarios with a faster-than-standard expansion of the universe also require larger DM couplings to meet relic density constraints, enhancing again the importance of displaced object searches~\cite{DEramo:2017ecx}. \vspace{.5cm}

\paragraph{Conversion-driven freeze-out regime}\label{sec:LLPsCDFO}
\paragraph*{}\vspace{.3cm}

\begin{figure}
    \centering
    \includegraphics[width=0.62\textwidth,trim={0.4cm 0.66cm 0.4cm 0.4cm},clip]{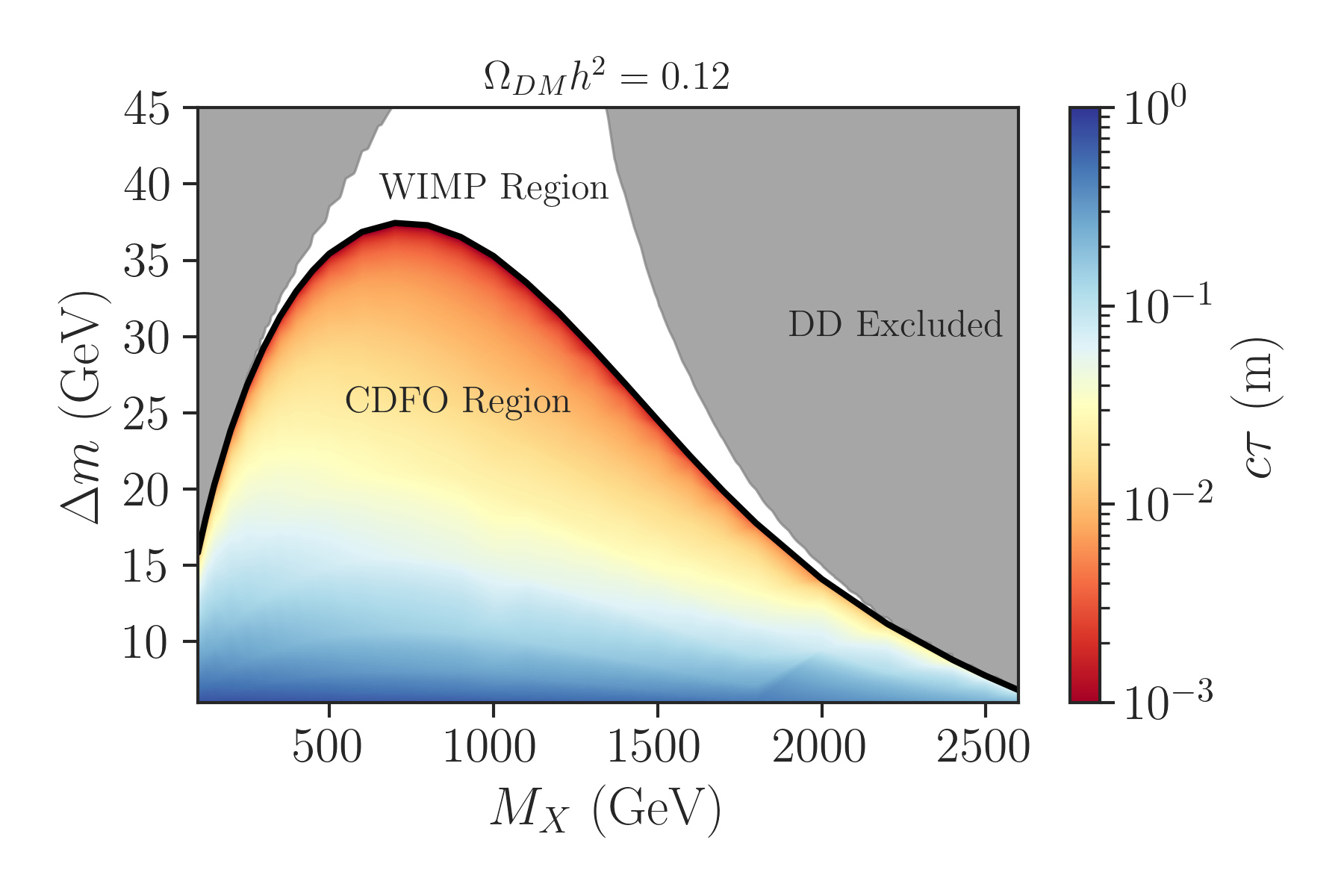}
    \caption{Representation of a typical cosmologically viable parameter space for $t$-channel DM scenarios featuring a bottom-philic scalar mediator~\cite{Garny:2021qsr}, adapted from~\cite{Heisig:2024xbh}. The black line separates the WIMP and CDFO regimes, where in the latter the mediator decay length is displayed via the colour code. We recall that in the WIMP regime decays are prompt.}
    \label{fig:CDFOlifetime}
\end{figure}

The cosmologically viable regions of the parameter space for CDFO scenarios is characterised by a relatively small mass splitting $\Delta m$ between the DM and mediator particles, below a few tens of GeV, and very weak DM couplings $\lambda$ of the order of $10^{-6}$~\cite{Garny:2017rxs}. This parameter space region borders the WIMP regime at larger $\Delta m$, where significantly stronger couplings are required to match the measured relic density. A striking feature of the CDFO scenario is its prediction of macroscopic mediator decay lengths, typically $1 \,\mathrm{mm} \lesssim c\tau \lesssim 1\,\mathrm{m}$ for two-body decays, or even larger for three-body-suppressed or four-body-suppressed decays. As an illustrative example, figure~\ref{fig:CDFOlifetime} shows this behaviour for a scenario with a bottom-philic scalar mediator. The shortest lifetimes occur for large $\Delta m$ near the boundary with the WIMP region, while lifetimes increase as $\Delta m$ approaches the kinematic threshold of the two-body decay defined by $\Delta m = M_f$ (with $f$ representing the SM fermion coupled to the mediator). In addition, for $\Delta m < M_f$, the mediator is typically detector-stable. The predicted lifetimes thus align well with the sensitivity range of LLP searches at the LHC, making this scenario particularly relevant for experimental exploration.

\begin{figure}
    \centering
    \includegraphics[width=0.55\textwidth,trim={0.4cm 0.4cm 0.4cm 0.4cm},clip]{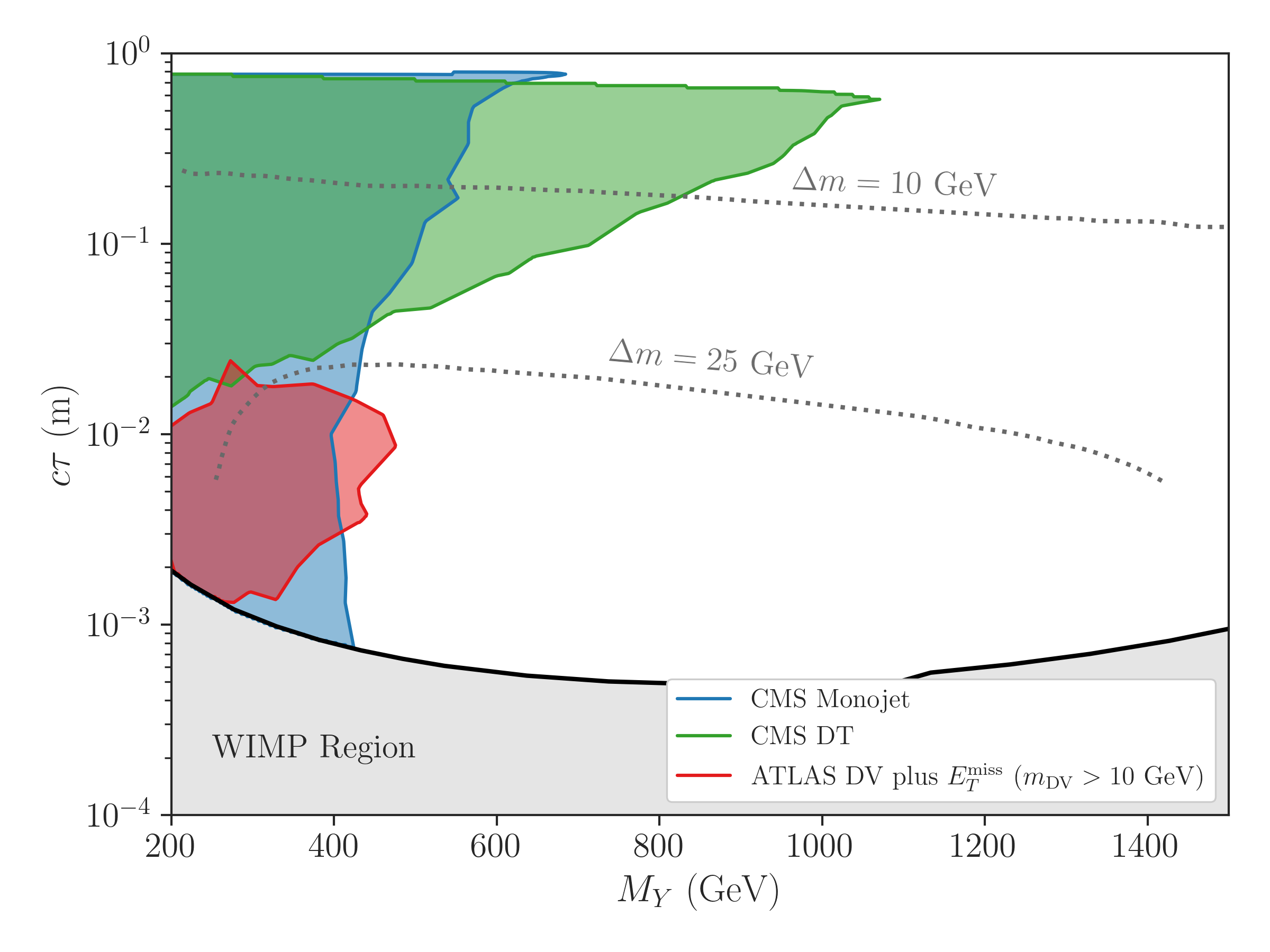}
    \caption{LHC constraints on the parameter space of a scenario featuring a bottom-philic mediator and DM production in the CDFO regime~\cite{Heisig:2024xbh}. We display constraints from LHC searches for disappearing tracks (green, \cite{CMS:2020atg}), displaced vertices (red, \cite{ATLAS:2017tny}) and monojet searches (blue, \cite{CMS:2021far}).} \label{fig:excCurves_mDV10}
\end{figure}

Within a quark-philic framework where the DM interacts with right-handed bottom quarks ($f = b$), different LLP searches are sensitive to specific ranges of mediator decay lengths. These include searches for heavy stable charged particles, disappearing tracks and displaced vertices, as well as searches for signals with missing transverse energy whose sensitivity could extend to longer lifetimes depending on their inclusiveness. Let us note that scenarios where $f=u,d,s,c$ yield a very similar phenomenology, but with a different location of the two-body threshold $\Delta m=M_f$. In contrast, for $f=t$, the mediator is effectively detector-stable throughout the entire CDFO parameter space as $\Delta m < M_t$, making HSCP searches alone well-suited to probing this case~\cite{Garny:2018icg}. Each class of searches covers different regions of the CDFO parameter space, as shown in figure~\ref{fig:excCurves_mDV10} which illustrates the constraints that could be imposed on the mediator mass and decay length for scenarios compliant with $\Omega h^2 = 0.12$. Bounds from CMS disappearing track searches~\cite{CMS:2020atg, CMS:2018rea} (green), the ATLAS displaced vertex$+\met$ search~\cite{ATLAS:2017tny} (red), and the CMS multijet+$\met$ search~\cite{CMS:2021far} (blue) are indicated in the figure, while those from HSCP searches are not shown. While these searches dominate at large $c\tau\gtrsim 10$~m or $\Delta m \lesssim M_b$, the corresponding region in parameter space is indeed omitted from the figure. In the detector-stable limit, scenarios with mediator masses up to approximately $1.3$~TeV would however turn out to be constrained~\cite{ATLAS:2019gqq, ATLAS:2022pib, CMS:2024nhn}.

For smaller decay lengths with $\Delta m > M_b$, disappearing track searches (green areas in figure~\ref{fig:excCurves_mDV10}) become most sensitive, the mass splitting dependence being highlighted in the figure through grey dotted isolines. However, care must be taken with the reinterpretation of the experimental disappearing track search results, focusing on colourless chargino decays to DM and an ultra-soft pion, to the CDFO signal. This requires accounting for differences originating from mediator hadronisation into neutral or charged $R$-hadrons, and from the specific properties of the $b$-jet emerging from the mediator decays (that is generally not ultra-soft for $\Delta m \gtrsim 10$~GeV). To solve this issue, we have made use of the implementation available in \lstinline{MadAnalysis 5}~\cite{Araz:2021akd} and enforced the presence of a charged $R$-hadron and $\Delta R(Y,b) > 0.2$ in order to avoid decay configurations where the $b$-jet is aligned with the track candidate. For decay lengths below approximately $1$~cm, displaced vertex and monojet searches provide the best sensitivity (red in figure~\ref{fig:excCurves_mDV10}). However, signal jets are usually soft and with a low invariant mass, yielding small signal efficiencies. Consequently, the alternative signature where an extra hard jet would stem from initial state radiation could be tested by conventional monojet searches (blue in figure~\ref{fig:excCurves_mDV10}). As it is not clear that events featuring displaced jets would be vetoed, we conservatively rejected, in our analysis, events exhibiting jets with a displacement larger than 2~mm and a transverse momentum $p_T > 20$~GeV. 

Despite all these efforts, the large mediator production cross section for $M_Y \lesssim 500\,\mathrm{GeV}$ and the potentially prominent LLP signatures, a substantial gap remains in the coverage of the parameter space for intermediate to small lifetimes ($\mathrm{mm} \lesssim c\tau \lesssim \mbox{few }\mathrm{cm}$), revealing an uncharted regime that we discuss further in section~\ref{sec:LLPsGaps}.

\begin{figure}
  \centering
  \includegraphics[width=0.55\textwidth]{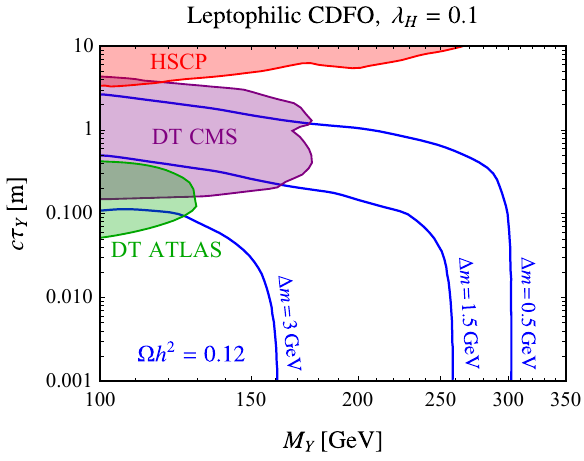}
  \caption{LHC constraints on the parameter space of a scenario featuring a muon-philic Majorana DM state and a scalar mediator, where the observed relic density is achieved within the CDFO regime~\cite{Junius:2019dci}. The blue contours correspond to isolines of correct relic abundance for different values of the mass splitting $\Delta m$, and we display constraints from LHC searches for HSCP (red, \cite{CMS:2016ybj, CMS:2015lsu}), and disappearing tracks (purple and green, \cite{CMS:2018rea, ATLAS:2017oal}). For this illustrative case a quartic coupling of $\lambda_H=0.1$ has been assumed between the scalar mediator and the SM Higgs boson.}\label{fig:lRCDV}
\end{figure}

We continue our discussion with leptophilic scenarios, that we illustrate with a setup involving a Majorana DM candidate coupled to a scalar slepton-like mediator~\cite{Junius:2019dci}. For an interaction strength $\lambda \lesssim 10^{-6}$, neglecting bound state formation effects~\cite{Garny:2021qsr, Binder:2023ckj} and ignoring Higgs-mediator couplings, the measured DM relic abundance can be achieved through CDFO production for mediator masses below $200~\text{GeV}$ and mass splittings $\Delta m < 2.6~\text{GeV}$. While such a setup is already stringently constrained by searches at the LHC, introducing a non-zero quartic coupling between the scalar mediator and the Higgs boson significantly enlarges the viable regions in the model parameter space. In figure~\ref{fig:lRCDV}, we consider such a quartic coupling $\lambda_H = 0.1$, and illustrate the constraints on the parameter space of a model including interactions with the right-handed muon. The results are as usual given in terms of the mediator lifetime and mass. The upper region of the plot corresponds to long-lived mediators with decay lengths of a few centimetres or more, or equivalently small values of the DM-mediator coupling together with a very compressed mass spectrum (as indicated by the blue contours). For $\lambda_H = 0.1$, the CDFO regime extends to mediator masses up to $300~\text{GeV}$, and necessitates mass splittings $\Delta m < 3~\text{GeV}$. We include in the figure the complementary constraints emerging from early LHC Run~2 searches for HSCP~\cite{CMS:2016ybj, CMS:2015lsu} (red) and disappearing tracks~\cite{ATLAS:2017oal, CMS:2018rea} (green and purple), that should get a little bit stronger by relying on more recent analyses like~\cite{CMS:2020atg, ATLAS:2019gqq}.

Similar to the quark-philic case, we observe a significant lack of experimental coverage in the region with small lifetimes, where the considered scenario predicts soft displaced leptons. Here, the small mass splitting renders the emitted leptons too soft to be efficiently reconstructed in existing displaced lepton searches. This contrasts sharply with leptophilic scenarios involving freezing-in DM in alternative cosmologies (see figure~\ref{fig:ctaumP}), where displaced lepton searches are highly effective. Nevertheless, it would be worthwhile to investigate whether recent DL searches like~\cite{CMS:2023bay} could probe this particular region of the CDFO parameter space. We further discuss this issue in section~\ref{sec:LLPsGaps}.

\begin{figure}
  \centering
  \includegraphics[width=0.48\textwidth,trim={0.0cm 0.1cm 0.0cm 0.0cm},clip]{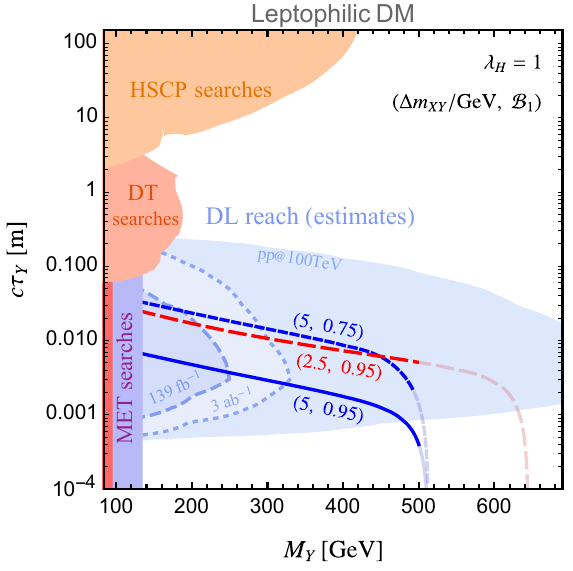}\hfill
  \includegraphics[width=0.48\textwidth,trim={0.0cm 0.55cm 0.0cm 0.0cm},clip]{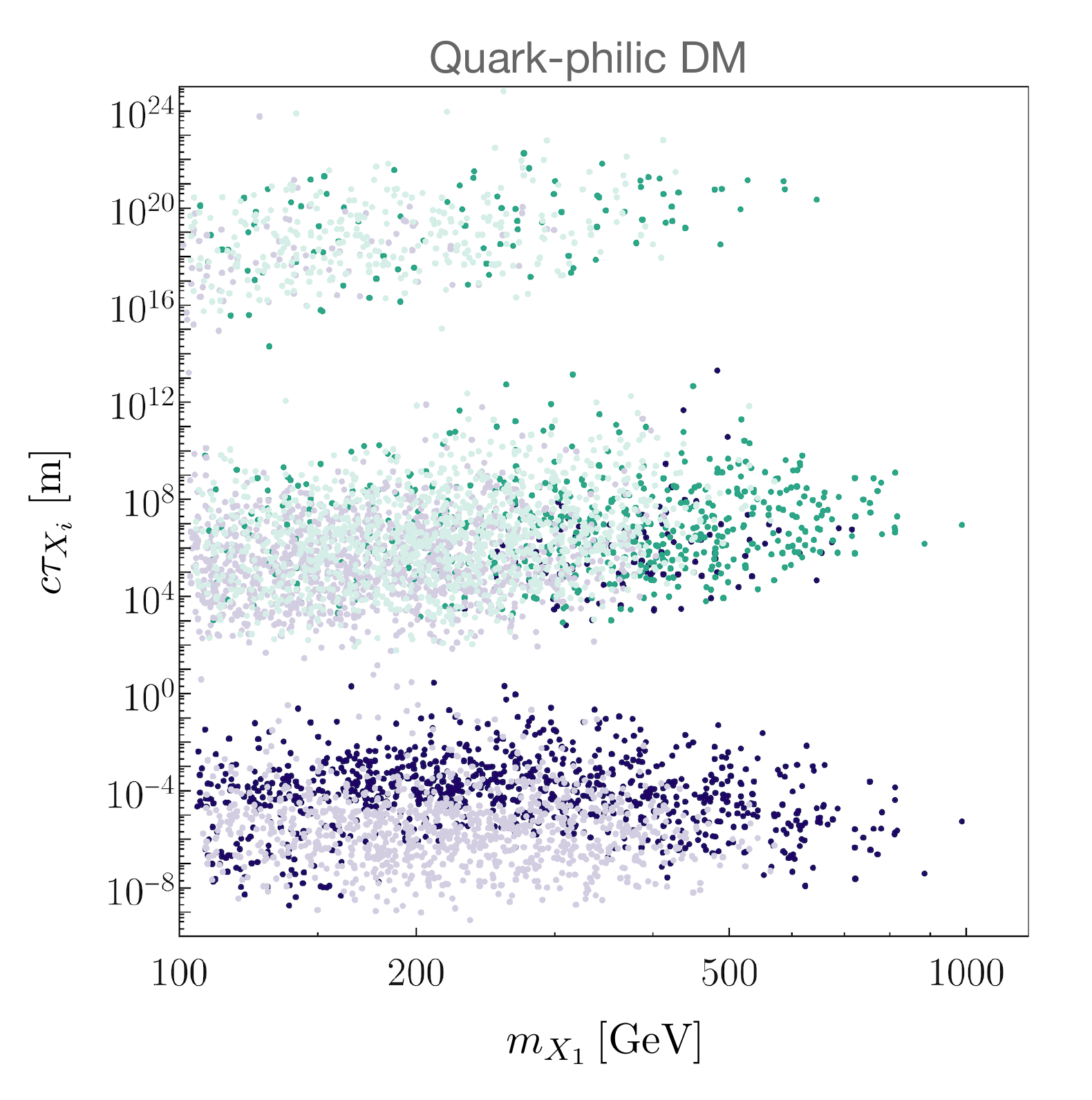}
  \caption{Constraints on CDFO flavoured DM scenarios, shown as a function of the proper lifetime and mass of the mediator (left, adapted from~\cite{Heisig:2024mwr}) and the heavier dark states (right, adapted from~\cite{Acaroglu:2023phy}). In the left panel, the blue and red lines represent slices in the parameter space of a leptophilic two-flavour model that simultaneously account for the observed DM abundance and successful leptogenesis (excluding the regions shown with faint colours). The solid shaded areas indicate current LHC constraints, while the transparent light blue areas illustrate projected sensitivities from dedicated searches. In the right panel, cosmologically viable points are shown for a three-flavour quark-philic model. The green and blue points represent the decay lengths of the heavier DM multiplet states $X_2$ and $X_1$, respectively, as functions of the mass of the heaviest state $X_1$.}\label{fig:FDM}
\end{figure}

CDFO realisations can also be achieved within non-minimal models~\cite{Acaroglu:2023phy, Heisig:2024mwr, Belanger:2021smw}. A particularly intriguing example is flavoured DM scenarios, where the DM state $X=(X_1, X_2, X_3)$ consists in a flavour multiplet whose lightest mass eigenstate $X_3$ serves as the DM candidate. Within these models, the CDFO mechanism has been explored for both quark-philic DM~\cite{Acaroglu:2023phy} and leptophilic DM~\cite{Heisig:2024mwr}. Notably, these frameworks offer the appealing prospect of addressing DM and baryogenesis within a unified framework, specifically through conversion-driven leptogenesis which provides additional motivation for experimental searches. In practice, CDFO DM production can manifest at colliders in multiple ways, potentially leading to signatures from either pair-produced long-lived mediators or from the production of long-lived heavier dark multiplet states $X_1$ or $X_2$. In the former case, the signatures resemble those of minimal models but with quantitative differences in the preferred lifetimes due to the presence of the additional states. For instance, in conversion-driven leptogenesis scenarios, the preferred decay lengths range from a few centimetres to millimetres, with $\Delta m \lesssim 5~\text{GeV}$ (see the left panel of figure~\ref{fig:FDM}). Moreover, these models predict soft, mildly displaced leptons as a challenging but intriguing target for LHC searches~\cite{Heisig:2024mwr}.

Alternatively, if the coupling matrix $\lambda$ exhibits significant hierarchies, a heavier multiplet state $X_i$ can interact more sizeably with the SM and become the LLP produced at the LHC. In such cases, an electrically neutral LLP would not generate detectable tracks but could lead to displaced vertices. Interestingly, the kinematic suppression inherent to the three-body decay via an off-shell mediator can result in large lifetimes for these states. The right panel of figure~\ref{fig:FDM} illustrates the dependence of the proper decay lengths of the heavier DM multiplet states $X_2$ (green) and $X_3$ (blue) on the mass of the heaviest state in a quark-philic setup~\cite{Acaroglu:2023phy}. The large lifetimes of these neutral $X_i$ states allow them to escape LHC detectors, making them ideal candidates for detection at dedicated experiments such as MATHUSLA~\cite{Curtin:2018mvb}. 

\subsubsection{Gaps in coverage}\label{sec:LLPsGaps}
The scenarios discussed in section~\ref{sec:LLPscoverage} illustrate the importance of LLP searches in testing various DM mechanisms. Despite the broad scope of the LLP search programme at the LHC, specific regions of the various parameter spaces considered remain uncovered by current search strategies. In particular, within the CDFO scenario, configurations with $c \tau \lesssim 1$~cm ($\Delta m \gtrsim 10$--$30$~GeV) are not covered because they fall outside the lifetime range addressed by disappearing track searches, and/or the LLP visible decay products are too soft to be detected in conventional displaced vertex searches. This gap in coverage is approximately represented by the white region shown in figure~\ref{fig:schematic}.

\begin{figure}
  \centering
  \includegraphics[width=0.525\textwidth,trim={0.45cm 0.61cm 0.4cm 0.4cm},clip]{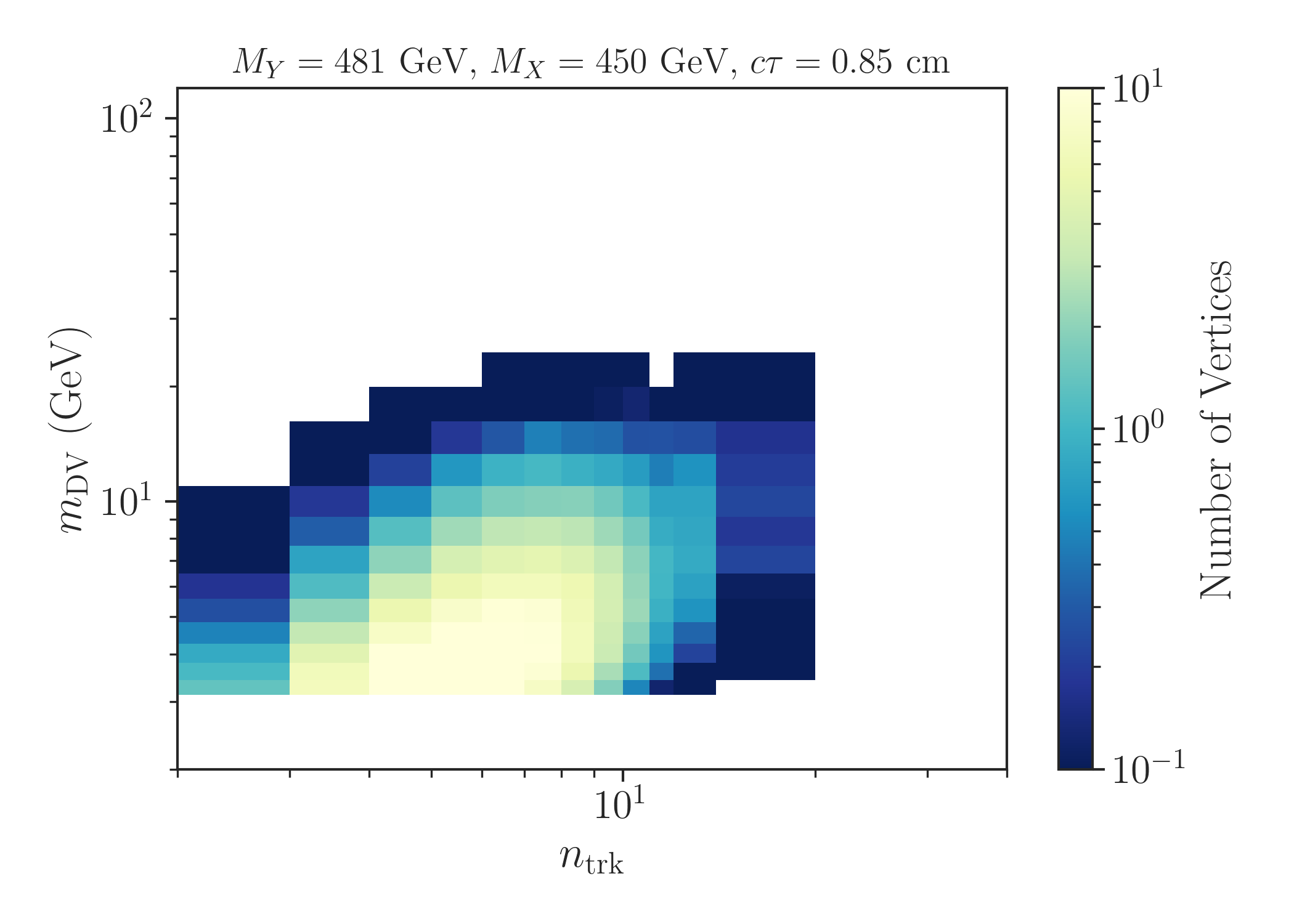}\hfill
  \includegraphics[width=0.47\textwidth,trim={0.4cm 0.4cm 0.4cm 0.4cm},clip]{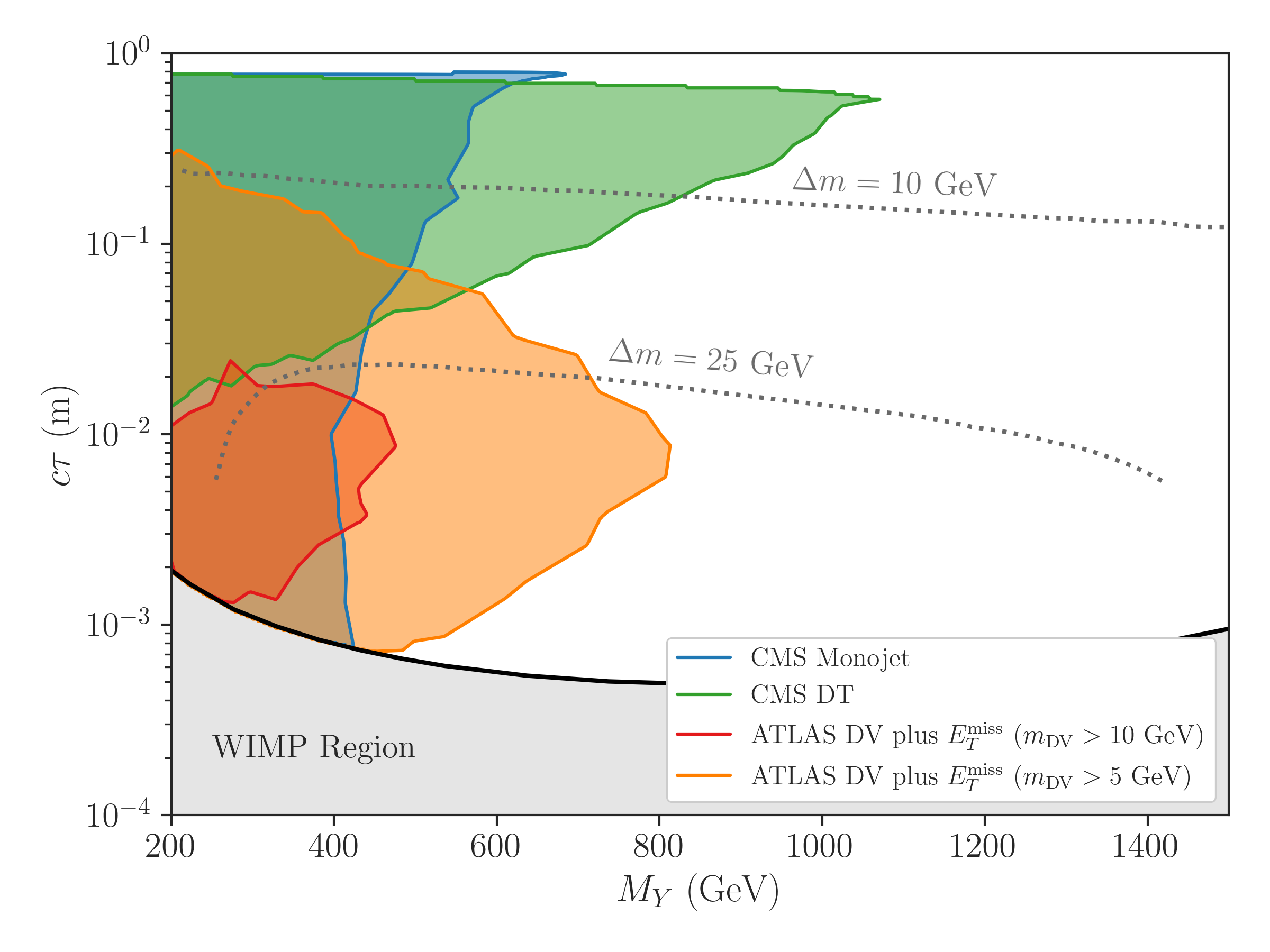}
  \caption{In the left panel, we show the two-dimensional distribution in the number of displaced vertices as a function of their invariant mass $m_\text{DV}$ and their number of tracks $n_\text{trk}$, for a quark-philic CDFO benchmark scenario with $M_{Y} = 481$~GeV, $\Delta m = 31$~GeV and $\lambda = 3.9 \times 10^{-7}$~\cite{Heisig:2024xbh}. In the right panel, we examine how the parameter space region excluded by the ATLAS DV search depends on the specific requirements on the displaced vertices, as detailed in the text.}\label{fig:mDVnTracks}
\end{figure}

The quark-philic scenario discussed above provides an example of this gap in coverage and how it could be addressed with minor modifications to existing search strategies. For instance, we could consider the ATLAS displaced vertex search~\cite{ATLAS:2017tny} (ATLAS-SUSY-2016-08), which relies on a $\met$ trigger and is sensitive to decay lengths between 4~mm and 30~cm. The signal region defined in this search requires at least one displaced vertex with five or more tracks and an invariant mass exceeding 10~GeV. In the left panel of figure~\ref{fig:mDVnTracks}, we present the distribution of the number of displaced vertices as a function of the number of tracks $n_{\rm trk}$ associated with it and of the invariant mass oof the vertex $m_{\rm DV}$, in the framework of a quark-philic CDFO benchmark model with $M_X = 450$~GeV, $\Delta m = 31$~GeV, and $c \tau = 0.85$~cm. Most vertices have $3 < n_{\rm trk} < 9$ and $m_{\rm DV} < 6$~GeV, making them too soft to satisfy the ATLAS analysis requirements. For this benchmark, the $m_{\rm DV}$ cut reduces the signal efficiency from approximately $6\%$ to $0.1\%$, therefore significantly diminishing the search sensitivity. Although the high $m_{\rm DV}$ threshold effectively suppresses SM backgrounds, it could likely be lowered without a substantial increase in background contamination. Indeed, the ATLAS search observed zero vertices with $n_{\rm trk} > 5$ and $m_{\rm DV} > 4$~GeV so that lowering the $m_{\rm DV}$ requirement, for instance to 5~GeV, could still maintain an effective SM background suppression. Assuming that the SM background associated with the modified selection $m_{\rm DV} > 5$~GeV remains negligible (\ie\ fewer than one vertex), we estimate in the right panel of figure~\ref{fig:mDVnTracks} the sensitivity improvement from reducing the invariant mass threshold from 10~GeV (red exclusion curve) to 5~GeV (orange exclusion curves), as further detailed in \cite{Heisig:2024xbh}. This demonstrates that a significant portion of parameter space becomes accessible when mildly soft displaced vertices are included, this region being in addition not covered by other LLP searches.

Similarly, the strategy employed in the mildly displaced soft track search from~\cite{ATLAS:2024umc} could also address a similar gap in the leptophilic CDFO case. Here, soft and mildly displaced leptons arise for moderate $\Delta m$ values so that, as illustrated in figure~\ref{fig:lRCDV}, scenarios with $\Delta m > 3$~GeV are currently unconstrained. Comparable gaps in coverage exist in non-minimal models as well, as seen in the right panel of figure~\ref{fig:FDM} and \cite{Heisig:2024mwr} for the leptophilic case, particularly in conversion-driven leptogenesis scenarios, and \cite{Acaroglu:2023phy} for the quark-philic case.

\begin{figure}
  \centering
  \includegraphics[width=0.47\textwidth,trim={0.25cm 0.2cm 0.0cm 0.4cm},clip]{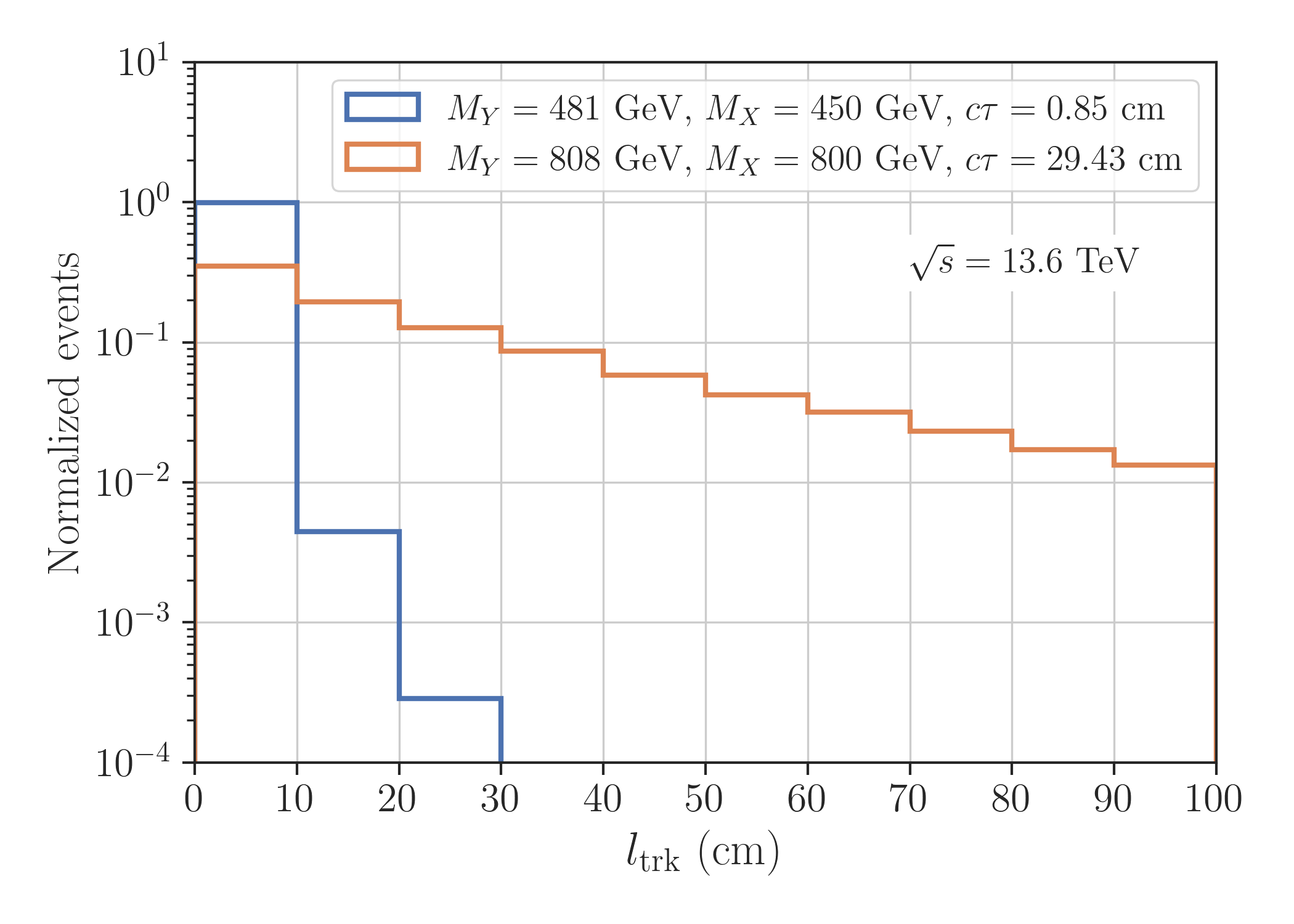}\hfill
  \includegraphics[width=0.47\textwidth,trim={0.45cm 0.2cm 0.0cm 0.4cm},clip]{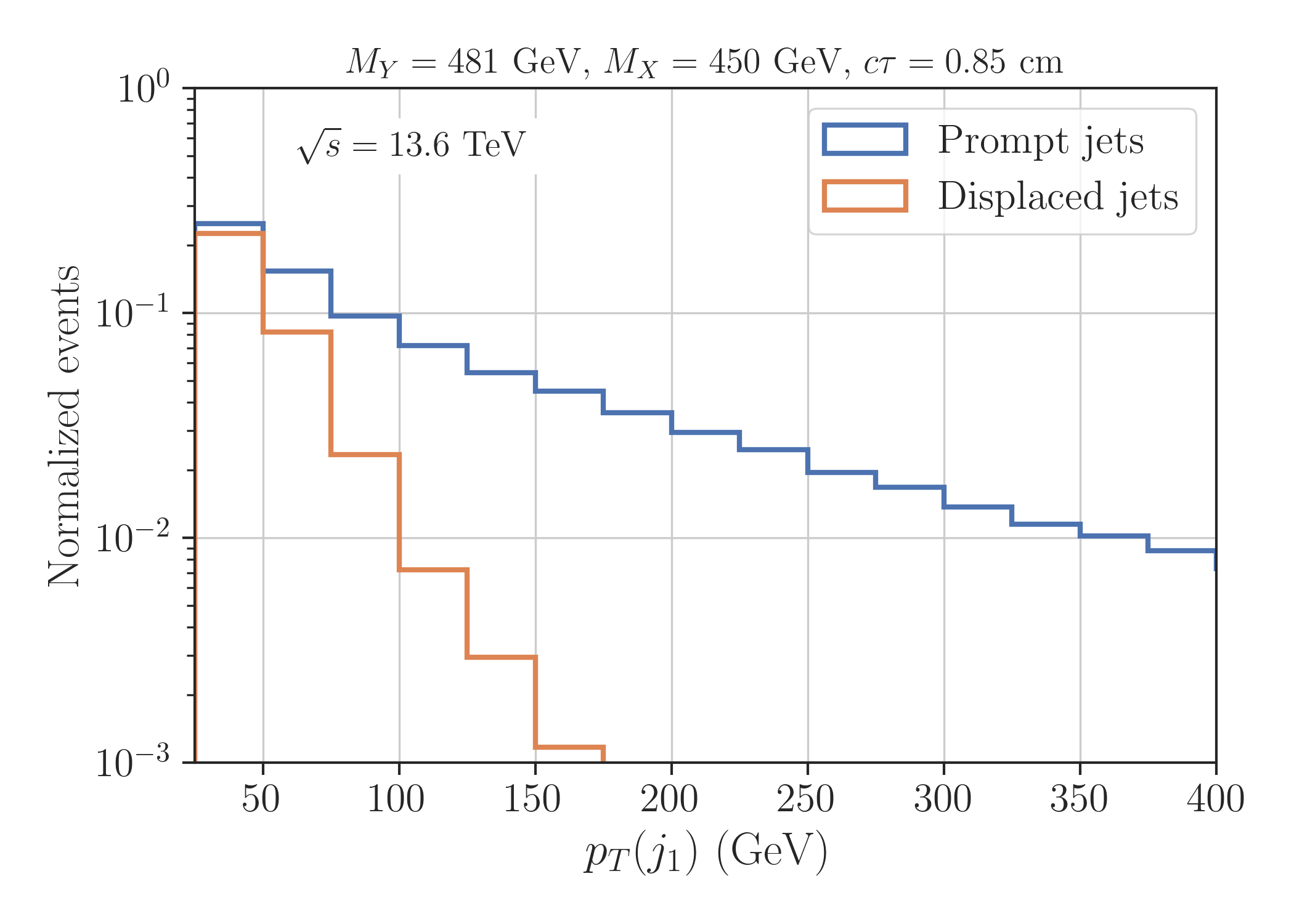}\\
  \includegraphics[width=0.47\textwidth,trim={0.45cm 0.6cm 0.0cm 0.2cm},clip]{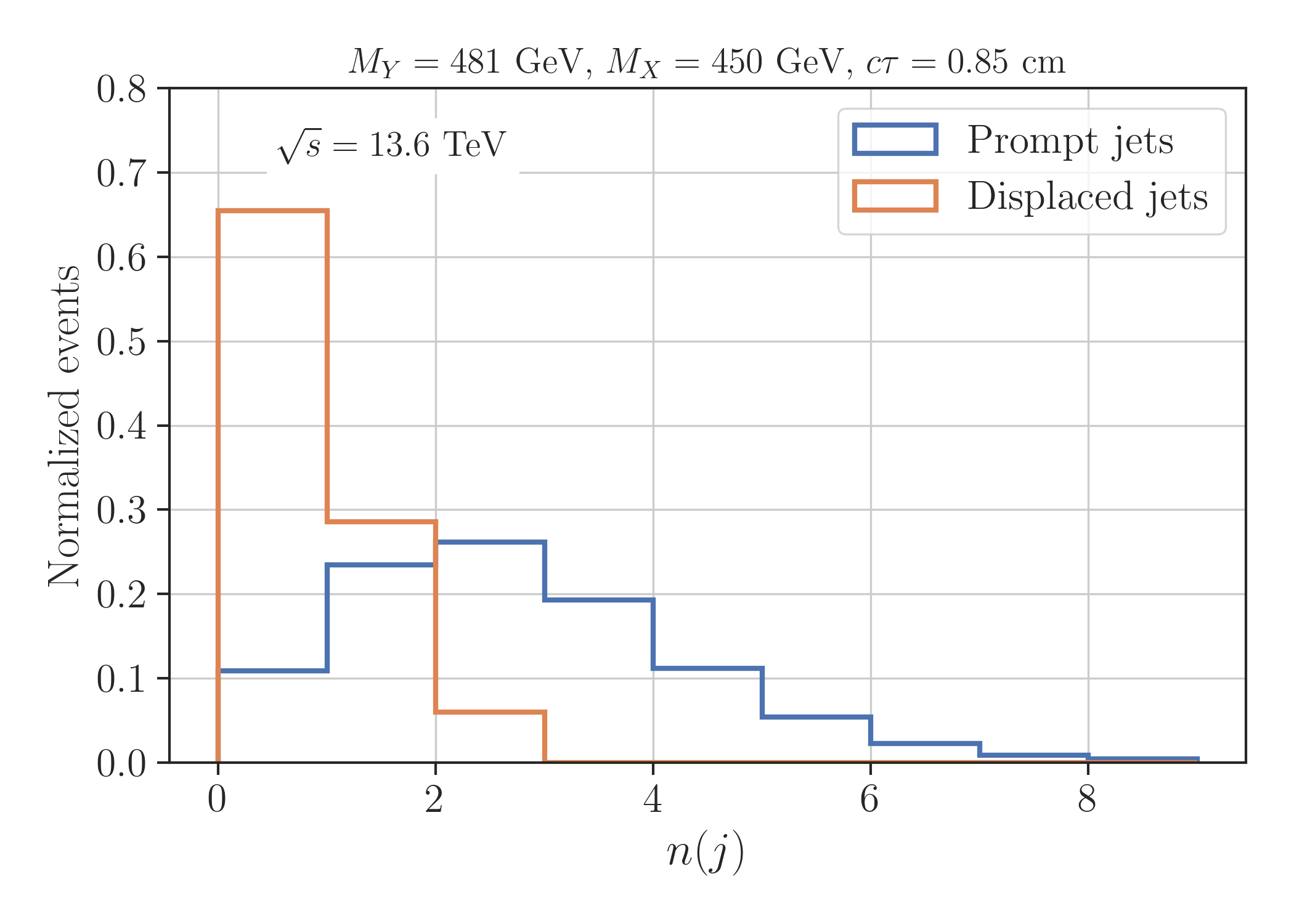}\hfill
  \includegraphics[width=0.47\textwidth,trim={0.45cm 0.6cm 0.0cm 0.2cm},clip]{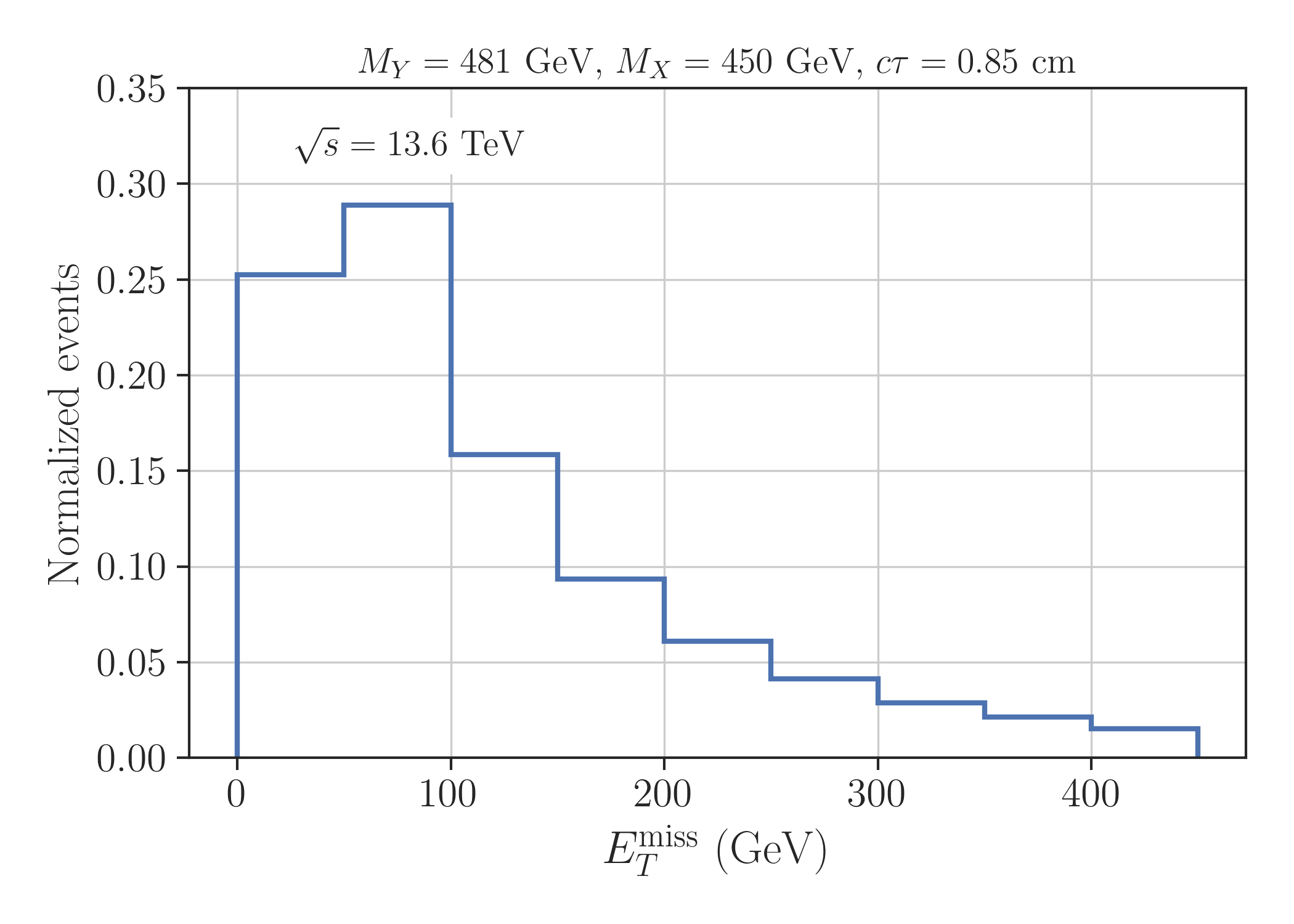}
  \caption{Kinematic properties of an LLP signal relevant to design new LLP searches at the 13.6\,TeV LHC, emerging from a typical quark-philic CDFO scenario. We consider the distribution in the charged $R$-hadron decay length (upper left) for a benchmark with $c \tau \simeq 1$~cm (blue) and $30$~cm (orange), the $p_T$ distributions of prompt (blue) and displaced (orange) jets with $p_T(j) > 25$~GeV and $|\eta(j)| < 5$ (upper right) as well as the associated prompt (blue) and displaced (orange) jet multiplicity spectrum (lower left), and the $\met$ distribution (lower right).} \label{fig:LLPdistrib1} 
\end{figure}

Beyond soft displaced objects, additional signatures could be explored to enhance background suppression. 
These include more sophisticated $\met$-based discriminators, timing information from the tracker or calorimeters, and the combination of the prompt track of the mediator with the displaced track of its decay products (yielding so-called `kinked tracks'). In figure~\ref{fig:LLPdistrib1}, we present some key distributions relevant for designing such searches, using the quark-philic model discussed earlier as an example. The upper-left panel shows the track length distributions for two benchmarks with $\Delta m = 31$~GeV ($c\tau \simeq 1$~cm) and $\Delta m = 8$~GeV ($c\tau \simeq 30$~cm). The colour-charged mediator, which forms charged $R$-hadrons with SM quarks, produces highly ionising tracks. In addition, those decaying within the tracker but traversing multiple inner layers give rise to disappearing or kinked tracks, depending on the softness of the decay products. Since $\Delta m < 40$~GeV is required by the CDFO mechanism, the displaced jets are in general relatively soft so that hard jets from initial state radiation may be necessary to satisfy trigger requirements. We refer to~\cite{Heisig:2024xbh} for further discussion.

\subsubsection{Advancing reinterpretation methods}\label{sec:RecastMaterial}
We conclude this chapter with some remarks on the material necessary for the reinterpretation of LLP searches (see also~\cite{LHCReinterpretationForum:2020xtr} and chapter~6 of~\cite{Alimena:2019zri}).

As emphasised in section~\ref{sec:LLPscoverage}, the scarcity of validated recast analyses is a major obstacle to assessing the current coverage of the LLP parameter spaces and to identifying the associated gaps. Therefore, it is pertinent to discuss potential measures to improve the situation, which requires adequate resources and collaboration. Reinterpretation tools are typically developed by small teams, often consisting of only a few researchers. In contrast, each experimental analysis is conducted by a dedicated team of analysts. Consequently, a single `reinterpreter' must handle the implementation and validation of numerous experimental analyses, frequently resulting in delays between the publication of an analysis and its incorporation into public reinterpretation frameworks. Consequently, by the time at which reinterpretation tools are updated, obtaining further information about the original analysis can become challenging, if not impossible. A more direct interaction between theorists and experimentalists could help alleviate these issues.

Another critical factor is access to data from experimental collaborations, which is essential for proper implementation and validation. For instance, implementing analyses into \lstinline{SModelS} requires acceptance times efficiency maps for each signal region for pure simplified models, given as a function of the simplified-model parameters. These can be either provided directly by the experimental collaborations, or derived via simulation-based recasting. Full-fledged recast codes, on the other hand, demand clear and unambiguous object definitions, a step-by-step description of the analysis logic, and truth-level versus reconstruction-level efficiencies. Moreover, recasting LLP searches presents additional challenges due to their reliance on non-standard objects, low-level detector inputs not reproducible in fast simulation, and the use of machine learning techniques for optimisation.\footnote{Machine-learning-based analyses have long been considered impossible to recast; see, however, \cite{Araz:2023mda} for a discussion of recent progress and guidelines for reusable machine-learning models in LHC analyses.}  

Significant strides have been made to address these challenges, with several recent examples demonstrating good practices. First, the ATLAS Run 2 search for LLPs decaying into hadronic jets in the calorimeter~\cite{ATLAS:2022zhj} provided six-dimensional efficiencies for an event to enter Region A (of the ABCD method). These efficiencies parametrise the output of a sequence of boosted decision trees and neural networks as functions of the LLP kinematics, decay type and decay position. An example code illustrating their usage is also included. Second, the CMS Run 3 search for LLPs decaying to a pair of muons~\cite{CMS:2024qxz} published truth-level signal efficiencies for the considered dimuon categories. These efficiencies are functions of the minimum muon transverse momentum $p_T$ and displacement $d_0$ in three intervals of the generated transverse decay length $L_{xy}$. The collaboration also provided detailed reinterpretation instructions. As a last example, the updated ATLAS Run 2 search for pairs of neutral LLPs in events with displaced jets and leptons~\cite{ATLAS:2024ocv} (extending~\cite{ATLAS:2022zhj}) went a significant step further. It included a BDT trained to compute the overall selection probability in the ABCD plane, using truth-level input of the decay position ($L_{xy}$, $L_z$), kinematics ($p_T$, $\eta$, $\met$), and Child ID. This approach effectively serves as a surrogate model for the complete analysis, incorporating detector effects. The trained BDTs were provided as pickle files alongside example code demonstrating their use. In each of the examples above, the accuracy and range of validity were thoroughly documented, making such material valuable for advancing recasting efforts.

Finally, reliable reinterpretation efforts also require detailed information about the probability models underlying the analyses~\cite{Cranmer:2021urp}. This is particularly crucial when combining signal regions in the statistical interpretation or performing combined fits to signal and control regions. Since 2019, ATLAS has been providing comprehensive \lstinline{HistFactory} models in \lstinline{JSON} format~\cite{ATLAS:2019oik}, mainly for supersymmetry and top quark analyses. More recently, CMS released its \lstinline{Combine} software~\cite{CMS:2024onh} along with data cards detailing early Higgs boson measurements. These developments represent significant progress, and we hope to see similar initiatives applied to LLP searches.

\newpage\section{Cosmology}\label{sec:cosmology}
\noindent \textit{Contributions from C. Arina, M.J. Baker, M. Becker, A. Belyaev, M. Blanke, L.M. Carpenter, E. Copello, B. Fuks, M. Garny, J.~Harz, J. Heisig A. Ibarra, S. Khalil, L. Lopez-Honorez, K. Mohan, A. Moreno Briceño, T. Murphy, L. Panizzi, D. Sengupta, W.~Shepherd, A. Thamm}\vspace{.2cm}

\subsection{Generalities}

In this section, we explore the phenomenology of dark matter in the $t$-channel models introduced in section~\ref{sec:models}, which encompass both minimal and non-minimal realisations. We outline the constraints and requirements that a viable dark matter candidate must satisfy from cosmological and astroparticle perspectives. We then assess the primary impact of these constraints on the parameter space of the model, providing explicit examples from the existing literature on $t$-channel models. Among minimal scenarios, a Majorana dark matter candidate with a scalar mediator is the most extensively studied case, particularly when it couples to right-handed third-generation quarks. As a benchmark, we will thus frequently refer to the \lstinline{S3M_tR} model.

This section is organised as follows. In section~\ref{sec:DMprod}, we discuss the mechanisms available to achieve the correct relic density in the early universe. Section~\ref{sec:searches} addresses the primary signatures for direct and indirect dark matter searches, as well as the constraints from early-universe physics. Finally, in section~\ref{sec:CosmoConsBench}, we examine the associated impact on the parameter space of selected benchmark models.

This section serves as a reference for understanding the complementarity between cosmological and astrophysical searches for dark matter, and collider-based searches. Each subsection is designed to convey the key concepts while providing references for readers seeking a deeper understanding of the topics discussed. For a more general overview of dark matter constraints, models, and features, we direct readers to recent reviews~\cite{Cirelli:2024ssz, Balazs:2024uyj}.

\subsection{Dark matter relic density} \label{sec:DMprod}
In $t$-channel mediator models, the DM relic density can be achieved through a variety of production mechanisms, as already sketched in section~\ref{sec:generalities_llp}. These include (by decreasing DM interaction strength) canonical WIMP freeze-out (FO)~\cite{Gondolo:1990dk, Griest:1990kh}, conversion-driven freeze-out (CDFO)~\cite{Garny:2017rxs, DAgnolo:2017dbv}, freeze-in (FI)~\cite{McDonald:2001vt, Asaka:2005cn, Hall:2009bx}, and superWIMP (SW)~\cite{Covi:1999ty, Feng:2003uy} production. Describing these mechanisms requires at least two coupled Boltzmann evolution equations for the DM $X$ and the mediator $Y$. A priori, these equations
should be solved at the level of the distribution functions taking the form:
\begin{equation}
  \frac{\d f_i}{\d t}={\cal C}(f_i,f_j)\,,
\label{eq:BEf}\end{equation}
where $\d/\d t$ is the total time derivative and ${\cal C}(f_i, f_j)$ is the collision operator. The distributions $f_i$ and $f_j$ are the distribution functions for particle species $i$ and $j$, which, for the purpose of DM relic abundance calculations, can be considered as functions of proper time and of the particle momenta. Non standard cosmology (early matter-dominated era, see
\eg~\cite{Chung:1998rq, Giudice:2000ex, Allahverdi:2020bys, Drees:2017iod, Co:2015pka, Calibbi:2021fld, Becker:2023tvd, Bernal:2022wck}) will essentially affect the left-hand side of this equation through a specific Hubble expansion rate, while particle physics processes affect its right-hand side. Unless stated otherwise, a standard cosmological history is assumed, with non-standard cosmology being discussed briefly in section~\ref{sec:nonstandardCosmo}.

In standard cosmology, the universe is radiation-dominated during DM production, with energy density $\rho=g_*\frac{\pi}{30}T^4$ and entropy density $s=g_{*S}\frac{\pi}{45}T^3$ satisfying $sa^3=\mathrm{constant}$. In these expressions, $g_*$ and $g_{*S}$ are the conventional effective numbers of relativistic degrees of freedom. Assuming a Boltzmann distribution for all particles involved in the DM production process and negligible variation of the number of relativistic degrees of freedom during DM production, the evolution equations can be integrated over the momentum of the particle species $i$, yielding~\cite{Edsjo:1997bg}:
\begin{equation}\label{eq:BEfull}
  \frac{\d Y_i}{\d x}=\frac{1}{3H  s^2}\frac{\d s}{\d x}\sum_{ij} \left[ \gamma_{ij} \left(\frac{Y_{i} Y_{j}}{Y_{i}^{\rm eq} Y_{j}^{\rm eq}}-1 \right) - \gamma_{j \to i }\left( \frac{Y_{j}}{Y_{j}^{\rm eq}}-\frac{Y_{i}}{Y_{i}^{\rm eq}} \right) \right]\,.
\end{equation}
Here the quantities $\gamma_{ij}$ and $\gamma_{j \to i}$ represent the reaction densities (described below), $Y_i=n_i/s$ are the comoving number density for the particle species $i=X,Y$ with $n_i$ being the number density, and $Y_{i}^{\rm eq}=n_i^{\rm eq}/s$ are the associated equilibrium comoving density with $n_i^{\rm eq}$ being the $i^\mathrm{th}$ species number density assuming kinetic equilibrium and zero chemical potential $\mu_i$. If kinetic equilibrium is maintained while the dark matter chemically decouples, as typical of freeze-out scenarios, $n_i/n_i^{\rm eq}= \exp(-\mu_i/T)$. Moreover, \eqref{eq:BEfull} relies on a standard and convenient practice that introduces a dimensionless time variable $x$ inversely proportional to the bath temperature $T$, thus defined by $x=M_X/T$ $(M_Y/T)$ for freeze-out (FI and SW) scenarios. Finally, we remind that $H$ is the Hubble rate and that the prefactor in \eqref{eq:BEfull} simplifies under the assumption of approximately constant $g_{*S}$, for which $\d s/\d x\simeq -3s/x$. This assumption will be implicit in the following (for simplicity), unless specified otherwise.

In order to write \eqref{eq:BEfull}, we have assumed that all particles involved in (co-)annihilation and conversion processes are in kinetic equilibrium at the time of DM production. For non-thermal DM production discussed in sections~\ref{sec:FI} and~\ref{sec:SW}, the relevant relation~\eqref{eq:BEdec} can be obtained from \eqref{eq:BEfull} without making any assumptions on the DM distribution, but after instead enforcing a negligible initial abundance for $X$ and suppressed couplings to $Y$. Equation~\eqref{eq:BEfull} is thus also perfectly suitable for evaluating the DM comoving number density $Y_X$ in FI and SW scenarios for which the loss terms for the dark matter particle $X$ are neglected.\footnote{For further discussion on kinetic decoupling, see \eg~\cite{Bringmann:2006mu, Binder:2017rgn, Garny:2017rxs}.} In addition, the semi-classical Boltzmann equation~\eqref{eq:BEfull} neglects various quantum and thermal effects, such as quantum coherence, quantum statistics, screening, multiple scatterings, and particle interactions with coherent condensates. A systematic framework for deriving quantum kinetic equations that incorporate these effects in relic density computations is provided by the Schwinger-Keldysh formalism (see, \eg, \cite{Beneke:2014gla, Drewes:2015eoa, Binder:2018znk, Becker:2023vwd, Ai:2023qnr, Kainulainen:2024etd,Becker:2025yvb}, as well as \cite{Kim:2016kxt, Biondini:2020ric, DEramo:2021lgb, Bouzoud:2024bom} for related computations in equilibrium and \cite{Kainulainen:2024etd} for a lattice-based approach). In particular for freeze-in scenarios quantum and thermal effects can alter the relic density significantly. This was demonstrated in~\cite{Biondini:2020ric} for a Majorana DM model and in~\cite{Becker:2023vwd} for a scalar DM model, respectively. The latter work, which employs non-equilibrium quantum field theory from first principles, quantifies the expected corrections to various approaches commonly used in the literature (freeze-in with only decays, including scattering with and without thermal masses, hard thermal loop approximation, \etc) and provides recommendations for accurate results in phenomenological studies. For freeze-out scenarios and in particular for co-annihilating configurations, significant impact is expected from the Sommerfeld effect and bound state formation, as discussed in section~\ref{sec:SE-BSF}. However, only small corrections are expected due to thermal effects, as detailed in the context of dark matter annihilation in, \eg, \cite{Beneke:2016jpw,Beneke:2014gla} and Sommerfeld and bound state formation in, \eg, \cite{vonHarling:2014kha, Kim:2016kxt, Binder:2020efn,Binder:2019erp, Biondini:2023zcz, Biondini:2024aan, Biondini:2025jvp}.

The $\gamma_{ij}$ quantities appearing in \eqref{eq:BEfull} represent the reaction densities for (co-)annihilation of the dark species $i$ and $j$, while $\gamma_{i\to j}= \gamma_{i\to jk}^{\rm dec}+\gamma_{ik\to jl}^{\rm scat}$ govern the conversion processes of the dark species $i$ to $j$ through (inverse-)decays or scatterings off SM particles $k,l$. For $2\to 2$ and $1 \to 2$ processes, the reaction densities respectively take the forms:
\be \label{eq:reaction-densities}\bsp
 \gamma_{ij\to kl} =& \int \d\phi_i\, \d\phi_j\, f_i^{\rm eq}(t,p_i)\, f_j^{\rm eq}(t,p_j) \int\d\phi_k\, \d\phi_l\, (2\pi)^4\delta^4(p_i\!+\!p_j\!-\!p_k\!-\!p_l)\, \big|{\cal M}_{ij\to kl}\big|^2\,,\\
 \gamma_{i\to j k} =& \int \d\phi_i\, \d\phi_j\, \d\phi_k\, f_i^{\rm eq}(t,p_i)\, (2\pi)^4\delta^4(p_i-p_j-p_k)\, \big|{\cal M}_{i\to j k}\big|^2 \,, 
\esp\ee
where $f_i$ and $p_i$ denote the $i^\mathrm{th}$ species distribution function and momentum, while $|{\cal M}|^2$ represents the relevant squared scattering amplitude averaged over the initial-state {\it and} final-state degrees of freedom, and $\d\phi_i=g_i \d^3p_i/(2E_i(2\pi)^3)$ stands for a one-body phase-space element. These reaction densities can be expressed in terms of thermally averaged cross sections and decay rates of the particle species $i$ in its rest frame.  For instance, for (co-)annihilation processes and for the $i\to jk$ decay contribution to a conversion processes,
\be\bsp
  \gamma_{ij}=&\ n_i^{\rm eq}\, n_j^{\rm eq}\, \langle\sigma_{ij}v_{ij}\rangle\,,\\
  \gamma_{i\to j}^{\rm dec}=&\ n_i^{\rm eq}\, \langle\Gamma_{i\to jk}\rangle =  n_i^{\rm eq}\, \Gamma_{i\to jk}\, \frac{K_1(M_i/T)}{K_2(M_i/T)} \,.
\esp\ee
Here $\langle  \sigma_{ij}v\rangle$ stands for the thermally averaged (co-)annihilation cross section entering the freeze-out relic abundance calculations, and $\langle\Gamma_{i\to jk}\rangle$ is the thermally averaged decay rate related to the process $i\to jk$ in the rest frame of particle $i$ (with $M_i > M_j + M_k$). Moreover, $K_1$ and $K_2$ are the modified Bessel functions of the second kind. 

\begin{figure}
    \centering
    \includegraphics[width=0.7\textwidth]{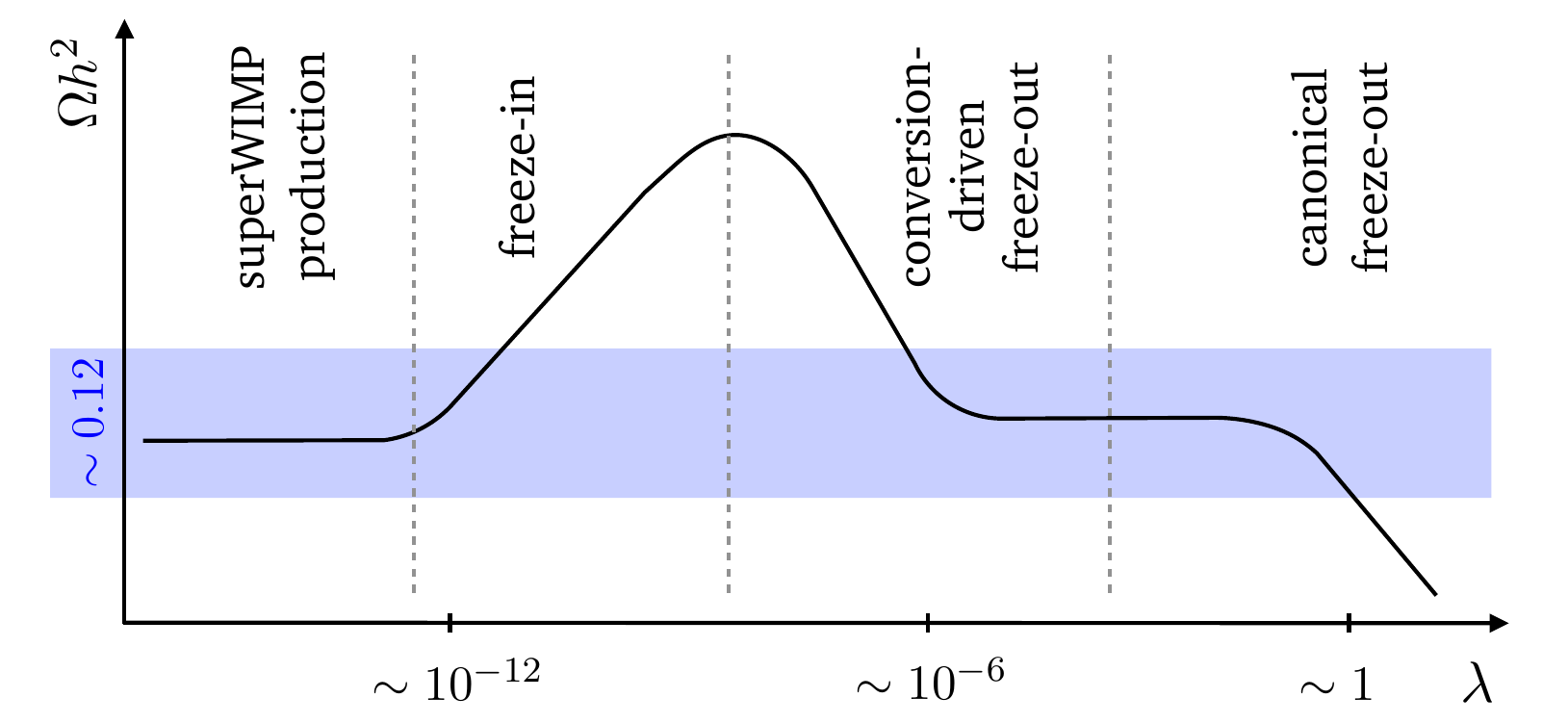}
    \caption{Schematic plot showing the dependence of the relic density, $\Omega h^2$, on the DM coupling $\lambda$ for a scenario with a relatively small mass splitting between the DM particle $X$ and the $t$-channel mediator $Y$. The blue band indicates the region for which $\Omega h^2=0.12$ and depends on the new physics masses $M_X$ and $M_Y$. The numbers for $\lambda$ indicate rough orders of magnitude, and are model-dependent. The four characteristic production regimes are further discussed in sections~\ref{sec:FO}, \ref{sec:conv}, \ref{sec:FI} and \ref{sec:SW}, respectively, in the order of decreasing coupling strength $\lambda$. This leads to a distinct phenomenology in each case.}
    \label{fig:schematicrelic}
\end{figure}
Equations~\eqref{eq:BEfull} represent general Boltzmann equations that can be simplified according to the dominant processes and the size of the DM coupling to the mediator. Depending on this coupling, certain terms can indeed be neglected or integrated out. Moreover, the size of the coupling strength largely determines the type of DM production mechanism in the early universe, as schematically depicted in figure~\ref{fig:schematicrelic} (see~\cite{Junius:2019dci} for an explicit example). From left to right in the plot, very feebly interacting DM particles achieve their relic density via the SW mechanism. As the coupling $\lambda$ increases, FI becomes active, followed by conversion-driven freeze-out, and finally standard freeze-out mechanism for couplings of $\mathcal{O}(1)$.

Returning to~\eqref{eq:BEfull}, during freeze-out, the DM particle $X$ can be assumed to remain in thermal and chemical equilibrium with the mediator $Y$, provided that the coupling is sufficiently strong. In this case, the conversion processes occur much faster than the Hubble rate, enforcing $n_X/n_Y = n_X^{\rm eq}/n_Y^{\rm eq}$. Under these conditions, the coupled set of equations~\eqref{eq:BEfull} reduces to a single equation describing the evolution of the DM abundance, driven by an effective annihilation cross section that accounts for all co-annihilation processes of $X$ and $Y$, weighted by the relevant Boltzmann suppression factors. For more details, see section~\ref{sec:FO} and references such as~\cite{Griest:1990kh, Gondolo:1990dk, Edsjo:1997bg}. Conversely, if the conversion processes between $X$ and $Y$ are inefficient, the full coupled set of Boltzmann equations must be solved. This situation is discussed in section~\ref{sec:conv}. 

The four primary DM production mechanisms are now described in the following subsections. For further discussion, see also~\cite{Heisig:2018teh, Lopez-Honorez:2022sge}.

\subsubsection{Canonical freeze-out} \label{sec:FO}
\begin{figure}
    \centering
    \includegraphics[width=0.8\textwidth]{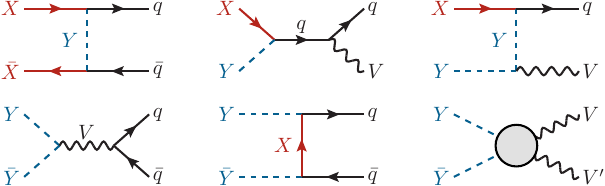}    
    \caption{Representative LO Feynman diagrams contributing to the DM annihilation cross section $\langle \sigma_\text{eff} v \rangle$. They include $t$-channel DM annihilations (top left), $XY$ co-annihilations (top central and right), and mediator pair-annihilations (bottom row) with the blob accounting for channels yielding SM electroweak bosons or gluons. The final state $q$ represents any quark flavour, and may include a sum over flavours depending on the model. For leptophilic models, $q$ can be replaced by a lepton $\ell$, and for models with self-conjugate DM, additional $YY$ annihilation diagrams must be added. Moreover, in cases where the S-wave contribution to $\langle \sigma_\text{ann} v \rangle (X\bar{X} \to q\bar{q})$ is helicity suppressed, NLO corrections become relevant (see the diagrams in figure~\ref{fig:indirect}). \label{fig:sigmav-ann}}
\end{figure}

DM production via standard freeze-out occurs if the interaction between the DM and the SM is strong enough to establish chemical equilibrium between the dark sector and the SM at early times. If chemical equilibrium is also maintained within the dark sector through conversion processes among its particle species, we can approximately write 
\be
   \frac{Y_{j}}{Y_{j}^{\rm eq}}=\frac{Y_{i}}{Y_{i}^{\rm eq}} \,,
\label{eq:chemeq}\ee
which allows us to simplify \eqref{eq:BEfull}. Under this assumption, the evolution of the total comoving number density $\tilde{Y} = \sum_i Y_i$ is governed by
\be
  \frac{\d \tilde{Y} }{\d x} = \sum_i \frac{\d Y_i}{\d x} =- \frac{c g_{\star, \text{eff}}^\frac{1}{2}}{x^2} \left \langle \sigma_\text{eff} v \right \rangle \left( \tilde{Y}^2 - \left( \tilde{Y}^{\rm eq} \right)^2 \right)\, ,
\label{eq:BEfreezeout}\ee
where $c = \sqrt{\pi / 45} \, M_\text{Pl} \, M_X$. In this last expression, the effective annihilation cross section and number of relativistic degrees of freedom are defined by
\be\bsp
  g_{\star,\text{eff}}^\frac{1}{2} =&\ \frac{g_{\star S}}{\sqrt{g_\star}} \left( 1 + \frac{T}{3 g_{\star S}} \frac{\d g_{\star S}}{\d T} \right) \,, \\
    \left \langle \sigma_\text{eff} v \right \rangle =&\ \sum_{i,j} \left \langle \sigma_{ij} v \right \rangle \frac{Y_i^{\rm eq}Y_j^{\rm eq}}{\tilde{Y}^{\rm eq} \tilde{Y}^{\rm eq}}  \,, 
\esp\label{eq:effDMCS}\ee
where $g_{\star S}$ is the effective number of relativistic degrees of freedom for entropy density. Here, we assume that the distribution functions of all particles can be expressed as a rescaling of their equilibrium distributions, $f_i = \alpha_i f_i^{\rm eq}$, and that these equilibrium distributions follow a Boltzmann distribution, \ie\ $f_i^{\rm eq} = \exp( -E_i / T)$. The annihilation and co-annihilation processes inherent to the $t$-channel DM models considered and contributing to $\left \langle \sigma_\text{eff} v \right \rangle$ are depicted in figure~\ref{fig:sigmav-ann}.

Since all dark sector particles heavier than the DM candidate $X$ are assumed to eventually decay into $X$, the DM relic density is determined by the solution to \eqref{eq:BEfreezeout}. Inspecting the effective annihilation cross section~\eqref{eq:effDMCS} reveals that annihilations of heavier dark sector particles $Y$ can significantly contribute to the total cross section if $M_Y \lesssim 1.2 \, M_X$, a scenario commonly referred to as a co-annihilation.

Within the standard freeze-out mechanism, the final DM relic density $\Omega h^2$ exhibits the characteristic \emph{increasing} trend as the coupling strength $\lambda$ is \emph{reduced}, as shown on the right-hand side of figure~\ref{fig:schematicrelic}. Within the co-annihilation regime, this trend levels off when the coupling $\lambda$ between the DM and the mediator becomes smaller than the gauge couplings relevant for the $t$-channel mediator. In this regime, DM freeze-out is dominated by efficient conversion to the mediator and subsequent freeze-out of mediator pair-annihilations, resulting in the plateau observed in figure~\ref{fig:schematicrelic}. When $\lambda$ becomes even smaller, the system transitions into the domain of conversion-driven freeze-out.

The canonical freeze-out picture can be affected by non-perturbative effects, the so-called Sommerfeld effect, as well as the formation and successive decay of bound states, as discussed in detail in section~\ref{sec:SE-BSF}. Specifically, for a mediator that is both coloured and electrically charged, long-range effects via gluon exchange impact the expected viable regions of the parameter space significantly~\cite{Harz:2018csl, Becker:2022iso}, as found in the case of exchanges of relatively light scalars~\cite{Harz:2017dlj,Harz:2019rro}.

\subsubsection{Conversion-driven freeze-out} \label{sec:conv}
The assumption of chemical equilibrium among the co-annihilating partners during freeze-out holds if conversions between $X$ and $Y$, such as $Y \to AX$ decays or $YB \to XA$ scatterings along with their inverse processes (with $A$ and $B$ denoting SM particles), remain fully efficient, \ie\ if $\Gamma_\text{conv} \gg H$. This is generally true if the DM coupling $\lambda$ and the mediator gauge interaction strengths are of comparable magnitude, and if $A$ and $B$ are comparably light SM particles. In such cases, conversion rates are not suppressed, while annihilation processes  initiate chemical decoupling due to their Boltzmann suppression from the initial-state heavy particles. As $\lambda$ decreases, two effects occur. First, direct annihilation processes involving $\lambda$ become negligible relative to mediator pair annihilation, while conversions remain efficient. In this regime, the relic density becomes independent of $\lambda$, corresponding to the plateau in figure~\ref{fig:schematicrelic}. However, further reduction of $\lambda$ eventually renders conversions inefficient, initiating chemical decoupling at $\Gamma_\text{conv} \sim H$. In this \emph{conversion-driven} freeze-out scenario~\cite{Garny:2017rxs, DAgnolo:2017dbv}, the relic density rapidly increases with decreasing $\lambda$, with typical values of $\lambda \sim 10^{-6}$ for couplings to light quarks, while for top-philic scenarios slightly larger couplings are required due to the involvement of massive top quarks in conversions~\cite{Garny:2018icg}. Since decays and scatterings contribute to conversions, the condition $\Gamma_\text{conv} \sim H$ constrains the decay rate $\Gamma_\text{dec} \lesssim H$ at freeze-out. For DM masses near the weak or TeV scale, this implies $c\tau_Y \gtrsim (10^{-3} - 1)\,\mathrm{m}$, indicating a long-lived mediator with significant collider implications~\cite{Garny:2017rxs} (see section~\ref{sec:LLP}).

Conversion-driven freeze-out requires solving coupled Boltzmann equations for the DM and mediator abundances. In addition, the weak strength of the DM couplings can induce deviations from kinetic equilibrium, necessitating to solve a momentum-dependent Boltzmann equation~\cite{Garny:2017rxs, DAgnolo:2017dbv}. However, for the models considered, these deviations have a minor quantitative impact~\cite{Garny:2017rxs}, and the system can often be described using the coupled rate equations~\eqref{eq:BEfull} for $i = X, Y$, after explicitly accounting for the conversion processes entering the rates $\gamma_{X \to Y}$ and $\gamma_{Y \to X}$. 
For $\lambda$ values where $\Gamma_\text{conv} \lesssim H$, annihilation processes proportional to $\lambda$ are typically negligible, leaving mediator pair-annihilation ($Y \bar{Y} \to \mathrm{SM}$) as the only relevant channel. In this case, the Boltzmann equations reduce to:
\be\label{eq:BEcdfo}\bsp
\frac{\d Y_X}{\d x}=&\ \frac{1}{H x s} \left[  
  \gamma_{Y \to X }\left( \frac{Y_{Y}}{Y_{Y}^{\rm
      eq}}-\frac{Y_{X}}{Y_{X}^{\rm eq}} \right) + \gamma_{\bar Y \to X }\left( \frac{Y_{\bar Y}}{Y_{\bar Y}^{\rm
      eq}}-\frac{Y_{X}}{Y_{X}^{\rm eq}} \right) \right]\,,\\
\frac{\d Y_Y}{\d x}=&\ \frac{-1}{H x s} \left[  \gamma_{Y\bar Y}
  \left(\frac{Y_{Y}Y_{\bar Y} }{Y_{Y}^{\rm eq}Y_{\bar Y}^{\rm eq} }-1 \right)+
  \gamma_{Y \to X }\left( \frac{Y_{Y}}{Y_{Y}^{\rm
      eq}}-\frac{Y_{X}}{Y_{X}^{\rm eq}} \right) \right]\,,
\esp\ee
where we assume the detailed balance condition $\gamma_{X \to Y} = \gamma_{Y \to X}$. The mediator $Y$ and its antiparticle $\bar{Y}$ are treated separately for clarity, and we have assumed a self-conjugate DM species $X$. The Boltzmann equation for $\bar{Y}$ is analogous, and under $CP$-conservation $Y_Y = Y_{\bar{Y}}$ and $\gamma_{\bar{Y} \to X} = \gamma_{Y \to X}$. This reduces the system to the two coupled equations given above (see~\cite{Heisig:2024mwr} for a $CP$-violating scenario in the context of baryogenesis). Furthermore, the mediator annihilation rate $\gamma_{Y \bar{Y}}$ is dominated by gauge interactions, including QCD for coloured $t$-channel mediators and electroweak interactions for leptophilic mediators. Accordingly, these processes are subject to Sommerfeld enhancement and bound-state formation effects, as discussed in section~\ref{sec:SE-BSF}~\cite{Garny:2021qsr, Becker:2022iso, Binder:2023ckj}. At very low temperatures, mediator pair-annihilation eventually becomes subdominant due to the double Boltzmann suppression related to the initial-state mediators. Thus, conversion-driven freeze-out arises from the interplay between the freeze-out of the conversion processes $\gamma_{ Y \to X }$ due to the small coupling strength $\lambda$, and deviations from equilibrium of the mediator itself due to a finite annihilation rate $\gamma_{Y\bar Y}$. Both are crucial in determining the DM relic abundance~\cite{Garny:2017rxs}.

The cosmologically viable CDFO region in the model parameter space requires the DM to be under-abundant, when assuming, regardless of $\lambda$, chemical equilibrium between the $X$ and $Y$ states. This condition implies relatively small mass splitting between the dark matter and the mediator particles. However, it cannot be met above a certain mediator mass for which $\gamma_{Y\bar Y}$ is too small to provide sufficient dilution of the dark sector particles, even for mass-degenerate $X$ and $Y$ particles. This limits the allowed mass range for the mediator to a few hundred GeV (a few TeV) for leptophilic~\cite{Junius:2019dci} (quark-philic~\cite{Garny:2021qsr}) models. CDFO scenarios therefore provide a well-bounded parameter space, that is likely fully testable with collider searches in the foreseeable future. This constitutes a difference from freeze-in or superWIMP scenarios that permit a broader mediator mass range.

Another notable feature of the CDFO mechanism is that, despite weak DM couplings, the DM particle remains thermalised, making the relic abundance independent of the initial conditions, akin to canonical freeze-out. In contrast, freeze-in and superWIMP scenarios require even weaker $\lambda$ such that DM never reaches thermal equilibrium during the entire evolution of the universe. CDFO has been studied in different minimal $t$-channel mediator models, including quark-philic~\cite{Garny:2017rxs, Garny:2018icg, Garny:2021qsr, Heisig:2024xbh} and leptophilic~\cite{Junius:2019dci} setups (where the Higgs-portal interactions of the mediator can become important), as well as in non-minimal models like flavoured DM~\cite{Acaroglu:2023phy, Heisig:2024mwr}. Notably, CDFO may simultaneously achieve baryogenesis~\cite{Heisig:2024mwr}, motivating further exploration.

\subsubsection{Freeze-in} \label{sec:FI}
Within the framework of freeze-in production of DM~\cite{McDonald:2001vt, Asaka:2005cn, Hall:2009bx}, we assume that the DM particle $X$ is so feebly coupled ($\lambda \lesssim 10^{-10}$) that it has never reached chemical or kinetic equilibrium (see figure~\ref{fig:schematicrelic}). Under this assumption and starting with a negligible initial abundance of dark matter, the evolution of the DM population is governed primarily by conversion processes $Y \to X$, which include both decays and scatterings. The Boltzmann equation for the evolution of DM abundance~\eqref{eq:BEfull} simplifies to
\begin{equation}
  \frac{\d Y_X}{\d x}=\frac{1}{H x s} \gamma_{Y \to X }\frac{Y_{Y}}{Y_{Y}^{\rm eq}} \,,
\label{eq:BEdec}\end{equation}
where we neglect $X \to Y$ contributions because the DM abundance is highly suppressed at early times. During freeze-in production, we assume that the mediator $Y$ remains in \emph{chemical and kinetic equilibrium}, as most of the DM production process occurs slightly before the freeze-out of the mediator $Y$ ($T \gtrsim M_Y$). Consequently, we can approximate $Y_Y \approx Y_Y^{\rm eq}$ all along the production process. This assumption holds for a freeze-in mechanism embedded in a renormalisable theory where production is dominated by infrared (IR) effects, in contrast to models featuring ultraviolet freeze-in~\cite{Elahi:2014fsa}. For the simplified $t$-channel DM models analysed in this report, freeze-in remains IR-dominated as long as $M_Y < T_\text{max}$, where $T_\text{max}$ is the maximum temperature of the universe. For scenarios with DM production occurring after the chemical decoupling of the mediator, we refer instead to section~\ref{sec:SW}. However, we must keep in mind that both contributions should generally be taken into account, as discussed, for example, in~\cite{Garny:2018ali, Decant:2021mhj}.

Interactions with the thermal plasma can play a crucial role in scenarios of freeze-in dark matter production. This was first demonstrated in~\cite{Biondini:2020ric} for a Majorana DM model, and later extended to scalar DM models in~\cite{Becker:2023vwd, Becker:2025lkc}. The latter works employed a first-principles non-equilibrium quantum field theory approach, specifically the closed-time path formalism, to quantify the corrections to standard approximations commonly used in the literature for freeze-in dark matter. These approximations include vacuum decays only, decays with thermal masses, decays and scatterings with thermal masses, and the hard thermal loop approximation. It was shown that using decays with vacuum masses alone can underestimate the relic abundance by up to 90\% in scenarios with small mass splittings where scattering processes, particularly those involving multiple soft scatterings, dominate. In contrast, for large mass splittings where decay processes dominate, the relic abundance is predicted correctly to within $\pm \mathcal{O}(10\%)$, the sign depending on the gauge coupling. For large gauge couplings, the relic abundance is typically underestimated by about 10\%, whereas for small gauge couplings it is overestimated by a similar margin. For further details, we refer the reader to~\cite{Becker:2023vwd, Becker:2025lkc}. Scattering processes such as $VY \to Xf$ may contribute slightly to the relic abundance, up to $\sim 25\%$, when $V$ is a gluon and $f$ is a coloured SM fermion. A more precise treatment of such thermal effects, particularly for small mass splittings, may nevertheless be necessary~\cite{Biondini:2020ric}. For simplicity, we focus here on mediator decay contributions during a radiation-dominated era to illustrate the main aspects of freezing-in DM production. In this case, \eqref{eq:BEdec} reduces to
\begin{equation}\label{eq:BEFI1}
  \frac{\d Y_X}{\d x}=\frac{1}{H x }  \Gamma_{Y \to X } \frac{K_1(x)}{K_2(x)} Y^{\rm eq}_{Y}\,,  
\end{equation}
where $\Gamma_{Y \to X}$ is the decay rate of $Y \to Xf$, and $x = M_Y / T$. For DM production via decay, which is also relevant for the superWIMP scenarios discussed in section~\ref{sec:SW}, it is convenient to define the dimensionless parameter
\begin{equation}
  R_\Gamma^{\rm prod}=\frac{M_0(T_{\rm prod}) \Gamma_{Y\to X}}{M_Y^2}\,,
\label{eq:Rgam}\end{equation}
where $M_0(T) = 45 M_{\rm pl} /(4 \pi^3 g_*(T))$ is the rescaled Planck mass entering the Hubble rate during a radiation-dominated era, $H = T^2 / M_0(T)$, evaluated at the DM production temperature $T_{\rm prod}$. In addition, $g_*(T)$ denotes the number of relativistic degrees of freedom at a temperature $T$. In the case of freezing-in DM, $T_{\rm prod} \simeq T_{\rm FI} \simeq M_Y / 3$. In this context, $R_\Gamma^{\rm prod} \equiv R_\Gamma^{\rm FI}$  approximately represents the ratio of the DM production rate to the Hubble rate at $T = M_Y$.

Integrating \eqref{eq:BEFI1} over time yields
\begin{equation}
    \Omega_X h^2|_\mathrm{FI,\,dec}=M_X\times\frac{135}{8\pi^3}\frac{g_Y}{g_*(T_{\rm FI})}R^\mathrm{FI}_\Gamma \frac{s_0h^2}{\rho_\mathrm{crit}}\,,
\label{eq:OmFIdec}\end{equation}
where $\rho_\mathrm{crit} = 3 M_{\rm pl}^2 H_0^2 / (8 \pi)$ is the critical energy density today, $s_0$ is the entropy density today, $h$ is the rescaled Hubble parameter, $h = H_0 / (100\,\text{km}\,\text{s}^{-1}\text{Mpc}^{-1}) \sim 0.7$, and $g_Y$ is the number of degrees of freedom of the mediator. Increasing the DM coupling $\lambda$ enhances $\Gamma_{Y \to X}$ (or equivalently $R^\mathrm{FI}_\Gamma$), increasing the DM production rate and resulting in a larger relic abundance. Finally, non-negligible scattering contributions may also be added to the decay terms in equations~\eqref{eq:BEFI1}--\eqref{eq:OmFIdec}.

In summary, the freeze-in production of DM is a cumulative effect of rare decay or scattering processes involving the thermalised mediator particle $Y$, which produces DM at a very low (sub-Hubble) rate. Consequently, the final DM abundance decreases strongly with a decreasing coupling strength $\lambda$, as shown in figure~\ref{fig:schematicrelic}. However, as $\lambda$ is reduced further, a plateau in the relic density is reached, as indicated at the leftmost end of figure~\ref{fig:schematicrelic}. We now turn to this extremely weakly coupled regime.

\subsubsection{SuperWIMP} \label{sec:SW}
For extremely small coupling strengths between the mediator and the DM particle, the superWIMP production mechanism~\cite{Covi:1999ty, Feng:2003uy} can account for the observed relic DM density, as shown in figure~\ref{fig:schematicrelic}. This mechanism typically unfolds in two distinct stages. First, at temperatures above or around the mediator mass, the mediator remains in thermal equilibrium with the SM thermal bath due to its gauge interactions, while the DM particle is decoupled because of its extremely weak interaction strength, assuming a negligible initial abundance. When the temperature drops below the mediator mass, mediator pair annihilation into SM particles becomes Boltzmann-suppressed and the mediator then eventually freezes out, similarly to the usual WIMP mechanism. At this stage, the DM abundance remains negligibly small because of its tiny interaction strength. The evolution of the mediator abundance during this stage is governed by conventional freeze-out via its gauge interactions, while conversion to DM is neglected. The corresponding Boltzmann equation is:
\begin{equation} \label{eq:SW}
  \frac{\d Y_Y}{\d x} = \frac{-1}{H x s} \left[  \gamma_{Y\bar Y} \left(\frac{Y_{Y}Y_{\bar Y} }{Y_{Y}^{\rm eq}Y_{\bar Y}^{\rm eq} }-1 \right)\right]\,, 
\end{equation}
with a similar equation for the antiparticle $\bar{Y}$. When the age of the universe becomes comparable to the lifetime of the mediator, the frozen-out distribution of mediator particles decays into DM. For each mediator particle one DM particle is therefore produced. For self-conjugate DM (\eg, a Majorana fermion or a real scalar), this implies
\begin{equation}
    Y_X^{\rm SW} = (Y_Y + Y_{\bar{Y}})|_{\rm FO}\,.
\end{equation}
These decays also produce SM particles, and impart a boost to the DM particles whose magnitude depends on the mass ratio between the mediator and the DM particle. This boost can result in a non-negligible DM velocity distribution, potentially leading to constraints that could be derived from structure formation. In particular, Lyman-$\alpha$ forest observations could provide very strong bounds~\cite{Garny:2018ali, Decant:2021mhj}, as discussed in section~\ref{sec:earlyCons}, and the long lifetime of the mediator can additionally result in collider signatures of long-lived particle searches and may affect Big Bang Nucleosynthesis (BBN)~\cite{Garny:2018ali}.

The scenario described above can be modified under certain conditions, particularly when non-perturbative effects are taken into account. For instance, the formation of bound states between pairs of mediator particles can significantly enhance their effective annihilation rate, $\gamma_{Y \bar{Y}}$~\cite{Garny:2021qsr, Becker:2022iso, Binder:2023ckj}. Specifically, for a mediator that is both coloured and electrically charged, bound states formed through gluon exchanges together with electromagnetic transition processes among the different bound state levels can prevent the mediator from freezing out completely. In such cases, the mediator abundance continues to be depleted even at temperatures much lower than its mass, and this depletion ceases only when the age of the universe becomes comparable to the mediator lifetime. When these effects are significant, the two stages described earlier become interconnected rather than distinct. Further details on this regime can be found in~\cite{Binder:2023ckj}. On the other hand, while bound states are also quantitatively relevant for leptophilic models, they do not qualitatively alter the evolution of the superWIMP mechanism as outlined above.

\subsubsection{Non-perturbative effects}\label{sec:SE-BSF}
The freeze-out of dark sector annihilations typically occurs when the annihilating particles are non-relativistic. This regime allows for significant non-perturbative effects, namely the Sommerfeld Effect (SE) and Bound State Formation (BSF), that affect the annihilation cross section through multiple exchanges of light states between the initial or final-state particles. Non-perturbative corrections in the final state are neglected here, since the annihilation products are much lighter than the dark sector particles so that they are relativistic. 

\begin{figure}
    \centering
    \includegraphics[width=0.4\textwidth]{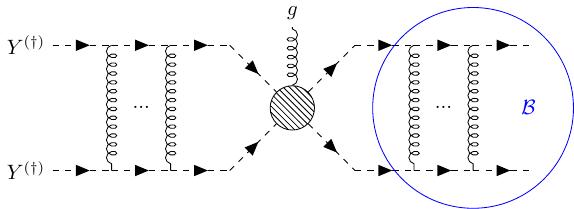}
    \caption{Diagramatic representation of bound state ($\cal B$) formation of two coloured particles via the emission of a gluon, and of the Sommerfeld Effect relevant for the initial state. Figure adapted from~\cite{Becker:2022iso}.\label{fig:BSF-diagram}}
\end{figure}

Simplified $t$-channel DM models always involve a dark sector particle $Y$ carrying SM charges. Consequently, annihilation involving two SM-charged dark sector particles is subject to corrections from multiple exchanges of (massless) gauge bosons. Each exchange of a massless gauge boson contributes parametrically with a factor of $\alpha/v$ where $\alpha$ is the relevant coupling constant and $v$ is the relative velocity between the two initial-state particles, which could be sizeable if $\alpha \sim v$. During the freeze-out of DM annihilations, $v \sim 0.1$. This thus necessitates the resummation of the ladder diagrams shown in figure~\ref{fig:BSF-diagram} if $\alpha \sim 0.1$, which is particularly relevant for mediator particles $Y$ charged under $SU(3)_C$ (as in the models considered). In the non-relativistic regime, this resummation reduces to solving the Schrödinger equation for the wave functions of the initial-state particles, incorporating a colour potential as discussed in~\cite{vonHarling:2014kha, Hisano:2003ec, Hisano:2004ds, Beneke:2012tg, Petraki:2015hla}. Since gluons are massless, the potential at energy scales far above the QCD confinement scale is Coulomb-like, and can be either attractive or repulsive depending on the representation of the initial state particles. In the context of simplified $t$-channel DM models, the dark sector particles are either singlet or lie in the (anti-)fundamental representation of $SU(3)_C$. Dark sector annihilation therefore leads to the following initial state colour configurations,
\be\bsp
    \mathbf{3} \times \mathbf{\bar{3}} &= \mathbf{1} + \mathbf{8} \,, \\
    \mathbf{3} \times \mathbf{3} &= \mathbf{\bar{3}} + \mathbf{6} \,.
\esp\ee
Among these, the $\mathbf{1}$ and $\mathbf{\bar{3}}$ configurations result in an attractive potential, allowing for particle-antiparticle and particle-particle bound states, respectively, while the $\mathbf{6}$ and $\mathbf{8}$ configurations are repulsive.

The Sommerfield effect affects both attractive and repulsive initial states by capturing the distortion of the incoming wave functions due to the effective colour potential, which are otherwise modelled as plane waves. For repulsive configurations, the cross section is reduced, while for attractive ones, it is enhanced. This effect can be incorporated by multiplying the perturbatively calculated cross section for each partial wave by the corresponding SE factor $S_l$~\cite{Cassel:2009wt, Iengo:2009ni},
\be
    \sigma_{ij} v_{ij} \rightarrow \sum_l \left( \sigma_{ij} v_{ij} \right)_l S_l \left( \frac{\alpha_\text{eff}}{v_{ij}} \right) \,,
\ee
where $l$ denotes the $l^\mathrm{th}$ partial wave.

Bound state formation contributes to the effective annihilation cross section by introducing an additional channel. A bound state forms via the emission of at least one gluon, and it can subsequently either decay into dark sector or SM particles, or be ionised by a gluon from the thermal bath. Only bound states that decay contribute to the effective annihilation cross section by depleting the dark sector. The related contribution in the case of a single bound state is given by~\cite{Ellis:2015vaa}
\be
  \left \langle \sigma_\text{eff} v \right \rangle \rightarrow \left \langle \sigma_\text{eff} v \right \rangle + \frac{\Gamma_\text{dec}}{\Gamma_\text{dec} + \Gamma_\text{ion}} \left \langle \sigma_\text{BSF} v \right \rangle\,,
\ee
where $ \langle \sigma_\text{BSF} v \rangle$ is the thermally averaged cross section for bound state formation under the emission of a gluon~\cite{Harz:2018csl}. The prefactor depends on the bound state decay and ionisation rates $\Gamma_\text{dec}$ and $\Gamma_\text{ion}$, and reflects that only bound state decays into SM particles deplete the dark sector. When excited bound states are included, transitions between different states $i$ and $j$ with rates $\Gamma_\text{trans}^{i \rightarrow j}$ must be considered, which yields~\cite{Binder:2021vfo, Garny:2021qsr,Binder:2023ckj}
\be
    \left \langle \sigma_\text{eff} v \right \rangle \rightarrow \left \langle \sigma_\text{eff} v \right \rangle + \sum_i R_i \left \langle \sigma_{\text{BSF},i} v \right \rangle \,,
\ee
with
\be 
    R_i = 1 - \sum_j \left( \delta_{ij} - \frac{\Gamma_\text{trans}^{i \rightarrow j}}{\Gamma^i_\text{ion} + \Gamma^i_\text{dec} + \sum_{j \neq i} \Gamma_\text{trans}^{i \rightarrow j} }\right)^{-1} \frac{\Gamma^j_\text{ion}}{\Gamma^i_\text{ion} + \Gamma^i_\text{dec} + \sum_{j \neq i} \Gamma_\text{trans}^{i \rightarrow j}} \,.
\ee
When summing the bound-state formation contributions to the cross section over all possible bound states in the final state, unitarity violation however arises~\cite{Flores:2024sfy, Beneke:2024nxh}, although a possible solution has been recently proposed in~\cite{Petraki:2025zvv}.

In summary, the annihilation cross section involving two possibly different coloured particles $Y_1$ and $Y_2$ must be modified to include two classes of non-perturbative effects, Sommerfeld effect and bound state formation effects,
\be
    \left \langle \sigma_{Y_1 Y_2} v\right \rangle \rightarrow \sum_l \left \langle S_l \left( \frac{\alpha_\text{eff}}{v} \right) \left( \sigma_{Y_1 Y_2} v \right)_l \right \rangle + \sum_i R_i \left \langle \sigma_{\text{BSF},i} v \right \rangle \,.
\ee

The Sommerfeld effect and bound state formation have been extensively studied in $t$-channel scenarios~\cite{Kim:2016kxt, Mitridate:2017izz, Harz:2017dlj, Harz:2018csl, Harz:2019rro, Becker:2022iso, Binder:2020efn, Binder:2019erp, Garny:2021qsr}. Their impact on the interpretation of the constraints existing in the parameter space is in particular explored in~\cite{Becker:2022iso} for freeze-out scenarios and in~\cite{Garny:2021qsr} for CDFO models.

\subsubsection{Altered cosmological histories} \label{sec:nonstandardCosmo}
The precise measurement of the cosmic abundance of light element relics provides one of the strongest constraints on the properties of the universe at the time of their creation, specifically when BBN occurred at temperatures of approximately $T \sim \mathcal{O}(1)\,\text{MeV}$~\cite{Kawasaki:2000en}. Experimental data indicates that the universe must have been in a period of radiation domination by the time of BBN, with the energy density of the relativistic particles in the thermal bath driving the expansion. At earlier times, however, the expansion history of the universe remains uncertain, as no data conclusively establishes the expansion history between inflation and BBN. This period could span up to 20 orders of magnitude in energy (given that the energy scale of inflation must be below $10^{16}$ GeV), during which various phenomena could have altered the expansion rate, impacting particle interactions and production~\cite{Giudice:2000ex, Chung:1998rq}. Among the possibilities are low-reheating temperature scenarios after inflation, early matter domination epochs~\cite{Moroi:1999zb, Coughlan:1983ci}, and faster-than-radiation eras such as kination~\cite{Barrow:1982ei, Ford:1986sy}. 
These phenomena could result from slow decays of the inflaton during reheating~\cite{Giudice:2000ex}, decays of heavy particles like moduli fields (scalar fields with generic equations of state), primordial black holes, quintessence models of dark energy, and more. For a comprehensive discussion, we refer the reader to the review~\cite{Allahverdi:2020bys}.

If the production of dark sector particles occurs during such altered cosmological histories, the evolution of their densities can change significantly, leading to predictions for model observables that differ by orders of magnitude~\cite{Co:2015pka}. For instance, in scenarios with low-reheating temperatures, the reheating phase after inflation may be prolonged, during which the inflaton oscillates around the minimum of its potential and decays into ultra-relativistic particles. This process generates a substantial entropy increase in the universe, causing a non-adiabatic evolution of thermodynamic quantities. In such cases, the relationship between the temperature and the scale factor is altered to the general form
\be
    T \propto a^{-k}\,,
\ee
where $k$ is a model-dependent real number that under adiabatic conditions would be $k=1$. If the production of a relic such as dark matter ceases during the reheating era, its abundance undergoes dilution until entropy is conserved again, for instance when the universe transitions back to a radiation-dominated phase. This dilution can span several orders of magnitude, drastically impacting predictions for dark sector particle masses and couplings.

For simplified $t$-channel DM models, low-reheating temperature scenarios have been studied in~\cite{Calibbi:2021fld, Becker:2023tvd}, focusing on freeze-in production for both Majorana and scalar DM models with scalar and vector-like fermion mediators, respectively. Ref.~\cite{Calibbi:2021fld} highlights that when the reheating temperature $T_\text{RH}$ is lower than the mediator mass $M_Y$, such that DM freezes in during reheating, entropy dilution significantly reduces the comoving number density. Reproducing the observed DM relic density for a fixed mass configuration then requires much stronger Yukawa couplings between DM, the mediator, and the SM fermions. This, in turn, shifts the collider signatures of the mediator from the long-lived particle regime to the prompt regime, making the interpretation of potential signals and constraints from colliders highly dependent on the reheating history. This result was extended in~\cite{Becker:2023tvd} to reheating scenarios with potentials of the form $V(\phi) \propto \phi^k$ ($k \geq 2$) and different inflaton decay channels into bosons and fermions, leading to distinct expansion histories and entropy dilution factors. Such scenarios are motivated by $\alpha$-attractor models of inflation~\cite{Starobinsky:1980te, Kallosh:2013maa, Kallosh:2013hoa}, which predict inflationary observables consistent with the constraints from the Planck 2018 analysis~\cite{Planck:2018jri}. 

As shown in section~\ref{sec:LLP}, these findings modify the predictions for LLP signatures, demonstrating the intricate dependence of the results on the reheating potential and the main inflaton decay channel. This arises from the power-law dilution of the relic abundance,
\be
    Y_X \sim \Gamma_Y \times
    \begin{cases} 
        \left(\dfrac{T_\text{rh}}{M_Y}\right)^{4k-1}\,&\text{Bosonic Reheating,}\\
        \left(\dfrac{T_\text{rh}}{M_Y}\right)^{\frac{9-k}{k-1}}\;&\text{Fermionic Reheating,}
    \end{cases} \label{eq:Dilution_FreezeIn}
\ee
which is true if $M_Y \gtrsim T_\text{RH}$. Although dilution of the relic abundance necessitates larger mediator decay widths in freeze-in scenarios, it reduces the DM-SM interaction strength for thermal freeze-out production. Since freeze-out during reheating has not been explored in the context of simplified $t$-channel DM models, we refer instead the reader to general discussions in {\it e.g.}~\cite{Bernal:2022wck,Silva-Malpartida:2023yks,Silva-Malpartida:2024emu}.

\subsection{Dark matter constraints} \label{sec:searches}
In this section, we describe the most relevant astrophysical and cosmological constraints that could be applied to the models studied. 

In sections~\ref{sec:ID} and~\ref{sec:DD}, we focus on astrophysical searches probing DM interactions at the present time, where DM is cold and clustered within halos in galaxies and galaxy clusters. The typical velocities $v$ involved range from approximately $10^{-5} c$ to $10^{-2} c$, with $v \sim 10^{-3} c$ in the Milky Way, and current detection methods primarily include indirect and direct detection approaches. Indirect detection relies on the annihilation of DM into SM particles in regions with high DM density, such as the Galactic Centre and dwarf spheroidal galaxies (dSphs). The SM particles produced through DM annihilations subsequently decay, shower, and hadronise to produce a flux of stable cosmic rays ($\gamma$, $e^{+}$, $\bar{p}$, and $\nu$), which can be detected by space-based and ground-based telescopes. Direct detection, on the other hand, relies on observing elastic scattering of DM off protons or neutrons within underground detectors, where the scattering produces a measurable recoil of the associated nucleus. However, these detection methods are most effective in probing the canonical freeze-out regime, where the DM annihilation or DM-nucleon scattering rates are sufficiently large. In contrast, scenarios that involve the freeze-in, superWIMP, or conversion-driven freeze-out mechanisms predict extremely small rates that are typically not testable with these methods. Instead, such scenarios often give rise to the interesting long-lived particle signals at colliders that have been discussed in section~\ref{sec:LLP}.

Cosmological constraints are presented in section~\ref{sec:earlyCons}. While constraints from the cosmic microwave background (CMB) apply broadly across all DM scenarios, constraints from Big Bang Nucleosynthesis and structure formation are particularly relevant for CDFO and freeze-in and superWIMP scenarios.

Several computational tools are available to automatically calculate DM observables and constraints such as those originating from the calculation of the relic density, as well as indirect and direct detection predictions. These tools generally perform calculations at LO, with some also offering a handle on NLO or loop-induced processes, and most of them facilitate automated testing of model parameter space against a wide range of experimental observables. The set of most comprehensive and widely used tools includes \lstinline{micrOMEGAs}~\cite{Alguero:2023zol}, \lstinline{MadDM}~\cite{Arina:2021gfn}, \mbox{\lstinline{DarkSUSY}~\cite{Bringmann:2022vra}} and \lstinline{GAMBIT}~\cite{GAMBITDarkMatterWorkgroup:2017fax}.

\subsubsection{Indirect detection}\label{sec:ID}
Predictions for present-day annihilation of DM in dense astrophysical environments, in the context of minimal simplified models coupling DM to fermions (quarks and/or leptons), depend strongly on the spin of the DM particles. The dominant tree-level annihilation process, $X X \to f \bar{f}$, is illustrated by the Feynman diagram shown in the top left corner of figure~\ref{fig:sigmav-ann}. 

\begin{figure}
  \centering
  \includegraphics[width=\textwidth]{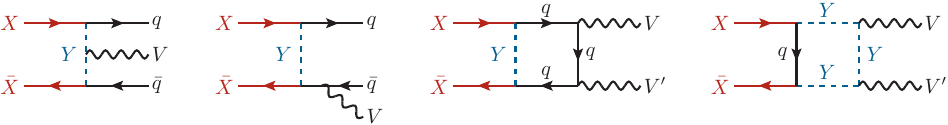}
  \caption{VIB and loop-induced processes contributing to DM indirect detection through the production of gamma-ray lines and other features in the gamma-ray spectrum.\label{fig:indirect}}
\end{figure} 

For Majorana and real scalar DM, this process is P-wave and D-wave suppressed, respectively, with the S-wave contribution being proportional to the fermion mass. This S-wave contribution is therefore negligible in the chiral limit or for top quarks in scenarios featuring DM and mediator masses above a few hundred GeV. For complex scalar DM, this suppression is still relevant, and it is primarily P-wave. There are however two processes that could enhance the detectability of such DM particles. Virtual internal bremsstrahlung (VIB), where the final-state quark pair or lepton pair is produced alongside a photon emitted by the internal $t$-channel propagator ($X X \to f \bar{f} \gamma$), can provide significant corrections to the tree-level annihilation cross section under certain mediator-DM spectrum configurations, overcoming P-wave suppression by orders of magnitude. Moreover, VIB yields a sharp spectral feature at the high-energy end of the gamma-ray spectrum, that could enhance the sensitivity of related experiments~\cite{Bringmann:2007nk, Bringmann:2011ye, Barger:2011jg, Toma:2013bka, Garny:2014waa, Giacchino:2013bta, Giacchino:2014moa, Giacchino:2015hvk, Garny:2015wea, Ibarra:2015nca, Colucci:2018qml}. Additionally, loop-induced annihilation processes, such as $X X \to \gamma \gamma$, $\gamma Z$, or $\gamma H$, produce distinctive gamma-ray lines that have the potential to directly reveal the DM mass~\cite{Bouquet:1989sr, Bergstrom:1989jr, Rudaz:1989ij, Bergstrom:1997fh, Bern:1997ng, Bertone:2009cb}. Representative Feynman diagrams for these two sets of processes are given in figure~\ref{fig:indirect}, and we emphasise that analytical expressions for the most studied models are available in the literature, such as in~\cite{Giacchino:2013bta, Giacchino:2014moa, Giacchino:2015hvk, Garny:2018icg}. Moreover, automated tools like \lstinline{MadDM} could be used to compute predictions for such higher-order processes for any DM model, provided that the Lagrangian is translated in the UFO format at NLO~\cite{Darme:2023jdn}.

In freeze-out scenarios where DM self-annihilation dominates and proceeds via S-wave processes (as for Dirac, real vector and most complex DM models except when it is a scalar), indirect detection signals are typically expected in the form of a continuum of photons, positrons, neutrinos and antiprotons. However, in regions where co-annihilation or bound-state formation plays an important role (typically when the mediator mass is less than about 1.2 times the DM mass), the connection between the relic density and indirect detection can weaken or break down entirely. In particular, if the relic density is set by processes involving the pair-annihilation of the co-annihilating partner (such as $Y Y$ annihilations to a pair of SM particles), no indirect detection signal is expected today. Nonetheless, in the mass and coupling ranges considered, annihilation cross sections near the canonical value of $10^{-26}~\text{cm}^3/\text{s}$ still arise in considerable parts of the parameter space.

The search for gamma-ray fluxes from DM annihilation in dwarf spheroidal galaxies, as performed by the Fermi-LAT satellite~\cite{Fermi-LAT:2016uux, McDaniel:2023bju}, is among the most sensitive methods to probe $t$-channel DM models, and it can exclude a large portion of the parameter space of minimal models~\cite{Arina:2023msd}. Measurements of cosmic-ray antiproton fluxes by the AMS-02 experiment at the International
Space Station~\cite{Cuoco:2017iax, Calore:2022stf, DelaTorreLuque:2024ozf} provide complementary constraints on DM annihilation in our galaxy, particularly for higher DM masses. While gamma-ray lines and VIB are higher-order contributions, they yield sharp spectral features that can be distinguished from smooth astrophysical foreground emission~\cite{Bringmann:2011ye}. These features lead to excellent experimental sensitivity from Fermi-LAT~\cite{Fermi-LAT:2015kyq} and other gamma-ray observatories, and allow us to probe cross sections well below $10^{-26}~{\rm cm}^3/{\rm s}$ for a wide range of DM masses. We should however keep in mind that theoretical predictions for these higher-order processes are however also often smaller than this benchmark value.

For DM models involving leptonic interactions, annihilations into neutrinos, either mono\-chro\-ma\-tic or as a continuum, are instead relevant. Although the observation of the associated signal is partially challenged by continuum photon emission~\cite{Queiroz:2016zwd} and the existing antiproton flux~\cite{Garny:2011cj}, neutrinos produced from DM annihilation can in principle be detected by current and future neutrino telescopes like IceCube~\cite{IceCube-Gen2:2020qha} and KM3NeT~\cite{KM3Net:2016zxf}. This has been particularly demonstrated for specific models such as secluded $t$-channel models and $U(1)_{L_\mu-L_\tau}$ gauge scenarios~\cite{Arguelles:2019ouk, BasegmezDuPree:2021fpo, Miranda:2022kzs}. 

Finally, gamma-ray line searches by HESS~\cite{HESS:2013rld, HESS:2016glm} and MAGIC~\cite{MAGIC:2022acl}, along with neutrino searches, provide sensitivity to very heavy DM, offering a complementary probe to collider experiments limited by the current centre-of-mass energies and luminosities.

\subsubsection{Direct detection}\label{sec:DD}
The standard direct detection signal considered in simplified $t$-channel models is the scattering of dark matter off nucleons. This signal varies significantly depending on the nature of the DM particle, the charge of the mediator, and its coupling to the SM particles. In this section, we illustrate the discussion by considering the case of fermionic (Majorana or Dirac) DM. Similar developments can easily be achieved for bosonic DM models. For a recent work on direct detection of t-channel models we refer to ~\cite{Arcadi:2023imv}. 
 
\begin{figure}
  \centering
  \includegraphics[width=0.9\textwidth]{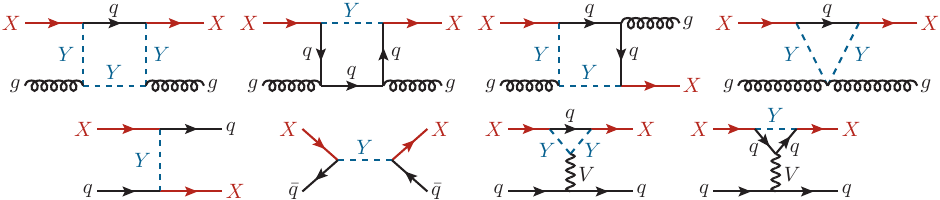}
  \caption{Illustrative Feynman diagrams contributing to DM-nucleon scattering, where we select as an example the case of fermionic DM. We include all diagrams that account for the scattering off gluons at one-loop and that are always present regardless of the model details (top row), tree-level scattering (relevant if the DM couples to an up or down quark) and the one-loop exchange of a photon, $Z$ boson or Higgs boson (bottom row).\label{fig:scattering_diagrams}}
\end{figure} 

For a Dirac DM particle, interactions with nucleons can be described by the representative Feynman diagrams shown in figure~\ref{fig:scattering_diagrams}, which depicts tree-level and one-loop contributions. These include the tree-level scattering process featuring the exchange of the coloured scalar mediator $Y$ (first two diagrams in the bottom row), as well as one-loop penguin diagrams mediated by the photon, the $Z$ boson, or the Higgs boson (two rightmost diagrams in the bottom row). While the tree-level contributions only arise for DM coupling to the up or down quark, the loop-induced subprocesses arise universally, regardless of the choice of SM fermion connecting the dark and SM sectors. Additionally, box and triangle diagrams involving two external gluons appear at one-loop, and are ubiquitous of a model where the DM couples to a coloured mediator. We now discuss below the effective Lagrangian interactions between the DM state and nucleons arising in each case.

If the DM particle is a Dirac fermion and couples at tree level to a first-generation quark, the dominant contribution to the scattering cross section comes from the $t$-channel exchange of the scalar mediator $Y$, as depicted in figure~\ref{fig:scattering_diagrams}. For the DM antiparticle $\bar{X}$, scattering off quarks occurs additionally via $s$-channel exchanges of the mediator. After a Fierz rearrangement of the corresponding matrix elements, these diagrams yield a vector interaction of the $X$ states with quarks, which translates into an effective vector interaction with nucleons $N$,
\be \label{eq:Leff_tree}
 \mathcal{L}_\text{eff,tree} = f_{V,\text{tree}}^{(N)}\,\bar{X} \gamma^\mu X \, \bar{N} \gamma_\mu N\,.
\ee
This leads to spin-independent interactions in the non-relativistic limit, with the effective couplings for protons and neutrons appearing in this expression, $f_{V,\text{tree}}^{(p)}$ and $f_{V,\text{tree}}^{(n)}$, being given by
\be\label{eq:fV_tree}
 f^{(p)}_{V,\text{tree}}= \begin{cases} 2 \, \lambda_V &\text{for coupling to }u_R \\ \lambda_V &\text{for coupling to } d_R\\ 3 \, \lambda_V &\text{for coupling to }(u_L, d_L)\end{cases}
 \,,\ \
  f^{(n)}_{V,\text{tree}}= \begin{cases} \lambda_V &\text{for coupling to }u_R \\ 2 \, \lambda_V &\text{for coupling to } d_R\\ 3 \, \lambda_V &\text{for coupling to }(u_L, d_L)\end{cases}\,,
\ee
with 
\be
     \lambda_V = \frac{\lambda^2}{ 8 \left( M_Y^2-M_X^2\right)}\,.
\ee
Due to vector current conservation, similar spin-independent tree-level contributions relevant for DM coupling to second-generation or third-generation quarks vanish, and they are (obviously) irrelevant for DM coupling solely to leptons. These diagrams also generate an axial-vector interaction of the form $f_{A,\text{tree}}^{(N)} \bar{X} \gamma^\mu \gamma_5 X \bar{q} \gamma_\mu \gamma_5 q$ that leads to spin-dependent interactions. The coupling constants of the axial-vector and the vector interaction terms are generally comparable in size in $t$-channel models. Therefore, given that the experimental sensitivity to spin-independent interactions is several orders of magnitude better than in the spin-dependent case, this axial-vector contribution can usually be neglected.

For a Majorana DM particle coupled to first-generation or second-generation quarks, the vector current vanishes at tree level. Spin-independent interactions then arise predominantly from higher-order effects, like interactions with gluons via loop diagrams such as those shown in figure~\ref{fig:scattering_diagrams}, or higher-order corrections to the conventional mediator exchanges present at tree level. These generate a higher-dimensional effective Lagrangian involving quark and gluon operators,
\be\bsp
  {\cal L}^{\rm{eff}}_q =&\  f_q\, \bar{X}X \, \mathcal{O}_{q}^{(0)} + \frac{g^{(1)}_q}{M_{X}} \, \bar{X} \big(i \partial^{\mu}\gamma^{\nu} + i \partial^{\nu}\gamma^{\mu}\big) X \,  \mathcal{O}_{q,\mu\nu}^{(2)} + \frac{g^{(2)}_q}{M_{X}^2}\, \bar{X}(i \partial^{\mu})(i \partial^{\nu}) X \, \mathcal{O}_{q,\mu\nu}^{(2)}\,,\\
  {\cal L}^{\rm eff}_{ g}=&\ f_G\, \bar{X}X \, \mathcal{O}_{g}^{(0)} +\frac{g^{(1)}_G}{M_{X}}\ \bar{X} \big(i \partial^{\mu}\gamma^{\nu} + i \partial^{\nu}\gamma^{\mu}\big) X \,  \mathcal{O}_{g,\mu\nu}^{(2)} + \frac{g^{(2)}_G}{M_{X}^2}\, \bar{X}(i\partial^{\mu}) (i\partial^{\nu})X \, \mathcal{O}_{g,\mu\nu}^{(2)} \,.
\esp\ee
Here, the operators $\cal O$ represent the scalar and higher-twist quark and gluon operators defined by
\be\bsp
  \mathcal{O}_q^{(0)} \equiv m_q {\bar q} q \,, &\qquad 
  \mathcal{O}^{(2)\mu\nu}_{q} \equiv \frac12 \bar{q}\left( \gamma^{\{\mu} iD_-^{\nu\}}  - \frac14 {g^{\mu\nu}} i\slashed{D}_- \right) q\,,\\
  \mathcal{O}_g^{(0)} \equiv G^A_{\mu \nu} G^{A \mu \nu}   \,, & \qquad
  \mathcal{O}^{(2)\mu\nu}_g \equiv -G^{A \mu\lambda} G^{A \nu}_{\phantom{A \nu} \lambda} + \frac14 g^{\mu\nu} (G^A_{\alpha\beta})^2\,,
\esp\ee
where we have introduced the standard shorthand notation
\be
  A^{\{\mu}B^{\nu\}}  = \frac12 (A^\mu B^\nu + A^\nu B^\mu)\qquad\text{and}\qquad
  D^\mu_{\pm} =D^\mu \pm \overleftarrow{D}^\mu\,,
\ee
with $D^{\mu}$ being the usual covariant derivative. These operators yield the nucleon matrix elements detailed in~\cite{Hill:2014yxa, Hisano:2015bma, Mohan:2019zrk}, with the analytical expressions for the corresponding Wilson coefficients $f_q$, $f_G$, $g_q^{(1)}$, $g_q^{(2)}$, $g_G^{(1)}$, and $g_G^{(2)}$ given in~\cite{Drees:1993bu, Hisano:2015bma, Mohan:2019zrk}. Since these Wilson coefficients are generated from higher-order contributions, the resulting spin-independent and spin-dependent cross sections are comparable. Consequently, constraints from spin-independent cross sections typically dominate when the DM and mediator masses are close, while constraints from spin-dependent rates dominate for DM masses comparably smaller than the mediator mass~\cite{Mohan:2019zrk}. Finally, although such contributions also exist for models featuring a Dirac DM fermion, the resulting Wilson coefficients are subdominant and thus often ignored.

If the DM does not couple to first-generation and second-generation quarks, the penguin diagrams in figure~\ref{fig:scattering_diagrams} become more significant. Regardless of the precise SM fermion species involved, the DM interaction with quarks necessarily induces, at the quantum level, DM-nucleon scattering due to penguin diagrams mediated by a photon, a $Z$ boson, or a Higgs boson. The photon-mediated diagram generates electromagnetic moments for the DM particle $X$, with the most relevant ones in the case of Dirac DM being the magnetic dipole moment $\mu_X$ and the charge radius $b_X$. These are described by the effective DM-photon Lagrangian\footnote{The same diagram also induces an anapole moment, corresponding to the effective operator $\bar{X} \gamma^\mu \gamma_5 X \partial^\nu F_{\mu \nu}$. However, this operator is suppressed in the non-relativistic limit by the square of the DM velocity, and is thus always subdominant in the $t$-channel models considered~\cite{Arina:2020mxo}.}
\be \label{eq:L_eff_gamma}
 \mathcal{L}_{\text{eff},\gamma} = \frac{\mu_X}{2} \bar{X} \sigma^{\mu \nu} X F_{\mu \nu} + b_X \bar{X} \gamma^\mu X \partial^\nu F_{\mu \nu} \,.
 \ee
The electromagnetic moments $\mu_X$ and $b_X$ are determined by matching the coefficients of this effective Lagrangian to the results of explicit loop diagram calculations. Moreover, in the case of Majorana DM, all vector currents vanish, rendering thus these interactions irrelevant.

Similarly, DM interactions with the $Z$ boson arise at one-loop, and induce effective vector interactions of the form
\be\label{eq:Leff_Z}
 \mathcal{L}_{\text{eff},Z} = f_{V,Z}^{(N)} \,\bar{X} \gamma^\mu X \, \bar{N} \gamma_\mu N \,,
\ee
where
\be
 f_{V,Z}^{(p)} = \left( 4 s_W^2-1 \right) \frac{G_F a_Z}{\sqrt{2}} \quad , \quad f_{V,Z}^{(n)} = \frac{G_F a_Z}{\sqrt{2}} \,.
\label{eq:f_VZ}\ee
Here, $G_F$ represents the Fermi constant, $s_W$ the sine of the electroweak mixing angle, and $a_Z$ is an effective form factor whose analytical expression can be found in~\cite{Ibarra:2015nca}. Furthermore, for DM coupling to the doublet of left-handed SM fermions, we have $a_Z^{\left(u_L, d_L\right)} = -a_Z^{\left(u_R\right)} - a_Z^{\left(d_R\right)}$ and $a_Z^{\left(\nu_L, e_L\right)} = -a_Z^{\left(e_R\right)}$. The DM effective coupling to the $Z$ boson scales as $(M_f/M_X)^2$, making this contribution subdominant compared to photon exchange, except in scenarios where DM couples to the third-generation fields $t_R$ or $(t_L, b_L)$. 

The last penguin diagram in figure~\ref{fig:scattering_diagrams}, \ie\ the Higgs-mediated one, induces a coupling of the DM state $X$ to the SM Higgs boson $h$, which in turn generates the effective DM-nucleon interaction
\be\label{eq:Leff_higgs}
  \mathcal{L}_\text{eff,Higgs} = f_{\text{S,Higgs}}^{(N)}\,\bar{X} X \, \bar{N} N\,,
\ee
where expressions for $f_{\text{S,Higgs}}^{(N)}$ are provided in~\cite{Ibarra:2015nca, Kumar:2013hfa}. Since the coupling of fermions to the Higgs boson is proportional to their mass, this contribution is subdominant for DM coupling to light SM fermions. However, for DM coupling to third-generation quarks, the $Z$-exchange contribution is always dominant, making the Higgs exchange negligible as well. Additional contributions to the DM-Higgs interaction may arise from quartic couplings involving the scalar mediator and the Higgs boson, but even with quartic couplings of $\mathcal{O}(1)$, the Higgs-mediated contribution remains subdominant in DM-nucleon scattering.

Finally, in general and for any coupling to a Standard Model fermion, the running of the $\lambda$ coupling from the electroweak scale down to the GeV scale significantly enhances direct detection cross section predicted values~\cite{Mohan:2019zrk}, often generating additional low-scale operators absent at the tree level~\cite{DEramo:2016gos}. 

In the parameter space of the $t$-channel mediator models discussed in this work, searches for spin-independent nucleon-DM scattering in data from the experiments  XENON1T~\cite{XENON:2018voc}, XENONnT~\cite{XENON:2023cxc}, and LZ~\cite{LZ:2022lsv} have achieved unprecedented sensitivity, imposing constraints even on loop-suppressed interactions. For models with couplings to first-generation quarks, next-generation experiments such as DARWIN~\cite{Baudis:2024jnk, Aalbers:2022dzr} are expected to explore much of the currently viable parameter space, as their reach on the spin-independent scattering cross section exceeds the neutrino background~\cite{Billard:2013qya}. Regarding spin-dependent proton scattering, the PICO-60 experiment~\cite{PICO:2017tgi} sets strong exclusion limits for DM masses below 500~GeV, even for Majorana DM~\cite{Arina:2020mxo}. In contrast, xenon-based detectors provide the strongest limits for spin-dependent neutron interactions. While this review was being finalised, impressive new results were released by PandaX-4T~\cite{PandaX:2024qfu} and LZ2025~\cite{LZ:2024zvo}. In particular, the LZ2025 exclusion limit for spin-independent DM-nucleus scattering improves upon that of XENONnT by more than an order of magnitude. These updated bounds are not included in the present analysis.

The sensitivity of direct detection experiments to all these operators is constrained by the minimum threshold energy required to produce detectable excitations in the detector material. For nucleon-DM scattering, the recoil energy is well below the detection threshold for sub-GeV DM masses~\cite{Essig:2011nj, Lin:2019uvt}. In such cases, electron-DM scattering offers a promising alternative, as the larger available energy allows triggering inelastic atomic processes that produce visible signals~\cite{Battaglieri:2017aum}. An expanding experimental programme aims to explore this, including Super-CDMS~\cite{SuperCDMS:2020ymb}, DAMIC~\cite{DAMIC:2019dcn}, SENSEI~\cite{SENSEI:2020dpa}, PandaX-II~\cite{PandaX-II:2021nsg}, DarkSide-50~\cite{DarkSide:2022knj}, and XENON1T-S2~\cite{XENON:2019gfn}.

\subsubsection{Early universe constraints }\label{sec:earlyCons}
The early universe provides a wealth of constraints on the DM properties and interactions, through in particular observations of the CMB, BBN and structure formation. 

The CMB anisotropies are a powerful tool for probing the history and content of the early universe, and their analysis constrains the DM abundance to $\Omega_X h^2 = 0.120 \pm 0.001$~\cite{Planck:2018vyg}. Moreover, DM candidates that annihilate or decay into SM particles can alter the reionisation history, thus affecting the CMB anisotropy spectrum. This results in an upper bound on the annihilation efficiency, $p_{\rm ann} \lesssim 3 \times 10^{-28}$ cm$^3$/s/GeV~\cite{Planck:2018vyg}, where $p_{\rm ann} = f_{\rm eff}\langle \sigma v \rangle / M_X$ with $f_{\rm eff}$ representing the fraction of energy from annihilation transferred to ionisation at redshifts relevant to CMB data. This bound, however, may not directly apply to $t$-channel scenarios where the DM annihilation cross section is P-wave or D-wave suppressed (see section~\ref{sec:ID}). For the scenarios considered here, we should instead rely on the fact that the DM particle itself cannot decay in the early universe, but the mediator can. CMB anisotropies can then constrain the lifetime of the decaying particles, with bounds such as $\tau > 10^{13}$~s~\cite{Slatyer:2016qyl}. However, stronger additional constraints also stem from Lyman-$\alpha$ forest data (see below). On the other hand, particles featuring shorter lifetimes can be constrained either by CMB spectral distortions or by BBN data (as discussed below too), and the CMB can finally also probe the effective number of relativistic degrees of freedom $N_{\rm eff}$ near the surface of last scattering with similar precision to BBN constraints. Unfortunately, for DM produced with relatively large momenta through the freeze-in or superWIMP mechanisms, the Planck 2018 $N_{\rm eff}$ bounds are always less stringent than those originating from the Lyman-$\alpha$ forest data analysis and applicable to DM free-streaming~\cite{Decant:2021mhj}.

CMB spectral distortions also test the nature of the dark sector. These distortions, which arise from non-standard energy injection (\eg\ from particle decays or annihilations) that disrupts the thermodynamic equilibrium between photons and free electrons after BBN ($z \lesssim 10^8$), can be probed and thus used as constraints on models~\cite{Lucca:2019rxf}. The FIRAS instrument on COBE measured the CMB energy spectrum, and it found it consistent with a perfect black-body spectrum at $T_0 = 2.725 \pm 0.002$~K~\cite{Fixsen:1996nj, Mather:1998gm}. For mediator decays with lifetimes $10^4$~s $< \tau < 10^{13}$~s, CMB spectral distortions provide stronger constraints than CMB anisotropies, though they are still weaker than those originating from BBN. Future experiments such as PIXIE or PRISM have nevertheless the potential to significantly improve the bounds~\cite{Lucca:2019rxf}.

Big Bang Nucleosynthesis, which occurred at $T_{\rm BBN} \sim 0.1\,$MeV, is another critical probe of dark matter. New particles can affect the primordial abundances of light nuclei by altering the Hubble rate or the entropy density of the universe. This leads to constraints on $N_{\rm eff}$, which are comparable to those originating from the CMB~\cite{Hufnagel:2017dgo}, but less stringent than the Lyman-$\alpha$ forest bounds discussed below. In addition, if the mediator has a lifetime longer than approximately $0.1$~s, its decay can induce non-thermal nuclear reactions during or after BBN, modifying standard predictions~\cite{Kawasaki:2004qu, Jedamzik:2007qk}. Such constraints are especially relevant for superWIMP production scenarios involving mediators with lifetimes $10^4$~s  $< \tau < 10^{13}$~s that decay into hadronic final states. For example, in top-philic scenarios, these constraints help close the existing gap between constraints stemming from Lyman-$\alpha$ forest probes and the collider limits~\cite{Garny:2018icg, Decant:2021mhj}.

Within the context of the $t$-channel models considered here, feebly interacting DM produced from decays or scatterings off an heavier mediator in the earlier universe can have very large momentum at the time of production compared to their mass, thereby affecting small scale structures and possibly leaving an imprint similar to warm DM (WDM) through free-streaming. The Lyman-$\alpha$ forest data, a typical tracer of small-scale structure clustering, provide stringent upper bound on the mass of thermal WDM, $M_{\rm wdm} \gtrsim 5.3$~keV~\cite{Irsic:2017ixq}. Translating these constraints to \emph{non-thermally} produced DM, via for example freeze-in or the superWIMP mechanism, requires costly hydrodynamical simulations or approximations using velocity dispersion or free-streaming scale comparisons~\cite{Schneider:2016uqi, Murgia:2017lwo, Bae:2017dpt, Ballesteros:2020adh, DEramo:2020gpr, Bode:2000gq, Heeck:2017xbu, Decant:2021mhj}. Another more advanced approach requires to determine the non-cold DM velocity distribution at the time of production, feed it to a Boltzmann solver such as~\cite{Lesgourgues:2011rh}, obtain the corresponding linear power spectrum or the associated transfer function, and compare those to the WDM results. Finally, we could also employ the area criterion~\cite{Schneider:2016uqi, Murgia:2017lwo, DEramo:2020gpr, Decant:2021mhj}. 

Remarkably, for freezing-in and superWIMP DM, the velocity distributions are similar in shape to the thermal WDM case but shifted, allowing a straightforward mapping of WDM bounds to constraints on $M_X$~\cite{Decant:2021mhj}. The resulting Lyman-$\alpha$ forest constraint can generally be expressed as~\cite{Decant:2021mhj}
\be
  M_X  \gtrsim  
  \begin{cases}
    15\, {\rm keV} \times \left( \frac{106.75}{g_*(T_{\rm FI})} \right)^{1/3}& \text{for freeze-in through decays},\\
    3.8\, {\rm GeV} \times \left(R_\Gamma^{\rm SW}/ 10^{-12}\right)^{-1/2}\times \left( \frac{106.75}{g_*(T_{\rm SW})}\right)^{1/3}  & \text{for superWIMP},\\
   \end{cases}
\label{eq:limsly}\ee
where $M_Y \gg M_X$ and $ M_Y \gg M_f$, $R_\Gamma$ is the dimensionless parameter introduced in \eqref{eq:Rgam} and proportional to the mediator decay rate and inversely proportional to its mass squared, and $g_*$ denotes the number of relativistic degrees of freedom at production. Notably, Lyman-$\alpha$ forest data can probe DM masses well above the naive keV scale in the case of superWIMP production, depending on the mediator's lifetime and mass.

\subsection{Cosmological constraints for benchmark models} \label{sec:CosmoConsBench}

In this section, we summarise the cosmological and astrophysical constraints that could be imposed on several of the benchmark models considered. We start by examining minimal $t$-channel DM models in scenarios where the observed relic density is reproduced through conventional freeze-out, conversion-driven freeze-out and through the freeze-in/superWIMP mechanism in sections~\ref{sec:CosmConstMinQuark}, \ref{sec:BMquarkconversiondriven} and \ref{sec:tRfreezeinSW},  respectively. The discussion is thus structured by distinguishing between the different DM production mechanisms responsible for explaining the relic density across distinct domains in the model parameter space. Leptophilic models are next investigated in section~\ref{sec:CosmConstMinLept}, before we finally turn in section~\ref{sec:cosmo_nonmin} on non-minimal models. More precisely, sections~\ref{sec:CosmConstNonMin_flav}, \ref{sec:CosmConstNonMin_frust}, \ref{sec:CosmConstNonMin_compo} and \ref{sec:CosmConstNonMin_nonab} are dedicated to models featuring multiple DM states $X_i$ (\eg\ flavoured DM), multiple mediators (\eg\ frustrated DM), and mixed $t$-channel and $s$-channel mediator scenarios.

It is important to note that the selection of benchmark models presented here is not exhaustive. The choices are primarily guided by the availability of existing results and do not aim to reflect theoretical preferences for any specific models. For minimal quark-philic models, most studies have focused on couplings to right-handed quarks of the first and third generations. Similarly, investigations of minimal leptophilic models have predominantly concentrated on couplings to muons.

\subsubsection{Minimal simplified models}\label{sec:minicosmo}

\paragraph{Minimal quark-philic models in the canonical freeze-out regime}\label{sec:CosmConstMinQuark}
\paragraph*{}\vspace{.3cm}

\begin{figure}
    \centering
    \includegraphics[width=.45\textwidth]{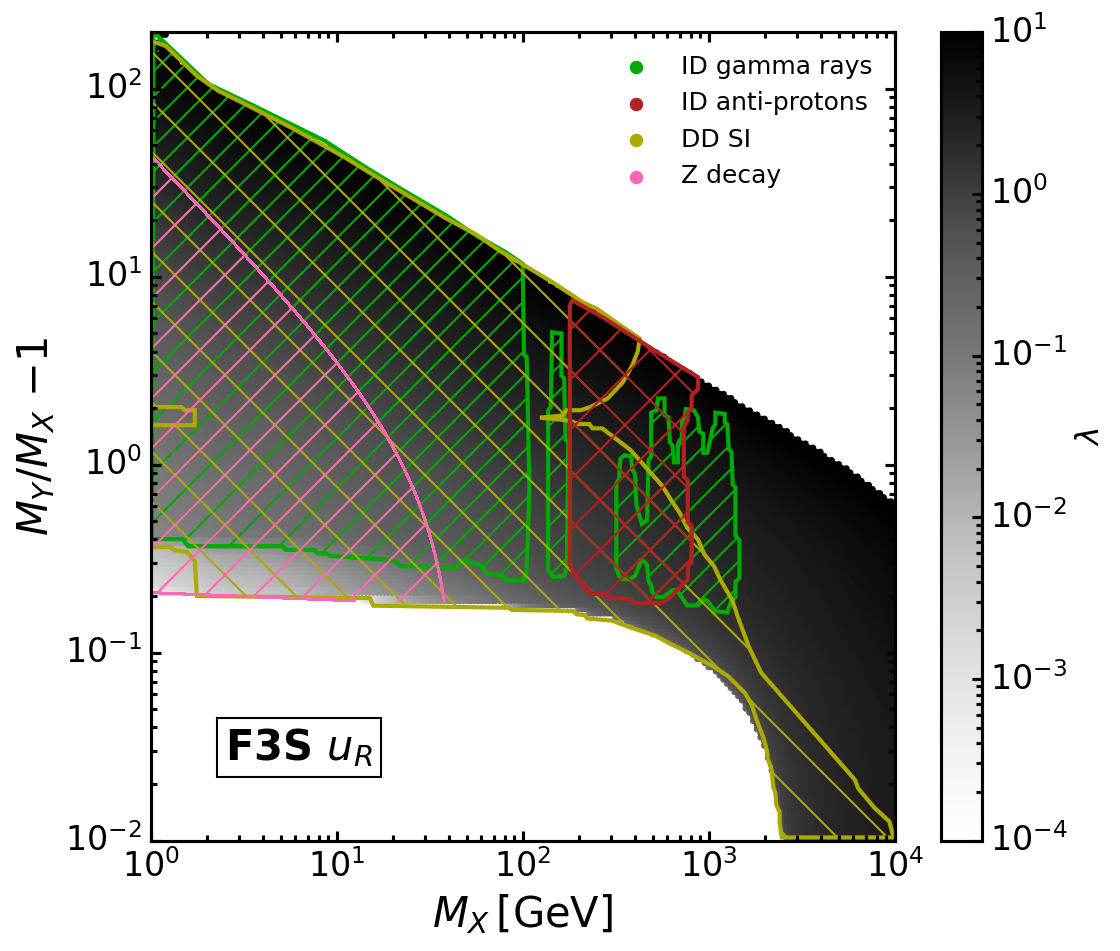}\hfill
    \includegraphics[width=.45\textwidth]{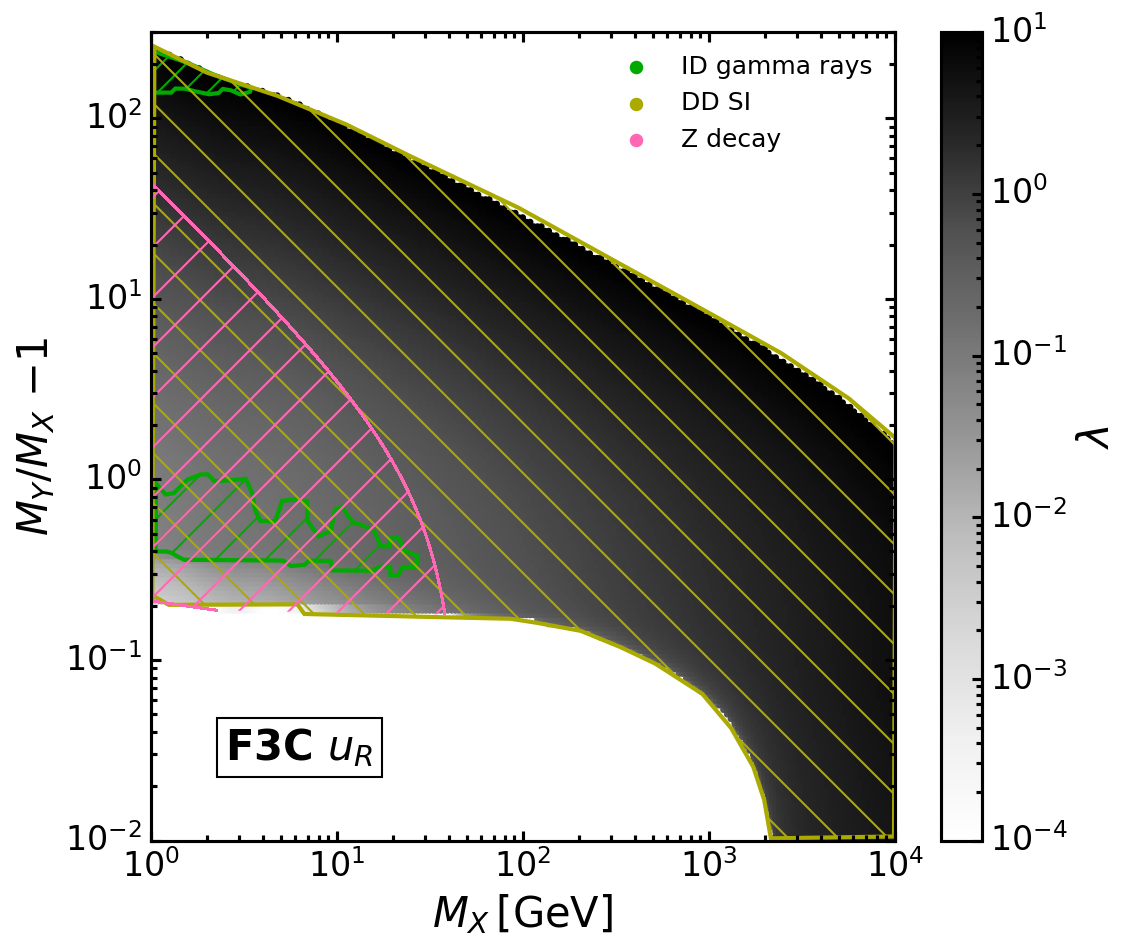}\\
    \includegraphics[width=.45\textwidth]{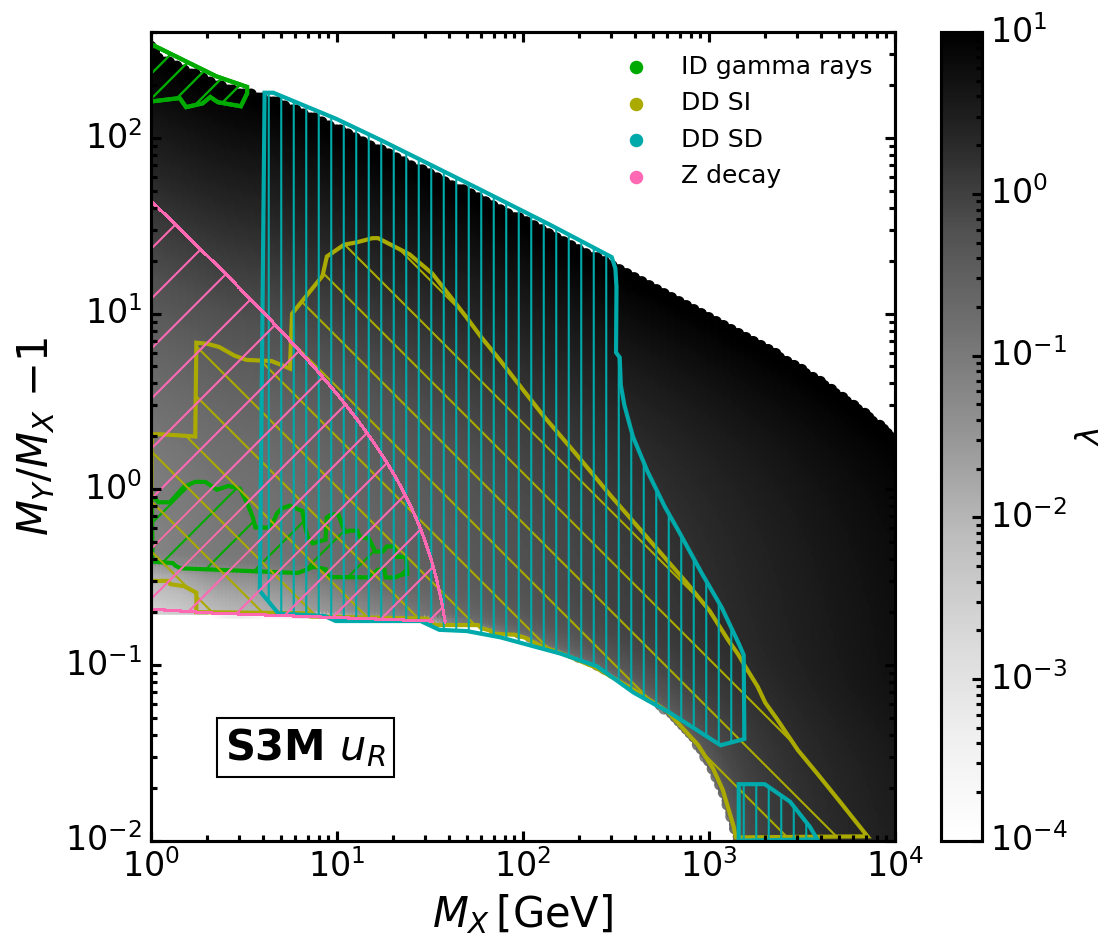}\hfill
    \includegraphics[width=.45\textwidth]{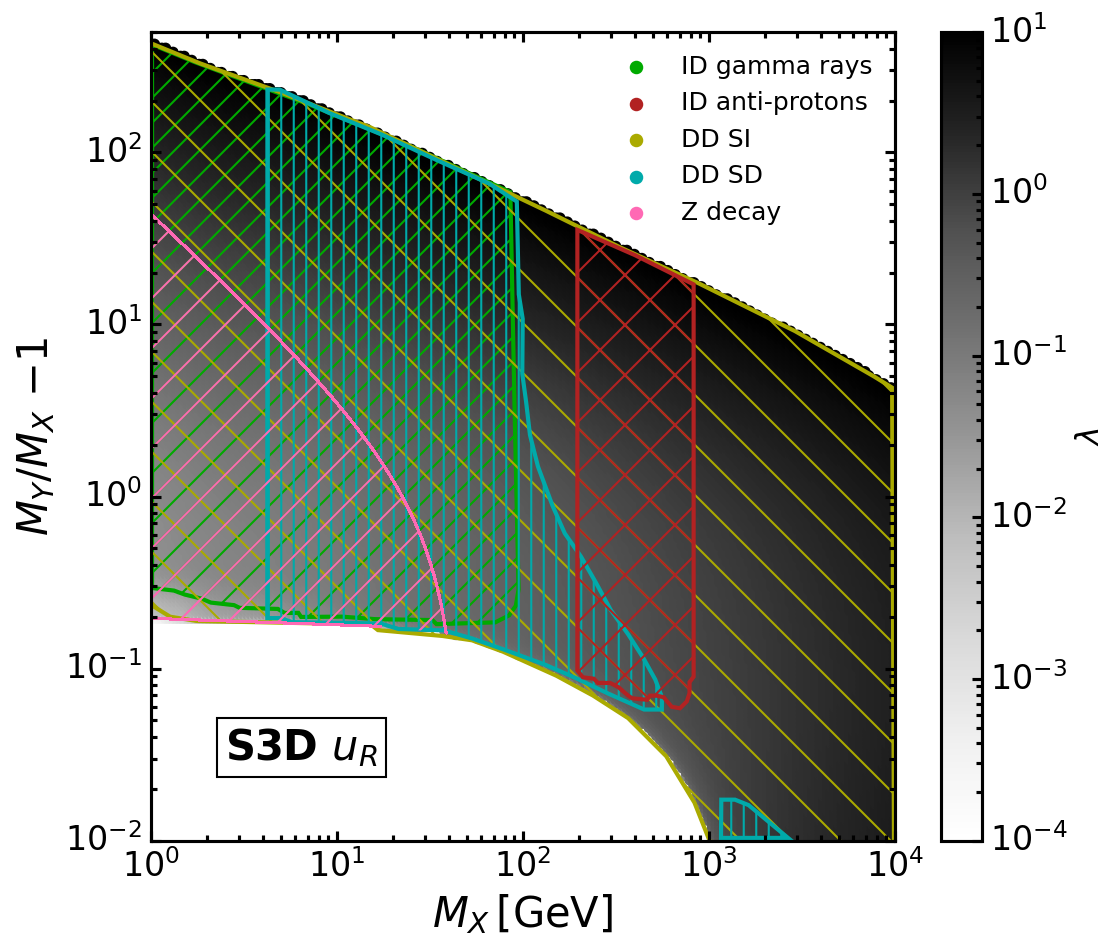}\\
    \includegraphics[width=.45\textwidth]{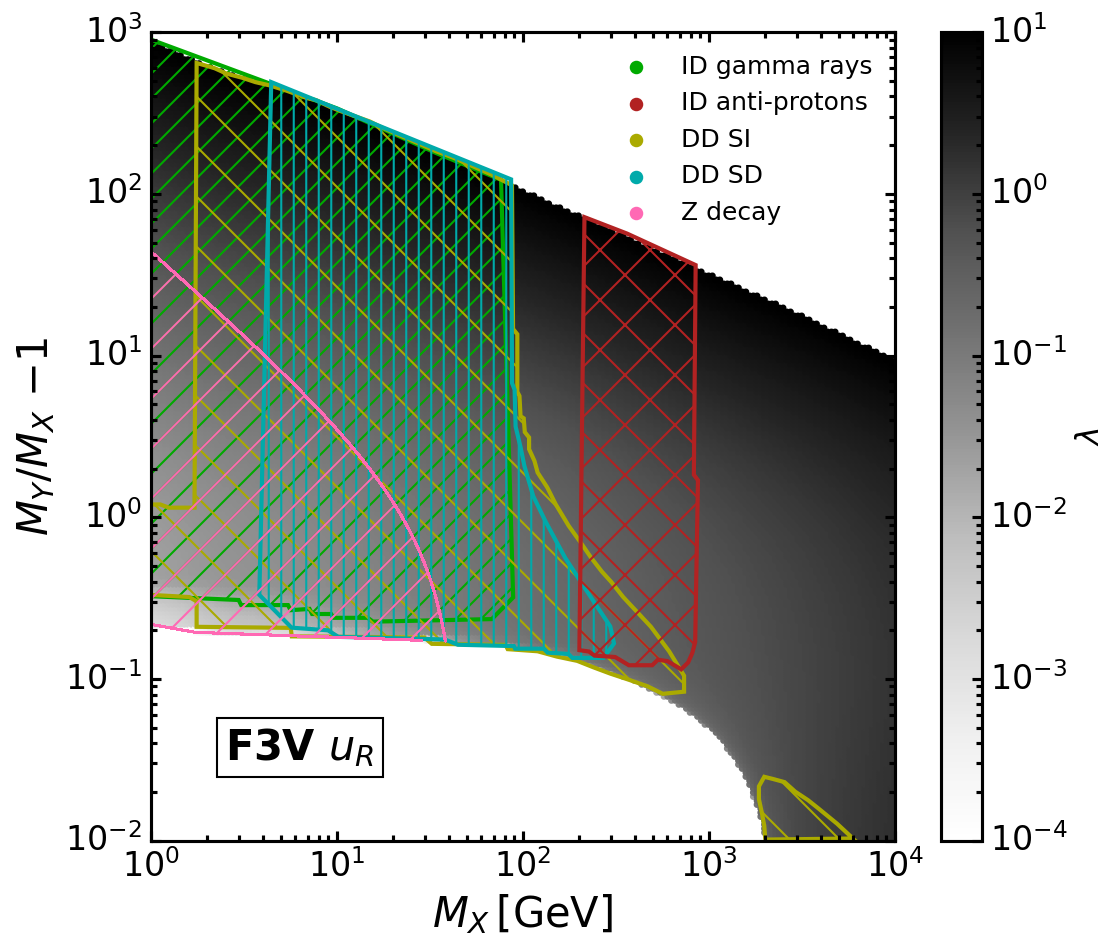}\hfill
    \includegraphics[width=.45\textwidth]{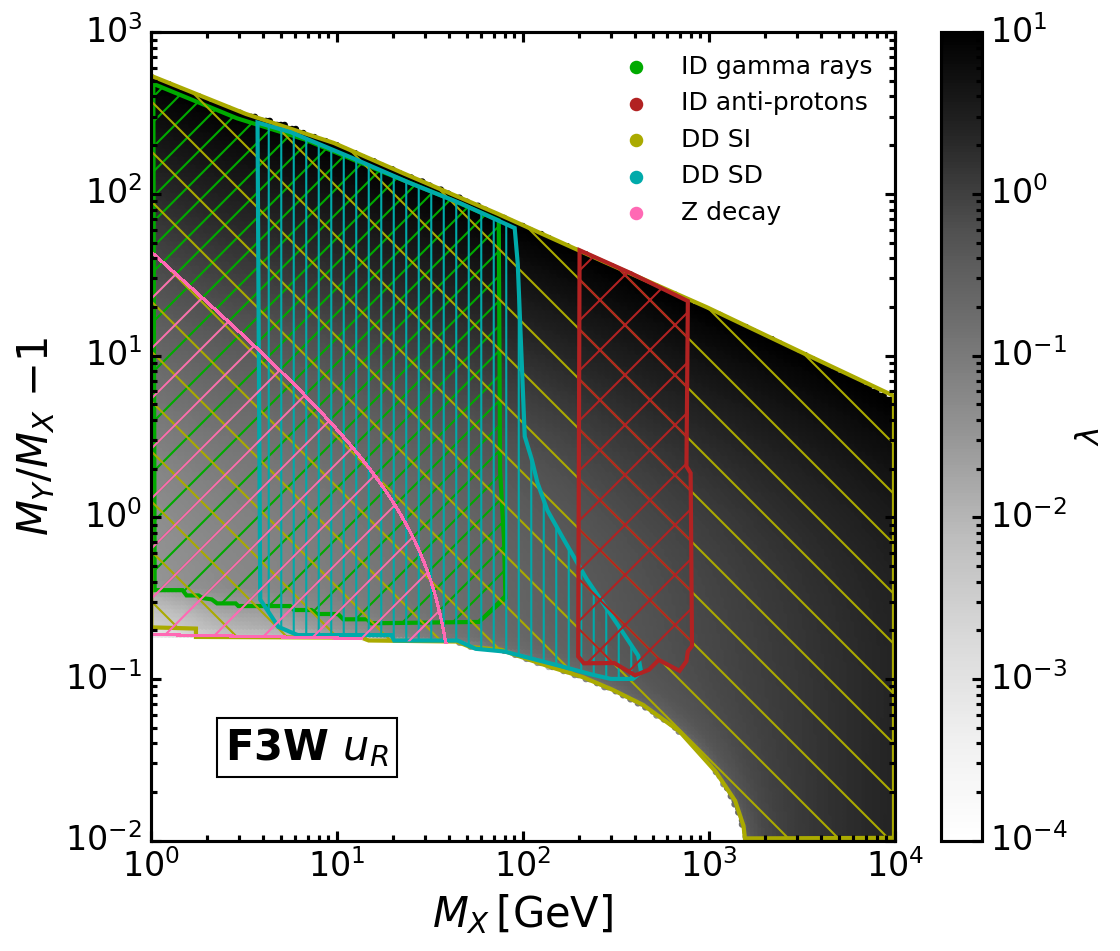}
    \caption{Constraints on minimal simplified $t$-channel DM models from cosmological and astrophysical observables, as well as from the measured $Z$-boson visible decay width. The coloured region in the $(M_X, M_Y/M_X - 1)$ plane represents scenarios that achieve $\Omega h^2 \simeq 0.12$, with the coupling value $\lambda$ indicated by the grey-scale colour map. The left (right) panels correspond to models with self-conjugate (complex) DM, featuring scalar (top row), fermionic (middle row), and vector (bottom row) DM particles. The hatched regions indicate exclusions from gamma-ray searches (`ID gamma rays', including gamma-line searches from Fermi-LAT~\cite{Fermi-LAT:2015kyq} and HESS~\cite{HESS:2018cbt} from the galactic centre and gamma-ray continuum searches in dSPhs by Fermi-LAT~\cite{Fermi-LAT:2016uux}), cosmic-ray antiproton searches (`ID antiprotons'), DM direct detection via spin-independent and spin-dependent interactions (`DD SI' and `DD SD,' respectively, including limits from LZ~\cite{LZ:2022ufs}, PICO-60~\cite{PICO:2017tgi}, CRESST-III~\cite{CRESST:2019jnq} and DarkSide-50~\cite{DarkSide:2018bpj}), and $Z$-boson visible decays (`Z decay', from~\cite{Zyla:2020zbs}). The blank upper region corresponds to scenarios requiring non-perturbative couplings, while the white region at the bottom represents the CDFO regime (see section~\ref{sec:BMquarkconversiondriven}). This figure is taken from~\cite{Arina:2023msd}, where further details can be found.\label{fig:cosmo3D}}
\end{figure}

Figure~\ref{fig:cosmo3D} illustrates the cosmologically viable regions of the parameter space for six minimal models described in section~\ref{sec:model_minimal} in which the DM candidate couples to the right-handed up quark $u_R$, three of them with a real DM candidate and three of them with a complex one. Specifically, these include self-conjugate scalar DM (\lstinline{F3S}, top left), Majorana DM (\lstinline{S3M}, central left), real vector DM (\lstinline{F3V}, bottom left), as well as complex scalar DM (\lstinline{F3C}, top right), Dirac DM (\lstinline{S3D}, central right), and complex vector DM (\lstinline{F3W}, bottom right) candidates. The results are displayed in the plane defined by the DM mass $M_X$ and the relative mass splitting $M_Y/M_X - 1$. Moreover, the grey-scale colour map indicates the value of the coupling $\lambda$ required to achieve the observed DM relic density, $\Omega h^2 \simeq 0.12$~\cite{Planck:2018vyg}. As the DM mass and mass splitting increase, the coupling value necessary to match the relic density also grows. Consequently, the white regions in the upper-right corners of the panels correspond to overabundant DM scenarios, where the annihilation cross section is insufficient within the perturbative regime of the coupling. Conversely, for small mass splittings, co-annihilation effects involving the mediator become increasingly significant. The white regions in the lower-left corners thus represent under-abundant scenarios, where mediators remain in chemical equilibrium with the thermal bath and dominate the annihilation process due to the large associated cross section. Although this condition persists for couplings $\lambda \geq 10^{-4}$, cosmologically viable solutions also exist for smaller couplings of the order of $10^{-6}$. In such cases, the relic density is determined via conversion-driven freeze-out~\cite{Garny:2017rxs}, in which the above chemical equilibrium breaks down due to semi-efficient conversion processes between the DM particle and the mediator. This regime, opening a new part of the parameter space, has not been explored in~\cite{Arina:2023msd}, despite that for $\lambda < 10^{-4}$ all astrophysical constraints are naturally evaded.

In the canonical freeze-out regime, direct detection constraints originating from spin-independent DM interactions with nuclei represent the strongest constraints across all six models. They are found to exclude the entire sampled parameter space for all three complex DM classes of scenarios, as shown in the right panels of figure~\ref{fig:cosmo3D}. This leaves the conversion-driven freeze-out region as the sole viable regime for these models. In contrast, for self-conjugate DM models, portions of the parameter space remain unaffected by direct detection bounds. Nevertheless, combining constraints from direct detection (both spin-independent and spin-dependent bounds), indirect detection via gamma-ray and cosmic-ray antiproton observations, and robust limits from $Z$-boson visible decay measurements excludes significant portions of the parameter space, as depicted in the left panels of figure~\ref{fig:cosmo3D}. For real scalar DM, these constraints exclude all scenarios with DM masses below $800\,\text{GeV}$ or mediator masses below $2\,\text{TeV}$. For Majorana DM and real vector DM, a similar exclusion pattern arises, except for isolated allowed regions featuring scenarios with $M_X \lesssim 4\,\text{GeV}$ and $100\,\text{GeV} \lesssim M_X \lesssim 200\,\text{GeV}$ respectively. In the latter cases, cosmic-ray antiproton data provide key constraints for $200\,\text{GeV} \lesssim M_X \lesssim 800\,\text{GeV}$, similar to the scalar DM case. Here, indirect detection bounds are particularly stringent for vector and Dirac DM, due to the S-wave nature of the annihilation process into quark pairs. For additional results on these models, we also refer to~\cite{Garny:2015wea}.

\begin{figure}
    \centering
    \includegraphics[width=0.49\textwidth]{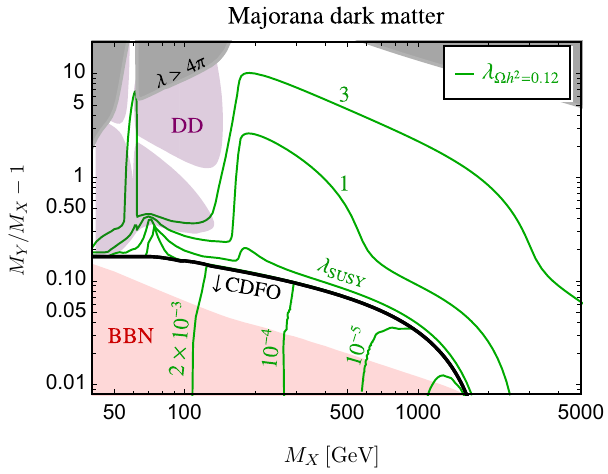}\hfill
    \includegraphics[width=0.49\textwidth]{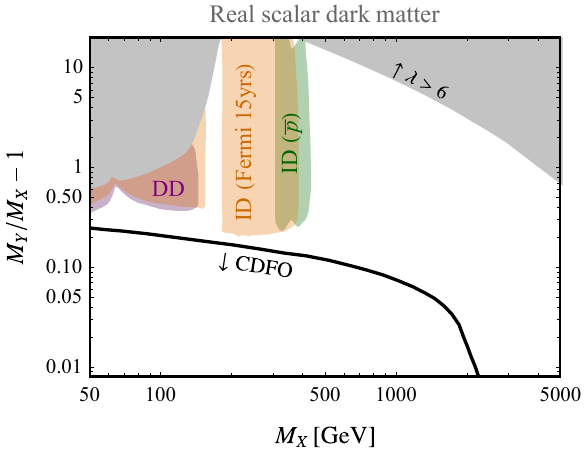}
    \caption{Cosmologically viable region of the parameter space of models featuring a DM state coupling to the right-handed top quark $t_R$ (with $\Omega h^2 = 0.12$), and constraints shown as functions of the DM mass $M_X$ and the relative mass splitting $M_Y / M_X - 1$. The left panel corresponds to Majorana DM, while the right panel depicts a scenario with real scalar DM. These plots are based on the results from~\cite{Garny:2018icg} and~\cite{Colucci:2018vxz}, respectively, and the direction detection (DD) constraints on both plots are derived from the XENON1T bounds. The thick black line separates the WIMP region (above) from the CDFO region (see section~\ref{sec:BMquarkconversiondriven}), and the shaded regions denote exclusions due to the various experimental and theoretical constraints discussed in the text.\label{fig:tRFO}}
\end{figure}

We now turn to simplified models in which the DM state interacts with the right-handed top quark $t_R$. In the left panel of figure~\ref{fig:tRFO}, we show the viable parameter space of a top-philic model featuring a Majorana dark matter state and a coloured scalar mediator (namely the \lstinline{S3M_tR} class of models), as obtained in~\cite{Garny:2018icg}. Constraints on the parameter space are displayed in terms of the DM mass $M_X$ and the mass splitting $M_Y / M_X - 1$ between the mediator and the DM particle. At each point in the parameter space, the Yukawa coupling strength $\lambda$ between the DM and the mediator is fixed to ensure that thermal freeze-out produces the observed DM relic density, $\Omega h^2 = 0.12$, the green contours being isolines of constant $\lambda$ value. 

The features at low DM masses arise due to resonant contributions from processes such as $XX \to h \to b\bar{b}$ via the loop-induced $XXh$ coupling, as well as from the co-annihilation channel $XY \to t \to Wb$. Additionally, $2 \to 3$ processes like $XX \to Wtb$, which are relevant below the $t\bar{t}$ threshold, and loop-induced processes like $XX \to gg$, are included in our calculations. For mediator annihilation $Y\bar{Y} \to gg$, Sommerfeld enhancement is accounted for, following~\cite{Ibarra:2015nca}, although bound-state effects are not included. In the lower-right region, below the black solid line, the DM relic density can only be explained via conversion-driven freeze-out, which will be discussed in section~\ref{sec:BMquarkconversiondriven}. In addition, the dark grey region at large relative mass splittings is excluded due to DM overproduction, and the purple-shaded region is excluded by direct detection constraints from XENON1T~\cite{XENON:2017vdw}. Here, the loop-induced $XXh$ and $XX gg$ couplings originating from box diagrams play a critical role as they mediate DM-nucleon scattering through partonic processes like $X g \to X g$. The `blind spot' in the direct detection constraints, visible at mass splittings of approximately 100\,GeV in the left panel of figure~\ref{fig:tRFO}, arises from destructive interference between these loop-induced contributions. Finally, indirect detection limits are relevant only within a very narrow band near the Higgs resonance at $M_X \simeq M_h / 2$, and are thus not shown here. In this region, DM annihilation predominantly proceeds via the loop-induced DM-Higgs interaction, as detailed in~\cite{Garny:2018icg}.

In the right panel of figure~\ref{fig:tRFO}, we move on with a study of the constraints that can be imposed on the parameter space of a simplified model with a real scalar DM candidate and a top-philic vector-like fermion mediator, as examined in~\cite{Colucci:2018vxz}. In the white region above the black solid line, the measured DM abundance can be achieved with WIMP-like couplings ($\lambda \sim 10^{-2}$ to ${\cal O}(1)$) through canonical freeze-out. Below this line, the DM relic density can instead be explained via the CDFO mechanism, which requires much weaker $\lambda$ coupling values. In the WIMP region, constraints from both direct and indirect DM searches exclude substantial portions of the cosmologically viable parameter space, as visible by the parameter space regions excluded when combining direct detection constraints from XENON1T~\cite{XENON:2017vdw}, indirect detection constraints from AMS-02 antiprotons~\cite{Cuoco:2017iax}, and the projected sensitivity of Fermi-LAT after 15 years of exposure~\cite{Fermi-LAT:2016afa}. Here, the related calculations necessitated a detailed treatment of radiative corrections, which was provided in~\cite{Colucci:2018qml}, and that accounts for the non-negligible mass of the final-state SM particles in the relevant processes. NLO processes are particularly important during freeze-out as the LO annihilation cross section is D-wave suppressed~\cite{Giacchino:2013bta}. In fact, for any $t$-channel DM models where the DM couples exclusively to the third generation, QCD corrections must be considered~\cite{Colucci:2018vxz}. For instance, in all scenarios where the DM couples to the right-handed top quark $t_R$, loop-induced DM annihilation into gluons dominates and determines the relic density below the $b$-quark threshold~\cite{Garny:2018icg}.

\begin{figure}
    \centering
    \includegraphics[width=0.5\textwidth]{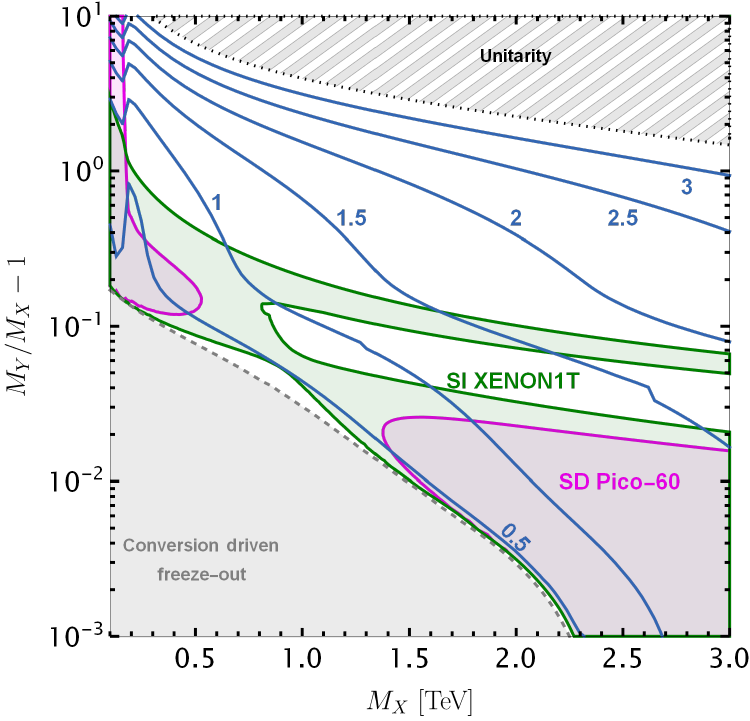}
    \caption{Constraints on a simplified DM model featuring three mass-degenerate mediators universally coupling to a Majorana DM candidate and all generations of SM right-handed up quarks. The blue solid lines are isocontours with fixed $\lambda$, as labelled. The black-shaded region represents parameter values where avoiding DM overproduction would violate perturbative unitarity. The green and magenta areas show exclusions from spin-independent and spin-dependent direct detection experiments, respectively, and the grey-shaded region corresponds to a DM relic density explained through the CDFO mechanism. The results incorporate the effects of bound states, including their excitations, on freeze-out dynamics. This plot is adapted from~\cite{Becker:2022iso}.\label{fig:uRuniversalFO}}
\end{figure}

Finally, we turn on the simplified models featuring universal couplings to all generations of SM fermions. In figure~\ref{fig:uRuniversalFO}, we present the exclusion limits on the parameter space for the DM model \lstinline{S3M_uni} with a SM-singlet Majorana DM candidate. This candidate couples universally to all SM up-type right-handed quarks via three mass-degenerate mediators, thus with a unique coupling strength $\lambda$. As above and following \cite{Becker:2022iso}, the constraints on the parameter space are shown in the plane defined by the DM mass and the mass splitting between the mediator and the DM particle. At each point in the plane, $\lambda$ is fixed to ensure that the observed DM relic density, $\Omega_\text{DM} = 0.12 \pm 0.005$, is not exceeded. The relic density calculation takes into account LO annihilation and co-annihilation processes, Sommerfeld effects, and bound state formation. In the grey-shaded region, freeze-out underproduces DM, requiring alternative production mechanisms to account for the observed DM abundance. Although such mechanisms are not explicitly addressed for this model, their phenomenology is expected to align with the CDFO and freeze-in/superWIMP regimes discussed in sections~\ref{sec:BMquarkconversiondriven} and \ref{sec:tRfreezeinSW}.  

The green-shaded region in figure~\ref{fig:uRuniversalFO} represents spin-independent direct detection constraints. For a Majorana DM candidate, all vector couplings to the SM neutral gauge bosons, including in particular those to the $Z$-boson that would otherwise lead to strong constraints, are identically zero. However, at the one-loop level, a DM-gluon coupling arises, yielding significant constraints at small mass splittings. These constraints are particularly sensitive for mass splitting values around the top quark mass ($M_Y-M_X\sim M_t$), where a resonant behaviour from top quarks running in the loop amplifies the effect. The magenta-shaded region corresponds to spin-dependent direct detection constraints, which mainly arise from tree-level interactions. Since spin-dependent limits are typically much weaker than spin-independent ones, they generally play a subdominant role unless the mass splitting is relatively large or the DM mass is small. However, spin-dependent constraints could become more significant in the future, especially with improved experimental sensitivities such as those projected for DARWIN, where they may even dominate for large mass splittings. Lastly we emphasise that figure~\ref{fig:uRuniversalFO} provides enough information to complementarily combine collider and cosmology exclusions, by appropriately incorporating the results from the collider section. For instance, the bounds determined in figures~\ref{fig:1stgen} and~\ref{fig:1stgenbis} can be directly applied in the large $\lambda$ regime where first-generation mediator production dominates the collider signal. 

\vspace{.5cm}

\paragraph{Minimal quark-philic models in the CDFO regime}\label{sec:BMquarkconversiondriven}
\paragraph*{}\vspace{.3cm}

\begin{figure}
  \centering
  \includegraphics[width=0.49\textwidth,trim=0.05cm 0cm 0cm 0cm,clip]{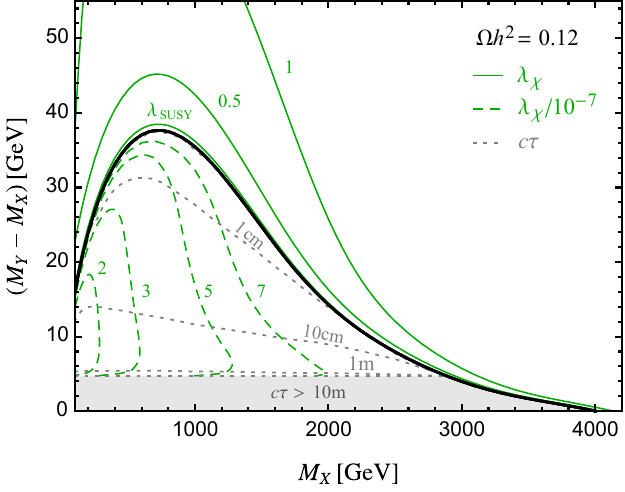}\hfill
  \includegraphics[width=0.51\textwidth,trim=0.cm 0.05cm 0cm 0cm,clip]{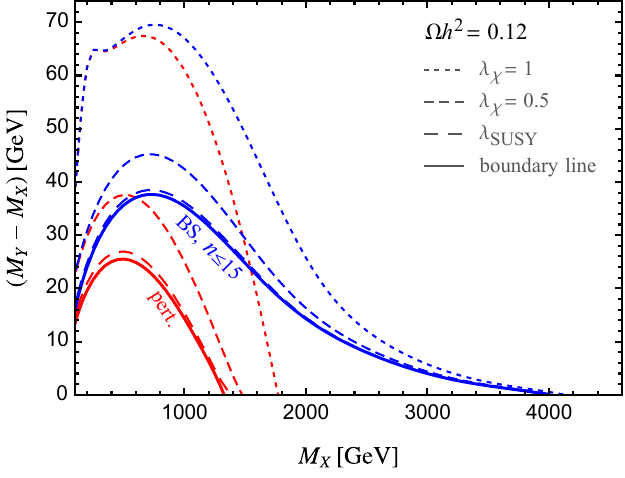}
  \caption{Parameter space regions of a bottom-philic Majorana fermion DM model~\cite{Garny:2021qsr} compatible with the observed relic density. The left panel shows the required couplings and the resulting decay length of the mediator, with the CDFO region lying below the thick black line. The results account for Sommerfeld enhancement and excited bound state effects. The right panel highlights the significance of bound state effects in this scenario, showing the CDFO contour’s boundary (solid lines) along with isolines of constant $\lambda$ values in the canonical freeze-out regime (dashed lines). Results including Sommerfeld enhancement and bound state effects are shown in blue, while those without these effects are shown in red.\label{fig:tRbR_conversiondriven}}
\end{figure}

In all models discussed in section~\ref{sec:CosmConstMinQuark}, the cosmologically viable parameter space that explains the relic density via canonical freeze-out is limited to regions featuring not too small mass splittings (up to a certain DM mass well in the TeV range). In the small mass splitting regime, the DM density can be explained instead by the CDFO mechanism~\cite{Garny:2017rxs, DAgnolo:2017dbv}, as described in section~\ref{sec:conv}. This region is left blank in figure~\ref{fig:cosmo3D} and marked accordingly in figures~\ref{fig:tRFO} and~\ref{fig:uRuniversalFO}. At the boundary between the canonical freeze-out and CDFO regimes, the required couplings decrease by several orders of magnitude, from a WIMP-like magnitude ($10^{-2}$ to $\mathcal{O}(1)$) above the boundary to approximately $10^{-6}$ deep in the CDFO region. Consequently, the lifetime of the mediator $Y$ increases sharply, rendering it long-lived with striking implications for its signatures at colliders that we have discussed in section~\ref{sec:LLP}. Since the CDFO scenario has been studied in detail in the literature for bottom-philic and top-philic models, the following discussion focuses solely on these cases. For mediators coupling to first-generation or second-generation quarks, the phenomenology is nevertheless expected to be very similar to the bottom-philic case.

For bottom-philic Majorana DM scenarios with a coloured scalar mediator (namely the \lstinline{S3M_bR} class of models), the constraints on the parameter space are shown in figure~\ref{fig:tRbR_conversiondriven}. In the region below the thick black line, the observed relic density can be explained within the CDFO scenario. The required values for the Yukawa coupling $\lambda$, of the order of $10^{-7}$, are indicated by green dashed contours. Isolines of constant decay length of the mediator are also shown, this time through grey dotted lines. In the bulk of the parameter space region above the kinematic threshold for the mediator's two-body decay, the decay length ranges from approximately 1~m (for small mass splittings) to 1~mm (near the upper boundary of the CDFO regime). Below this threshold, it becomes significantly larger, often exceeding the dimensions of typical LHC detectors. Unlike in the canonical freeze-out regime, the CDFO parameter space is not subject to strong constraints from direct or indirect detection due to the very weak DM-SM coupling, rendering the LHC searches the unique probes for this scenario. On the cosmological side, the dilution of dark sector particles is driven solely by mediator-pair annihilation. Thermal decoupling is a prolonged process, extending deeper into the non-relativistic regime than in the conventional freeze-out case. As a result, non-perturbative effects, such as Sommerfeld enhancement and bound state formation, play a particularly significant role~\cite{Garny:2021qsr}. These effects substantially extend the size of the CDFO regions in the parameter space, as illustrated in the right panel of figure~\ref{fig:tRbR_conversiondriven}. The red lines represent LO tree-level results, while the blue lines incorporate Sommerfeld enhancement and the effects of excited bound states, including states up to a principal quantum number of $n=15$.

For top-philic models, the phenomenology exhibits distinctive features by virtue of the large top mass $M_t$. The regions of the parameter space relevant for the CDFO regime corresponds to the one below the solid black line in figure~\ref{fig:tRFO}. Here, we focus on the Majorana DM case explicitly studied in~\cite{Garny:2018icg}, as the corresponding CDFO region for scenarios with real scalar DM solely extends to slightly higher masses and does not exhibit any different feature. A key difference from the bottom-philic scenario is that the two-body decay of the mediator is kinematically forbidden throughout the entire CDFO region. Since $M_Y-M_X < M_W + M_b$, the mediator's decay instead proceeds via very suppressed four-body channels. Consequently, the decay length of the mediator is large compared to the dimensions of the LHC detectors across the entire CDFO parameter space, despite the larger $\lambda$ coupling values relevant for top-philic scenarios. However, the large lifetimes of the mediator also imply that certain regions of the parameter space are inconsistent with successful BBN. These regions are shaded in red in the figure. Contours of constant coupling strength are additionally shown in green, ranging from values of approximately $10^{-3}$ for $M_X \lesssim M_t$ to $10^{-6}$ for $M_X \gg M_t$. While these results include Sommerfeld enhancement, bound state effects are not accounted for in this analysis. As demonstrated in the bottom-philic case, such effects are expected to further enlarge the CDFO parameter space.\vspace{.5cm}

\paragraph{Minimal quark-philic models in the freeze-in and superWIMP regimes}\label{sec:tRfreezeinSW}
\paragraph*{}\vspace{.3cm}
For couplings smaller than those considered in the CDFO regime, DM does not thermalise in the early universe and we must rely on freeze-in and superWIMP DM production to explain the measured relic density, as detailed in sections~\ref{sec:FI} and \ref{sec:SW}. In the class of $t$-channel mediator models under consideration, both production mechanisms are present, though their relative importance can vary significantly depending on the specifications of the model and the benchmark point in the parameter space. The following discussion focuses on the case of top-philic fermionic DM. For mediators coupling to first-generation or second-generation quarks, the cosmologically viable parameter space is expected to be similar for large mediator masses ($M_Y \gg M_t$) but differs for $M_Y \lesssim 1$~TeV. Moreover, scalar DM scenarios are expected to lead to qualitatively similar results~\cite{Calibbi:2021fld}.

\begin{figure}
  \centering
  \includegraphics[width=0.5\textwidth]{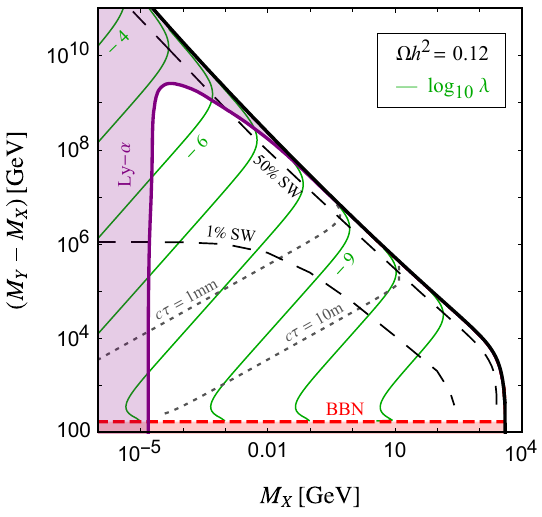}
  \caption{Regions of the top-philic Majorana DM model compatible with DM production through the freeze-in and superWIMP regime, and existing cosmological constraints as studied in~\cite{Decant:2021mhj}. The plot highlights the interplay between these two DM production processes across the parameter space (long dashed lines). Relevant constraints from structure formation (purple) and BBN (red) are also indicated. In addition, isolines of constant $\lambda$ values and of constant mediator decay length are shown in green and grey, respectively.\label{fig:tRFI}}
\end{figure}

In figure~\ref{fig:tRFI}, we display the cosmologically viable regions of the parameter space associated with the top-philic Majorana DM setup considered. The results, based on~\cite{Decant:2021mhj}, are shown in the $(M_X, M_Y - M_X)$ plane as a function of the coupling value $\lambda$. They assume no additional contributions to the DM abundance prior to infrared freeze-in (\eg\ from post-inflationary reheating processes) and a reheating temperature above the mediator mass scale $T_\text{RH} \gg M_Y$. The measured relic abundance is reproduced in the regions of the parameter space extending to the left of the solid black curve, whereas to the right of this curve, superWIMP production alone would exceed the observed relic density. In addition, long-dashed black lines indicate the relative contributions of freeze-in and superWIMP production to $\Omega h^2$, with the freeze-in contribution increasing as one moves away from the overproduction boundary, and thin green lines are isolines of constant $\lambda$ values. For masses around the TeV scale that are typically relevant for collider searches, $\lambda$ ranges between $10^{-8}$ and $10^{-12}$. Those small values imply large mediator lifetimes, as shown through the light dashed lines demonstrating that TeV-scale mediators have macroscopic decay lengths. 

Cosmological constraints are indicated by the different coloured regions. The purple area is excluded by structure formation constraints via Lyman-$\alpha$ observations, and the red area is inconsistent with successful BBN. Here, all the results include Sommerfeld enhancement and bound-state effects, although they only account for the impact of the ground bound state~\cite{Decant:2021mhj}. Higher bound-state excitations are expected to be significant as they reduce the superWIMP contribution, thereby expanding the cosmologically viable region of the parameter space towards larger DM masses~\cite{Binder:2023ckj}. Finally, we should keep in mind that thermal effects, which are particularly relevant in the freeze-in regime, have only been studied recently in~\cite{Becker:2023vwd}, as noted in section~\ref{sec:DMprod}. The full parameter space incorporating these effects has yet to be mapped.

The above results change if the assumption $T_\text{RH} \gg M_Y$ is relaxed to consider lower reheating temperatures. For simplified $t$-channel DM models, such scenarios with low reheating temperatures have been investigated in~\cite{Brooijmans:2020yij, Calibbi:2021fld, Becker:2023tvd}. These studies examine freeze-in production for both Majorana and scalar DM models with scalar and vector-like fermion mediators, respectively. If $T_\text{RH} < M_Y$, meaning that DM freezes in during reheating, the entropy dilution that reduces the comoving number density leads to larger Yukawa couplings between DM, the mediator, and the SM particle with which they interact in order to reproduce the observed DM relic density. These larger couplings shorten the lifetime of the long-lived mediator, bringing it into a range testable via displaced-vertex searches, as discussed in section~\ref{sec:LLP}.\vspace{.5cm}

\paragraph{Minimal quark-philic models: discussion and conclusion}\label{sec:cosmosummary}
\paragraph*{}\vspace{.3cm}

We can summarise our findings and discussion with the following general remarks. In minimal models, the quark flavour to which the DM couples has only a minor effect on the relic density across all DM production mechanisms, provided that the DM mass and the DM-mediator mass splitting are significantly larger than the corresponding quark mass. This condition is typically satisfied for most flavour choices. However, in the top-philic case, parts of the parameter space exhibit unique features due to the top quark large mass and, in some cases, its substantial Yukawa coupling to the SM Higgs boson. In the canonical freeze-out regime, direct detection experiments impose the strongest constraints on DM, particularly for couplings to first-generation quarks. Here, these experiments exclude significant portions, or even all, of the cosmologically viable parameter space. In contrast, models featuring couplings to third-generation quarks are less constrained by direct detection. Astrophysical observations thus provide a valuable complement to collider searches for freezing-out DM, as they remain sensitive to very heavy DM candidates that are hard (or even impossible) to be efficiently produced at colliders. In the CDFO and freeze-in/superWIMP case, conventional DM searches through direct and indirect detection offer limited prospects due to the extremely weak coupling required. However, certain regions of the parameter space predicting a very long-lived mediator are constrained by Big Bang Nucleosynthesis. These constraints apply to specific realisations of top-philic CDFO models and to portions of the parameter space related to the freeze-in/superWIMP regime that is also further constrained by cosmological structure formation, particularly Lyman-$\alpha$ forest observations. We additionally point out that models where DM couples to second-generation quarks remain relatively under-explored. Nevertheless, their astroparticle phenomenology is expected to broadly resemble that of models involving first-generation quarks. This contrasts with collider physics, where charm quarks play a distinctive role due to factors such as charm tagging, uncertainties in parton distribution functions, and other collider-specific considerations. 

In the $t$-channel mediator models considered in this work, DM is assumed to be a singlet under the SM gauge group. Relaxing this assumption introduces a broader class of models featuring $t$-channel mediators. Examples include in particular the Minimal Dark Matter framework~\cite{Cirelli:2005uq}, where DM lies in a non-trivial $SU(2)_L$ multiplet. Among the most studied realisations are the $SU(2)_L$ doublet (`higgsino' DM) and triplet (`wino' DM) scenarios. Detailed phenomenological studies can be found, for instance, in~\cite{Nagata:2014wma, Dessert:2022evk, Arina:2014xya, Cirelli:2014dsa, Beneke:2019qaa}, and also in \cite{Agin:2023yoq, Agin:2024yfs, Chakraborti:2024pdn, Martin:2024ytt} in light of concurring excesses in related LHC searches. The DM observables and related constraints in the case of models with non-trivial $SU(2)_L$ multiplets differ significantly from the gauge-singlet case. For instance, the small mass splitting among states within a multiplet, typically in the 100 MeV range, often leads to LLP signatures driven by phase-space suppression like in higgsino and wino models. While these specific models are not discussed in detail here, section~\ref{sec:LLP} addresses analogous signatures in the considered $t$-channel mediator models, where long lifetimes are instead induced by the small couplings required in the CDFO, freeze-in or superWIMP regimes. To maintain focus, we have indeed restricted the previous discussion to gauge-singlet models, and we refer instead to \cite{Cirelli:2024ssz} for a comprehensive overview of electroweak multiplet models. Specific non-minimal explorations are then conducted in the last part of this section, all highlighting interesting phenomenology not covered in the minimal framework.\vspace{.5cm}

\paragraph{Minimal leptophilic models}\label{sec:CosmConstMinLept}
\paragraph*{}\vspace{.3cm}

In this section, we explore leptophilic minimal models, focusing specifically on scenarios with a Majorana DM candidate and a scalar mediator with couplings to right-handed muons (\ie\ the \lstinline{S1M_muR} class of models). The phenomenology of models with couplings to right-handed electrons or taus is expected to be qualitatively similar, provided the DM and mediator masses satisfy $M_X, M_Y-M_X \gg M_\tau$. For a comprehensive overview of these models, we refer to~\cite{Bai:2014osa, Chang:2014tea, Agrawal:2014ufa, Garny:2015wea, Baker:2018uox, Belanger:2018sti, Barducci:2018esg, Junius:2019dci, Calibbi:2021fld, Liu:2021mhn, Biondini:2022ggt, Jueid:2023zxx, Jueid:2020yfj, Horigome:2021qof, Liu:2013gba}.

\begin{figure}
  \centering
  \includegraphics[width=0.46\textwidth]{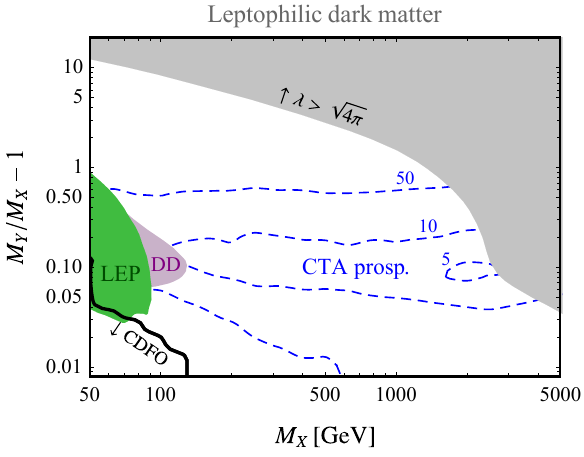}\hfill
  \includegraphics[width=0.44\textwidth]{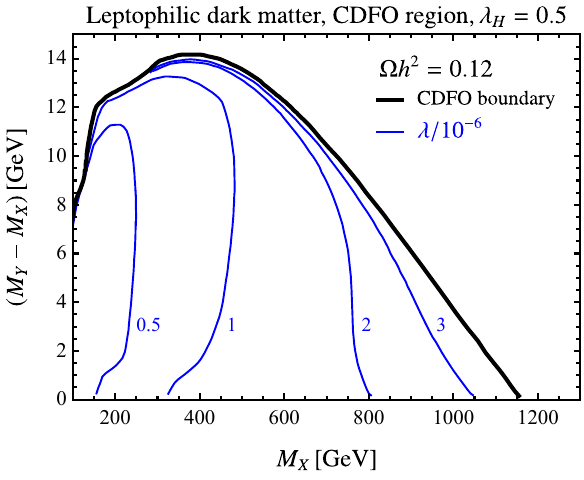}
  \caption{Viable regions of the parameter space of a simplified \lstinline{S1M_muR} leptophilic DM model where a Majorana DM particle couples to right-handed muons. In the left panel (adapted from \cite{Garny:2015wea}), results assume that the relic abundance is achieved through thermal freeze-out, while the region where the measured abundance could be reproduced within conversion driven freeze out production region is shown in the lower left corner. Projected constraints from indirect and direct detection are shown through the dashed blue and purple areas, while LEP limits are given in green. The figure in the right panel is dedicated to the CDFO regime and is adapted from \cite{Junius:2019dci}. Here, the Higgs-portal coupling is set to $\lambda_H=0.5$, and the blue lines are isolines of constant $\lambda/10^{-6}$ values.\label{fig:muR-freeze-out}}
\end{figure}

The canonical freeze-out regime provides an interesting starting point to examine these models. The left panel of figure~\ref{fig:muR-freeze-out} illustrates the viable parameter space and projected constraints, as presented in~\cite{Garny:2015wea}, under the assumption that the DM relic abundance is achieved through thermal freeze-out. In the upper-right region (grey shading), DM is overproduced unless couplings become non-perturbatively large ($\sqrt{4\pi} < \lambda$). Conversely, the lower-left region (below the thick black line) leads to under-abundant DM via canonical freeze-out, while conversion-driven freeze-out provides instead viable solutions discussed later. Compared to quark-philic models, non-collider constraints on leptophilic models are generally weaker. The annihilation of leptophilic DM produces less hadronic activity, resulting in fewer photons and significantly fewer antiprotons, which limits the detectability of a signal in gamma-ray and cosmic-ray detectors respectively. Additionally, the DM-nucleon scattering cross section is loop-suppressed, reducing direct detection prospects. For the considered case of Majorana DM, the direct detection cross section is further suppressed by its velocity dependence. Despite these limitations, experimental progress has allowed partial probing of the parameter space. For instance, a projected LZ constraint from 2015 excludes regions with mass splittings around 10\% for DM masses below roughly 100 GeV, as shown by the purple shaded area in the left panel of figure~\ref{fig:muR-freeze-out}. Indirect detection constraints remain elusive under reasonable assumptions such as an Einasto DM density profile due to their velocity suppression, though enhancements in gamma-ray flux relative to the Einasto profile could exclude additional regions, as indicated by the blue contour lines in the figure. In contrast, direct and indirect detection become more significant for Dirac fermion or complex scalar DM where the relevant cross sections are not velocity suppressed, as studied in~\cite{Baker:2018uox, Cermeno:2022rni}.

For small mass splittings, the CDFO regime becomes relevant, requiring very weak DM couplings of ${\cal O}(10^{-6})$. This region of the parameter space is represented in the lower-left corner of the left panel of figure~\ref{fig:muR-freeze-out}, with the thick black line marking the transition from the WIMP regime to CDFO regime. The measured relic density is still reproduced, but the required coupling value drops by several orders of magnitude. In the model with a scalar mediator considered here, the mediator pair-annihilation cross section depends on electroweak contributions from $\gamma/Z$ exchanges and potential Higgs-portal interactions, with their relative importance impacting the position of the boundary of the CDFO regime. While the Higgs portal contributions are taken vanishing in the left panel of the figure, introducing a sizeable Higgs-portal coupling, such as $\lambda_H = 0.5$, significantly expands the viable CDFO parameter space. This is illustrated in the right panel of the figure, where we can consider CDFO scenarios with DM masses ranging up to above 1 TeV~\cite{Junius:2019dci}. The green thick line in the figure denotes the new boundary of the CDFO regime, extending hence the relevant region of the parameter space significantly beyond 1~TeV. Additionally, blue thin lines represent isolines of constant DM-SM-mediator coupling values (multiplied by $10^6$). It is important to note that non-perturbative effects, including Sommerfeld enhancement and bound state formation, could substantially further enlarge the region of the parameter space relevant for the CDFO regime by modifying the mediator pair-annihilation cross section. While these effects have not been included in the results of figure~\ref{fig:muR-freeze-out}, they are expected to follow the trends observed for quark-philic models~\cite{Garny:2021qsr}. Additionally, the small coupling in the CDFO regime ensures that direct and indirect detection constraints remain negligible. Testing this scenario at colliders, however, is possible via LLP signatures, as discussed in section~\ref{sec:LLP}.

Finally, the freeze-in and superWIMP regimes arise for even smaller couplings, where DM production occurs out of equilibrium. These scenarios, characterised by extremely weak couplings, are not testable via direct or indirect detection experiments. Instead, the long-lived nature of the mediator $Y$ leads to LLP signatures at colliders, as detailed in section~\ref{sec:LLP}. Although the cosmologically viable regions of the parameter space of leptophilic models in the freeze-in/superWIMP regime have not been fully mapped, as for quark-philic models, they are expected to qualitatively resemble the latter case, albeit with different mass scales due to weaker mediator-SM interactions~\cite{Calibbi:2021fld}.

\subsubsection{Non-minimal models}\label{sec:cosmo_nonmin}

\paragraph{Flavoured dark matter}\label{sec:CosmConstNonMin_flav}
\paragraph*{}\vspace{.3cm}

The presence of multiple DM flavours significantly influences early-universe cosmology and shapes the experimental constraints on these models. In flavoured DM frameworks, it is common to consider three generations of DM particles $X_i$ with a mass hierarchy $M_{X_1} > M_{X_2} > M_{X_3}$. The lightest flavour $X_3$ is assumed to be stable and constitutes the observed DM relic density, while the heavier flavours $X_{1,2}$, as well as the mediator $Y$, are unstable and decay into $X_3$ and SM particles. The cosmological evolution of such models depends on the representations of the $X_i$ and $Y$ states and on the structure of the coupling matrix $\lambda$ linking the SM fermions to the dark sector. Furthermore, in the context of the DMFV models introduced in section~\ref{sec:DMFV}, the matrix $\lambda$ not only governs the texture of the interactions but also determines the DM mass spectrum, as the mass matrix $M_{X,ij}$ can be expressed in terms of a spurion expansion in $\lambda^\dagger \lambda$.

\begin{figure}
  \centering
  \includegraphics[width=0.48\textwidth]{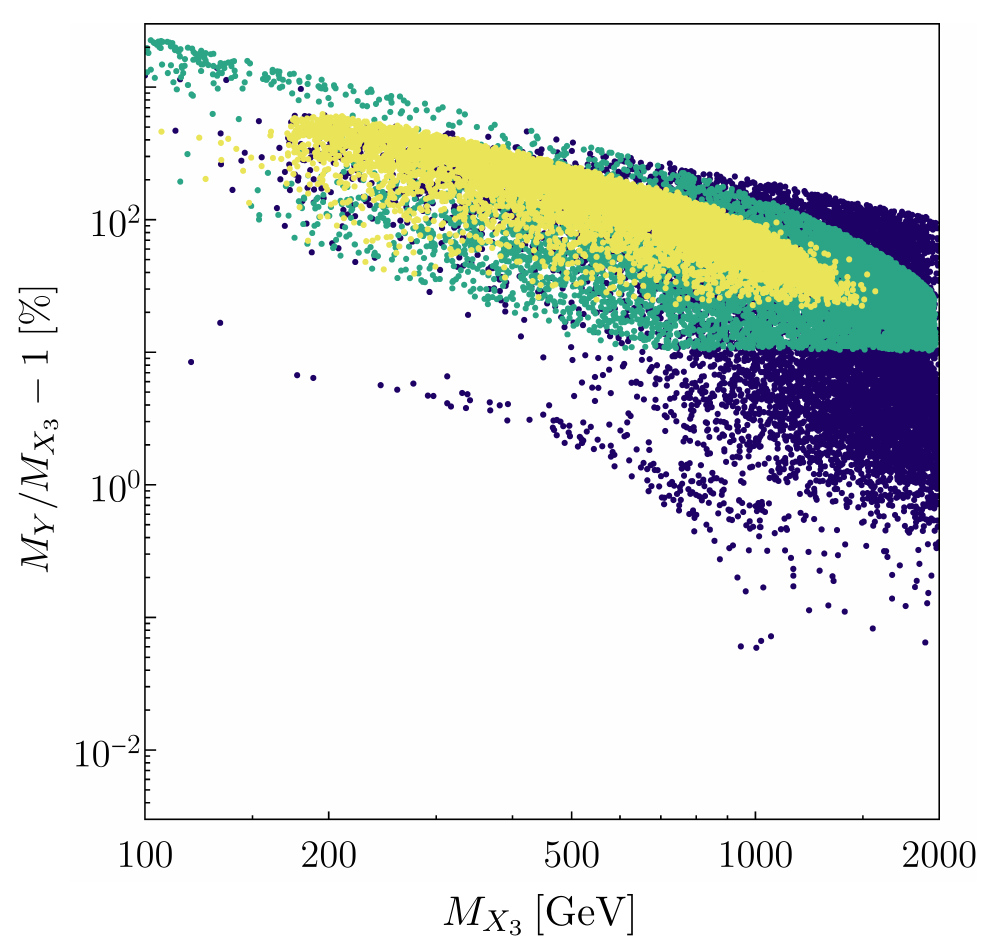}\hfill
  \includegraphics[width=0.48\textwidth]{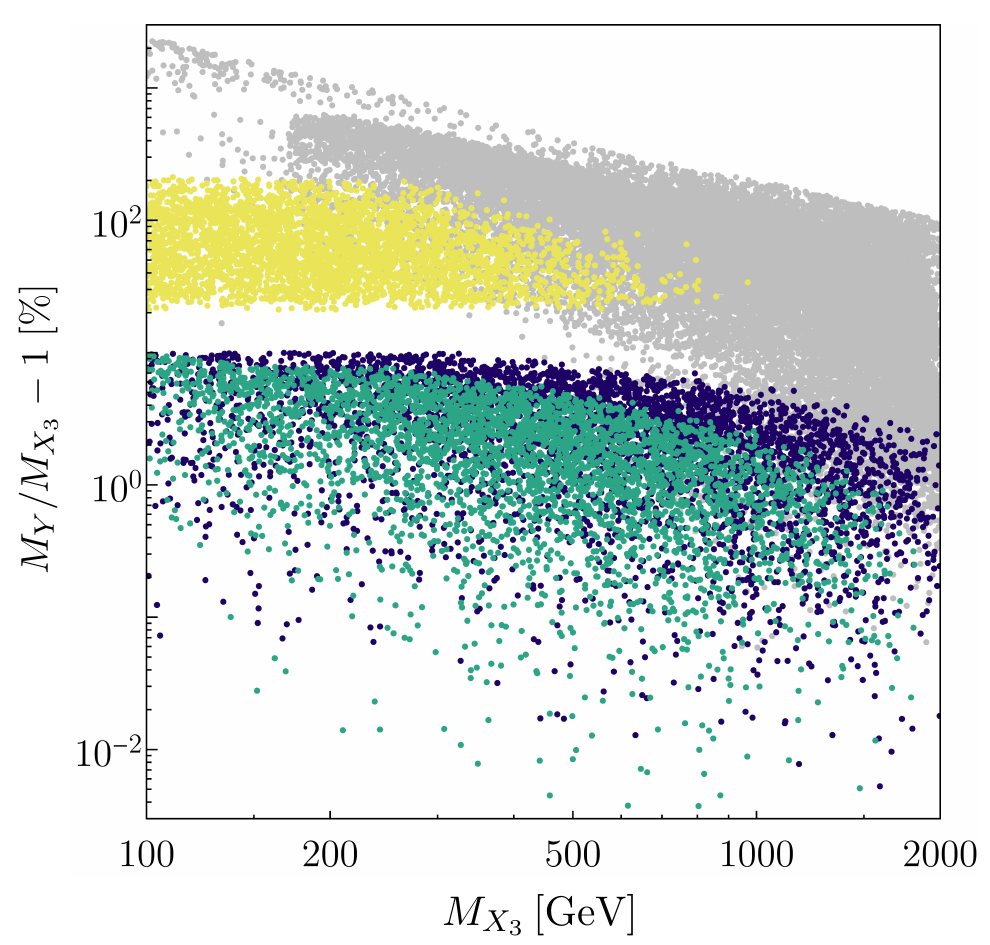}
  \caption{Viable regions of the parameter space of flavoured Majorana DM scenarios~\cite{Acaroglu:2023phy} in which the dark matter couples to right-handed up quarks. Constraints from the observed dark matter relic abundance, direct and indirect detection experiments, and flavour data are shown as a function of the DM mass $m_{X_3}$ and the mass splitting between the dark matter and the mediator $\Delta m_{Y3}$. On the left panel, the \mbox{(co-)annihilation} freeze-out regime is considered, with QDF scenarios shown in green, SFF ones in yellow, and generic ones in blue. On the right panel, the CDFO regime is examined instead, with conversions between $X_3/X_2$ (yellow), $X_3/Y$ (blue), and combined $X_3/X_2/Y$  (green).\label{fig:flavDM-const}}
\end{figure}

Flavoured DM has been explored in the regimes of both canonical freeze-out and conversion-driven freeze-out. Early studies~\cite{Agrawal:2014aoa, Chen:2015jkt, Blanke:2017tnb, Jubb:2017rhm, Blanke:2017fum, Acaroglu:2021qae, Acaroglu:2022hrm} of DMFV models have largely focused on two limiting scenarios of freeze-out. In the first called single-flavour freeze-out (SFF), the flavour $X_3$ is sufficiently separated in mass from the other states $X_{1,2}$ and $Y$ such that co-annihilation effects are negligible. In the second that was named quasi-degenerate freeze-out (QDF), the $X_i$ flavours are all nearly degenerate, and their combined annihilations contribute equally to the freeze-out process. More recent analyses, however, have incorporated general co-annihilation effects into the study of both quark-flavoured and lepton-flavoured DM \cite{Acaroglu:2023phy, Acaroglu:2023cza}. Furthermore, in addition to canonical freeze-out, the case of CDFO has been studied in a flavoured Majorana DM model coupled to right-handed up-type quarks $u_{Ri}$~\cite{Acaroglu:2023phy}. The inclusion of heavier dark flavours $X_{1,2}$ and the additional coupling parameters $\lambda_{ij}$ have been found to relax the constraints on the viable regions of the parameter space imposed by the relic abundance condition in comparison to unflavoured models. Notably, an inverse mass hierarchy within the dark sector  (with the $\eta$ parameter of \eqref{eq:flavDM_eta} being greater than zero), where the lightest state $X_3$ has the weakest coupling to visible matter, offers a compelling scenario. Here, the relic density relies predominantly on the annihilation of the heavier DM flavours and/or the mediator, and this configuration naturally evades direct and indirect detection experiment bounds. The latter indeed probe only the interactions of the $X_3$ state, and DM is thus hidden from these searches.

Figure~\ref{fig:flavDM-const} displays the allowed regions of the parameter space for scenarios featuring flavoured Majorana DM coupled to the right-handed up quarks $u_{Ri}$. On the left panel of the figure, the standard (co-)annihilation freeze-out scenario is examined, while the right panel depicts various CDFO scenarios~\cite{Acaroglu:2023phy}. Accounting for co-annihilation effects (blue points) significantly broadens the size of the viable region of the parameter space compared to the SFF (yellow points) and QDF (green points) benchmark cases. However, small mass splittings between the $X_3$ state and the mediator $Y$ remain excluded in the canonical freeze-out scenario. In the CDFO regime, where semi-efficient annihilation between $X_3$ and $Y$ occurs, this region of the parameter space becomes in contrast viable (green, blue). Moreover, for small splittings between $X_3$ and the heavier flavours, the conversion between these states can also become semi-efficient, enabling additional viable parameter space regions for CDFO scenarios involving the heavier dark flavours (yellow). Intriguingly, flavoured DM models with a CDFO realisation offer the possibility of simultaneously generating the baryon asymmetry of the universe through $CP$-violating $\lambda$ couplings~\cite{Heisig:2024mwr}.

The typical constraints that could be imposed on flavoured DM arise from a combination of direct and indirect detection experiments, electroweak precision measurements, LHC searches, and flavour physics. In the latter case, the most stringent limits typically stem from neutral meson mixing observables for quark-flavoured DM \cite{Agrawal:2014aoa, Blanke:2017tnb, Jubb:2017rhm, Blanke:2017fum, Acaroglu:2021qae} and from radiative decays such as $\ell_i \to \ell_j \gamma$ for lepton-flavoured DM \cite{Chen:2015jkt, Acaroglu:2022hrm, Acaroglu:2023cza}. However, these flavour observables constrain the structure of the coupling matrix $\lambda$ rather than the overall mass or coupling scale. In addition, direct and indirect detection constraints are often relaxed relative to minimal non-flavoured models. This occurs due to the extra parametric freedom provided by the flavour structure of the coupling matrix $\lambda$, which can allow cancellations between tree-level and loop-level contributions, thereby suppressing the DM-nucleon scattering cross section~\cite{Agrawal:2014aoa, Blanke:2017tnb}. Consequently, flavoured DM serves as a concrete realisation of the xenophobic DM paradigm~\cite{Feng:2013vod}. Finally, constraints from indirect detection are found generally weaker than those from direct detection experiments. \vspace{.5cm}

\paragraph{Frustrated dark matter}\label{sec:CosmConstNonMin_frust}
\paragraph*{}\vspace{.3cm}

The fDM framework~\cite{Carpenter:2022lhj} introduced in section~\ref{sec:fDM} describes a family of non-minimal models in which a fermionic dark matter state $X$ couples via a Yukawa-like interaction to a pair of mediators $\{\varphi, \psi\}$. At least one of these mediators interacts with the Standard Model, and to satisfy gauge invariance, they carry the same quantum numbers. While in section~\ref{sec:collider_frustrated} we provided a detailed review of the collider implications the fDM framework, we now focus on its cosmological consequences. Frustrated DM indeed exhibits intriguing astrophysical phenomenology, combining compatibility with the observed relic density and promising prospects for detection via both direct and indirect searches for DM. 

By construction, $2 \to 2$ interactions between the DM and the SM occur only at one-loop order. Consequently, for most DM masses, DM annihilation is dominated by $XX$ scattering to at least four SM particles, which effectively reduce to $t$-channel $XX$ annihilations to a pair of mediators when the mediators can be produced nearly on-shell. For perturbative Yukawa-like DM-mediator couplings $\lambda_X \lesssim \sqrt{4\pi}$, efficient DM annihilation requires these $t$-channel processes to be kinematically accessible; otherwise, DM tends to be overabundant under the assumption of a standard cosmological history. However, the fDM framework allows for a wide range of mediator masses and SM couplings, enabling the construction of numerous specific realisations with large experimentally viable regions in the model parameter space. The regions of the fDM parameter space that can produce a viable thermal relic are, in principle, testable via indirect searches for DM annihilation in the cosmos. While loop-suppressed $2 \to 2$ annihilation processes leading to photons or a $\gamma Z$ final state are possible, the dominant experimental signature of the models is typically an apparent excess in the continuum spectra of gamma or cosmic rays originating from DM annihilation into quarks and/or charged leptons. For fDM scenarios at or below the TeV scale, the strongest constraints thus come from \emph{Fermi}-LAT searches for gamma-ray production in nearby dwarf spheroidal galaxies. On the other hand, direct detection is fully loop-suppressed since there are no tree-level diagrams for fDM scattering off SM particles. Nonetheless, constraints from direct detection experiments such as XENON1T can be quite stringent. In scenarios where mediators carry weak hypercharge, spin-independent DM-nucleon scattering via off-shell photon exchange then provides the most robust constraints, that can also be interpreted as upper limits on the DM magnetic dipole moment~\cite{Arina:2020mxo}. In the absence of such processes, box diagrams involving mediators and SM fields can sometimes generate detectable direct detection signals~\cite{Mohan:2019zrk} that may lead to competitive or even exceed limits from indirect detection. More conclusive quantitative statements require, however, specifying a particular fDM construction.

In section~\ref{sec:collider_frustrated}, we proposed a benchmark scenario that was ideal for LHC exploration. In this scenario, the DM is a Dirac fermion, and the mediators are QCD colour sextets. The sole renormalisable model of this type features a colour-sextet scalar $\varphi$ coupling to pairs of like-sign quarks, while the colour-sextet fermion $\psi$ is sequestered from the SM, interacting only through usual gauge interactions. In the simplest case, this can be achieved by imposing a $\mathbb{Z}_2$ symmetry under which $\psi$ and $X$ are odd while the SM states are even. Moreover, to avoid scenarios with stable colourful fermions, we impose the hierarchy $M_{\psi} > M_X$. Furthermore, experimental constraints from direct searches and measurements of flavour-changing neutral currents in neutral meson transitions~\cite{Babu:2008rq, Babu:2013yca} motivate a scenario in which the scalar mediator preferentially couples to third-generation quarks. Specifically, we thus assume that both quarks to which the mediator couples are of up-type, with dominant $\varphi qq$ couplings to $ut$ and $tt$ systems. Current constraints favour TeV-scale mediators, which in turn suggest TeV-scale DM masses ($M_X \gtrsim M_{\varphi}$) to ensure efficient annihilation.

\begin{figure}
  \centering
  \includegraphics[width=0.7\textwidth]{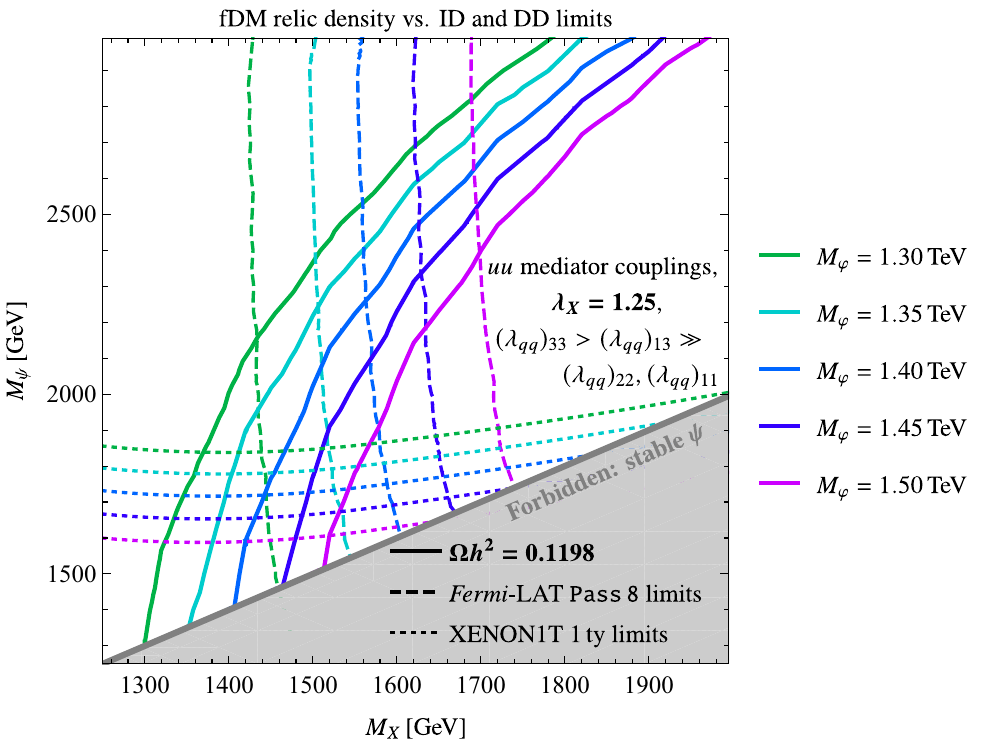}
  \caption{Limits from direct and indirect searches (XENON1T and Fermi-LAT) on a fDM realisation featuring Dirac DM and colour-sextet mediators. Here the scalar is taken to couple to pairs of like-sign up-type quarks, preferentially to the third generation.\label{fig:frusDMIDlim}}
\end{figure}

To illustrate the cosmological constraints that could be imposed on such an fDM setup, figure~\ref{fig:frusDMIDlim} shows contours of $\Omega h^2 = 0.1198$ in the $(M_X, M_{\psi})$ plane. We consider different values of $M_{\varphi}$, and a DM-mediator coupling $\lambda_X = 1.25$ that is relatively large but remains within the perturbative regime. The figure also includes Fermi-LAT \texttt{Pass\,8} (dashed lines) and XENON1T 1 ton-year (dotted lines) constraints. Here, the Fermi-LAT limits act as lower bounds on $M_X$, while the XENON1T limits correspond instead to a lower bound on $M_{\psi}$. In the chosen scenario, the indirect detection limits are derived primarily from $2 \to 4$ annihilation processes as calculated with \lstinline{MadDM}, while the direct detection bounds stem from one-loop off-shell photon-mediated DM-nucleon scattering, an effective field theory analysis, and our own calculation of the DM magnetic dipole moment~\cite{Arina:2020mxo}. In this specific realisation, frustrated DM remains a viable thermal relic across certain ranges of $M_X$ and $M_{\psi}$ values for all the considered mediator masses $M_{\varphi}$. While heavier mediator and DM mass spectrum correspond to regions of the parameter space less constrained by indirect detection, these setups turn out to be complementarily probed, or even excluded, by ongoing direct detection experiments. It is however important to emphasise that these limits do not directly apply to all fDM models. The mass scales, interaction strengths, and experimental constraints can shift depending on the mediator's SM charge assignments or other model specifics. For instance, if the DM is a Majorana fermion instead of a Dirac fermion, the limits from direct detection are significantly altered, as a Majorana fermion does not possess a magnetic (and an electric) dipole moment.\vspace{.5cm}

\paragraph{Composite dark matter}\label{sec:CosmConstNonMin_compo}
\paragraph*{}\vspace{.3cm}

\begin{figure}
    \centering
    \includegraphics[width=0.75\textwidth]{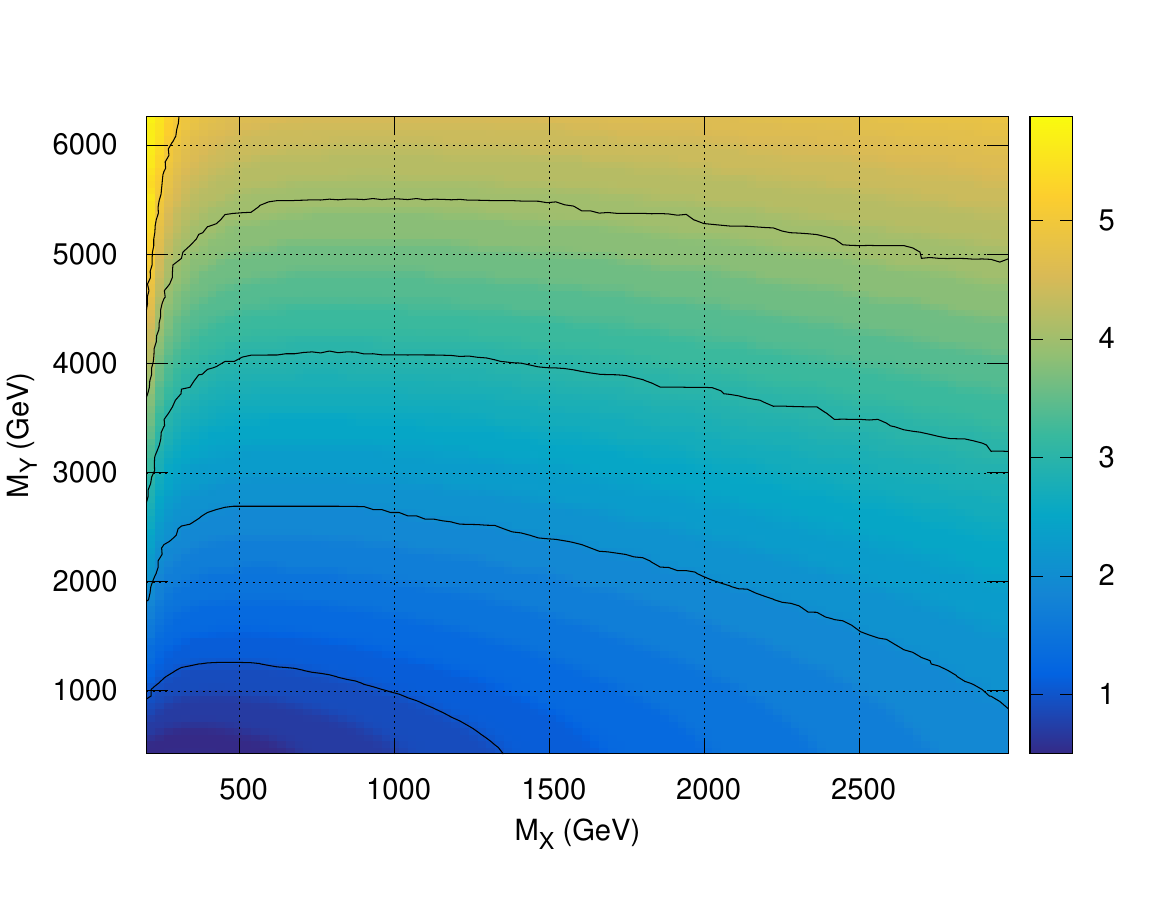}
    \caption{Constraints on the parameter space of a composite $t$-channel DM scenario featuring one DM state $X$ and one mediator state $Y$ that couples to the top quark with a strength $\lambda_t$. Results are given in the $(M_X, M_Y)$ plane, and the value of the coupling $\lambda_t$, given through the colour map, is determined in order to reproduce the observed relic density $\Omega_{CDM}h^{2} = 0.1186$. The black lines correspond to isolines of constant coupling values with $\lambda_{t}=1$, $2$, $3$, $4$ and $5$.}\label{fig:1}
\end{figure}

Another compelling possibility for exploring $t$-channel dark matter models beyond minimal frameworks arises from composite scenarios such as  those introduced in section~\ref{sec:compositeDM}, that lead to a new physics particle spectrum including both even and odd new states, one of the latter playing the role of the DM. In the present section, we consider two distinct realisations of such composite models, and we discuss their implications for dark matter observables in terms of the model parameters. The results shown below highlight an interesting and non-trivial interplay between the couplings and the mass parameters, offering insights into the underlying dynamics of composite models and their implications for DM phenomenology.

In the first case that we consider, the only non-vanishing free parameters of the model comprise a single coupling $\lambda_t$ that appears in the Lagrangian~\eqref{eq:lagtopcompoDM}, the DM mass $M_X$, and the mediator mass $M_Y$. To estimate the DM relic density including NLO effects, we employ \lstinline{MadDM} with the model implementation documented in  section~\ref{sec:compositeDM}. We perform a parameter scan in which $M_X$ ranges between 200 GeV and 3000 GeV, $M_Y$ varies between 200~GeV and 6.5 TeV and $\lambda_t$ varies between $0.1$ and $6$. Compatibility with the latest Planck collaboration results for the relic density, $\Omega_\mathrm{CDM} h^2 = 0.1186 \pm 20\%$, determines the allowed parameter space. The analysis reveals that a large viable region in the parameter space, with an increasing number of possible combinations for $M_X$, $M_Y$, and $\lambda_t$, as visible from the results shown in figure~\ref{fig:1}. In this figure, we display via a colour code the $\lambda_t$ coupling value required to obtained the measured relic density for different mass configurations.

\begin{figure}
  \centering
  \includegraphics[width=0.49\textwidth]{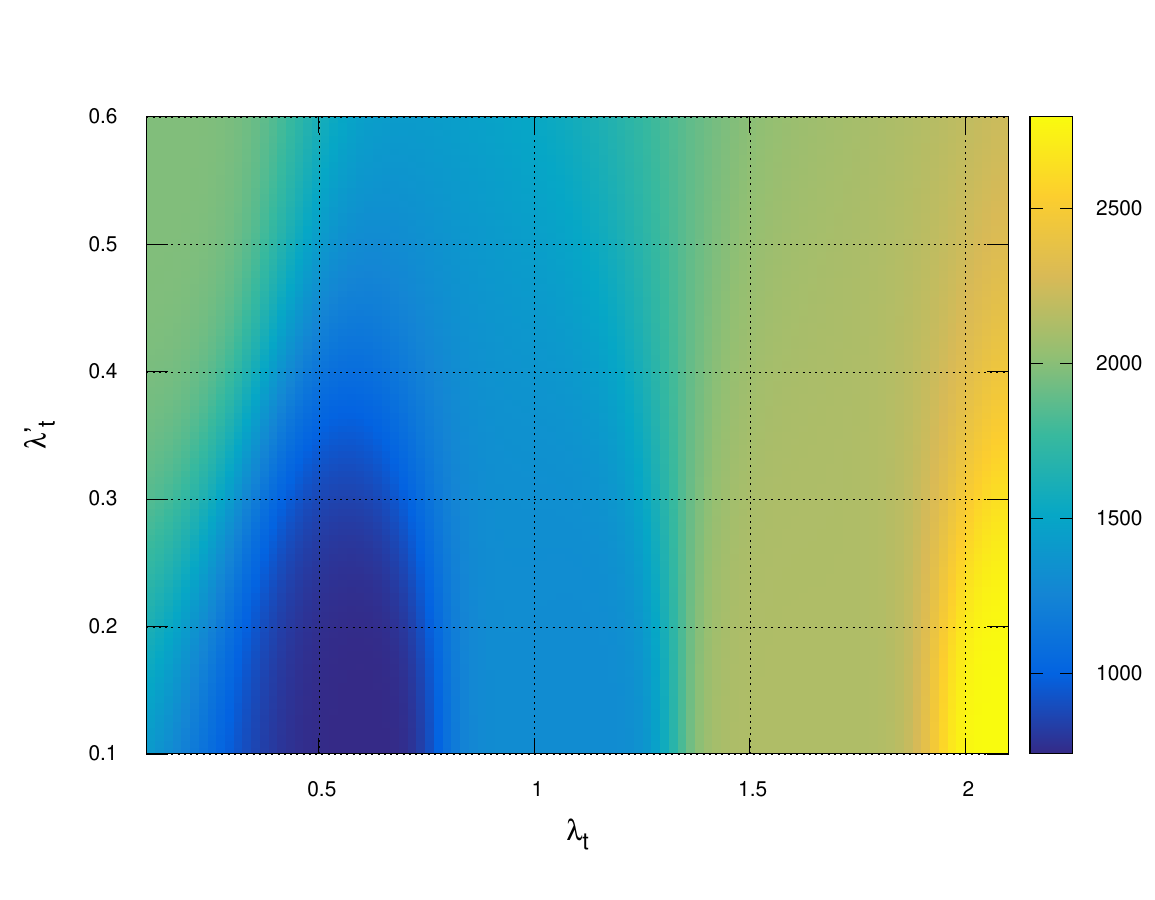}\hfill
  \includegraphics[width=0.49\textwidth]{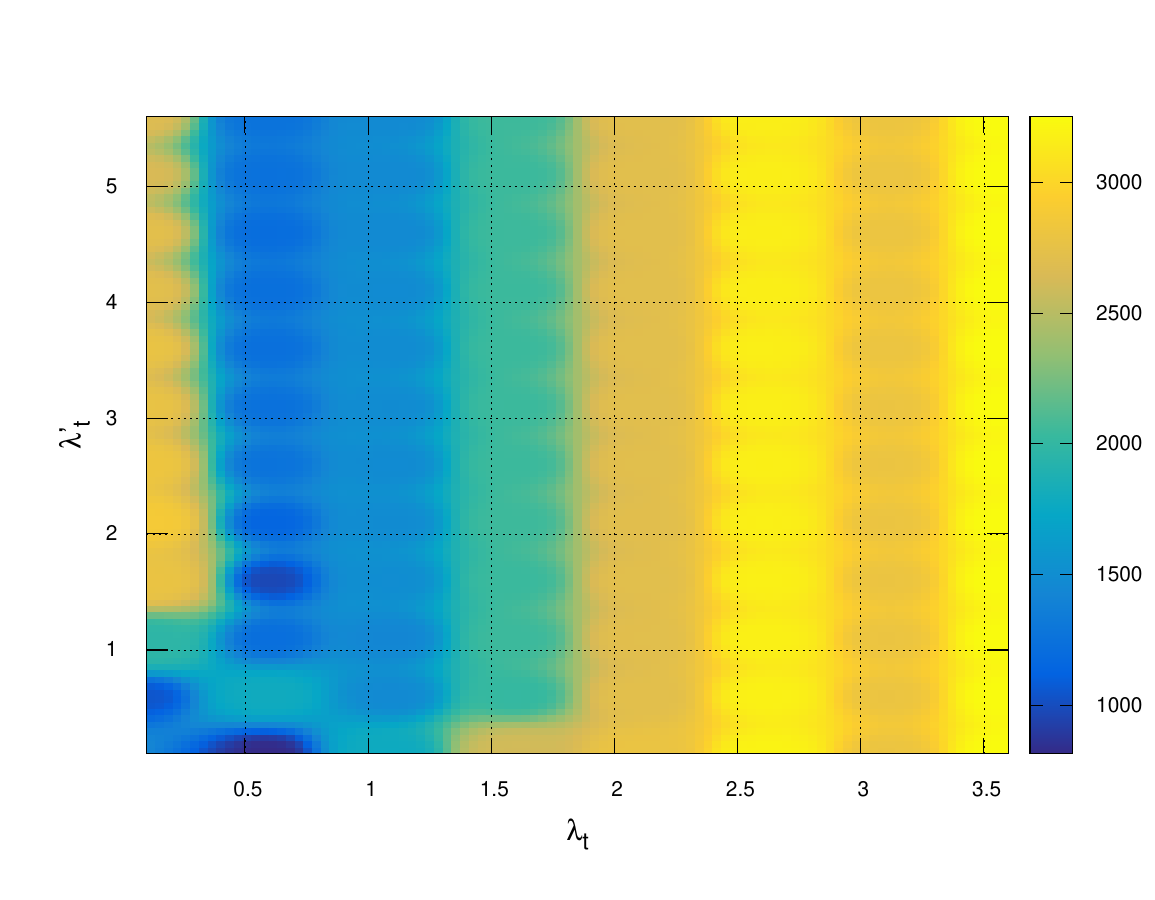}\\
  \includegraphics[width=0.49\textwidth]{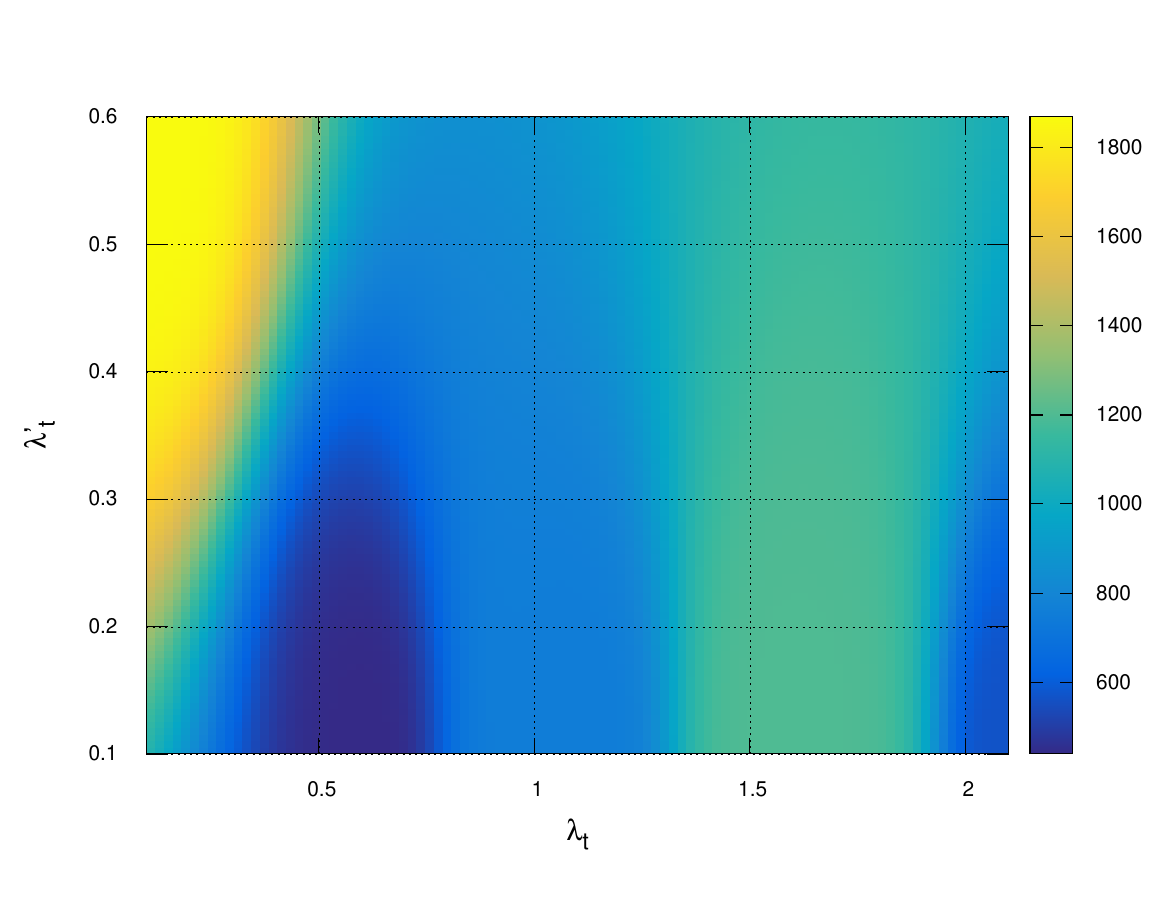}\hfill
  \includegraphics[width=0.49\textwidth]{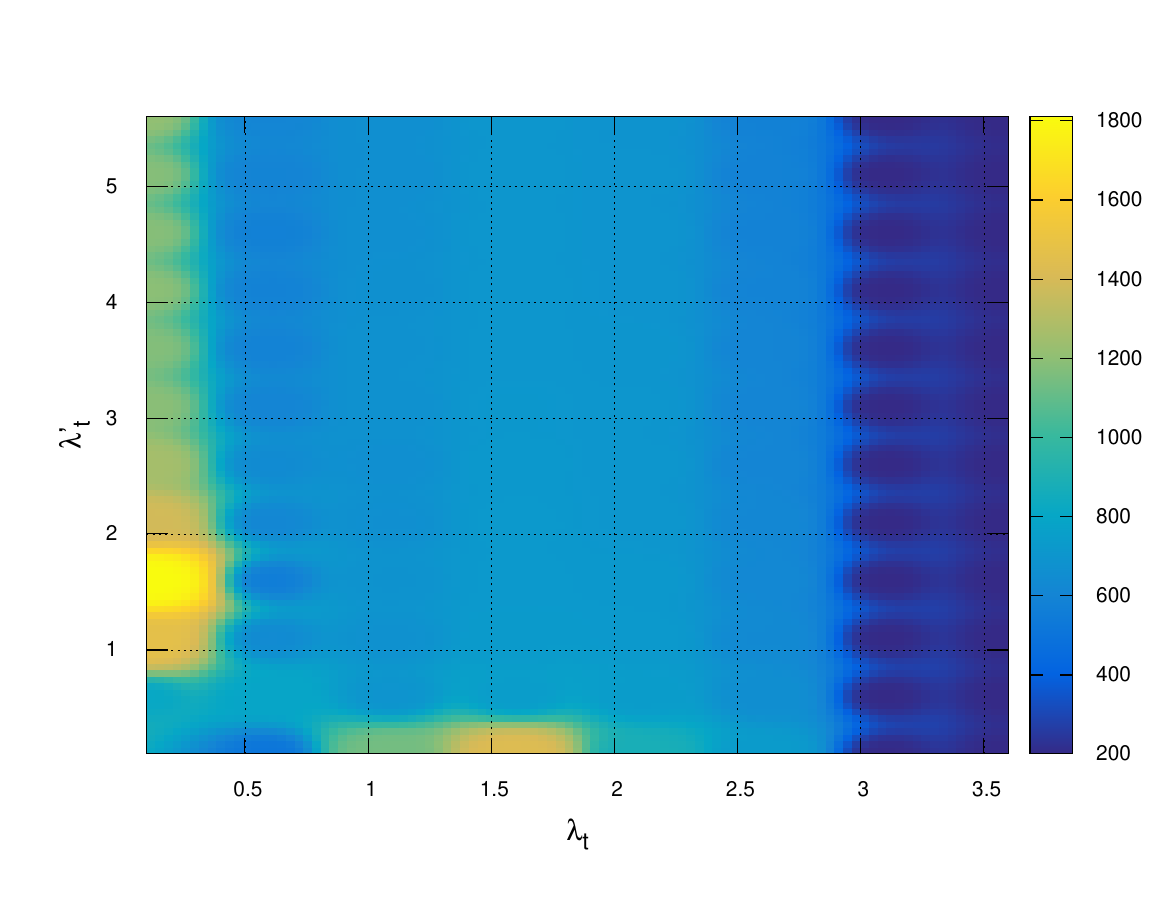}\\
  \includegraphics[width=0.49\textwidth]{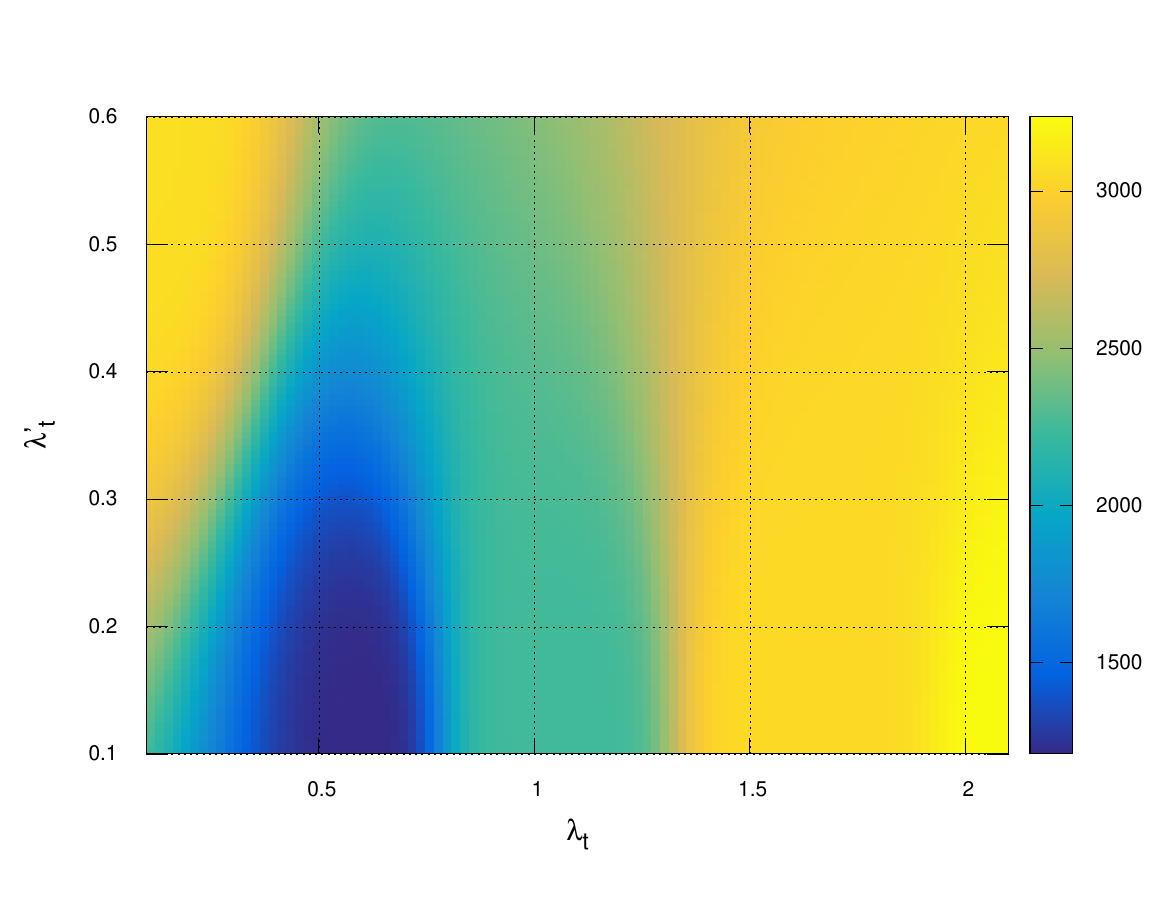}\hfill
  \includegraphics[width=0.49\textwidth]{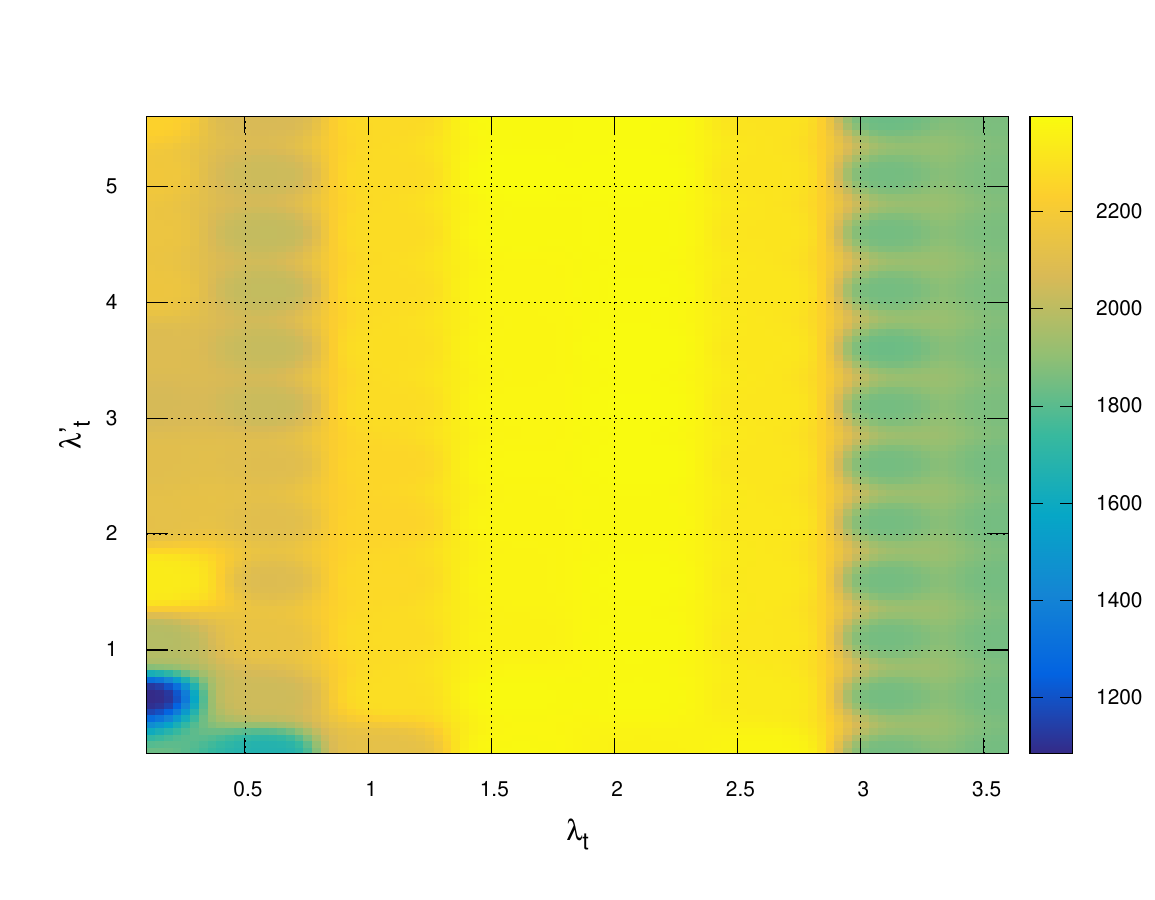}
  \caption{Constraints on the parameter space of a composite $t$-channel DM scenario featuring one DM state $X$, two mediator states $Y$ (that is $\mathbb{Z}_2$ odd) and $Y'$ (that is $\mathbb{Z}_2$-even), and the two new physics couplings $\lambda_t$ and $\lambda'_t$ of the Lagrangian~\eqref{eq:lagtopcompoDM}. Results are given in the $(\lambda_t, \lambda'_t)$ plane, and the value of the masses in GeV, represented through the colour code, are determined so that the relic density matches observations. The top, central and bottom rows respectively address the masses $M_X$, $M_Y$ and $M_{Y'}$, and we consider the mass hierarchies $M_{Y'}<M_{X}<M_{Y}$ (left) and $M_{X}<M_{Y'}$, $M_{Y}$ (right).}\label{fig:2}
\end{figure}

We next focus on a second class of scenarios in which the model is extended to include the two couplings $\lambda_t$ and $\lambda'_t$ of the Lagrangian~\eqref{eq:lagtopcompoDM}, the two mediator masses $M_Y$ and $M_{Y'}$, and the DM mass $M_X$. We perform a parameter space scan in which $M_X$ is varied between 200 GeV and 3000 GeV, $\lambda_t$ and $\lambda'_t$ between $0.1$ and $6$, and $M_Y$ and $M_{Y'}$ between 200 GeV and 3500 GeV. Two specific mass hierarchies are considered, $M_{Y'} < M_X < M_Y$ and $M_X < M_{Y'}, M_Y$, and the results are given, in the left and right columns of figure~\ref{fig:2} respectively, in the coupling plane $(\lambda_t, \lambda'_t)$. The colour code represents the masses of the different states required to reproduce the relic density as observed by the Planck collaboration for any given coupling configuration. When $M_{Y'}<M_{X}<M_{Y}$, $\lambda_t$ can reach values up to $2.1$, while $\lambda'_t$ remains below $0.6$, regardless of the values of $M_X$, $M_Y$, and $M_{Y'}$. Parameter combinations show a concentration of viable scenarios featuring small $\lambda_t$ values (around $0.5$) and $\lambda'_t$ values (between $0.1$ and $0.3$) when $M_X < 1000$ GeV, $M_Y < 1500$ GeV, and $M_{Y'} < 600$ GeV. Spectra featuring larger DM masses ($M_X \sim 2000$ GeV) and mediator masses ($M_Y \sim 3000$ GeV or $M_{Y'} \sim 1800$ GeV) are also viable, but in this case they require combinations of higher $\lambda_t$ values (around $2$) with smaller $\lambda'_t$ values (below $0.4$ or between $0.1$ and $0.6$). On the other hand, when $M_X < M_{Y'}, M_Y$, $\lambda'_t$ can reach larger values than $\lambda_t$, with $\lambda'_t$ going up to $5.5$ and $\lambda_t$ up to $3.6$, independent of the masses $M_X$, $M_Y$, and $M_{Y'}$. For $M_X < 400$ GeV, $\lambda_t$ values cluster around $3$ to $3.5$, while $\lambda'_t$ remains below $5.5$. For mediator masses $M_Y > 2500$ GeV or $M_{Y'} > 2000$ GeV, viable combinations include $\lambda_t > 2$ with $\lambda'_t$ below $5.5$ and lower values of both couplings. For these composite models a fully phenomenological investigation taking into account the relevant dark matter constraints from astroparticle searches is still missing, and thus deserves future studies. \vspace{.5cm}

\paragraph{Non-Abelian dark sector portal} \label{sec:CosmConstNonMin_nonab}
\paragraph*{}\vspace{.3cm}

\begin{figure}
  \centering
  \raisebox{0.30cm}{\includegraphics[width=0.23\textwidth]{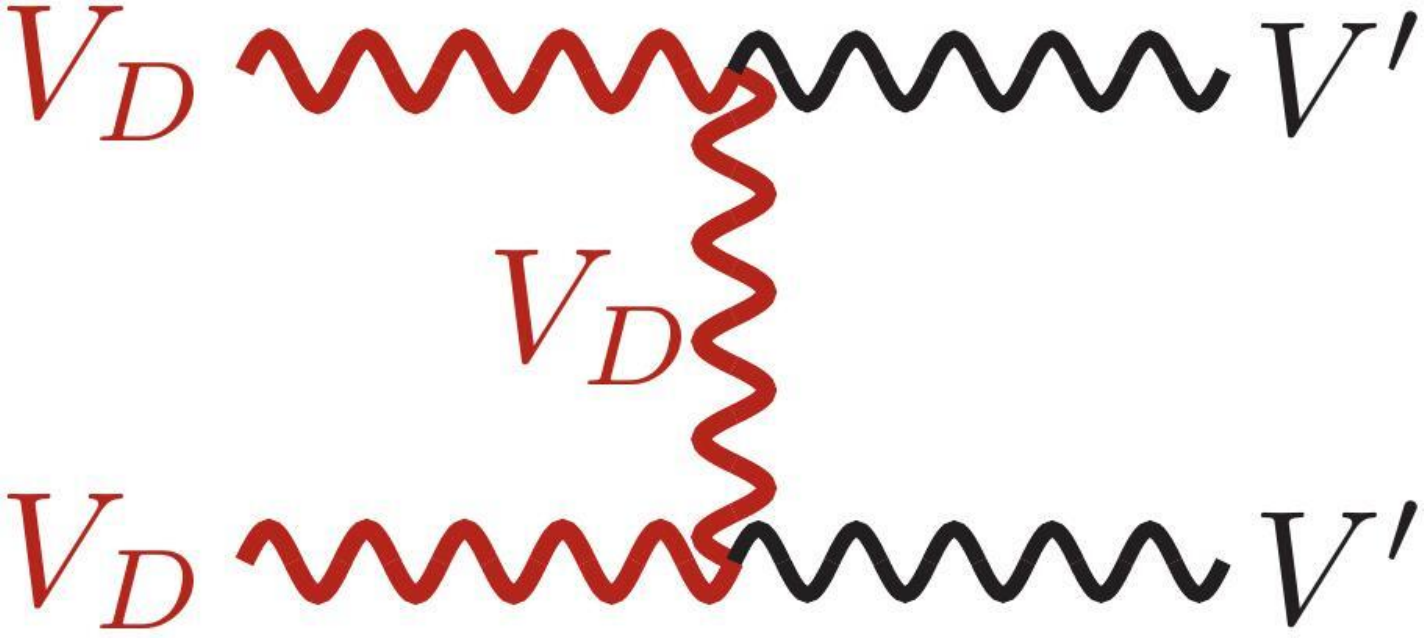}}\hfill
  \includegraphics[width=0.32\textwidth]{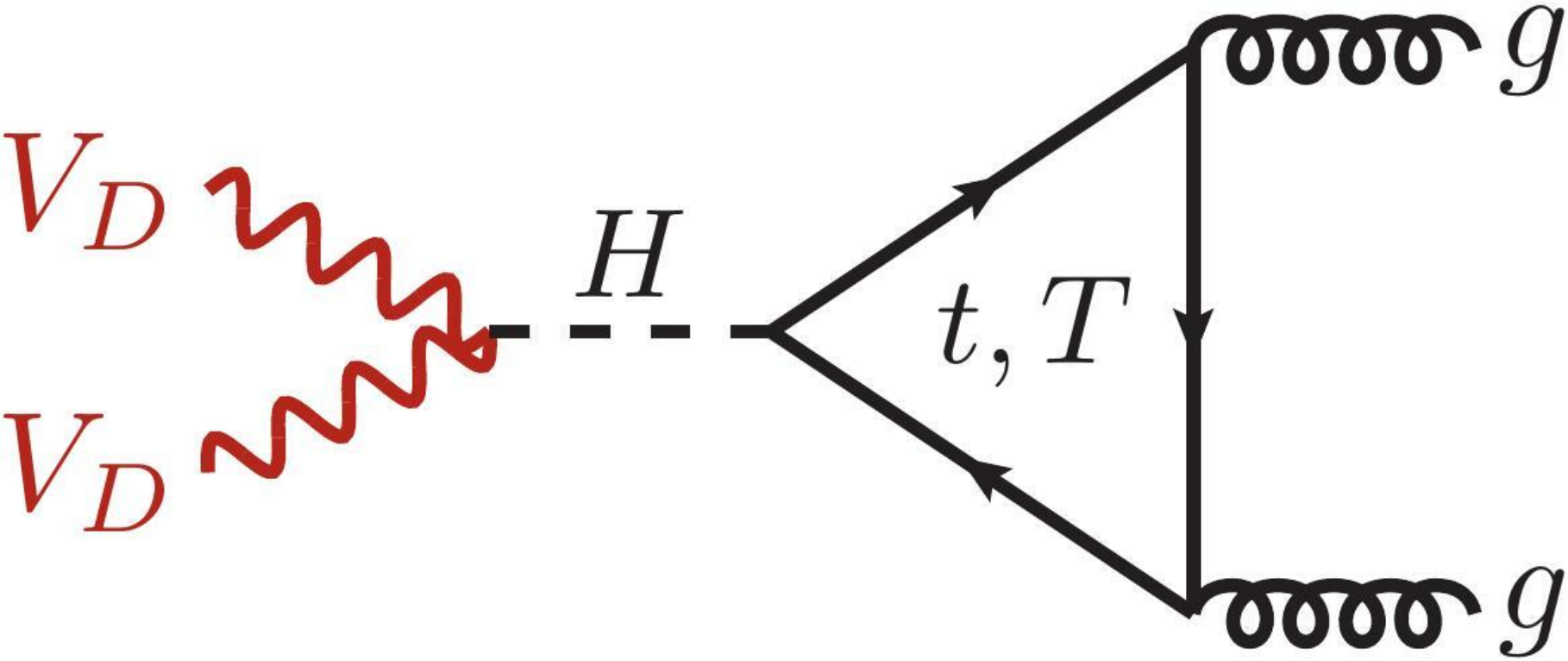}\hfill
  \raisebox{0.32cm}{\includegraphics[width=0.37\textwidth]{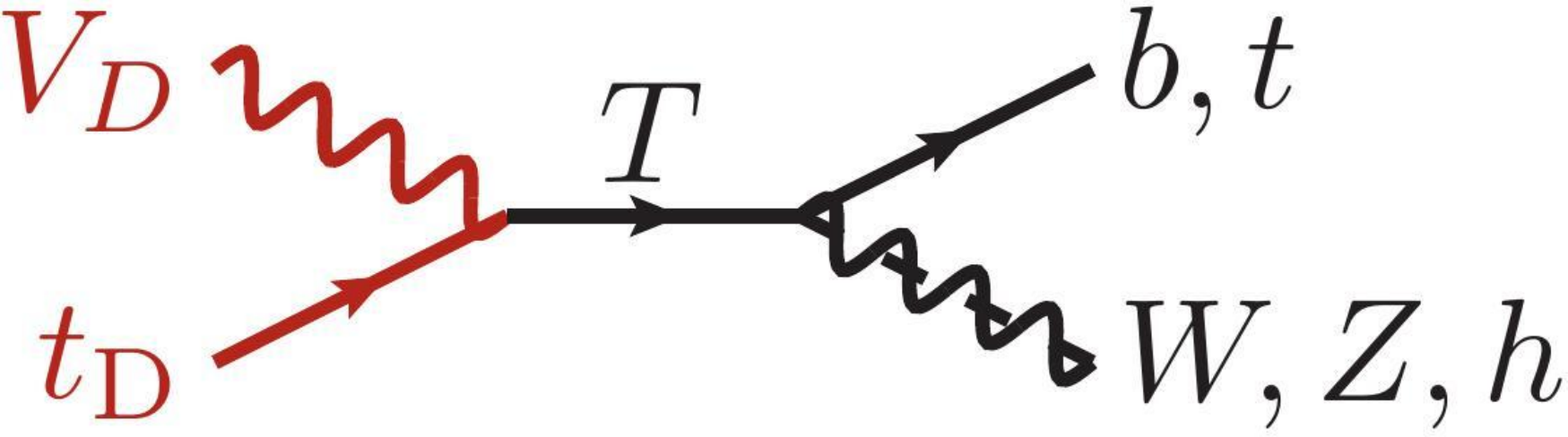}}\\[.6cm]
  \includegraphics[width=0.25\textwidth]{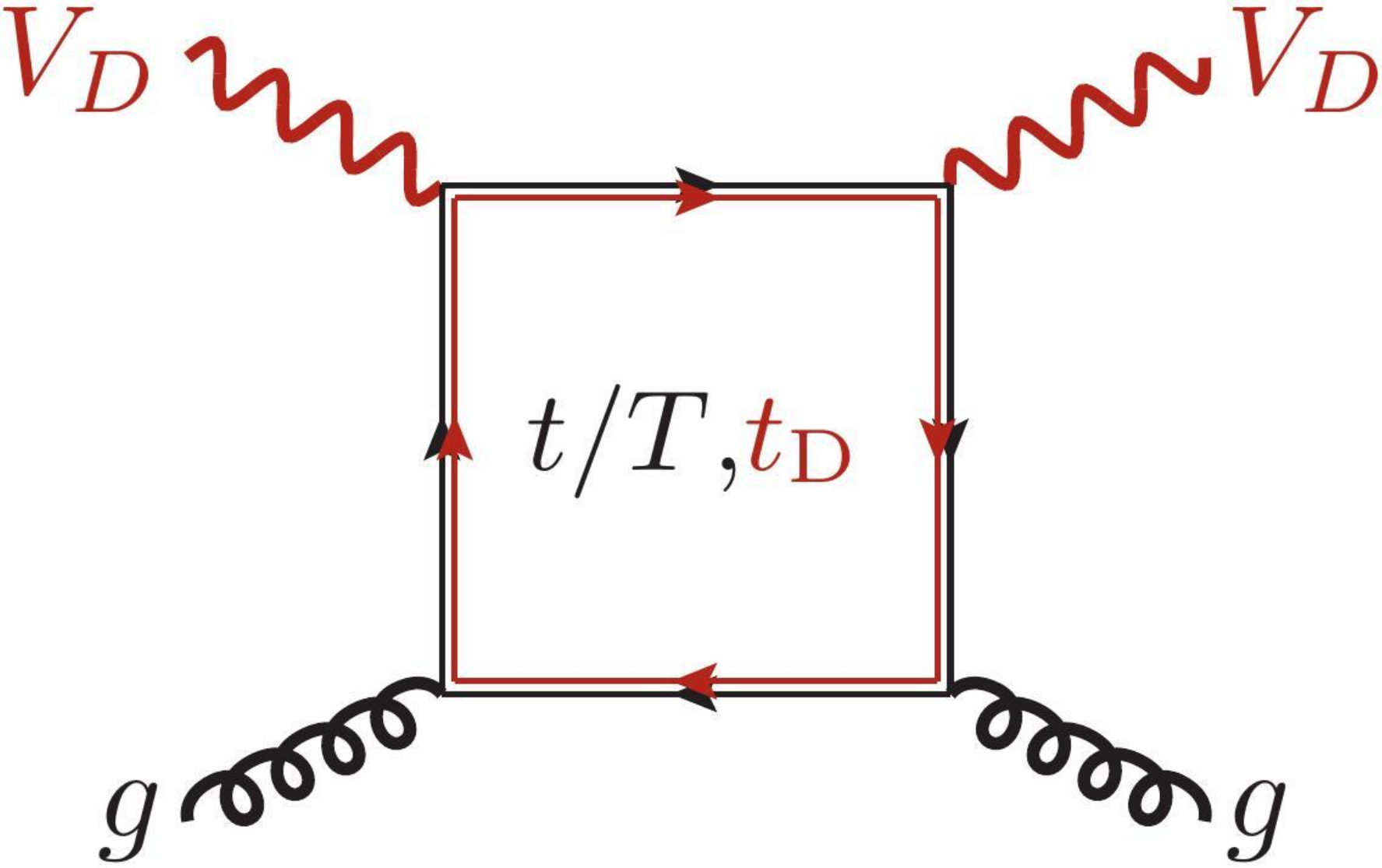}\hspace{.5cm}
  \includegraphics[width=0.25\textwidth]{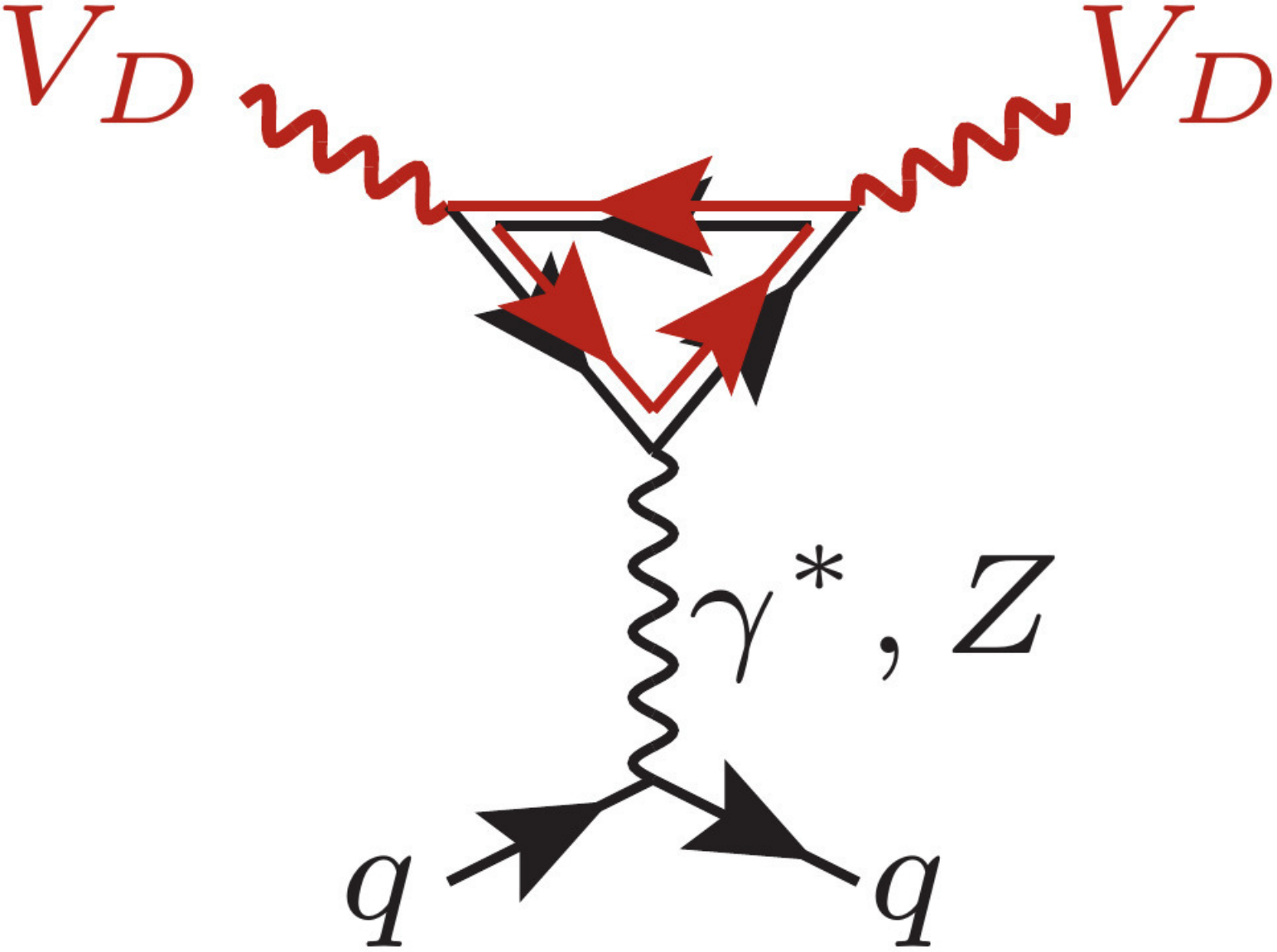}
  \caption{Representative Feynman diagrams for $t$-channel and resonant contributions to DM annihilation and DM-mediator co-annihilation processes (top), and processes relevant for direct detection experiments (bottom).\label{fig:FPVDM_diags_2}}
\end{figure}

In this section, we consider the simplest realisation of the FPVDM model described in section~\ref{sec:nonAbDS}, assuming that new vector-like fermions interact with only one SM flavour that we take to be the top quark, and that no mixing occurs between the SM Higgs boson $h$ and the new scalar $H$. Moreover, the hierarchy in the fermion sector follows $m_t < m_{t_{D}} \leq m_T$, and $H$ can have any mass value allowed by data. We test the above setup against multiple DM observables, importing our implementation within \lstinline{micrOMEGAs}~\cite{Belanger:2020gnr}. The relic density is determined by the interplay of annihilation and co-annihilation processes, some of which being represented in the top row of figure~\ref{fig:FPVDM_diags_2}. Indirect detection constraints are tied to DM annihilation rates during the CMB epoch, excluding regions of parameter space where energy injection into the SM plasma in the early universe is inconsistent with data. In the procedure that we follow to extract bounds on the model, both relic density and indirect detection observable predictions are tested against Planck data~\cite{Planck:2018vyg}. Finally, direct detection constraints are also assessed, and in the considered FPVDM scenario they are associated with processes such as those in the bottom row of figure~\ref{fig:FPVDM_diags_2}. Limits are this time determined by confronting our predictions against the rsults of the XENON1T collaboration~\cite{XENON:2018voc}.

\begin{figure}
  \centering
  \includegraphics[width=0.55\textwidth]{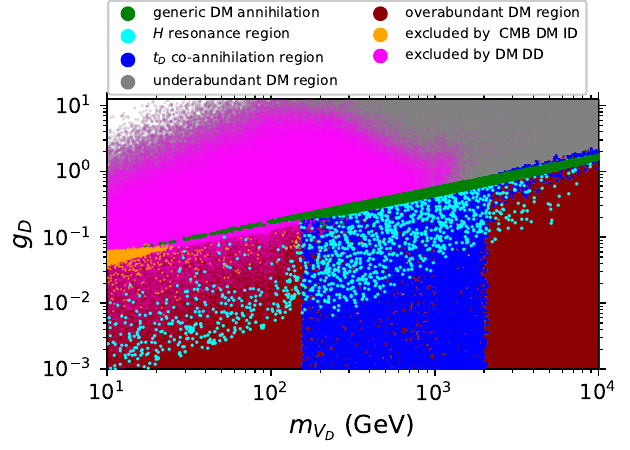}
  \caption{Excluded and allowed parameter space regions of the considered FPVDM realisation, as obtained from a full five-dimensional scan of the parameters in \eqref{eq:fpvdm_prm}. Results are projected in the  $(m_{V_D}, g_D)$ plane and include scenarios compatible with observations by virtue of $t$-channel annihilations (green), $H$-funnel resonant contributions (cyan) and co-annihilations (blue), as well as under-abundant (grey) and overabundant (red) DM setups. Constraints from indirect and direct detection are additionally displayed through the orange and magenta regions, while white areas represent a non-perturbative regime.\label{fig:FPVDM_results}}
\end{figure}

The regions of the parameter space compatible within 5\% with the relic density as measured by the Planck collaboration are shown by the green, cyan, and blue areas in figure~\ref{fig:FPVDM_results}. In this figure, the results of a comprehensive scan of the parameter space of the model are projected in the $(m_{V_D}, g_D)$ plane to highlight their dependence on the dark gauge boson mass and coupling. These regions correspond to a relic density driven by DM annihilations dominated by $t$-channel diagrams, resonantly-enhanced $H$ contributions and DM-$t_D$ co-annihilations, respectively. Generic DM annihilations induced by the $t$-channel diagrams set a lower limit on the dark gauge coupling $g_D$ as a function of $m_{V_D}$, while scenarios exhibiting an $H$-resonant enhancement allow $g_D$ to be reduced by up to two orders of magnitude. Furthermore, the strong DM-$t_D$ co-annihilation channel permits even lower values of $g_D$ for moderately heavy DM. For $m_{V_D} > 2$~TeV, however, the co-annihilation mechanism saturates, while $H$-resonant annihilation requires larger $g_D$ values for increasing DM masses to maintain the observed relic density. As a result, the region with $m_{V_D} \gtrsim 2$~TeV typically corresponds to overabundant DM (indicated by the dark red area in the figure), except for scenarios with large $g_D$ values involving $V_D V_D \to V'V'$ or $H$-funnel annihilations. Finally, regions with $m_{V_D} \lesssim 2$~TeV are additionally partially excluded by direct and/or indirect detection constraints, as indicated by the magenta and orange points, respectively.

\begin{figure}
  \centering
  \includegraphics[width=0.48\textwidth]{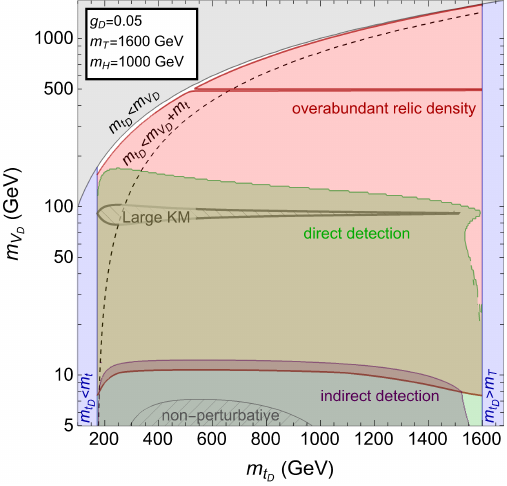}
  \hfill
  \includegraphics[width=0.48\textwidth]{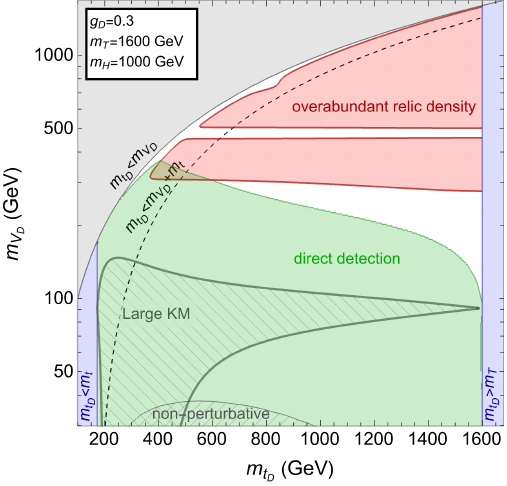}
  \caption{Excluded regions of the considered FPVDM realisation by cosmological observables projected onto the $(m_{t_D}, m_{V_D})$ plane for $m_T = 1600$ GeV, $m_H = 1000$ GeV and $g_D =$ 0.05 (left) or 0.3 (right). Non-perturbative and large kinetic-mixing regions are also shown as hatched areas.
  \label{fig:FPVDM_results2}}
\end{figure}

To facilitate comparison with the LHC bounds discussed in section~\ref{sec:collider_nonabelian}, we present in figure~\ref{fig:FPVDM_results2} the cosmological constraints projected onto the $(m_{t_D}, m_{V_D})$ plane for two benchmark scenarios. Both benchmarks share the parameter values $m_H = 1000$ GeV and $m_T = 1600$ GeV, but differ in the choice of the gauge coupling $g_D$, which is set to 0.05 and 0.3 in the left and right panel of the figure respectively. This projection clearly illustrates how smaller gauge couplings significantly restrict the allowed region of the parameter space, leaving only regions where dark matter annihilation is enhanced by $H$-resonance effects or co-annihilation processes. The correct relic density as observed by the Planck collaboration is achieved at the boundary of the overabundant regions.

Our results demonstrate that the simplest FPVDM realisation that connects a vector DM candidate to SM fermions via a non-Abelian $SU(2)_D$ gauge group without requiring a Higgs portal at tree level, thus involving dark sector interactions with a single SM fermion, has significant potential to explain DM phenomena. Furthermore, alternative realisations involving interactions with other SM fermions could address a large set of observed anomalies. For instance, if the vector-like fermion interacts with the SM leptonic sector, it might contribute to explaining the $(g-2)_\mu$ anomaly~\cite{Muong-2:2021ojo} and open novel opportunities for future $e^+e^-$ colliders~\cite{Aicheler:2012bya, ILC:2013jhg, An:2018dwb, FCC:2018evy}. Moreover, non-minimal scenarios, incorporating scalar sector mixing, extra vector-like partners, or interactions within the same vector-like representation, could expand the scope of FPVDM, offering rich prospects for both phenomenological and experimental studies, as well as insights into the complementarity of collider and non-collider observables.

\newpage\section{Conclusion}\label{sec:conclusion}
In this report that has been designed following work achieved in the context of the LHC Dark Matter Working Group, we have explored the rich phenomenology of $t$-channel dark matter models. These range from minimal simplified scenarios, where the dark sector comprises a single dark matter candidate and one mediator interacting with a specific SM state, to complex and non-minimal constructions inspired by flavoured dark matter, compositeness, frustrated dark matter, and gauged dark sectors. We have examined dark matter production in the early universe, considering canonical freeze-out as well as the freeze-in, superWIMP, and conversion-driven freeze-out mechanisms, and studied the resulting cosmological and astrophysical implications. At the same time, we have investigated collider signatures, distinguishing between scenarios where new particles decay promptly (except for the stable dark matter candidate) and those featuring long-lived particles on collider scales. Our results highlight the theoretical diversity of $t$-channel models and emphasise the intricate interplay between collider, cosmological, and astrophysical studies. In addition, they demonstrate the need for complementary efforts to constrain or validate these theoretical frameworks, particularly in the event of a discovery.

Looking ahead, this whitepaper underscores the importance of a holistic approach to dark matter research, combining theoretical model building, high-precision Monte Carlo simulations, and detailed phenomenological studies leveraging data from both colliders and cosmology. Furthermore, we advocate for continued collaboration between theorists and experimentalists, not only to maximise the potential of existing data but also to develop innovative search strategies capable of probing both minimal and non-minimal models. 

For this purpose, this whitepaper has been designed to provide a comprehensive and up-to-date reference that can serve as a baseline for future theoretical and experimental investigations. Given the wide range of possible signals, we deliberately refrain from proposing specific benchmark scenarios. This choice is intended to avoid biasing future searches toward a limited set of cases while potentially overlooking other relevant possibilities. Instead, we focus on providing model-independent parametrisations that help identify key features of minimal scenarios and enable efficient parametric scans. This strategy has been applied, for instance, to analyse all collider signals emerging from the considered simplified $t$-channel models using a single set of Monte Carlo simulations, with datasets appropriately re-weighted to explore different configurations characterised by varying couplings. Numerical tools for collider and cosmological analyses, including recast efficiencies, cross section tables and simulated samples for individual contributions in these simplified scenarios, are available upon request.  

All scenarios discussed in this work, including non-minimal ones, are presented using a consistent notation wherever possible. This uniform approach aims to facilitate future analyses, enable robust reinterpretations of experimental results, and ultimately pave the way for groundbreaking discoveries. While this report does not propose specific benchmarks, we recommend that practical implementations of these guidelines in concrete experimental analyses include a harmonised approach across LHC experiments when selecting scenarios to report search results. 

\begin{acknowledgements}
We thank Simone Tentori for providing illustrative Feynman diagrams, Marco Drewes for highlighting current and future developments relevant for freeze-in and thermal corrections calculations, and Sam Junius for assisting in adapting the results of \cite{Calibbi:2021fld}. We are also grateful to the CERN IT division for maintaining the computing resources that have enabled us to run the simulations necessary in this report.

AB acknowledges support from the STFC Consolidated Grant ST/ L000296/1, the Leverhulme Trust RPG-2022-057 Grant, as well as from the Royal Society  International Exchanges Grant IES/R1/211138.
AL is supported by FAPESP grants no.\ 2018/25225-9 and 2021/01089-1.
AM is supported in part by the Strategic Research Program High-Energy Physics of the Research Council of the Vrije Universiteit Brussel and by the iBOF `Unlocking the Dark Universe with Gravitational Wave Observations: from Quantum Optics to Quantum Gravity' of the Vlaamse Interuniversitaire Raad. 
BF, MDG and TM have been partly supported by Grant ANR-21-CE31-0013 (project DMwithLLPatLHC) from the French \emph{Agence Nationale de la Recherche}.
CA has been partially supported by the F.R.S.-FNRS under the `Excellence of Science' EOS be.h project no. 30820817.
CP and WN are supported by an SA-CERN excellence bursary.
DK thanks the generous support of the Wolfson Foundation and the Royal Society to allow him to spend his sabbatical year at the University of Glasgow.
DT is supported by the U.S. Department of Energy, Office of Science.
EC, JuHa and MaBe were supported by the Emmy Noether grant `Baryogenesis, Dark Matter and Neutrinos: Comprehensive analyses and accurate methods in particle cosmology' (HA 8555/1-1, Project No. 400234416) funded by the Deutsche Forschungsgemeinschaft (DFG, German Research Foundation) and Cluster of Excellence `Precision Physics, Fundamental Interactions, and Structure of Matter' (PRISMA$^+$ EXC 2118/1) funded by the Deutsche Forschungsgemeinschaft (DFG, German Research Foundation) within the German Excellence Strategy (Project No. 390831469).
JaHe acknowledges support from the Alexander von Humboldt Foundation through the Feodor Lynen Research Fellowship for Experienced Researchers and the Feodor Lynen Return Fellowship during the early stage of this work.
JaHe and MoBl have been supported by the Deutsche Forschungsgemeinschaft (DFG, German Research Foundation) under the grant 396021762 - TRR 257.
JB is supported by the Natural Sciences and Engineering Research Council of Canada (NSERC).
KB is supported by the U.S.\ Department of Energy, Office of Science, Office of High Energy Physics program under Award Number DE-SC0020244.
KM was supported in part by the National Science Foundation under Grant No.~PHY-2310497.
LoCa acknowledges support from the National Natural Science Foundation of China~(NSFC) under the grants No.~12035008 and No.~12211530479.
LoCo and DK wish to thank the CNRS-Africa Residential Research School programme for funding the CHACAL2024 school, where some of the initial ideas for this project were developed.
LLH is supported by the Fonds de la Recherche Scientifique F.R.S.-FNRS through a research associate position, and acknowledges support of the FNRS research grant number J.0134.24, the ARC program of the Federation Wallonie-Bruxelles and the IISN convention No.~4.4503.15.
LMC is supported in part by Grant DE-SC0024179 from the United States Department of Energy (DOE). 
LP's work is supported by by ICSC – Centro Nazionale di Ricerca in High Performance Computing, Big Data and Quantum Computing, funded by the European Union – NextGenerationEU.
MaBe is supported in part by the Italian MIUR Departments of Excellence grant 2023-2027 `Quantum Frontiers' and by Istituto Nazionale di Fisica Nucleare (INFN) through the Theoretical Astroparticle Physics (TAsP) project.
SK is supported in part by the CHIST-ERA project OpenMAPP, grant number ANR-23-CHRO-0006. 
\end{acknowledgements}

\newpage 

\appendix

\section{\hspace{1.2cm} Implementation of the ATLAS-EXOT-2018-06 search in MadAnalysis 
5}\label{app:atlas_exot_2018_06}
As sketched in section~\ref{sec:collider_generalities}, the ATLAS-EXOT-2018-06 analysis~\cite{ATLAS:2021kxv} is sensitive to a signal comprising a not too large number of jets, the leading one being very energetic (with a transverse momentum larger than 150~GeV), no leptons (electrons, muons, and taus) or photons, and a significant amount of missing transverse energy well separated from the jet activity. It exploits 139 fb$^{-1}$ of LHC collision data at a centre-of-mass energy of 13 TeV, recorded during the period 2015-2018 by the ATLAS collaboration, and hence updates previous analyses conducted with 3.2 fb$^{-1}$~\cite{ATLAS:2016bek} and 36.1 fb$^{-1}$~\cite{ATLAS:2017bfj} of data using less sophisticated signal selections. It is thus relevant for probing the $t$-channel models that we explore in this whitepaper. In this section, we report on the validation of its implementation in the \lstinline{MadAnalysis 5} framework, facilitated by the substantial additional data made available via \lstinline{HepData}~\cite{hepdata.102093} by the ATLAS collaboration. This includes detailed cut-flow tables and exclusion curves for given benchmark scenarios, as well as digitised information on the figures.\footnote{See the webpage \url{https://www.hepdata.net/record/ins1847779}.}

\subsection{Description of the analysis}
The signal topology exploits jets and missing energy while vetoing leptons. Jets are reconstructed by clustering particles with the anti-$k_T$ jet algorithm~\cite{Cacciari:2008gp} with a radius parameter of $R=0.4$. Only jets with transverse momentum $p_T > 20$ GeV and pseudo-rapidity $|\eta|<2.8$ are considered. Moreover, the \lstinline{MV2} $b$-tagging algorithm~\cite{ATLAS:2016gsw} is used to identify $b$-jets, defined as jets with $p_T>30$ GeV and $|\eta|<2.5$ originating from $b$-quark fragmentation with an average efficiency of 60\%. Electron candidates must satisfy $p_T > 7$ GeV, $|\eta| <2.47$, and `Loose' track selection criteria~\cite{ATLAS:2019qmc}, which requires their longitudinal impact parameter to be less than $0.5$ mm. Overlaps between identified electrons and jets with $p_T > 30$ GeV are resolved by discarding non-$b$-tagged jets within $\Delta R < 0.2$ of any identified electron, and by removing electrons within $\Delta R < 0.4$ of any remaining jets. Muon candidates must pass a `Medium' identification selection~\cite{ATLAS:2016lqx}, and have $p_T > 7$ GeV, $|\eta| < 2.5$, and a longitudinal impact parameter smaller than $0.5$~mm. Jets with $p_T > 30$ GeV and fewer than three associated tracks with $p_T > 500$~MeV are then discarded if they are within $\Delta R < 0.4$ of an identified muon. Hadronically-decaying tau leptons are reconstructed from jets with $p_T > 10$~GeV and $|\eta|<2.5$, and they must satisfy `Loose' identification requirements, have a transverse momentum larger than 20 GeV after energy-scale corrections, and be associated with either one or three charged tracks~\cite{ATLAS:2014rzk, ATLAS:2015xbi}. Tau candidates within $\Delta R < 0.2$ of an electron or muon are removed, as are any jets within $\Delta R = 0.2$ of a reconstructed tau-lepton. Finally, the missing transverse momentum $\ptmiss$ is reconstructed from the negative vector sum of the transverse momenta of all reconstructed objects with $p_T > 20$ GeV and $|\eta| < 4.5$.

Event preselection requires a significant amount of missing energy, $\met > 200$ GeV, and an energetic leading jet with $p_T > 150$ GeV and $|\eta| < 2.4$. Moreover, up to three additional jets with $p_T > 30$ GeV and $|\eta| < 2.8$ are allowed. Additionally, the missing transverse momentum $\ptmiss$ must be well separated from the four leading jets by an angle of $\Delta \phi(j, \ptmiss) > 0.4$ for events with $\met > 250$ GeV and $\Delta \phi(j, \ptmiss) > 0.6$ for events with $\met \leq 250$ GeV, such a separation criterion helping to reduce the multijet background contributions.

\begin{table}
  \centering\renewcommand{\arraystretch}{1.3}\setlength{\tabcolsep}{12pt}
    \begin{tabular}{c | c c }
      Cuts & Exclusive SRs & Inclusive SRs\\ \hline
      \multirow{6}{*}{Preselection} & \multicolumn{2}{c}{$\met > 150$~GeV} \\
      & \multicolumn{2}{c}{Lepton veto}\\
      & \multicolumn{2}{c}{$N(j)\in [1, 4]$}\\
      & \multicolumn{2}{c}{$\Delta \Phi(j_i, \ptmiss) >0.4\ \ (\forall i)$}\\
      & \multicolumn{2}{c}{$p_T(j_1) > 150$~GeV}\\
      & \multicolumn{2}{c}{$\met > 200$~GeV} \\
      \hline
      Bin~0 & \texttt{EM0}: $200$~GeV $< \met < 250$~GeV & \texttt{IM0}: $\met > 200$~GeV\\
      Bin~1 & \texttt{EM1}: $250$~GeV $< \met < 300$~GeV & \texttt{IM1}: $\met > 250$~GeV\\
      Bin~2 & \texttt{EM2}: $300$~GeV $< \met < 350$~GeV & \texttt{IM2}: $\met > 300$~GeV\\
      Bin~3 & \texttt{EM3}: $350$~GeV $< \met < 400$~GeV & \texttt{IM3}: $\met > 350$~GeV\\
      Bin~4 & \texttt{EM4}: $400$~GeV $< \met < 500$~GeV & \texttt{IM4}: $\met > 400$~GeV\\
      Bin~5 & \texttt{EM5}: $500$~GeV $< \met < 600$~GeV & \texttt{IM5}: $\met > 500$~GeV\\
      Bin~6 & \texttt{EM6}: $600$~GeV $< \met < 700$~GeV & \texttt{IM6}: $\met > 600$~GeV\\
      Bin~7 & \texttt{EM7}: $700$~GeV $< \met < 800$~GeV & \texttt{IM7}: $\met > 700$~GeV\\
      Bin~8 & \texttt{EM8}: $800$~GeV $< \met < 900$~GeV & \texttt{IM8}: $\met > 800$~GeV\\
      Bin~9 & \texttt{EM9}: $900$~GeV $< \met < 1000$~GeV & \texttt{IM9}: $\met > 900$~GeV\\
      Bin~10 & \texttt{EM10}: $1000$~GeV $< \met < 1100$~GeV & \texttt{IM10}: $\met > 1000$~GeV\\
      Bin~11 & \texttt{EM11}: $1100$~GeV $< \met < 1200$~GeV & \texttt{IM11}: $\met > 1100$~GeV\\
      Bin~12 & \texttt{EM12}: $\met > 1200$~GeV & \texttt{IM12}: $\met > 1200$~GeV\\
    \end{tabular}
  \caption{Selection cuts defining the different signal regions of the ATLAS-EXOT-2018-06 analysis~\cite{ATLAS:2021kxv}.}
  \label{tab:atlas_exot_met_selection}
\end{table}

Next, the ATLAS collaboration implements a twofold analysis strategy, including signal regions (SRs) with either an inclusive or an exclusive selection on the missing transverse energy. Inclusive $\met$ signal regions (denoted by names starting with \texttt{IM}) are used for model-independent interpretations of the search results, while exclusive signal regions (denoted by names starting with \texttt{EM}) are used for model-dependent interpretations. In the first series of 13 signal regions (\texttt{EM0}, \texttt{EM1}, $\ldots$, \texttt{EM12}), the analysis considers an exclusive missing transverse energy selection defined by $E_\mathrm{min} < \met < E_\mathrm{max}$. The different thresholds range from 200 GeV to 1200 GeV, as shown in Table~\ref{tab:atlas_exot_met_selection} which also includes all preselection cuts and an extra cut on the missing transverse momentum ($\met > 150$~GeV) allowing us to be consistent with generator-level cuts biasing event generation to the phase space region of interest (as implemented in the ATLAS simulations). In the second set of 13 signal regions (\texttt{IM0}, \texttt{IM1}, $\ldots$, \texttt{IM12}), the analysis instead considers an inclusive missing transverse energy selection defined by $\met > E_\mathrm{threshold}$, with the different thresholds again ranging from 200 GeV to 1200 GeV, as given in Table~\ref{tab:atlas_exot_met_selection}.

\subsection{Validation of the implementation}

The validation of our implementation of the ATLAS-EXOT-2018-06 search in \lstinline{MadAnalysis 5} has been achieved by focusing on top squark production and decay in the $R$-parity-conserving Minimal Supersymmetric Standard Model (MSSM). We have considered two different final states, corresponding to the processes
\begin{equation}\label{eq:atlas_exot_signals}
 p p \to \tilde{t}_1 \tilde{t}_1^* \to (c \tilde{\chi}^{0}_{1}) (\bar{c} \tilde{\chi}^{0}_{1})\qquad\text{and}\qquad
 p p \to \tilde{t}_1 \tilde{t}_1^* \to ( b f\bar{f}^\prime \tilde{\chi}^{0}_{1}) (\bar{b} f\bar{f}^\prime \tilde{\chi}^{0}_{1})\,,
\end{equation}
and we have computed limits and cut-flows for different choices of the stop and neutralino masses, with all other superpartners being decoupled.

Hard-scattering signal event generation has been achieved with \lstinline{MadGraph5_aMC@NLO}~\cite{Alwall:2014hca} (version 3.4.2), while the simulation of supersymmetric particle decays, parton showering, and hadronisation has been performed with \lstinline{Pythia}~\cite{Sjostrand:2014zea} (version 8.2). This event generation procedure relies on the MSSM implementation in \lstinline{MadGraph5_aMC@NLO} described in \cite{Duhr:2011se}, which makes use of \lstinline{FeynRules}~\cite{Christensen:2009jx, Alloul:2013bka} and its UFO interface~\cite{Degrande:2011ua, Darme:2023jdn}. Moreover, we have merged event samples including up to two additional hard partons in the final state, following the MLM prescription as implemented in \lstinline{MadGraph5_aMC@NLO}~\cite{Mangano:2006rw, Alwall:2008qv} with a merging scale set to one quarter of the stop mass. To match the statistics of the reference cut-flows provided by the ATLAS collaboration on \lstinline{HepData}, we simulated 100,000 events before merging, which resulted in samples of about 90,000 merged events.

\begin{table}
  \centering\renewcommand{\arraystretch}{1.3}\setlength{\tabcolsep}{12pt}
  \begin{tabular}{l|cc|cc|c}
    \multirow{2}{*}{Cut} & \multicolumn{2}{c|}{ATLAS} & \multicolumn{2}{c|}{\lstinline{MadAnalysis 5}} & \multirow{2}{*}{$R [\%]$}\\
      & $N_\mathrm{events}^\mathrm{ATLAS}$ & $\varepsilon_\mathrm{ATLAS} [\%]$ & $N_\mathrm{events}^\mathrm{MA5}$ & $\varepsilon_\mathrm{MA5} [\%]$ & \\ \hline
      $\met > 150$~GeV & $39598$ & $100$ & $89529$ & $100$ & $-$\\
      Lepton veto & $37547$ & $94.82$ & $85417$ & $95.41$ & $0.62$\\
      $N(j)\in [1, 4]$ & $35412$ & $89.43$ & $76195$ & $85.11$ & $4.38$\\
      $\Delta \Phi(j_i, \ptmiss) >0.4\ \ (\forall i)$ & $33319$ & $84.14$ & $69253$ & $77.35$ & $8.07$\\
      $p_T(j_1) > 150$~GeV & $23134$ & $58.42$ & $47157$ & $52.67$ & $9.84$\\
      $\met > 200$~GeV & $18801$ & $47.48$ & $39183$ & $43.77$ & $7.81$\\
      \hline
      \texttt{EM0} & $4488$ & $11.34$ & $8509$ & $9.50$ & $16.23$\\
      \texttt{EM1} & $3789$ & $9.57$  & $7946$ & $8.88$ & $7.21$\\
      \texttt{EM2} & $2857$ & $7.21$  & $6226$ & $6.95$ & $3.61$\\
      \texttt{EM3} & $2111$ & $5.33$  & $4621$ & $5.16$ & $3.19$\\
      \texttt{EM4} & $2618$ & $6.61$  & $5847$ & $6.53$ & $1.21$\\
      \texttt{EM5} & $1352$ & $3.41$  & $2895$ & $3.23$ & $5.28$\\
      \texttt{EM6} & $712$  & $1.80$  & $1501$ & $1.67$ & $7.22$\\
      \texttt{EM7} & $393$  & $0.99$  & $719$  & $0.80$ & $19.19$\\
      \texttt{EM8} & $204$  & $0.52$  & $408$  & $0.46$ & $11.54$\\
      \texttt{EM9} & $122$  & $0.31$  & $207$  & $0.23$ & $25.80$\\
      \texttt{EM10}& $58$   & $0.15$  & $124$  & $0.14$ & $6.67$\\
      \texttt{EM11}& $42$   & $0.11$  & $77$   & $0.09$ & $18.18$\\
      \texttt{EM12}& $55$   & $0.14$  & $103$  & $0.11$ & $21.43$\\
    \end{tabular}
    \caption{Cut-flow associated with the ATLAS-EXOT-2018-06 analysis, for the signal emerging from the process $p p \to \tilde{t}_1 \tilde{t}_1^* \to ( b f\bar{f}^\prime \tilde{\chi}^{0}_{1}) (\bar{b} f\bar{f}^\prime \tilde{\chi}^{0}_{1})$ and a spectrum with $(m_{\tilde{t}_1}, m_{\tilde{\chi}^{0}_{1}}) = (450, 443)$~GeV. We compare our predictions for the the number of generated events surviving each cut and with the associated efficiencies with the information provided by ATLAS on \lstinline{HepData}.}
    \label{tab:atlas_cutflow}
\end{table}

The information available on \lstinline{HepData} provides cut-flow information for several benchmark scenarios. We begin our validation by focusing on the second process of eq.~\eqref{eq:atlas_exot_signals}, considering a scenario with a compressed spectrum featuring a stop mass of $m_{\tilde{t}_1} = 450$~GeV and a neutralino mass of $m_{\tilde{\chi}^{0}_{1}} = 443$~GeV. We have generated events as introduced above, and then analysed the produced sample by applying all the analysis preselection cuts and assessing how many signal events ($N_\mathrm{events}$) would populate the different exclusive bins in missing transverse energy. Our results are displayed, for each individual cut, in table~\ref{tab:atlas_cutflow}, that reports the number of events surviving each cut and the associated cumulative efficiency $\varepsilon$. We include both the predictions provided by ATLAS ($N_\mathrm{events}^\mathrm{ATLAS}$ and $\varepsilon_\mathrm{ATLAS}$) and those predicted using our implementation in \lstinline{MadAnalysis 5} ($N_\mathrm{events}^\mathrm{MA5}$ and $\varepsilon_\mathrm{MA5}$), along with the relative difference $R$ between the two,
\begin{equation}
  R = \left| \frac{\varepsilon_\mathrm{ATLAS} - \varepsilon_\mathrm{MA5}}{\varepsilon_\mathrm{ATLAS}} \right|\,.
\end{equation}
We observe a good agreement between our predictions and the ATLAS ones, with $R$ values ranging from 1\% to 25\%. The largest discrepancies are associated with signal regions in which the number of Monte Carlo events populating the bin is small, especially for the ATLAS predictions, indicating that a large numerical uncertainty must be accounted in the comparison.

\begin{figure}
  \centering
  \includegraphics[width=.48\textwidth]{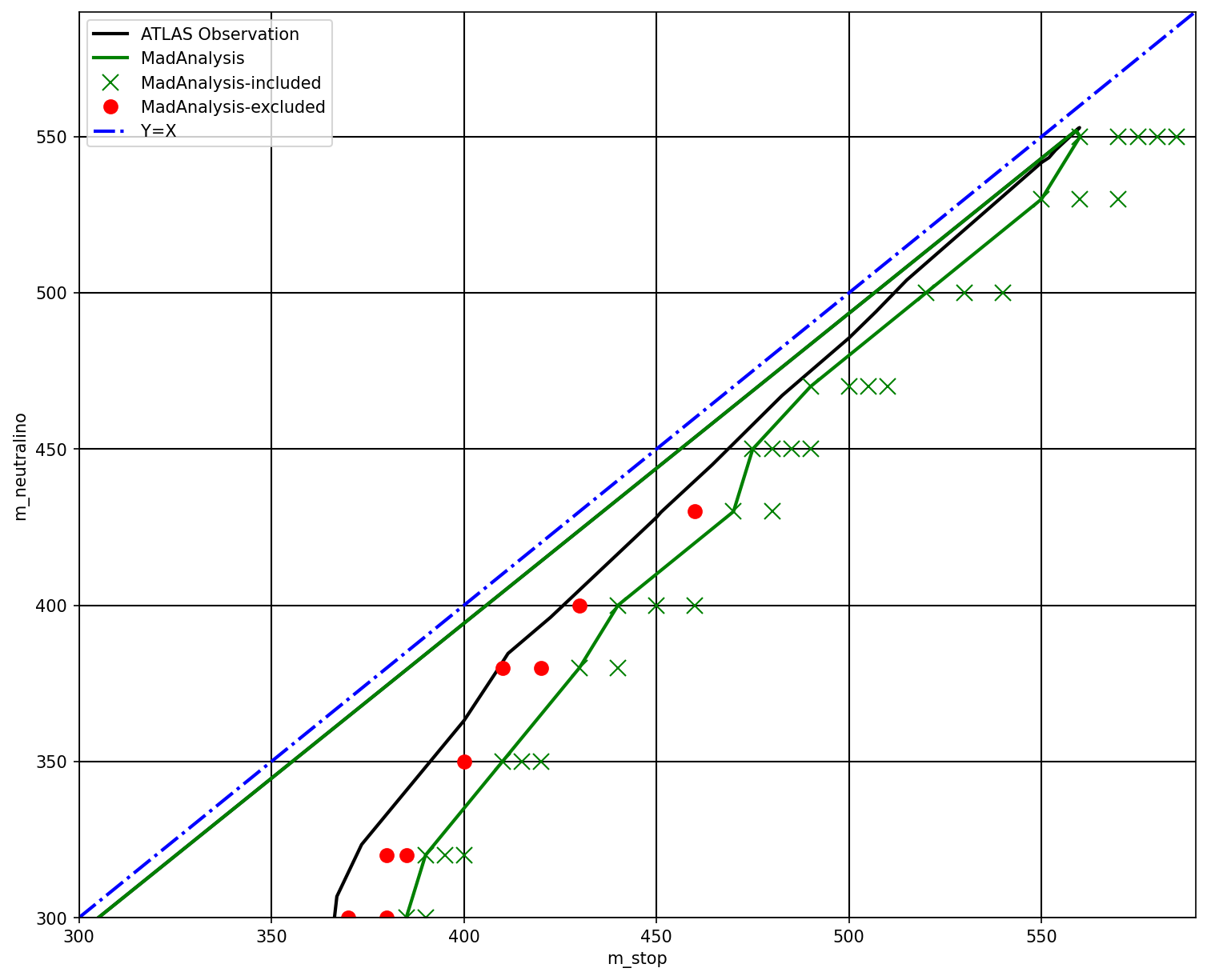}\hfill
  \includegraphics[width=.48\textwidth]{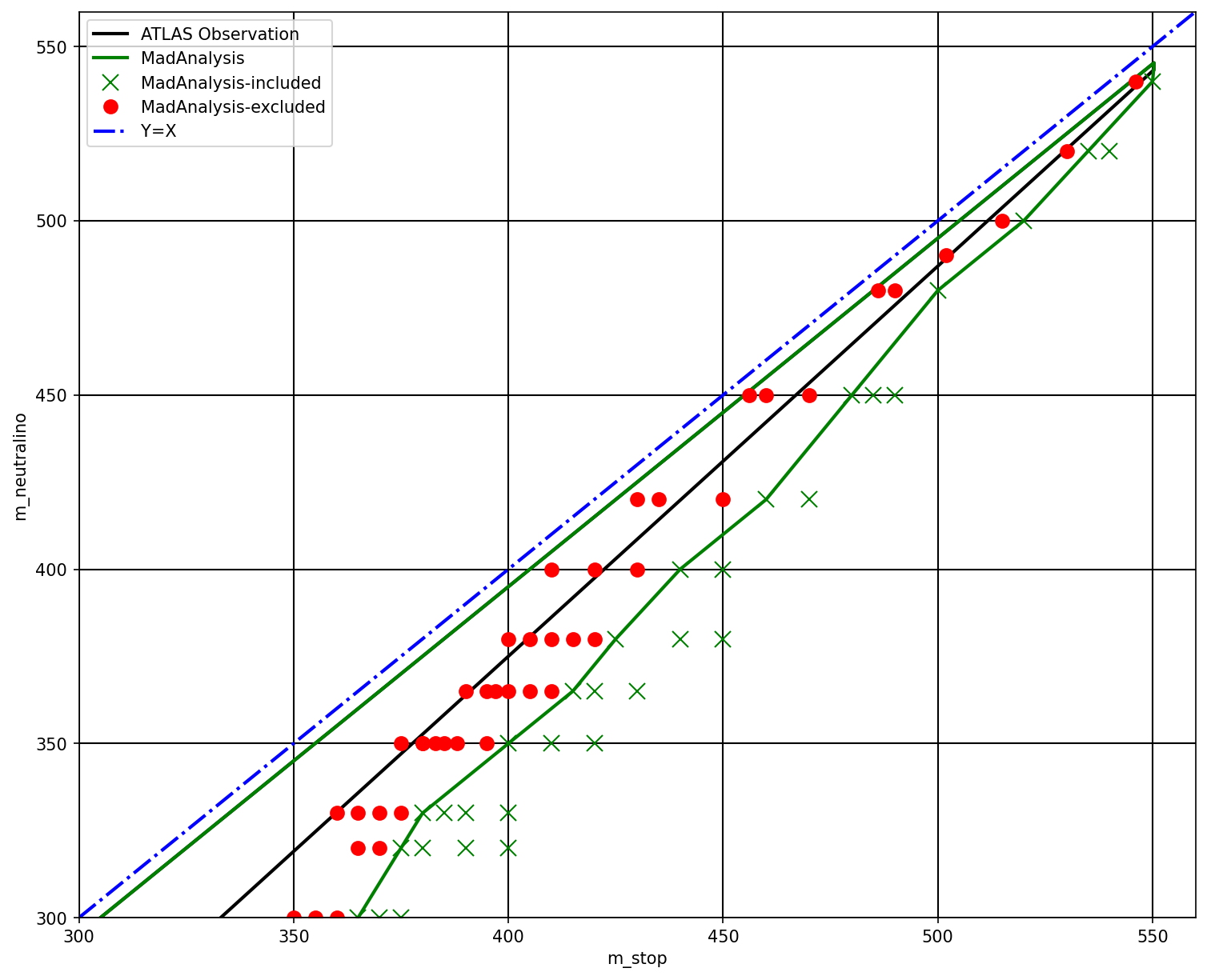}
  \caption{Excluded region at the 95$\%$ CL, displayed in the $(m_{\tilde{t}_1}, m_{\tilde{\chi}^{0}_{1}})$ mass plane for the $p p \to \tilde{t}_1 \tilde{t}_1^* \to (c \tilde{\chi}^{0}_{1}) (\bar{c} \tilde{\chi}^{0}_{1})$ (left) and $p p \to \tilde{t}_1 \tilde{t}_1^* \to ( b f\bar{f}^\prime \tilde{\chi}^{0}_{1}) (\bar{b} f\bar{f}^\prime \tilde{\chi}^{0}_{1})$ (right) processes. The green crosses and red dots respectively correspond to scenarios allowed and excluded when using our implementation of the ATLAS-EXOT-2018-06 analysis in \lstinline{MadAnalysis 5}. The excluded parameter space regions are thus delineated by the green contours, that could be compared to the official ATLAS exclusions (black).}
  \label{fig:atlas_exot_valid}
\end{figure}

To further validate our implementation, we also determine exclusion contours at the 95\% confidence level (CL) for the two simplified models corresponding to the two processes of eq.~\eqref{eq:atlas_exot_signals}. Our results are presented in the stop mass versus neutralino mass plane in figure~\ref{fig:atlas_exot_valid} for the $p p \to \tilde{t}_1 \tilde{t}_1^* \to (c \tilde{\chi}^{0}_{1}) (\bar{c} \tilde{\chi}^{0}_{1})$ process (left panel) and $p p \to \tilde{t}_1 \tilde{t}_1^* \to ( b f\bar{f}^\prime \tilde{\chi}^{0}_{1}) (\bar{b} f\bar{f}^\prime \tilde{\chi}^{0}_{1})$ process (right panel). We explore scenarios with stop and neutralino masses varying between 300 and 600~GeV, and superimpose the exclusion contours obtained with \lstinline{MadAnalysis 5} (green) with the official ones provided by the ATLAS Collaboration (black). An excellent degree of agreement is observed, with the excluded mass configurations agreeing at the level of a few percent. This therefore validates our implementation.

\newpage 
\bibliographystyle{JHEP}
\bibliography{99_dmtchan}

\end{document}